\documentclass[preprintnumbers, floatfix, letterpaper, twocolumn,aps,prd,epsfig,nofootinbib,natbib,longbibliography]{revtex4-2}
\usepackage{graphicx}
\usepackage{epstopdf}
\usepackage{latexsym}
\usepackage{amssymb}
\usepackage{amsmath}
\usepackage{color}
\usepackage{mathrsfs}
\usepackage{xparse}
\usepackage{bbding}
\usepackage{enumitem}
\usepackage{pifont}
\usepackage[center]{subfigure}
\usepackage[
            pdfstartview=FitH,
            bookmarksnumbered=true,
            bookmarksopen=true,
            colorlinks,
            linkcolor=blue,
            anchorcolor=green,
            citecolor=blue
            ]{hyperref}
\begin{document}

%%%%%%%%%%%%%%%%%%%%%%%%%%%%%%%%%%%%%%%%%%%%%%%%%%%%%%%%%%%%%%%
  \renewcommand\arraystretch{2}
 \newcommand{\bq}{\begin{equation}}
 \newcommand{\eq}{\end{equation}}
 \newcommand{\bqn}{\begin{eqnarray}}
 \newcommand{\eqn}{\end{eqnarray}}
 \newcommand{\nb}{\nonumber}
 \newcommand{\lb}{\label}
 
\newcommand{\La}{\Lambda}
\newcommand{\va}{\scriptscriptstyle}
\newcommand{\be}{\nopagebreak[3]\begin{equation}}
\newcommand{\ee}{\end{equation}}

\newcommand{\ba}{\nopagebreak[3]\begin{eqnarray}}
\newcommand{\ea}{\end{eqnarray}}

\newcommand{\la}{\label}
\newcommand{\n}{\nonumber}
\newcommand{\su}{\mathfrak{su}}
\newcommand{\SU}{\mathrm{SU}}
\newcommand{\U}{\mathrm{U}}

\newcommand{\R}{\mathbb{R}}

 \newcommand{\cb}{\color{blue}}
    \newcommand{\cc}{\color{cyan}}
        \newcommand{\cm}{\color{magenta}}
\newcommand{\rc}{\rho^{\scriptscriptstyle{\mathrm{I}}}_c}
\newcommand{\rd}{\rho^{\scriptscriptstyle{\mathrm{II}}}_c} 
\NewDocumentCommand{\evalat}{sO{\big}mm}{%
  \IfBooleanTF{#1}
   {\mleft. #3 \mright|_{#4}}
   {#3#2|_{#4}}%
}
\newcommand{\PRL}{Phys. Rev. Lett.}
\newcommand{\PL}{Phys. Lett.}
\newcommand{\PR}{Phys. Rev.}
\newcommand{\CQG}{Class. Quantum Grav.}
 %%%%%%%%%%%%%%%%%%%%%%%%%%%%%%%%%%%%%%%%%%%%%%%%%%%%%%%%%%%%%%%

  %%%%%%%%%%%%%%%%%%%%%%%%%%%%%%%%%%%%%%%%%%%%%%%%%%%%%%%%%%%%%%%

\title{Understanding  quantum black holes from quantum reduced loop gravity}
  
\author{Wen-Cong Gan${}^{a, b, c}$}
\email{Wen-cong$\_$Gan1@baylor.edu}

\author{Geeth  Ongole  ${}^{a}$}
\email{Geeth$\_$Ongole1@baylor.edu}

\author{Emanuele Alesci ${}^{b, c}$}
\email{emanuele.alesci@gmail.com}

\author{Yang An ${}^{b, c}$}
\email{anyangpeacefulocean@zju.edu.cn}

\author{Fu-Wen Shu${}^{a, d, e}$}
\email{shufuwen@ncu.edu.cn}

\author{Anzhong Wang${}^{a}$ \footnote{Corresponding author}}
\email{Anzhong$\_$Wang@baylor.edu; Corresponding author}

\affiliation{${}^{a}$ GCAP-CASPER, Physics Department, Baylor University, Waco, Texas 76798-7316, USA\\
     ${}^{b}$Institute for Theoretical Physics \& Cosmology, Zhejiang University of Technology, Hangzhou, 310023, China\\
		${}^{c}$ United Center for Gravitational Wave Physics (UCGWP),  Zhejiang University of Technology, Hangzhou, 310023, China\\
 ${}^{d}$ Department of Physics, Nanchang University,
No. 999 Xue Fu Avenue, Nanchang, 330031, China\\
${}^{e}$ Center for Relativistic Astrophysics and High Energy Physics, Nanchang University, No. 999 Xue Fu Avenue, Nanchang 330031, China}

\date{\today}

\begin{abstract}

 {We systematically study the top-down model of loop quantum black holes (LQBHs), recently derived by Alesci, Bahrami and Pranzetti (ABP). Starting from the full theory of loop quantum gravity, ABP constructed a model with respect to coherent states peaked around spherically symmetric geometry, in which both holonomy and inverse volume corrections are taken into account, and shown that the classical singularity used to appear inside the Schwarzschild black hole is replaced by a regular transition surface.} To understand the structure of the model, we first derive several well-known LQBH solutions by taking 
proper limits. These  include the B\"ohmer-Vandersloot and Ashtekar-Olmedo-Singh models, which were all obtained by the so-called bottom-up 
polymerizations within the framework of the minisuperspace  quantizations. Then, we study the ABP model, and find that the  inverse volume
corrections become important only when the  {radius of the two-sphere is} of the Planck size. For macroscopic black holes, the minimal  {radius} obtained  at the transition surface is always much larger than the Planck scale, and hence these corrections are always sub-leading. The  transition surface divides the whole spacetime into two regions, and in one of them the spacetime is asymptotically  Schwarzschild-like, while in the other region, 
the asymptotical behavior sensitively depends on the ratio of two spin numbers involved in the model, and can be divided into three different classes. 
In one class, the spacetime in the 2-planes orthogonal to the two spheres is asymptotically flat, and in the second one it is not even conformally flat, while 
in the third one  it can be asymptotically conformally flat by properly choosing the free parameters of the model. In the latter, it is asymptotically de Sitter. 
However, in any of these three classes, sharply in contrast to the models obtained by the bottom-up approach,  the spacetime is already geodesically 
complete, and no additional extensions are needed in both sides of the transition surface. In particular,  {identical} multiple  black hole and white hole structures  do not exist.

\end{abstract}

\maketitle

\section{Introduction} 
 \renewcommand{\theequation}{1.\arabic{equation}}\setcounter{equation}{0}
The resolution of General Relativity (GR) singularities is a well established result in Loop Quantum Gravity (LQG) \cite{Thiemann:2007zz},
 and is ultimately due to the presence of a minimum area implied by the quantum nature of the gravitational field.
The studies of the cosmological singularity carried out  in the last decades represent the first applications of LQG to cosmology \cite{Bojowald:2001xe,Bojowald:2002gz}
 in a well established research area  now called Loop Quantum Cosmology (LQC) \cite{Ashtekar:2011ni}, which is already at the stage of predicting observable consequences \cite{Ashtekar:2016wpi,Agullo:2016hap,Ashtekar:2021izi}.
LQC is built on first performing a classical symmetry reduction and then importing from the full theory a quantum structure adapted to the reduced system,  namely the polymer quantization \cite{Corichi:2007tf}.

The LQC success in identifying the resolution of the Big Bang singularity naturally shifted the effort to study  black  hole interiors \cite{Ashtekar:2005qt,Modesto:2005zm} with LQC techniques: {\em classical symmetry reduction and polymer quantization of the resulting minisuperspace}.  {However, in both contexts} the quantization procedure leaves several ambiguities: LQC needs to import from the full theory the area gap, and part of the quantum degrees of  {freedom are} lost once the classical symmetry reduction is performed. In fact,  in LQG the quantum states of the  {gravitational} field are spinnetwork states labeled by spins (SU(2) quantum numbers, the eigenvalues of the geometrical operators, such as the area and volume operators, etc.) and graphs on 3-dimensional manifolds with vertices locating the quanta space and realizing arbitrary quantum spaces. On the other hand, in LQC dealing with classically homogenous models the Hilbert space can't accommodate graphs and the polymer quantization employed is not sensible to the SU(2) representations.  {These} ambiguities have been fixed \cite{Ashtekar:2007em}
 in the cosmological setting with the evolution from the $\mu_0$ to the $\bar{\mu}$ scheme \cite{Ashtekar:2006wn},  while for black holes, although there are many proposals \cite{BV07,Campiglia:2007pb,Brannlund:2008iw,Modesto:2008im,Chiou:2008nm,Chiou:2008eg,Perez:2012wv,Gambini:2013exa,GOP14,GP14,Haggard:2014rza,Joe:2014tca,Corichi:2015xia,Olmedo:2017lvt,Cortez:2017alh,CR17,AP17,BMM18,RMD18,BCDHR18,AOS18a,AOS18b,Assanioussi:2019twp,Bodendorfer:2019nvy,BMM19,MDR19,AAN20,AO20,Ashtekar20,Zhang:2020qxw,Gambini:2020nsf,GSSW2020,Kelly:2020uwj,Liu:2020ola,BMM21,SG21,Giesel:2021dug,Garcia-Quismondo:2021xdc,BMM21,GOP21,LFSW21,HL22,ZMSZ22,RD22},  their LQC treatment is still evolving.  In both schemes the origin of the ambiguities is rooted in the fact that there is no fixed prescription to obtain the LQC Hilbert space from the LQG one and the fundamental property of LQG, namely the existence of space quanta, can only be imported. Now if the introduction of a minimum volume as external input is enough to solve the singularity, details of the evolution deeply depend on the amount of structure imported ad hoc from the full theory.

Recently,  a new technique (Quantum Reduced Loop Gravity - QRLG) aimed to disentangle those ambiguities was proposed by Alesci, Bahrami and Pranzetti (ABP), the so-called top-down approach \cite{LP22}. QRLG is based on the tentative of reverting the reduction-quantization process to implement a quantum symmetry reduction. Performing gauge fixing to adapt the full quantization to the symmetry compatible coordinates, QRLG allows to study the homogeneous spacetimes as coherent states of the full theory retaining all the quantum degrees of freedom of LQG. In this sense, QRLG doesn't need an external area gap or an ad-hoc Hilbert space, because it just uses the full LQG Hilbert space. QRLG program has been successfully applied to cosmology \cite{Alesci:2013xd} and a direct link to LQC has been unveiled \cite{Alesci:2016gub}. However,
 the inclusion of new degrees of freedom also opens the possibility for new scenarios as the replacement of the big bounce scenario \cite{Ashtekar:2006rx} with  the emergent bouncing one \cite{Alesci:2016xqa}. 
The application of QRLG to the interior of a black hole \cite{ABP18,ABP19} has been recently performed and showed a completely new possibility. The black hole singularity is replaced by a bounce followed by an expanding Universe that could be asymptotically de Sitter \cite{ABP20}.

 In this paper, we shall study the ABP model in detail and confirm several major conclusions obtained in \cite{ABP19,ABP20}, and meanwhile  clarify some silent points.  In particular, the article is organized as follows. In Sec. II, we provide a brief review of the ABP model \cite{ABP18,ABP19,ABP20}, by paying particular attention to its semi-classical limit conditions, which are essential 
in order to understand the physical implications of the model. In Sec. III, we first consider its classical limit, whereby the physical interpretation of quantities of the ABP model become clear, and then obtain  
the B\"ohmer-Vandersloot (BV) \cite{BV07} and Ashtekar-Olmedo-Singh (AOS) models \cite{AOS18a,AOS18b}   by taking proper limits and replacements. In doing so, we look for the possible relation among these models. Although formally
we can obtain all these models, they all fall to the case where the semi-classical limit conditions of the ABP model are not satisfied. As a result, these models  
 cannot be embedded properly into the ABP model. However, we do find that such derivation is helpful in understanding
the structure of the ABP model. In Sec. IV, we study the ABP model without the inverse volume corrections  in detail, by first showing that such corrections become important only when the curvature becomes the order
of the Planck scale. The subsequent detailed analysis shows that the minimal  {radius of the two-sphere obtained at the transition surface is } always much larger than the Planck scale for macroscopic black holes. As a result, the inverse volume corrections should be always sub-leading for such black holes. In Sec. V, we confirm this by focusing only on the cases with $\gamma = 0.274$ obtained by the considerations of 
black hole entropy \cite{ABBDV10}, and $j_x$ and $j$ given by Eq.(\ref{eq2.5a}) below, obtained  by demanding that the spatial manifold triangulation remain consistent on both sides of the black hole horizons  \cite{ABP20}. Our main results are summarized in Sec. VI, while in Appendix \ref{app-a}, we provide some properties of the Struve functions.

In this paper, we shall use $\ell_p, m_p, \tau_p$ to  {denote}, respectively, the Planck length, mass and time. In all the numerical plots, we  shall use them as the units.
For example, when plotting a figure with $m =1$ we always mean  $m/m_p =1$, and so on. 

\section{Effective Hamiltonian of Internal spherical Black Hole Spacetimes} 
 \renewcommand{\theequation}{2.\arabic{equation}}\setcounter{equation}{0}

Spherically symmetric spacetimes inside black holes can be written in the form
\begin{equation}
\label{eq2.1}
ds^2= - N(\tau)^2d\tau^2+ \Lambda(\tau)^2 dx^2 + R(\tau)^2d\Omega^2,
\end{equation}
where $N(\tau)$ is the lapse function and $d\Omega^2 \equiv d\theta^2+\sin^2\theta d\phi^2$. Clearly, the above metric is invariant under the following transformations
\begin{equation}
\label{eq2.1a}
 \tau = \xi(\tau'), \quad x = a_0 x' + b_0,
\end{equation}
where $\xi(\tau')$ is an arbitrary function of $\tau'$ and $a_0$ and $b_0$ are arbitrary constants.

\subsection{Classical Spherical Spacetimes and Canonical Variables}

 It should be noted that, instead of using the canonic variables ($\Lambda, R$) and their momentum conjugates ($P_{\Lambda}, P_{R}$), one often uses  
  ($p_b, b, p_c, c$) \cite{AOS18a}, which can be obtained by comparing the gravitational connection $A^i_a\tau_i dx^a$ and the spatial triads
 $E_i^a\tau^i \partial_a$,   given in \cite{AOS18a,ABP20}, and yield 
 \bqn
 \lb{eq2.13}
 p_c &=&  R^2, \quad
 p_b =   L_0 R\Lambda, \quad
 b =   -\frac{\gamma G}{R}P_{\Lambda},\nb\\
 c &=&   -\frac{\gamma G L_0}{R}\left(P_R-\frac{\Lambda P_{\Lambda}}{R}\right),
 \eqn
where $L_0$ is a constant, and related to ${\cal{L}}_0$ introduced in \cite{ABP20} by $L_0 = 2 {\cal{L}}_0$. Note that in writing down the above expressions we assumed $p_c >0$. With the choice  of  the lapse function   \cite{AOS18a,AOS18b}  
 \bq
 \lb{eq2.14a}
 N_{cl}  = \gamma b^{-1} {\text{sgn}}(p_c)\left|p_c\right|^{1/2} = - \frac{R^2}{GP_{\Lambda}}, 
 \eq
 we find that  the metric (\ref{eq2.1}) takes the form  
  \bq
 \lb{eq2.14ca}
 ds^2 = - \frac{\gamma^2 p_c(T)}{b^2(T)} dT^2 + \frac{p_b^2(T)}{L_0^2 p_c(T)} dx^2 + p_c(T) d\Omega^2,
 \eq
  where  \footnote{It should be noted that the parameter $m$ used in \cite{BV07,AOS18a,AOS18b} corresponds to $Gm$ introduced  in this paper.}
  \bq
  \lb{eq2.14caa}
  T \equiv \frac{\tau}{2Gm}+\log(2Gm).
  \eq
Then, the corresponding classical Hamiltonian is given by
\bqn 
  \lb{eq2.14b}
 H_{cl}[N_{cl}] &\equiv&N_{cl} \mathcal{H}_c\nb\\
&=& -\frac{1}{2G \gamma}\left(2c\;p_c+\left(b+\frac{\gamma^2}{b}\right)p_b\right)\nb\\
 &=& \frac{L_0 R^2}{G P_\Lambda}\Bigg(\frac{G P_{\Lambda} P_{R}}{R}-\frac{G  P_{\Lambda}^2 \Lambda}{2 R^2}+\frac{\Lambda}{2G}\Bigg).
 \eqn

 \subsection{Quantum Black Holes in QRLG}

Within the framework of QRLG, starting   from a partial gauge fixing of the full LQG Hilbert space, ABP
 \cite{ABP18,ABP19,ABP20} studied the interior of a Schwarzschild black hole,  and derived
 an effective Hamiltonian by including the inverse volume and coherent state sub-leading corrections, which  differs crucially from the ones introduced previously in the minisuperspace models. 
 In particular, by fixing  the quantum parameters associated with the structure of coherent states   through geometrical considerations, the authors found that the post-bounce interior geometry 
  sensitively depends on the  value of the Barbero-Immirzi parameter $\gamma$, and that  the value $\gamma \simeq 0.274$,  deduced from  the SU(2) black hole entropy calculations in LQG \cite {ENPP10,ABBDV10},
gives rise to  an asymptotically de Sitter geometry in the interior region \footnote{Note that, instead of using the  SU(2)  black hole entropy as done in \cite {ENPP10,ABBDV10}, if one uses the U(1)  black hole entropy arguments, the parameter
 $\gamma$  was found to be $\gamma \simeq 
0.2375$ \cite{KM04}.}.

\begin{figure}[h!]
\includegraphics[height=4.8cm]{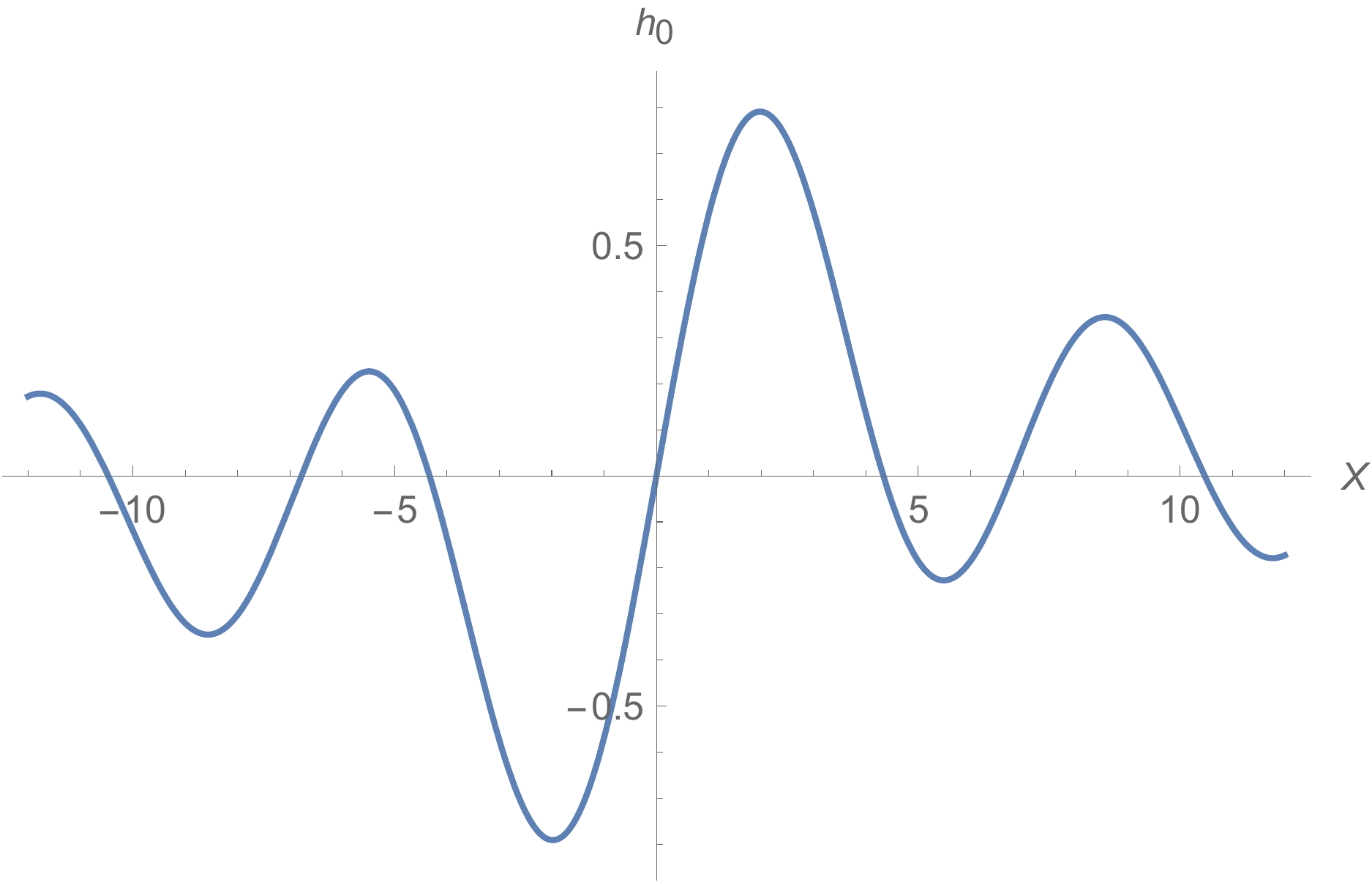}\\
{(a)}\\
\vspace{.5cm}
\includegraphics[height=4.8cm]{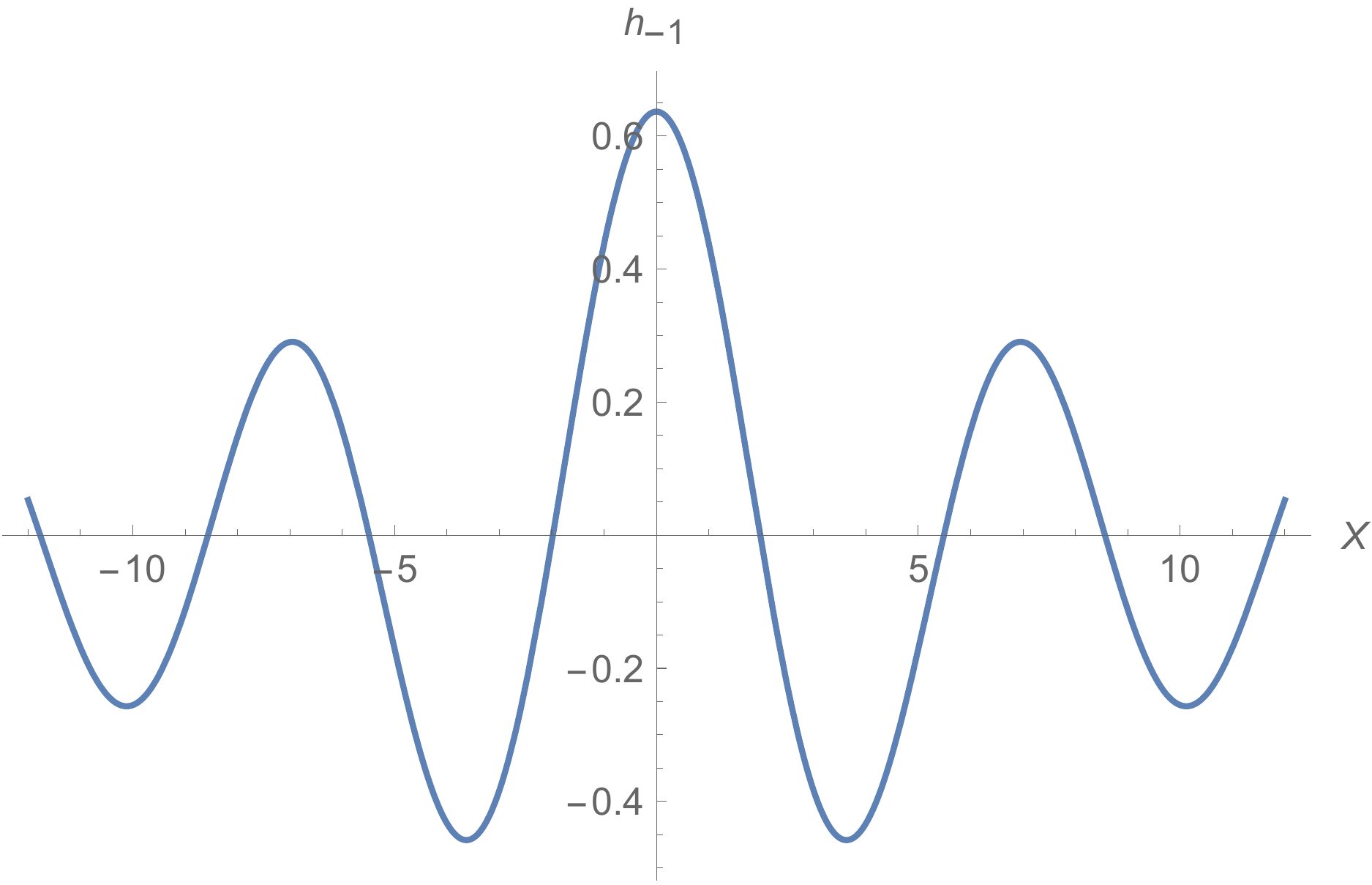}\\
{(b)}\\
\vspace{.5cm}
\caption{The  Struve functions $h_0[X]$ and $h_{-1}[X]$. } 
\label{fig1}
\end{figure}

Introducing the following parameters 
\bqn
\label{eq2.2}
A&\equiv&2{\ell_p}^2\left(\frac{{\ell_p}^2 \gamma^2}{\beta^2}-\frac{4 \gamma^2}{\delta_x}+\frac{4(3-\nu)  \gamma^2}{\delta}\right),\nb\\
B&\equiv&{\ell_p}^2\left(\frac{{\ell_p}^2 \gamma^2}{\beta^2}-\frac{8 \gamma^2}{\delta_x}+\frac{8(3\nu-1) \gamma^2}{\delta}\right),\nb\\
C&\equiv&2{\ell_p}^2\left(\frac{{\ell_p}^2 \gamma^2}{\alpha^2}+\frac{12 \gamma^2}{\delta_x}-\frac{4(1+\nu) \gamma^2}{\delta}\right),
\eqn
and the functions
\bqn
\lb{eq2.3}
 X&\equiv& \alpha \gamma G \left(\frac{P_\Lambda}{R^2}\right),\quad
Y \equiv \beta\gamma G \left(\frac{P_R}{R\Lambda}-\frac{P_\Lambda}{R^2}\right),\nb\\
Z&\equiv&8\gamma^2 \cos\left(\frac{\alpha}{R}\right)\sin^2\left(\frac{\alpha}{2R}\right),
\eqn
we find that   the  effective Hamiltonian of the ABP model can be cast in the form
\bqn
\label{hamiltonian}
 \mathcal{H}_{int}^{IV+CS}= -\frac{\mathcal{L}_0 R^2 \Lambda}{2\alpha^2 \gamma^2 G}\mathcal{C}(\tau),
\eqn
where
\bqn
\lb{Cfactor}
&& \mathcal{C}(\tau)\equiv  \frac{\alpha} {\beta} \sin[Y] \Bigg\{\left(1+\frac{A}{R^2}\right)  \pi h_0[X]\nb\\
&& ~~~~~~~~~~~~~~~~~~~~~~~ + 2\left(1+\frac{B}{R^2}\right)  \sin[X]\Bigg\}\nb\\
&&~~~~~~~~~~ +Z+ \left(1+\frac{C}{R^2}\right) \pi \sin[X] h_0[X],
\eqn
and $\mathcal{L}_0$ denotes the length of the fiducial cell with $x \in\left[-\mathcal{L}_0, \mathcal{L}_0\right]$, and ${\ell_p}$  is the Planck length with ${\ell_p} \equiv \sqrt{\hbar G/c^3}$, 
while $G$ and $c$ are the  {Newton's} constant and the speed of light, respectively.  
The super indices ``IV" and ``CS" stand for, respectively, the inverse volume and coherent state, while
the dimensionless  parameters $\delta, \; \delta_x$ and $\nu$ are the spread parameters, characterizing the coherent state corrections. The terms proportional to the constants $A, B$ and $C$ 
characterize the  inverse volume corrections and are subdominant \cite{ABP20}. The function $h_0[X]$ denotes the  zeroth-order Struve function 
and its series expansion reads  \cite{AS72}
\bq
\lb{eq2.4}
h_0[z] = \frac{2}{\pi}\left(z - \frac{z^3}{1^2\cdot 3^2} + \frac{z^5}{1^2\cdot 3^2\cdot 5^2} - ... \right).
\eq
In Fig. \ref{fig1}, we plot out the  Struve function $h_0$ together with $h_{-1}$, as the latter will appear in the dynamical equations. In general, the $\nu$-th order Struve functions are defined by
Eq.(\ref{A.1}) in Appendix A, in which some of their properties are also given. For more details, we refer readers to  \cite{AS72}.

In terms of the spin numbers 
$j$ and $j_x$, the parameters $\alpha$ and $\beta$ are given by
\bq
\lb{eq2.5}
\alpha \equiv 2\pi \sqrt{\gamma j_x}\; {\ell_p}, \quad \beta \equiv 4\sqrt{\frac{8\pi\gamma}{j_x}}\; j \;  {\ell_p},
\eq
where $j_x$ denotes the averaged spin number of all plaquettes that tessellate the 2-sphere $S^2$ spanned by $(\theta, \phi$), while $j$ is the averaged spin number associated with the links dual to the plaquettes in both ($\theta, x$) and  ($\phi, x$) planes.  It must be noted that this effective Hamiltonian is valid only in the semi-classical limits \cite{ABP20}
\bq
\lb{eq2.25v1}
j, \; j_x \gg 1.
\eq
 To understand further the geometrical meaning of $j$ and $j_x$, we introduce
the coordinate lengths along $x, \; \theta,\; \phi$ directions by $\epsilon_x, \; \epsilon_{\theta}, \; \epsilon_{\phi}$, respectively. Due to the spherical symmetry, we have 
$\epsilon_{\theta} = \epsilon_{\phi} \equiv \epsilon$. Then, we introduce two new quantities ${\cal{N}}$ and  ${\cal{N}}_x$, in terms of which  
$\epsilon$ and $\epsilon_x$ can be written as
\bq
\lb{eq2.5c1}
\epsilon \equiv \frac{2\pi}{{\cal{N}}}, \quad \epsilon_x \equiv \frac{{\cal{L}}_0}{{\cal{N}}_x}, 
\eq
where ${\cal{N}}^2/2$ is the total number of the plaquettes on $S^2$, and ${\cal{N}}_x$ denotes the total number of plaquettes in the $x$ direction for a given fiducial 
length ${\cal{L}}_0$. The effective Hamiltonian (\ref{hamiltonian}) was obtained under the assumption 
\bq
\lb{eq2.5d1} 
 {\cal{N}}, \; {\cal{N}}_x \gg 1 \;\;\; \text{or} \;\;\; \epsilon, \; \epsilon_x \ll 1.
\eq
 To find the relations between (${\cal{N}}, \; {\cal{N}}_x$) and ($j, j_x$), we can calculate the area of a given $S^2$ and the volume
of a given spatial three-surface spanned by $x, \theta, \phi$, which are given, respectively, by
\bqn
\lb{eq2.5ea}
 A(R) &=& 4\pi R^2 = 8\pi \gamma \ell_p^2\sum_{p\in S^2}\tilde{j}^p_x \simeq 8\pi   \gamma \ell_p^2 \left(\frac{{\cal{N}}^2}{2} j_x\right), ~~~~~~~ \\
 \lb{eq2.5eb}
 V(\Sigma) &=& 8\pi {\cal{L}}_0 \Lambda R^2 \simeq 4 \left(8\pi \gamma \ell_p^2\right)^{3/2} j \sqrt{j_x} {\cal{N}}_x {\cal{N}}^2, 
\eqn
where $\tilde{j}^p_x$ is the spin number associated with the link dual to the given plaquettes $p$ on $S^2$. In the limit $ {\cal{N}} \gg 1$, the sum of $\tilde{j}^p_x$
in Eq.(\ref{eq2.5ea}) 
was approximated by the average spin $j_x$ of a single cell times the total number of the plaquettes in $S^2$.  In the last step of Eq.(\ref{eq2.5eb}), the average spin number $j$ is associated with the links dual to the plaquettes in both ($x, \theta$)- and  ($x, \phi$)-planes. Therefore, we find
\bqn
\lb{eq2.5f1}
 {\cal{N}} = \frac{R}{\sqrt{\gamma\ell_p^2}}\left(\frac{1}{\sqrt{j_x}}\right), \quad  
  {\cal{N}}_x = \frac{{\cal{L}}_0 \Lambda}{4\sqrt{8\pi \gamma\ell_p^2}}\left(\frac{\sqrt{j_x}}{j}\right). ~~~~~~~~~~
 \eqn
 Inserting Eq.(\ref{eq2.5f1}) into Eq.(\ref{eq2.5c1}), we obtain
 \bq
\lb{eq2.5g1}
\epsilon = \frac{\alpha}{R}, \quad \epsilon_x = \frac{\beta}{\Lambda}, 
\eq
where $\alpha$ and $\beta$ are defined by Eq.(\ref{eq2.5}).

It should be noted that the understanding of the  {geometrical meaning} of ${\cal{N}}, {\cal{N}}_x, j$ and $j_x$ is important for our following discussions, 
especially when we consider some specific models within the framework of QRLG. As to be seen below, both of the semi-classical limit conditions
(\ref{eq2.25v1}) and (\ref{eq2.5d1}) must be fulfilled, in order to have the effective Hamiltonian (\ref{hamiltonian}) valid. These also provide the keys for us to understand the
semi-classical structures of black holes in the framework of LQG.

We further note that, by demanding that the spatial manifold triangulation remain consistent on both sides of the black hole horizons, ABP found \cite{ABP20}
\bq
\lb{eq2.5a}
j = \gamma j_x,  
\eq
for which we have 
\bq
\lb{eq2.5b}
\eta \equiv \frac{\alpha }{\beta} =\frac{\sqrt{2\pi}}{8\gamma},
\eq
 as can be seen from Eq.(\ref{eq2.5}).  
 Then, in the effective Hamiltonian (\ref{hamiltonian}) five new parameters
 $$
 (\gamma, j; \nu,  \delta, \delta_x) \;\;\; \text{or} \;\;\; 
 (\gamma,  \alpha; \nu, \delta, \delta_x),  
 $$
 are present  in addition to $G, c, \hbar$, where ($\nu,  \delta, \delta_x$) are related to the inverse volume corrections. 
  One of the purposes of this paper is to understand their effects on the local and global properties of the  spacetimes.

 It should be noted  that the two spin numbers  $j$ and $j_x$ used in this paper, which are consistent with those used in  \cite{ABP20}, are different from the ones ($\hat j, \hat j_0$)  introduced in \cite{ABP19}
 \footnote{Note that, instead of using ($j, j_0$) as those adopted in  \cite{ABP19}, here we use the symbols  with hats, in order to distinguish them from the ones used in this paper.}. In particular, we have
 \bq
\lb{eq2.5g2}
\hat j = \sqrt{8\pi}\; j, \quad  \hat j_0 = \frac{\pi}{2} j_x.  
\eq

To write down the corresponding    dynamic equations  for the effective Hamiltonian (\ref{hamiltonian}), using the gauge freedom (\ref{eq2.1a}), ABP chose the lapse function $N(\tau)$ as
\bq
\lb{eq2.7b}
N(\tau) = - \frac{2\alpha\gamma}{mGW},
\eq
where $m$ is a mass parameter, and $W$ is defined as
\bqn
\lb{eq2.3aa}
W&=&\pi h_0[X]+2\sin[X].
\eqn
  Taking $\hbar \rightarrow 0$, it reduces to 
\bq
\lb{eq2.7c}
N_c \equiv \; \lim_{\hbar \rightarrow 0}{N} = - \frac{R^2}{2mG^2 P_{\Lambda}},  
\eq
which corresponds to the classical limit, and $m$ represents the  mass of the Schwarzschild black hole. 
Taking Eq.(\ref{eq2.14caa})  into account, we find that 
\bq
\lb{eq2.7ca}
N_c^2 d\tau^2=N_{cl}^2 dT^2, \quad N_{cl}=2Gm N_c,
\eq
where  $N_{cl}$ and $N_c$ are given, respectively,  by Eqs.(\ref{eq2.14a}) and \eqref{eq2.7c}. 

%Thus,  $N_{cl}=(2Gm) N_c$.

Then, the smeared effective Hamiltonian of Eq.(\ref{hamiltonian}) with the choice of the lapse function (\ref{eq2.7b})  is given by  
\bqn
\label{hamil1}
H_{int}^{IV+CS}[N] &\equiv&   N(\tau) \mathcal{H}_{int}^{IV+CS} 
=   \frac{\mathcal{L}_0 R^2 \Lambda}{\alpha \gamma m G^2 W} {\cal{C}}(\tau). ~~~~~~~
\eqn
Hence,  the corresponding  {dynamical equations can be cast in the form}
\begin{eqnarray}
\lb{eq2.8a}
-2 G m\frac{z}{\ell}R'&=&\frac{R \cos[Y]}{W} {\cal{D}}, \\ 
\lb{eq2.8b}
-2 G m\frac{z}{\ell}P_{\Lambda}'&=&\frac{R P_R\cos[Y]}{\Lambda W} {\cal{D}}, 
\eqn
\begin{widetext}
\bqn
\lb{eq2.8c}
-2 G m\frac{z}{\ell}\frac{\Lambda'}{\Lambda}&=&- \frac{ \cos[Y]}{W}{\cal{D}}  
+\frac{1}{W^2}\Bigg\{\pi h_{-1}[X]\Bigg[2\left(1+\frac{C}{R^2}\right) \sin^2[X] -Z\Bigg]+\cos[X] \Bigg[\left(1+\frac{C}{R^2}\right)  \pi^2 h_0^2[X] -2Z \Bigg]\nb\\
&& +\frac{2\pi\alpha(A-B)}{\beta R^2}\sin[Y]\Big(\sin[X] h_{-1}[X]-\cos[X] h_0[X]\Big)\Bigg\},\\
\lb{eq2.8d}
-2 G m\frac{z}{\ell}P_{R}'&=&\frac{R P_R- 2\Lambda P_\Lambda}{R W} \cos[Y]{\cal{D}} +\frac{2\pi\Lambda P_\Lambda}{ R W} \sin[X] h_{-1}[X]\left(1+\frac{C}{R^2}\right)  \nb\\  
&&
 +\frac{2\pi\Lambda }{R W}  h_{0}[X]\Bigg\{\left(\frac{C}{\alpha \gamma G}\right) \sin[X]  + P_ \Lambda \cos[X]\left(1+\frac{C}{R^2}\right) \Bigg\}\nb\\
&&+\frac{2\Lambda \sin[Y]}{R W }\Bigg\{ \frac{\alpha \pi}{\beta}P_\Lambda h_{-1}[X]\left(1+\frac{A}{R^2}\right)+\frac{A}{\beta \gamma G} \pi h_0[X]+  \frac{2B}{\beta\gamma G} \sin[X]
+\frac{2   \alpha}{\beta}  P_\Lambda \cos[X] \left(1+\frac{B}{R^2}\right)\Bigg\} \nb\\
&& -\frac{4\gamma \Lambda}{G W}\Bigg\{\sin\left(\frac{\alpha}{R}\right)-\sin\left(\frac{2\alpha}{R}\right)\Bigg\},
\end{eqnarray}
\end{widetext}
where 
\bqn
\lb{eq2.9}
{\cal{D}}(X) \equiv \left(1+\frac{A}{R^2}\right)\pi h_0[X] +2\left(1+\frac{B}{R^2}\right)\sin[X], ~~~~~~~~
\eqn
and a prime denotes the ordinary derivative with respect to $z$, with $z \equiv \exp(-\tau/\ell)$, where $\ell$ is a constant and has the length dimension.
The function $h_{-1}[X] \left(\equiv dh_{0}[X]/dX\right)$ denotes the  Struve function of order $-1$. In Appendix A, we present some basic properties of these functions, and for
other properties of them, we refer readers to \cite{AS72}.

\section{Some Known Loop Quantum Black Holes as Particular Limits of the ABP Model} 
 \renewcommand{\theequation}{3.\arabic{equation}}\setcounter{equation}{0}

To understand the  quantum reduced loop black hole (QRLBH) spacetimes with both of the holonomy and  inverse volume corrections, in this section  let us first consider some limits of the parameters involved,
and derive several well-known spacetimes. In doing so, we can gain a better understanding of the QRLBH spacetimes and their relation with other models.  
 
\subsection{Classical Limit}\lb{cla limit}

The classical limit is obtained by taking $\hbar \rightarrow 0$, that is, by setting  $\ell_{p} = 0$,  which leads to 
\bqn
\lb{eq2.9a}
&&  A = B = C  = 0,\nb\\
&& \mathcal{D}\simeq W \simeq 4X, \quad Z \simeq  \frac{2 \gamma^2\alpha^2}{R^2}.
\eqn
Then, Eqs.(\ref{eq2.8a}) - (\ref{eq2.8d})
reduce  respectively to 
\bqn
\lb{eq2.10a}
-2 G m\frac{z}{\ell}R'&=&R, \\ 
\lb{eq2.10b}
-2 G m\frac{z}{\ell}P_{\Lambda}'&=&\frac{R P_{R}}{{\Lambda}}, \\  
\lb{eq2.10c}
-2 G m\frac{z}{\ell}\frac{\Lambda'}{\Lambda}&=&-\frac{G^2 P_{\Lambda}^2+R^2}{2 G^2 P_{\Lambda}^2},\\
\lb{eq2.10d}
-2 G m\frac{z}{\ell}P_{R}'&=&3 P_R -\frac{2\Lambda P_{\Lambda}}{R}+\frac{\Lambda R}{G^2 P_{\Lambda}},
\eqn
while the effective Hamiltonian (\ref{hamiltonian}) reduces to (\ref{eq2.14b}) with $L_0 = 2{\cal{L}}_0$. 
Then, from the   Hamiltonian constraint $\mathcal{H}_c=0$, we  find the following two useful expressions
\bqn
\lb{eq2.11aa}
\frac{R P_R}{\Lambda} &=& \frac{G^2P_{\Lambda}^2 - R^2}{2G^2 P_{\Lambda}},\\
\lb{eq2.11bb}
\frac{\Lambda P_{\Lambda}}{R} &=& 2 P_R + \frac{R\Lambda}{G^2P_{\Lambda}}.
\eqn
Inserting them into Eqs.(\ref{eq2.10b}) and (\ref{eq2.10d}), respectively, we obtain two new equations for $P_{\Lambda}'$ and  $P_{R}'$, and together with the other two, they can be  cast in the forms 
\bqn
\lb{eq2.11a}
-2 G m\frac{z}{\ell}R'&=&R, \\ 
\lb{eq2.11b}
-2 G m\frac{z}{\ell}P_{\Lambda}'&=&\frac{G^2 P_{\Lambda}^2-R^2}{2G^2 P_{\Lambda}}, \\  
\lb{eq2.11c}
-2 G m\frac{z}{\ell}\frac{\Lambda'}{\Lambda}&=&-\frac{G^2 P_{\Lambda}^2+R^2}{2 G^2 P_{\Lambda}^2},\\
\lb{eq2.11d}
-2 G m\frac{z}{\ell}P_{R}'&=&-\frac{G^2 P_{\Lambda}P_{R}+\Lambda R}{G^2 P_{\Lambda}}.
\eqn
 Now, the above equations can be solved in sequence, that is, we first solve Eq.(\ref{eq2.11a}) to find $R(z)$, and then substituting it into Eq.(\ref{eq2.11b}), we can find $P_{\Lambda}(z)$. 
Once $R(z)$ and $P_{\Lambda}(z)$ are given, we can substitute them into Eq.(\ref{eq2.11c}) to find $\Lambda(z)$. Then, we can find $P_{R}(z)$  either by integrating 
Eq.(\ref{eq2.11d}) explicitly  or by using the Hamiltonian constraint $\mathcal{H}_c=0$. In the first approach, we shall have four integration constants, but only three of them are independent, as
the  Hamiltonian constraint $\mathcal{H}_c=0$ must be satisfied, which will relate  one of the four constants to the other three. Therefore, a simpler way is to solve   $\mathcal{H}_c=0$ directly to find $P_R$,
 once $R, P_{\Lambda}$ and $\Lambda$ are found from Eqs.(\ref{eq2.11a})-(\ref{eq2.11c}). However, to illustrate what we mentioned above, let us first integrate the above four equations directly to get  
 \bqn
\lb{eq2.12a}
R&=&c_0 e^{\frac{\tau}{2Gm}}, \\ 
\lb{eq2.12b}
P_{\Lambda}&=&\mp \frac{\sqrt{c_1 G^2 e^{\frac{\tau }{2 G m}}-{c_0}^2 e^{\frac{\tau }{G m}}}}{G},  \nb\\ 
\lb{eq2.12c}
\Lambda&=&c_2 e^{-\frac{\tau }{4 G m}} \sqrt{c_1 G^2-c_0^2 e^{\frac{\tau }{2 G m}}}, \\  
\lb{eq2.12d}
P_{R}&=&c_3 e^{-\frac{\tau }{2 G m}}\pm\frac{c_0 c_2}{G },
\eqn
 where $c_{n}$'s are the four integration constants. As noticed above, only three of them are independent. In fact,
substituting the above expressions into the   Hamiltonian constraint $\mathcal{H}_c=0$ we find that 
\bqn
\lb{eq2.12e}
c_1 c_2 G=\mp 2 c_0 c_3.
\eqn
 On the other hand, from Eq.(\ref{eq2.7b}), we find
 \bqn
\lb{eq2.12f}
N   &=& - \frac{R^2}{2mG^2 P_{\Lambda}} \nb\\
&=& \pm\frac{c_0^2 e^{\frac{\tau }{G m}}}{2 G m \sqrt{c_1 G^2 e^{\frac{\tau }{2 G m}}-c_0^2 e^{\frac{\tau }{G m}}}}.
\eqn
Thus, we finally obtain
\bqn
\lb{eq2.12g}
ds_c^2 &=& - N^2 d\tau^2 + \Lambda^2dx^2 + R^2d\Omega^2\nb\\
&=&-\frac{dR^2}{\frac{G^2 c_1}{c_0 R}-1}+c_0^2 c_2^2 \left(\frac{G^2 c_1}{c_0 R}-1\right)dx^2 + R^2d\Omega^2.\nb\\
\eqn 
Clearly, using the gauge residual (\ref{eq2.1a}), we can always absorb the factor $c_0^2 c_2^2$ into $x$ by setting 
$a_0 \equiv (c_0 c_2)^{-1}$. Then, the metric essentially depends only on one independent combination, $G^2 c_1/c_0$,  of the parameters, which is related to   the mass of the black hole via the relation
\bq
\lb{eq2.12ga}
m \equiv \frac{c_1 G}{2c_0}. 
\eq
 
 It should be noted that the integration constants $c_n$'s can be also determined by the   boundary conditions
 \bq
 \lb{eq2.12gb}
 R=2 G m, \quad \Lambda=0, \quad P_{\Lambda}=0,\; (\tau=0), 
 \eq
and the Hamiltonian constraint   at the horizon $\tau=0$, which will be elaborated in more detail below, when we try to solve the  field equations (\ref{eq2.8a}) - (\ref{eq2.8d}) numerically for the general case. 
In the current case,  it can be shown that the above conditions together with the Hamiltonian  constraint lead to  
\bq
\lb{eq2.12h}
c_0=2 G m, \quad c_1=\frac{c_0^2}{G^2}, \quad c_2=\frac{1}{c_0}, \quad c_3 =\mp\frac{1}{2G},
\eq
 so the classical metric finally  takes its standard form 
 \bqn
\lb{eq2.12i}
ds_c^2 &=& \left(1-\frac{2Gm}{R}\right)^{-1}dR^2-\left(1-\frac{2Gm}{R}\right)dx^2 \nb\\
&& + R^2d\Omega^2. ~~~~
\eqn

\subsection{B\"ohmer-Vandersloot Limit}\lb{bv limit}

Following the so-called $\bar\mu$ scheme in LQC \cite{Ashtekar:2006wn}, B\"ohmer-Vandersloot (BV) \cite{BV07} considered the case in which the physical area of the closed loop is equal to the minimum area gap predicted
by  LQG 
\bq
\lb{eq2.5za}
\Delta = 2\sqrt{3} \pi \gamma \ell^2_{\text{p}}.
\eq
For example, the holonomy loop in the ($x, \theta$)-plane leads to
\bq
\lb{eq2.5zb}
A_{x\theta} = \delta_b\delta_c p_b, 
\eq
while the one in the  ($\theta, \phi$)-plane leads to
\bq
\lb{eq2.5zc}
A_{\theta\phi} = \delta_b^2  p_c, 
\eq
where the new variable $b,\; c$ and their moment conjugates $p_b,\; p_c$ are related to the ABP variables through Eq.(\ref{eq2.13}), which can be written 
in the form
\bqn
\lb{eq2.5zd}
p_b &=& L_0 \Lambda R, \quad b = - \alpha^{-1} RX,\nb\\
p_c &=&   R^2, \quad c = - \beta^{-1} {L_0} \Lambda Y,
\eqn
where $X$ and $Y$ are defined in Eq.(\ref{eq2.3}). Then, setting 
\bq
\lb{BVc}
A_{x\theta}  = \Delta = A_{\theta\phi}, 
\eq
will lead to
\bq
\lb{eq2.5ze}
\delta_b = \sqrt{\frac{\Delta}{p_c}}, \quad
\delta_c = \frac{\sqrt{\Delta {p_c}}}{p_b}.
\eq
 Making the replacements 
 \bq
\lb{eq2.5zf}
b \rightarrow \frac{\sin(\delta_b b)}{ \delta_b}, \quad c \rightarrow \frac{\sin(\delta_c c)}{ \delta_c},
\eq
in the classical  lapse function $N_{\text{cl}}$ (\ref{eq2.14a}) and Hamiltonian $H_{\text{cl}}$ (\ref{eq2.14b}), 
we obtain
\bqn
\lb{eq2.5zh1}
N_{\text{BV}} &=& \frac{\gamma\delta_b \sqrt{p_c}}{\sin(\delta_b b)},\\
\lb{eq2.5zh2}
H^{\text{eff}}_{\text{BV}}[N] &=& - \frac{1}{2\gamma G}\Bigg[2 \frac{\sin(\delta_c c)}{ \delta_c} p_c\nb\\
&& + \left(\frac{\sin(\delta_b b)}{ \delta_b} + \frac{\gamma^2\delta_b}{\sin(\delta_b b)}\right)p_b\Bigg].
\eqn

It is remarkable to note that the above effective Hamiltonian can be obtained from the ABP Hamiltonian without the inverse volume corrections presented in the last subsection. 
In fact, making the following approximation
\bq
\lb{eq2.7a}
h_0[X] \rightarrow \frac{2}{\pi} \sin[X], \quad \cos[\epsilon]\sin^2\left[\frac{\epsilon}{2}\right] \rightarrow \frac{\epsilon^2}{4},
\eq
where $\epsilon$ is defined in Eq.(\ref{eq2.5g1}), we find that \footnote{It should be noted that Eq.(\ref{eq2.2}) tells that physically the conditions  $A = B = C = 0$ imply that: (a) the parameters $\alpha$ and $\beta$ defined in terms of  
the spin numbers $j$ and $j_x$ [cf. Eq.(\ref{eq2.5})] must satisfy the condition $\alpha, \; \beta \gg \ell_p$;  and (b) the spread dimensionless parameters $\delta_x$ and $\delta$ appearing in the quantum reduced
coherent states \cite{ABP20} must satisfy the condition $\delta, \; \delta_x \gg \gamma^2$. Both conditions are consistent with the semi-classical approximation of the effective Hamiltonian \cite{ABP20}. Further considerations
of these conditions are presented in Section IV given below. }
\bqn
\lb{eq2.14c}
A &=& B = C = 0,\nb\\
W &\simeq& 4 \sin[X],\quad  \mathcal{D}\simeq    4 \sin[X],   \nb\\
\frac{\mathcal{D}}{W} &\simeq&  1, \quad Z \simeq 2 \gamma^2 \left(\frac{\alpha}{R}\right)^2,\nb\\
h_{-1} &\simeq& \frac{2}{\pi} \cos[X].
\eqn
Then, substituting the above into the effective Hamiltonian (\ref{hamiltonian}), we
shall obtain precisely  the BV Hamiltonian (\ref{eq2.5zh2}) with
\bqn
\lb{eq2.25i}
\delta_b=\frac{\alpha}{R}=\frac{\alpha}{\sqrt{p_c}}, \quad
\delta_c=\frac{\beta}{\Lambda L_0}=\frac{\beta  \sqrt{p_c}}{p_b}.
\eqn
Comparing them with those given by Eq.(\ref{eq2.5ze}), we find that 
\bqn
\lb{eq2.25ii}
&& \alpha^\text{(BV)} = \beta^\text{(BV)} = \sqrt{\Delta},  
\eqn
which immediately leads to
\bqn
\lb{eq2.25iii}
 j^{\text{(BV)}} &=& \sqrt{\frac{3}{128\pi}}  \simeq 0.0864 \simeq 0.313 j^{\text{(BV)}}_x> \gamma j^{\text{(BV)}}_x,\nb\\
 j^{\text{(BV)}}_x &=& \frac{\sqrt{3}}{2\pi} \simeq 0.275.
\eqn
Therefore, the BV Hamiltonian is precisely the  limit of the effective ABP Hamiltonian,\footnote{In the BV limit, $N(\tau) \rightarrow \frac{N_{\text{BV}}}{2Gm}$ because $d\tau=2Gm dT$. Thus, we have
 $H_{int}^{IV+CS}[N] \rightarrow \frac{H^{\text{eff}}_{\text{BV}}[N]}{2Gm}$.} provided that: 
\begin{itemize}
\item the inverse volume corrections vanish, $A= B= C = 0$; 

\item 
 the Struve functions $h_0[X]$ and $h_{-1}[X]$ are replaced respectively by $(2/\pi)\sin[X]$ and $(2/\pi)\cos[X]$; and 
 
 \item  the spin parameters $j_x$ and $j$ are chosen as those given by Eq.(\ref{eq2.25iii}). 
 \end{itemize}
It is clear that  the last condition is in sharp conflict with the semi-classical limit requirement of Eq.(\ref{eq2.25v1}).

In addition, as $T \rightarrow - \infty$, BV found the following asymptotic behaviors
\bqn
\lb{eq2.14d}
b &\simeq& \bar b, \quad p_b \simeq \bar p_b e^{-\bar\alpha T}, \nb\\
c &\simeq& \bar c e^{-\bar \alpha T}, \quad p_c \simeq \bar p_c,
\eqn
where $ \bar b, \bar p_b, \bar c, \bar p_c$ and $\bar\alpha > 0$  are constants, given by  [cf. Eqs.(64) - (69) in  \cite{BV07}]
\bqn
\lb{bv1}
&&2\sin(\bar\delta_b \bar b)-\sin(\bar\delta_b \bar b)^2=\frac{\Delta \gamma^2}{\bar p_c},\\
\lb{bv2}
&&\bar{\alpha}=-\cos(\bar\delta_b \bar b)+\cot(\bar\delta_b \bar b),\\
\lb{bv3}
&&\sin (\bar\delta_b \bar b)-\left(\bar\delta_b \bar b+\frac{\pi }{2}\right) \Big[\cos (\bar\delta_b \bar b)-\cot (\bar\delta_b \bar b)\Big]-2=0,\nb\\
\eqn
with 
\bqn
\lb{bv3a}
\bar\delta_b=\frac{\sqrt{\Delta}}{\sqrt{\bar p_c}}, \quad
\bar\delta_c=\frac{\sqrt{\Delta\; \bar p_c}}{\bar p_b},\quad \bar\delta_c \bar c = - \frac{\pi}{2}. 
\eqn
 Then, from Eqs.(\ref{eq2.5ze}) and (\ref{eq2.5zh1}) we find that asymptotically
\bq
\lb{eq2.14e}
N_{\text{BV}} \simeq \bar{N} \equiv \frac{\gamma\sqrt{\Delta}}{\sin(\bar\delta_b \bar b)}.
\eq
Hence, the spacetime is asymptotically described by the metric
\bqn
\lb{eq2.14f}
ds^2 &=& - {N}_{\text{BV}}^2 dT^2 + \frac{p_b^2}{L_0^2 p_c} dx^2 + p_c d\Omega^2 \nb\\
&\simeq& \left(\frac{\bar t_0}{\bar t}\right)^{2}\left(- d\bar t^2 + d\bar x^2\right) + \bar p_cd\Omega^2,
\eqn
where  
\bqn
\lb{eq2.14g}
 d\bar t = e^{\bar\alpha T} dT, \quad \bar x = \frac{\bar p_b}{{\bar{N}}L_0 \sqrt{\bar p_c}} x, \quad \bar t_0 \equiv \frac{\bar N}{\bar\alpha}.
\eqn
Loop quantum black holes do not satisfy the classical Einstein's equations. However, in order to study the loop quantum gravitational effects (with respect to GR),  we introduce the effective energy-momentum tensor  $T^{\;\text{eff}}_{\mu\nu}$
by  $T^{\;\text{eff}}_{\mu\nu} \equiv  G_{\mu\nu}$ \footnote{It should be noted that the Einstein field equations usually read as $ G_{\mu\nu} = (8\pi G/c^4) T_{\mu\nu}$, while in this paper we drop the factor $8\pi G/c^4$, as this will not affect
our analysis and conclusions.}, which takes the form
\bqn
\lb{eq2.5g}
T^{\;\text{eff}}_{\mu\nu} &\simeq& \rho u_{\mu} u_{\nu} + p_{\bar x} \bar x_{\mu}\bar x_{\nu} + p_{\bot}\left(\theta_{\mu}\theta_{\nu} + \phi_{\mu}\phi_{\nu}\right), ~~~~
\eqn
 in the current case, where $u_{\mu} = (\bar t_0/\bar t)\delta^{\bar t}_{\mu}$, $\bar x_{\mu} =  (\bar t_0/\bar t)\delta^{\bar x}_{\mu}$, $ \theta_{\mu} = \sqrt{p_c} \delta^{\theta}_{\mu}$, $\phi_{\mu} = \sqrt{p_c} \sin\theta  \delta^{\phi}_{\mu}$, 
 and
\bqn
\lb{eq2.14h}
\rho &\simeq& \frac{1}{\bar{p}_c}, \;\; p_{\bar x}  \simeq  -\frac{1}{\bar{p}_c}, \;\;
 p_{\bot} \simeq -\frac{1}{\bar{t}_0^2}.
 \eqn
 From the above it is clear that the spacetime corresponds to a spacetime with a homogeneous and isotropic perfect fluid  only when  $\bar{t}_0 = \sqrt{\bar{p}_c}$. When $\bar{t}_0 \not= \sqrt{\bar{p}_c}$, the radial pressure 
 is different from the tangential one, despite the fact that they are all constants. The latter (with $\bar{t}_0 \not= \sqrt{\bar{p}_c}$)   can be interpreted as the charged Nariai solution \cite{Bousso97}. 
 In addition, we also have
\bqn
\lb{eq2.14i}
&& {\cal{R}}  \simeq 2 \left(\frac{1}{\bar{p}_c}+\frac{1}{\bar{t}_0^2}\right), \nb\\
&& R_{\mu\nu} R^{\mu\nu} \simeq 2 \left(\frac{1}{\bar{p}_c^2}+\frac{1}{\bar{t}_0^4}\right), \nb\\
 && R_{\mu\nu\alpha\beta} R^{\mu\nu\alpha\beta} \simeq 4 \left(\frac{1}{\bar{p}_c^2}+\frac{1}{\bar{t}_0^4}\right), \nb\\
&& C_{\mu\nu\alpha\beta} C^{\mu\nu\alpha\beta} \simeq \frac{4 \left(\bar{p}_c+\bar{t}_0^2\right){}^2}{3 \bar{t}_0^4 \bar{p}_c^2}.
\eqn
It is remarkable to note that, even when $\bar{t}_0 = \sqrt{\bar{p}_c}$, the spacetime is still not conformally flat. So,  it must not be the de Sitter space. In fact, as 
 noticed by BV \cite{BV07}, it is the Nariai space \cite{Nariai99,Bousso02}.

On the other hand, from Eqs.(\ref{bv1})-(\ref{bv3}), BV found the following solutions
\bqn
\lb{bv4}
&& \bar b\simeq 0.156, \;\;\; \bar p_c \simeq 0.182 \ell_p ^2, \;\;\; \bar{\alpha} \simeq 0.670,\nb\\  
&& \frac{\bar c}{\bar p_b} \simeq - 2.290 m_p ^2, \;\;\; \bar N \simeq 0.689 \ell_p,
\eqn
from which we find that
\bqn
\lb{bv5}
\bar t_0 =  \frac{\bar N}{\bar{\alpha}} \approx 1.029 \ell_p \not=\sqrt{\bar p_c} \left(\approx 0.427 \ell_p\right). 
\eqn
Therefore, the solution is asymptotically approaching to the charged Nariai solution \cite{Bousso97}, instead of the Nariai solution \cite{Nariai99}. 

It should be noted that in the above calculations, BV took   $\gamma \approx 0.2375$ in the expression $\Delta = 2 \sqrt{3} \pi \gamma \ell_p^2$. Instead,  if we  take $\gamma \approx 0.274$ \cite{ABP20}  
 we find 
 \bqn
 \lb{bv6}
 && \bar N \approx 0.854 \ell_p, \quad \bar p_c \approx 0.279 \ell_p^2, \; (\gamma \approx 0.274),  \nb\\
 &&  \bar t_0 \equiv \frac{\bar N}{\alpha} \approx 1.275 \ell_p \not= \sqrt{\bar p_c} \left(\approx 0.529 \ell_p\right),
 \eqn
 that is, even in this case the spacetime is still not asymptotically Nariai, but the charged Nariai  \cite{Bousso97}.

\subsection{Ashtekar-Olmedo-Singh Limit}\lb{aos-limit}

From the analysis of the BV limit, it becomes clear that from the general ABP model, the  AOS limit \cite{AOS18a,AOS18b}   can be obtained by the replacements
\bqn
\lb{eqAOS1}
&& h_0[X] \rightarrow \frac{2}{\pi} \sin[X], \quad h_{-1}[x] \rightarrow \frac{2}{\pi} \cos[X],\nb\\
&&  \cos[\epsilon]\sin^2\left[\frac{\epsilon}{2}\right] \rightarrow \frac{\epsilon^2}{4},
\eqn
so that 
\bqn
\lb{eqAOS2}
W &\simeq& 4 \sin[X],\quad  \mathcal{D}\simeq    4 \sin[X],   \nb\\
\frac{\mathcal{D}}{W} &\simeq&  1, \quad Z \simeq 2 \gamma^2 \left(\frac{\alpha}{R}\right)^2.
\eqn
In addition, we must also set
%\footnote{\textcolor{red}{\bf In \cite{AOS18b}, $\delta_b, \; \delta_c$ depend on Dirac observable $m$, which is constant \textit{only} on physical trajectory. However, solutions \eqref{AOS2b} can only be obtained when $\delta_b, \; \delta_c$ are constants in whole phase space.}}
\bqn
\lb{eqAOS2a}
 && A = B = C = 0, \nb\\
 && \delta_b, \; \delta_c = {\text{Constant.}}
\eqn
Then, the resultant lapse function and effective Hamiltonian will be precisely given by the same form as  Eqs.(\ref{eq2.5zh1}) and (\ref{eq2.5zh2}) but with different $\delta_b, \; \delta_c$.
With the above in mind, AOS found the following solutions \cite{AOS18b}
\bqn
\lb{AOS2b}
&& \sin\left(\delta_c c\right) = \frac{2a_0e^{2T}}{a^2_0 + e^{4T}}, \nb\\
&& \cos\left(\delta_b b\right) = b_0 \frac{b_+ e^{b_0T} - b_-}{b_+ e^{b_0T} + b_-}, \nb\\
&& p_b = - \frac{GmL_0e^{-b_0T}}{2b_0^2}\left(b_+ e^{b_0T} + b_-\right){\cal{A}},\nb\\
&& p_c = 4(Gm)^2\left(a_0^2 + e^{4T}\right)e^{-2T}, 
\eqn
where $m$ is an integration constant, related to the mass parameter as noticed previously, and 
\bqn
\lb{AOS2c}
{\cal{A}} &\equiv& \Big[2\left(b_0^2 + 1\right)e^{b_0T} - b_-^2 - b_+^2 e^{2b_0T}\Big]^{1/2}, \nb\\  
a_0 &\equiv&  \frac{\gamma \delta_c L_0}{8Gm}, \quad b_0 \equiv \left(1 + \gamma^2 \delta_b^2\right)^{1/2}, \nb\\
b_{\pm} &\equiv& b_0 \pm 1,
\eqn
with
\bqn
\lb{AOS2d}
&& \delta_b b \in \left(0, \pi\right), \quad \delta_c c \in \left(0, \pi\right),\nb\\
&& p_b \le 0, \quad p_c \ge 0, \quad - \infty < T < 0.
\eqn
In terms of $p_b$ and $p_c$, the metric takes the form
\bq
\lb{AOS2e}
ds^2 = - N_{\text{AOS}}^2 dT^2 + \frac{p_b^2}{|p_c| L_0^2} dx^2 + |p_c|d\Omega^2,
\eq
where \footnote{In the AOS limit, $N(\tau) \rightarrow \frac{N_{\text{AOS}}}{2Gm}$ because $d\tau=2Gm dT$. Thus, we have $H_{int}^{IV+CS}[N] \rightarrow \frac{H^{\text{eff}}_{\text{AOS}}[N]}{2Gm}$.}
\bqn
\lb{AOS2ea}
N_{\text{AOS}} &=& \frac{\gamma \delta_b\; \text{sgn}\left(p_c\right)\left|p_c\right|^{1/2}}{\sin\left(\delta_b b\right)}\nb\\
 &=& \frac{2Gm}{{\cal{A}}}e^{-T}\left(b_+e^{b_0T} + b_-\right) \left(a_0^2 + e^{4T}\right)^{1/2}.\nb\\
\eqn

From Eq.(\ref{AOS2b}), it can be seen that the transition surface is located at $\partial p_c\left({\cal{T}}\right)/\partial T = 0$, which yields
\bq
\lb{AOS2f}
{\cal{T}} = \frac{1}{2}\ln\left(\frac{\gamma \delta_c L_0}{8Gm}\right) < 0.
\eq
There exist two horizons, located respectively at
\bq
\lb{AOS2g}
T_{\text{BH}}  = 0, \quad T_{\text{WH}} = - \frac{2}{b_0}\ln\left(\frac{b_0 + 1}{b_0 -1}\right), 
\eq
at which we have ${\cal{A}}(T) = 0$, where $ T = T_{\text{BH}}$ is the location of the black hole horizon, while $T = T_{\text{WH}}$ is the location of the white hole horizon. 
In the region ${\cal{T}} < T < 0$, the 2-spheres are all trapped, while in the one  $T_{\text{WH}} < T < {\cal{T}}$, they are all anti-trapped.  Therefore, 
the region ${\cal{T}} < T < 0$ behaves like the internal of a black hole, while the one $T_{\text{WH}} < T < {\cal{T}}$ behaves like the  internal of a white hole.

The extension across the  {black hole}  horizon can be obtained by the following replacements \cite{AOS18a,AOS18b}
\bqn
\lb{AOS2h}
&& b \rightarrow i b, \quad p_b  \rightarrow i p_b, \nb\\
&& c \rightarrow c, \quad p_c  \rightarrow  p_c.
\eqn
Then, AOS found that the corresponding Penrose diagram consists of infinite diamonds along the vertical direction, alternating between black holes and white holes,
but the spacetime singularity used appearing at $p_c = 0$ now is replaced by a non-zero minimal surface with 
\bqn
\lb{AOS2i}
p^{\text{min}}_c = p_c({\cal{T}}) > 0,
\eqn
where ${\cal{T}}$ is given by Eq.(\ref{AOS2f}).

To completely fix the values of $\delta_b$ and $\delta_c$, AOS required that on the transition surface ${\cal{T}}$, the physical areas of $A_{x\theta}$ and $A_{\theta\phi}$ be
equal to the area gap $\Delta$ \cite{AOS18a,AOS18b}
\bqn
\lb{eqAOS3a}
 && 2\pi \delta_c \delta_b \left|p_b({\cal{T}})\right|  = \Delta, \\
 \lb{eqAOS3b}
 && 4\pi  \delta_b^2 p_c({\cal{T}})  = \Delta. 
\eqn

It is interesting to note that, substituting Eq.(\ref{eq2.25i}) into the above equations, we find that
\bq
\lb{eqAOS4}
2\pi \alpha\beta = \Delta, \quad 4\pi \alpha^2 = \Delta,
\eq
which are all independent of $p_b$ and $p_c$ and given by
\bq
\lb{eqAOS5}
  \alpha = \frac{1}{2} \beta = \sqrt{\frac{\Delta}{4\pi}} = \sqrt{2\sqrt{2}\; \gamma}\; \ell_p.  
\eq
Comparing it with Eq.(\ref{eq2.5}) we find that
\bqn
\lb{eqAOS6}
 j^{\text{(AOS)}} &=&  \frac{1}{4 \pi^{3/2}} < \frac{1}{2}, \quad   j^{\text{(AOS)}}_x = \frac{1}{\sqrt{2}\; \pi^2} < \frac{1}{2},\nb\\
  {j}^{\text{(AOS)}} &=&  \sqrt{\frac{\pi}{8}}\; {j^{\text{(AOS)}}_x} \simeq 0.6265 {j^{\text{(AOS)}}_x} > \gamma {j^{\text{(AOS)}}_x}, ~~~~
\eqn
from which we find that such given  $j$ and $j_x$ do not satisfy the semi-classical limit conditions (\ref{eq2.25v1}) either. Therefore,  
 the AOS model cannot be realized in the framework of QRLG either, although it can be obtained formally by the approximations
(\ref{eqAOS2}) and  (\ref{eqAOS2a}) from the ABP model.

  \section{Quantum Reduced Loop Black Holes without Inverse Volume Corrections} 
  \lb{rlqg1}
 \renewcommand{\theequation}{4.\arabic{equation}}\setcounter{equation}{0}

Setting   the three constants $A, B$ and $C$ to zero,   the effective Hamiltonian (\ref{hamiltonian}) reduces to the one given in \cite{ABP19}, but  with the replacement of the
constants $\alpha$ and $\beta$ by 
\bq
\lb{eq2.5c}
\alpha \equiv \sqrt{8\pi \gamma} \; \ell_p \; \sqrt{\hat j_0}, \quad \beta = \frac{\sqrt{8\pi \gamma}\;\ell_p\; \hat j}{\sqrt{\hat j_0}}, \;\;\; 
\eq
where now $\hat j_0$ and $\hat j$ denote  the quantum numbers associated respectively with the longitudinal and angular links 
of the coherent states, as mentioned in Section II.  The relations between ($j, j_x$) and $(\hat j, \hat j_0$) are given explicitly by Eq.(\ref{eq2.5g2}).  Without causing any confusion, in the rest of
this section we shall drop the hats from   $(\hat j, \hat j_0$):
$$
(\hat j, \hat j_0) \quad \rightarrow \quad (j,  j_0),
$$
 unless some specific statements are given. 

It is interesting to note that dropping the terms that are proportional to the constants $A,\; B$ and $C$ defined in Eq.(\ref{eq2.2})  is  physically equivalent to assuming that
\bq
 \lb{eq2.23}
 \frac{A}{R^2},   \; \frac{B}{R^2},  \; \frac{C}{R^2}  \ll 1,
 \eq
as can be seen from the effective Hamiltonian given by Eq.(\ref{hamiltonian}). 
Before proceeding further, let us first pause here for a while and consider the above limits. In particular, from  
Eqs.(\ref{eq2.5}) and (\ref{eq2.5a}), we find $\alpha \sim \beta \sim \sqrt{j} \ell_p$, where ``$\sim$'' means ``being the same order''. 
On the other hand, introducing the spread parameters  $\delta_i$ via the relations \cite{ABP20}
\bqn
\la{deltas}
\delta_r&=&\frac{\pi^2 \ell_{\va P}^2 R^2}{\alpha^4 (\sin{\theta})^2} \delta_x\,,\;\;\;
\delta_\theta = \frac{\pi^2 \ell_{\va P}^2 R^2}{\alpha^2\beta^2 (\sin{\theta})^2} \delta\,,\nb\\
\delta_\varphi&=&\frac{\pi^2 \ell_{\va P}^2 R^2}{\alpha^2\beta^2 } \frac{\delta}{\nu},
\eqn
we find that the  terms appearing in the expressions of $A,\; B$ and $C$ behave, respectively, as
\bqn
 \lb{eq2.23a} 
&& \ell_p^2\left(\frac{\ell_p^2 \gamma^2}{\beta^2}\right) \sim \frac{\ell_p^2 \gamma^2}{j},\;\;\; 
 \ell_p^2\left(\frac{\gamma^2}{\delta_x}\right) \sim \frac{\gamma^2 \pi^2 R^2}{j^2  \sin^2(\theta) \delta_r},\nb\\
&& 
 \ell_p^2\left(\frac{(3-\nu) \gamma^2}{\delta}\right) \sim \frac{\pi^2 \gamma^2 R^2}{j^2  \sin^2(\theta) \delta_\theta}-\frac{\pi^2 \gamma^2 R^2}{j^2   \delta_\varphi}. ~~~
\eqn
Thus, the conditions (\ref{eq2.23})  imply 
\bq \lb{eq2.23d}
(i) \;\; \frac{\ell_p}{R} \ll 1, \quad (ii)\;\;  j \delta_{i}  \gg 1, \;\;\; (i = r, \theta, \varphi).
\eq
 Condition (ii) is required by the effective Hamiltonian approach \cite{ABP20}, while condition (i)
tells us that the effects of the inverse volume corrections are negligible when the geometric radius of the two-spheres (with $\tau,  x =$ Constant) is much large than the Planck length.

With the above in mind,  let us now turn to consider the effective  Hamiltonian given by Eq.(\ref{hamiltonian}) with 
\bq
\lb{ABC}
A=B=C=0. 
\eq
It was shown \cite{ABP19} that the classical singularity  of the Schwarzschild black hole now is replaced by a quantum bounce at $R = R_{\text{min}} > 0$, at which all the physical quantities, such as the Ricci scalar ${\cal{R}}$, 
Ricci squared $R_{\mu\nu} R^{\mu\nu}$, Kretschmann scalar $R_{\mu\nu\alpha\beta} R^{\mu\nu\alpha\beta}$, and Weyl squared $C_{\mu\nu\alpha\beta} C^{\mu\nu\alpha\beta}$, remain finite. In addition,
at the black hole horizons, the quantum effects become negligible for macroscopic black holes. 

A remarkable feature of this class of spacetimes is that the spacetime on the other side of the bounce is not asymptotically a white hole, as normally expected from the minisuperspace considerations \cite{Ashtekar20}. Instead, depending
on the values of $\eta$, defined by
\bq
\lb{eq2.5daa}
\eta \equiv \frac{\alpha}{\beta} = \frac{j_0}{j},
\eq
the spacetime has three different asymptotical limits, as $\tau \rightarrow -\infty$.  

In this section, we shall provide a more detailed study over the whole parameter
space. To this goal, let us consider the three cases $\eta = 1$, $\eta < 1$ and $\eta > 1$, separately.

  \begin{figure}[h!]
\includegraphics[height=7.5cm]{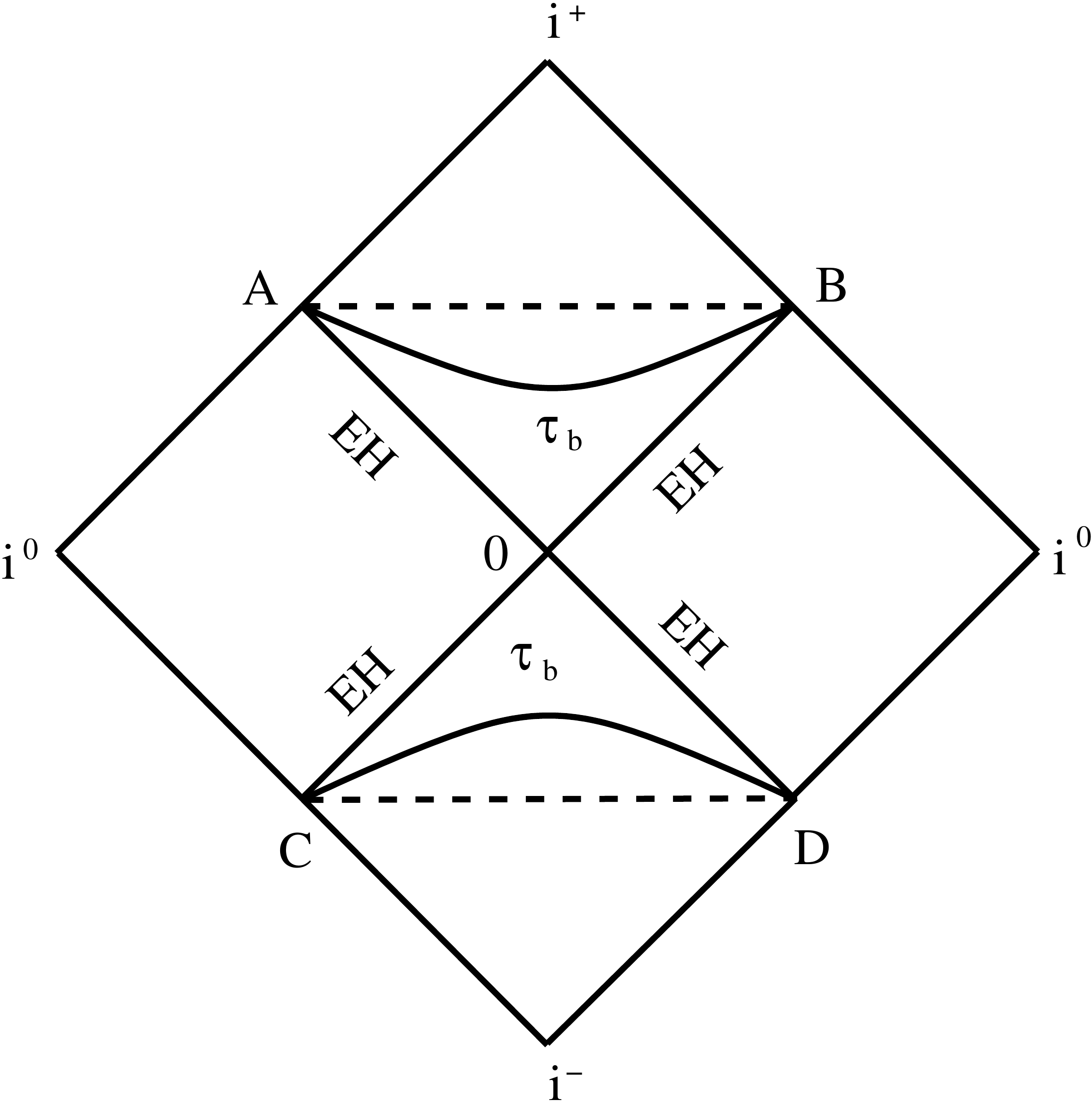}
\caption{The  Penrose diagram for the loop quantum spacetimes without the inverse volume corrections in the case $\eta = 1$. The curved lines denoted by $\tau_b$ are the transition surfaces (throats), and
the straight lines AD and CB  are the locations of the black hole horizons. The dashed lines AB and CD are the locations of the classical singularities of the Schwarzschild black and white holes,
which now are all free of singularities. } 
\label{fig2}
\end{figure}

\subsection{$\eta = 1$}

 In this case from Eq.(\ref{eq2.5daa}) we find that $j=j_0$. Then, as $\tau \rightarrow -\infty$, we have
  \bqn
  X &\simeq& -\pi, \quad Y \simeq -\pi, \quad
   W \simeq -\pi h_0[\pi], \nb\\
  \frac{P_\Lambda}{R^2} &\simeq& -\frac{\pi}{\alpha \gamma G}, \quad \frac{P_R}{R \Lambda} \simeq -\frac{2\pi}{\alpha \gamma G}.
  \eqn
  Hence, the metric coefficients have the following 
asymptotical behavior \cite{ABP19} \footnote{We found that the numerical factor, $31.49$, of $\Lambda$ weakly depends on the mass parameter $m$. For example, it is respectively  $31. 55, 31.77, 32.63$ for $m/m_p  = 10^6, 10^5, 10^4$. On the other hand, the numerical factors of $N(\tau)$ and $R(\tau)$ are very insensitive to $m$. In particular,  they are the same up to the third digital for $m/m_p  = 10^{12}, 10^6, 10^5, 10^4$.}
\bqn
\lb{eq2.5d}
 N(\tau)  &\simeq& - \frac{2\gamma \sqrt{8\pi \gamma} \; \ell_p \; \sqrt{ j_0}}{mG \left(-\pi h_0[\pi]\right)}\simeq 0.886\frac{\sqrt{j}\; \ell_p}{m G},\nb\\
  \Lambda(\tau) &\simeq& 31.49 \left(\frac{m G}{\sqrt{j}\; \ell_p}\right)^{1/3}, \nb\\
R(\tau) &\simeq& 0.0504\left(\frac{j^2\ell_p^4}{m G}\right)^{1/3} \exp\left(- \frac{\tau}{2mG}\right).
\eqn
Thus, the metric takes the following asymptotical form
\bq
\lb{eq2.5e}
ds^2 \simeq - d\bar\tau^2 + d\bar x^2 + R^2d\Omega^2,
\eq
which has a topology $R^2 \times S^2$, and the ($\bar\tau, \bar x$)-plane is flat, where $\bar\tau \equiv - N(\tau \rightarrow -\infty) \tau$ and $\bar x \equiv \Lambda(\tau \rightarrow -\infty) x$. Then, the low
half plane $ - \infty < \tau < 0$ and $-\infty < x < \infty$ is mapped to the upper half plane $ 0 < \bar\tau < \infty$ and $-\infty < \bar x < \infty$, and
the corresponding Penrose diagram is given by Fig. \ref{fig2}.

It should be noted that the spacetime is not vacuum as $\tau  \rightarrow -\infty$, despite the fact that the ($\bar\tau, \bar x$)-plane is asymptotically flat. This can be seen clearly by writing the metric (\ref{eq2.5e}) 
in terms of the timelike coordinate $R$
\bq
\lb{eq2.5f}
ds^2 \simeq - \left(\frac{R_0}{R}\right)^2 dR^2   + d\bar x^2 + R^2d\Omega^2,
\eq
where $R_0 \equiv 2 \sqrt{j}\; \ell_p$. For the metric (\ref{eq2.5f}), we find that the corresponding effective energy-momentum tensor can still be cast in the form
of   Eq.(\ref{eq2.5g}), but with 
 $u_{\mu} = (R_0/R)\delta^{R}_{\mu}$, $\bar x_{\mu} =  \delta^{\bar x}_{\mu}$, $\theta_{\mu} = R \delta^{\theta}_{\mu}$, $\phi_{\mu} = R\sin\theta  \delta^{\phi}_{\mu}$, and
\bqn
\lb{eq2.5h}
\rho &\simeq& \frac{1}{R^2}+\frac{1}{R_0^2}, \nb\\
p_{\bar x} &\simeq&  -\frac{1}{R^2}-\frac{3}{R_0^2}, \nb\\
 p_{\bot} &\simeq& -\frac{1}{R_0^2}.
 \eqn

 The commonly used three energy conditions are {\it the weak, dominant and strong energy conditions} \cite{HE73}. For   $T^{\;\text{eff}}_{\mu\nu}$ given by Eq.(\ref{eq2.5g}), they can be expressed respectively as
\begin{itemize} 
\item  the weak energy condition (WEC):
\bq
\lb{eq2.5ha}
\rho \geq 0, \;\;\; \rho+p_{\bar x} \geq 0, \;\;\; \rho+p_{\bot} \geq 0,
\eq
 
 \item the dominant energy condition (DEC): 
\bq
\lb{eq2.5hb}
\rho \geq 0, \;\;\;  - \rho \le  p_{\bar x} \le \rho,  \;\;\; -\rho \le p_{\bot} \le \rho,
\eq

\item the  strong energy condition (SEC):
\bq
\lb{eq2.5hc}
 \rho+p_{\bar x} \geq 0, \;\; \rho+p_{\bot} \geq 0,  \;\; \rho+p_{\bar x} + 2p_{\bot}\geq 0.
\eq
\end{itemize}
Clearly, Eq.(\ref{eq2.5h}) does not satisfy any of these conditions, but the energy density and the two principal pressures do approach  constant values that are inversely proportional
to $R_{0}^2 \propto   \ell_p^{2}$, that is, the spacetime curvature approaches to the Planck scale. 
On the other hand, we also find
\bqn
\lb{eq2.5i}
&& {\cal{R}} \simeq  \frac{2}{R^2}+\frac{6}{R_0^2}, \nb\\
&& R_{\mu\nu} R^{\mu\nu} \simeq 2 \left(\frac{1}{R^4}+\frac{4}{R^2 R_0^2}+\frac{6}{R_0^4}\right), \nb\\
 && R_{\mu\nu\alpha\beta} R^{\mu\nu\alpha\beta} \simeq 4 \left(\frac{1}{R^4}+\frac{2}{R^2 R_0^2}+\frac{3}{R_0^4}\right), \nb\\
&& C_{\mu\nu\alpha\beta} C^{\mu\nu\alpha\beta} \simeq \frac{4}{3 R^4}.
\eqn
It is interesting to note that the last expression of the above equation shows that asymptotically the spacetime is  conformally flat, while the Ricci,  Ricci squared and Kretschmann scalars are approaching to their Planck values.

 \begin{table}%\scriptsize%[h!]
\caption{
The initial values $P_{R}(\tau_i)$ obtained from the effective Hamiltonian constraint (\ref{eq2.5k}) and the choice of the initial values of the other three variables given by Eq.(\ref{eq2.5j}), 
and its corresponding relativistic values  $P_{R_c}(\tau_i)$, for different choices of $\tau_i$.
Results are calculated with $m=10^{12} m_p , \; j=j_0=10$.}
\begin{tabular}{|c|c|c|c|c|c|c|c|c|c|}
\hline
$\tau_i/\tau_{p} $    & -0.01 & -0.02 & -0.05 & -0.1  & -1    & -10   & -100  & $-10^3$ & $-10^4$ \\ \hline
$P_{R_c}$ & 0.500 & 0.500 & 0.500 & 0.500 & 0.500 & 0.500 & 0.500 & 0.500   & 0.500   \\ \hline
$P_R$     & 0.506 & 0.500 & 0.501 & 0.500 & 0.500 & 0.500 & 0.500 & 0.500   & 0.500   \\ \hline
\end{tabular}
\label{tab:initial}%
\end{table}

 \begin{table}
        \caption{
        The initial values $P_{R}(\tau_i)$ obtained from the effective Hamiltonian constraint (\ref{eq2.5k}) and the choice of the initial values of the other three variables given by Eq.(\ref{eq2.5j}), 
and its corresponding relativistic values  $P_{R_c}(\tau_i)$, for different choices of $j$ with $j_0 = j\;  $(or $\eta = 1$). Results are calculated with $m=10^{12} m_p , \; \tau_i = - 10\; \tau_{p} $.}
\begin{tabular}{|l|l|l|l|l|l|l|l|l|l|l|l|l|}
    \hline
        $j$ & 10 & $10^3$ & $10^5$ & $10^7$ & $10^8$ & $10^9$ & $10^{10}$ & $10^{11}$ & $10^{12}$  \\ \hline
        $P_{R_c}$ & 0.500 & 0.500 & 0.500 & 0.500 & 0.500 & 0.500 & 0.500 & 0.500 & 0.500  \\ \hline
        $P_R$ & 0.500 & 0.500 & 0.500 & 0.500 & 0.500 & 0.500 & 0.500 & 0.500 & 0.501  \\ \hline
    \end{tabular}
    \label{tab:initial-2}%
\end{table}

\begin{table}
    \caption{ The initial values $P_{R}(\tau_i)$ obtained from the effective Hamiltonian constraint (\ref{eq2.5k}) and the choice of the initial values of the other three variables given by Eq.(\ref{eq2.5j}), 
and its corresponding relativistic values  $P_{R_c}(\tau_i)$, for different choices of $m$. Results are calculated with $j =10, \; \tau_i = - 10\; \tau_{p} $.}
    \begin{tabular}{|l|l|l|l|l|l|l|l|l|l|l|l|}
    \hline
        $m/m_p$ & 10 & $10^2$ & $10^3$ & $10^5$  & $10^{10}$  & $10^{12}$  & $10^{14}$ \\ \hline
        $P_{R_c}$ & 0.176 & 0.474 & 0.497 & 0.500 & 0.500 & 0.500 & 0.500    \\ \hline
        $P_R$ &0.051 & 0.474 & 0.497 & 0.500 & 0.500 & 0.500 & 0.500  \\ \hline
    \end{tabular}
 \label{tab:initial-3}%
\end{table}

To study this class of solutions in more details, we need first to specify the initial conditions, which are often imposed near the black hole horizons \cite{BV07,AOS18a,AOS18b,ABP19}, as normally it is 
expected that the quantum effects for macroscopic black holes should be negligible \cite{Ashtekar20}, and the spacetime can be well-described by the Schwarzschild black hole spacetime. So, near the horizon, 
say, $\tau = \tau_i \simeq \tau_{H}$, we can take the initial values of ($\Lambda, P_{\Lambda}$) and ($R, P_{R}$) as their corresponding relativistic values,
 ($\Lambda_c, P_{\Lambda_c}$) and ($R_c, P_{R_c}$).  However, there is a caveat with the above prescription of the initial conditions, 
 that is, before carrying out the integrations of  the effective Hamiltonian equations, 
 we do not know if the corresponding model indeed has negligible quantum gravitational effects near the black hole horizons even for macroscopic black holes.
Therefore, a consistent way to choose the initial conditions should be: {\em First choose the initial conditions for any three of the four variables, $\left(R, \Lambda, P_R, P_{\Lambda}\right)$, and then obtain the initial condition for the fourth variable
 through the  Hamiltonian constraint ${\cal{H}}^{\text{IV+CS}}_{\text{int}} =  0$}.  The choice of the initial conditions for the first three variables clearly are arbitrary, which form the complete phase space ${\cal{D}}$
 of the initial conditions of the theory. However, in order to study quantum effects, one can choose them as their
 corresponding relativistic values.

   \begin{widetext}

     \begin{figure}[h!]
 \begin{tabular}{cc}
\includegraphics[height=4cm]{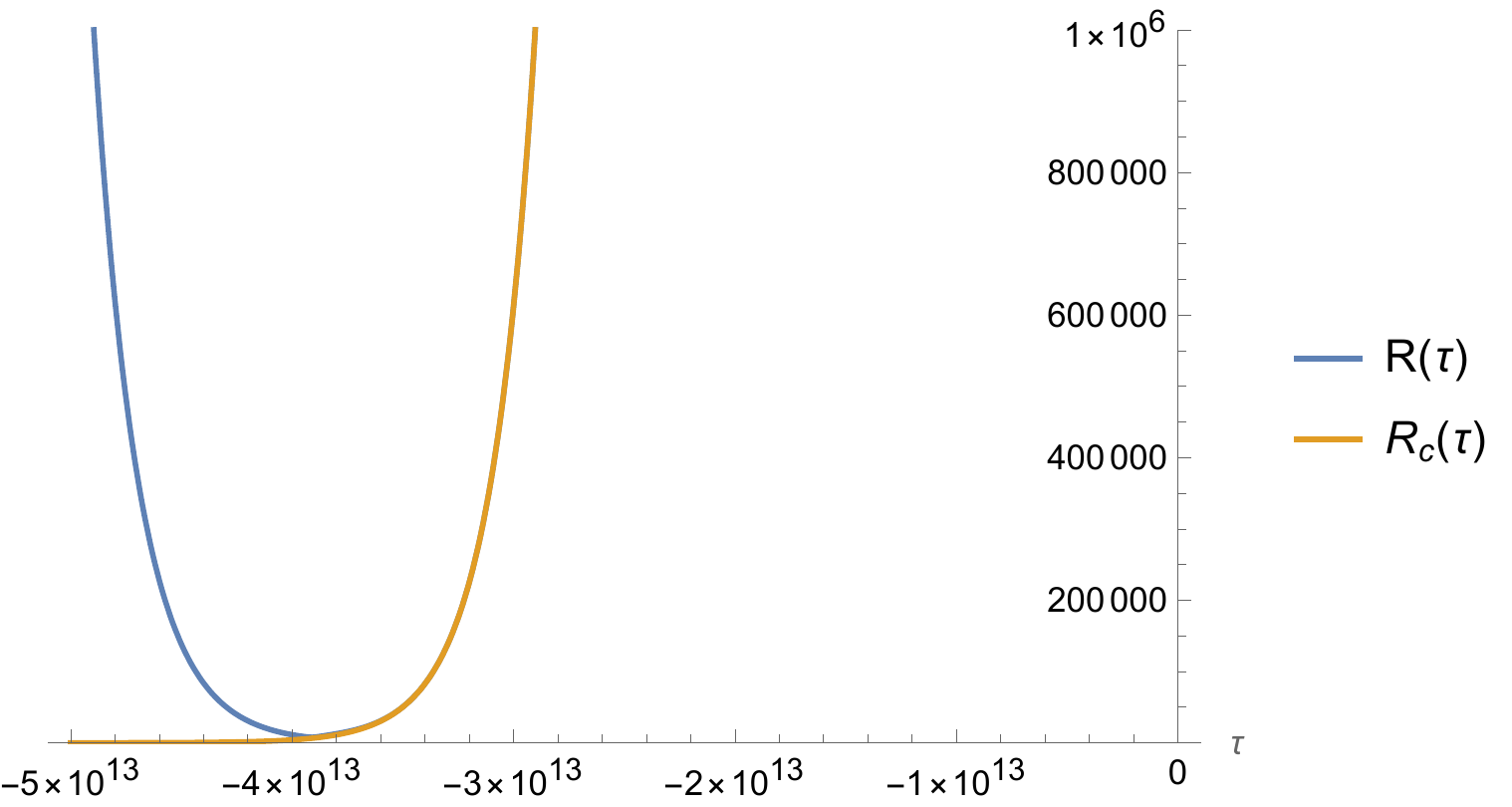}&
\includegraphics[height=4cm]{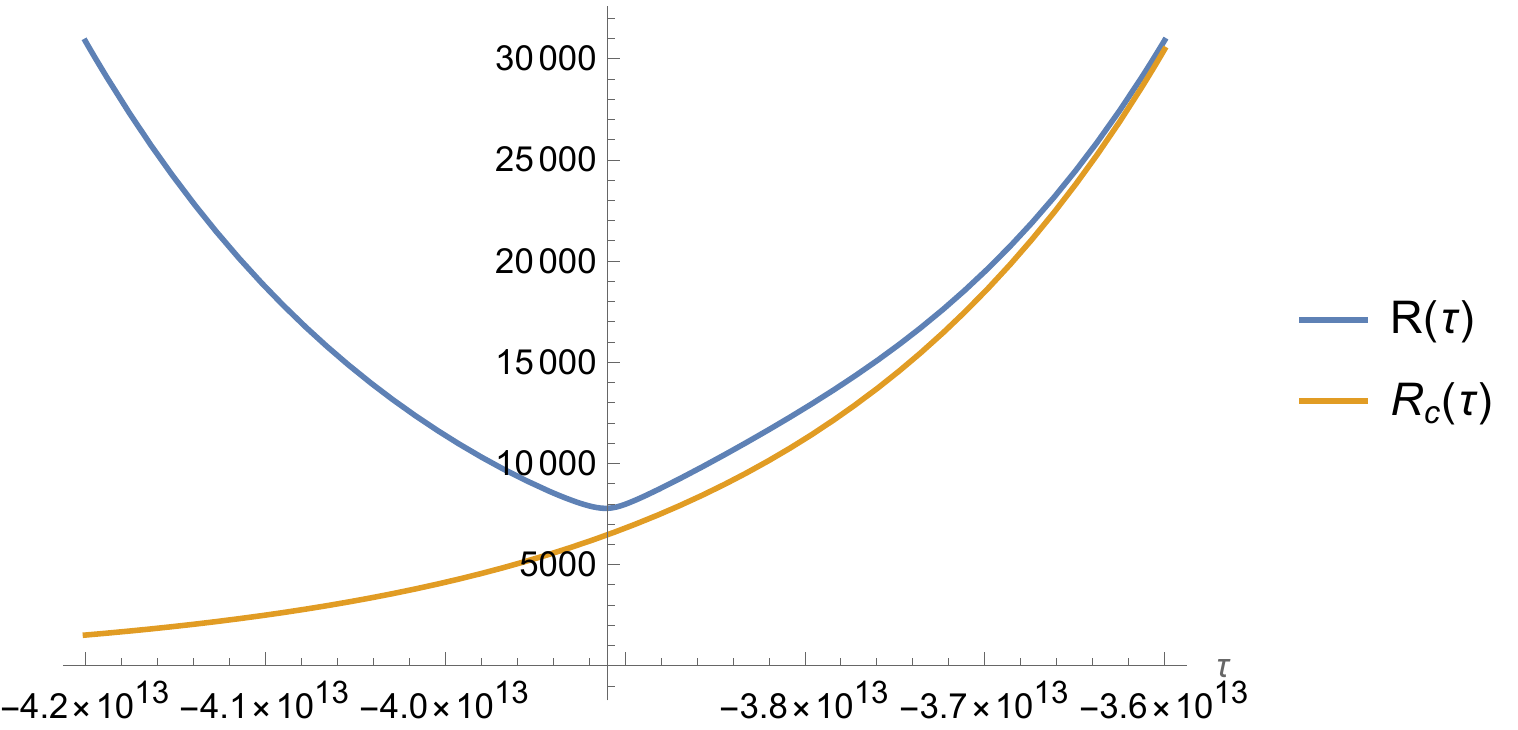}\\
	(a) & (b)  \\[6pt]
\includegraphics[height=4cm]{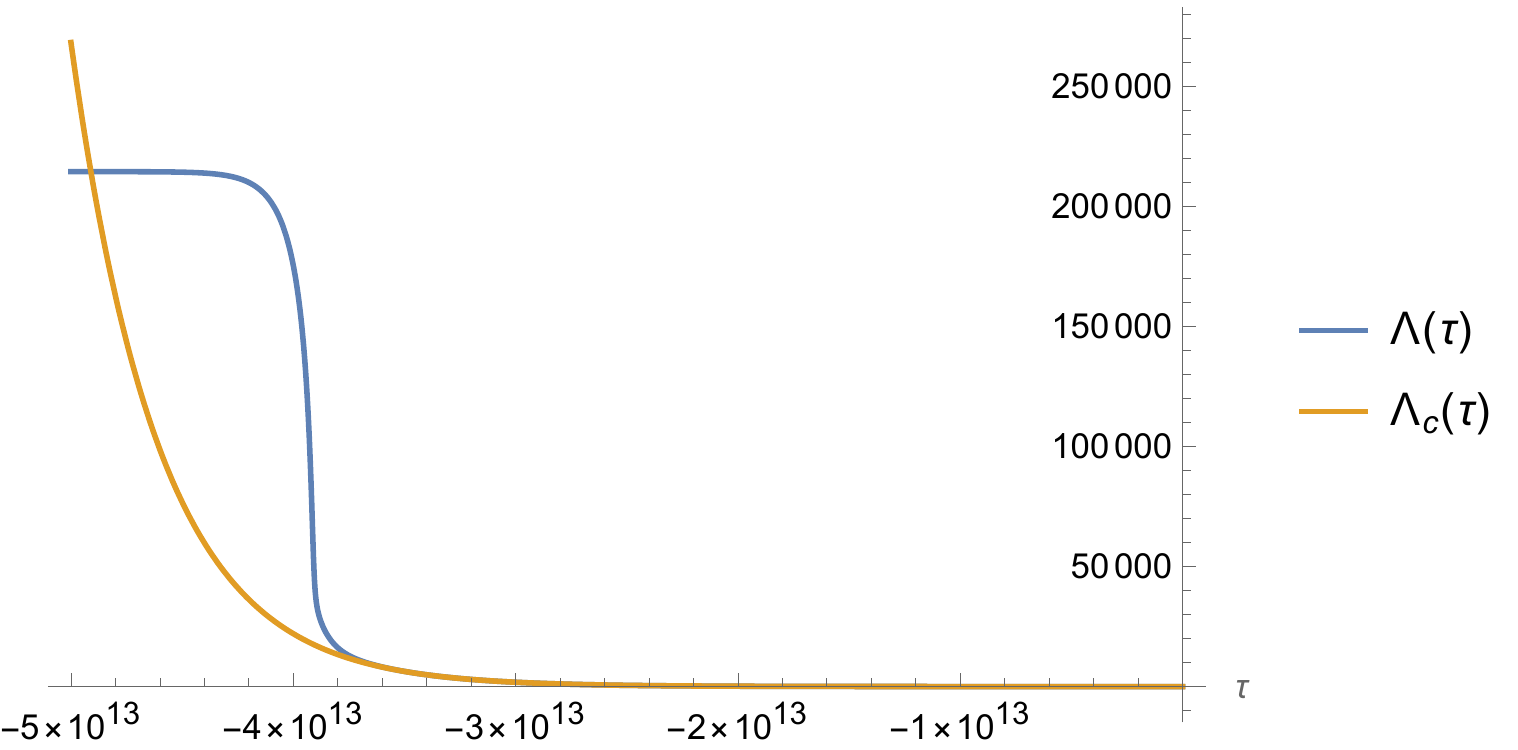}&
\includegraphics[height=4cm]{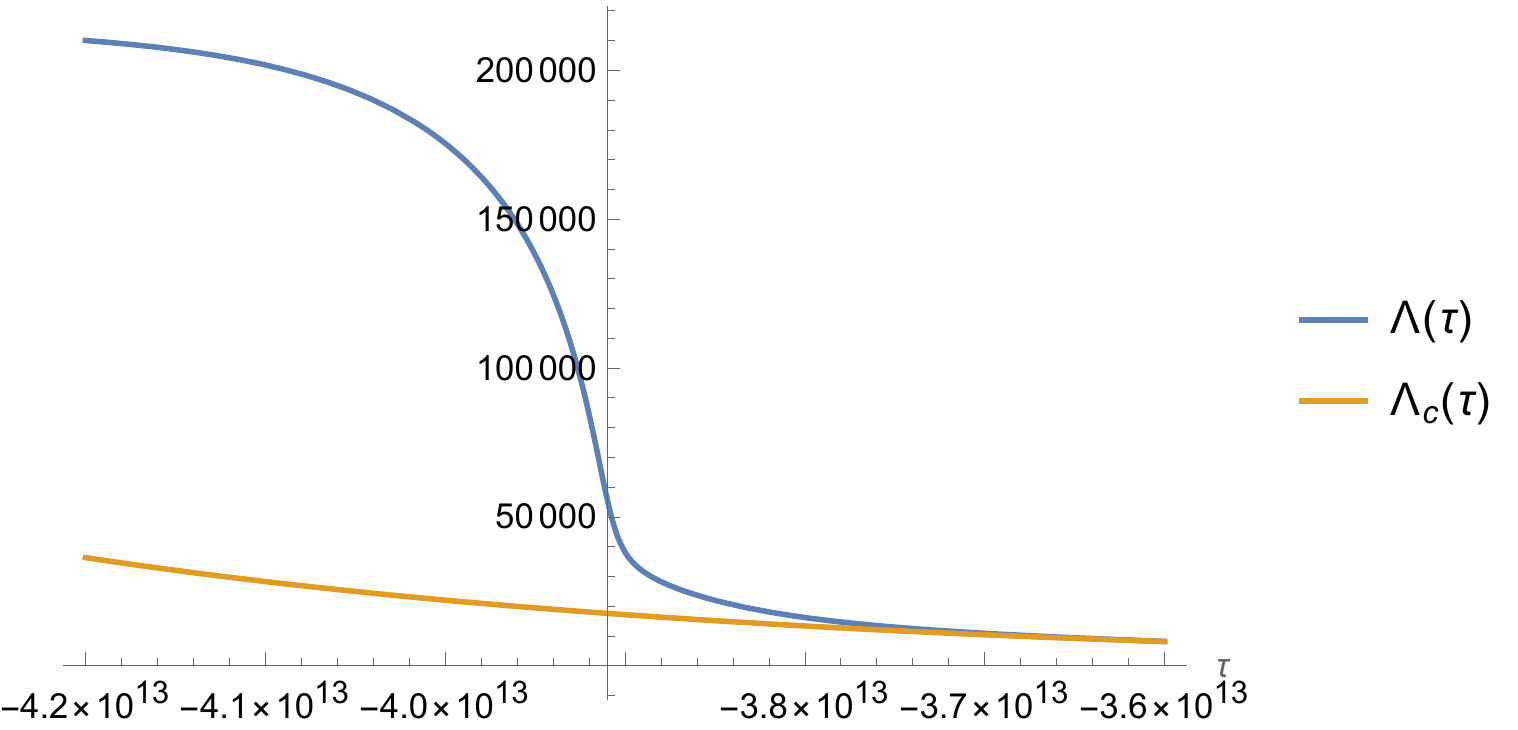}\\
	(c) & (d)  \\[6pt]
	\includegraphics[height=4cm]{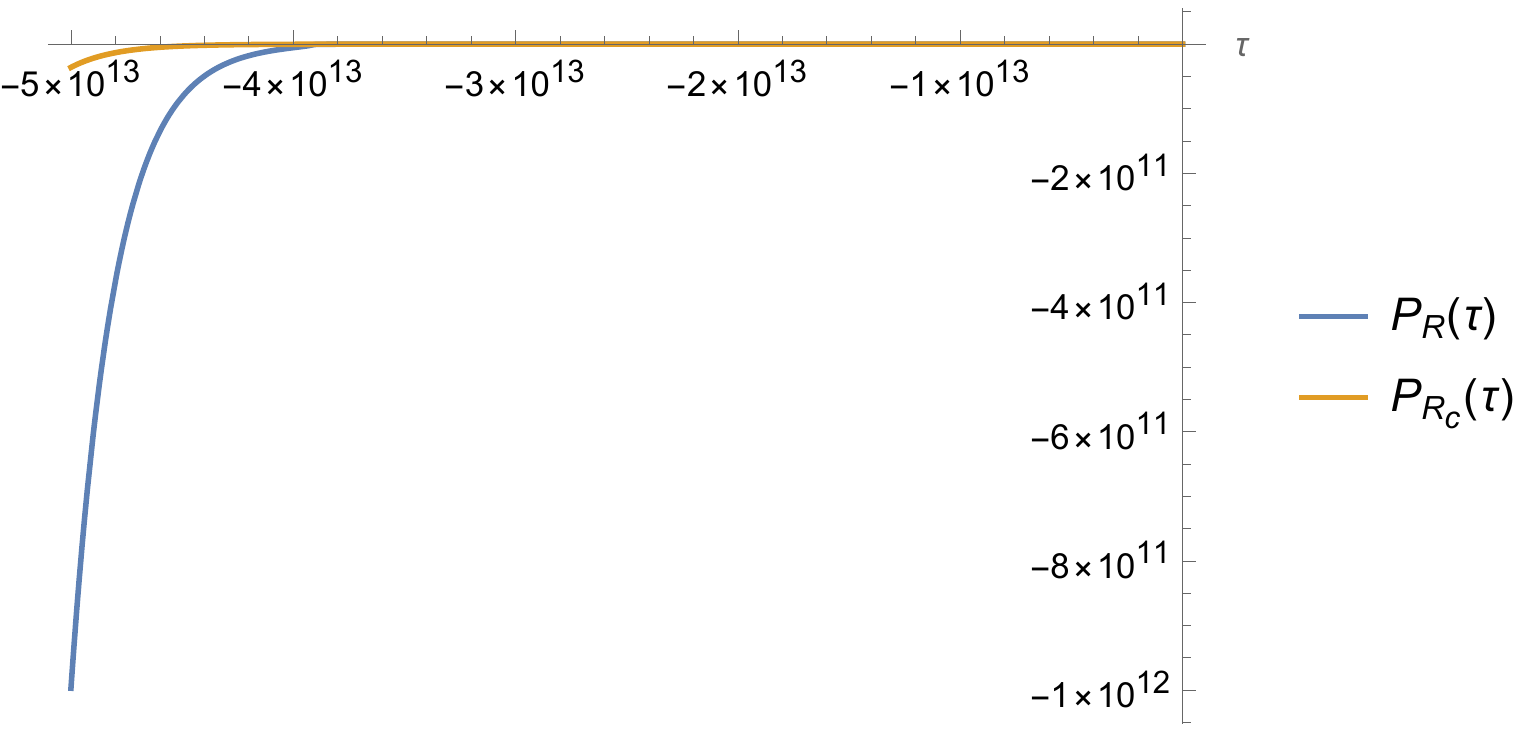}&
\includegraphics[height=4cm]{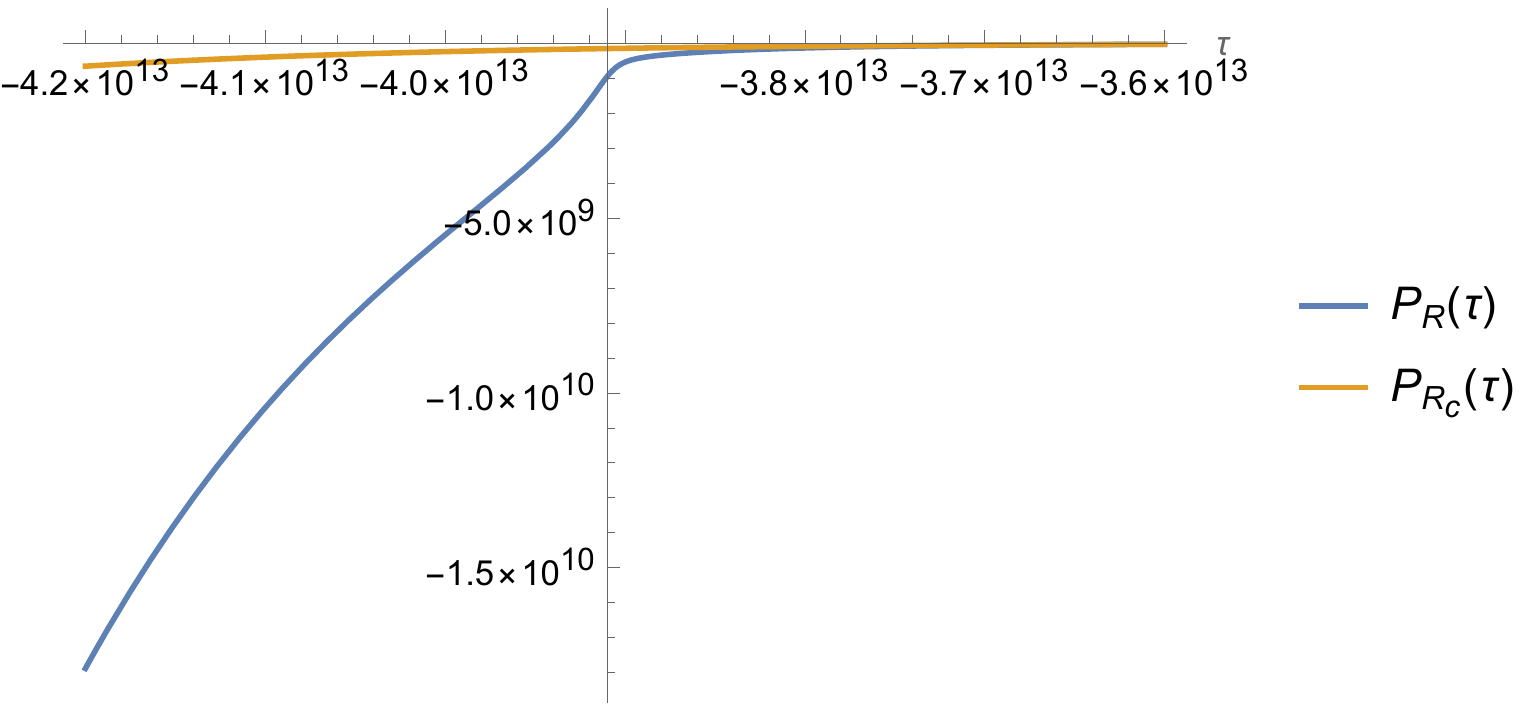}\\
	(e) & (f)  \\[6pt]
	 	\includegraphics[height=4cm]{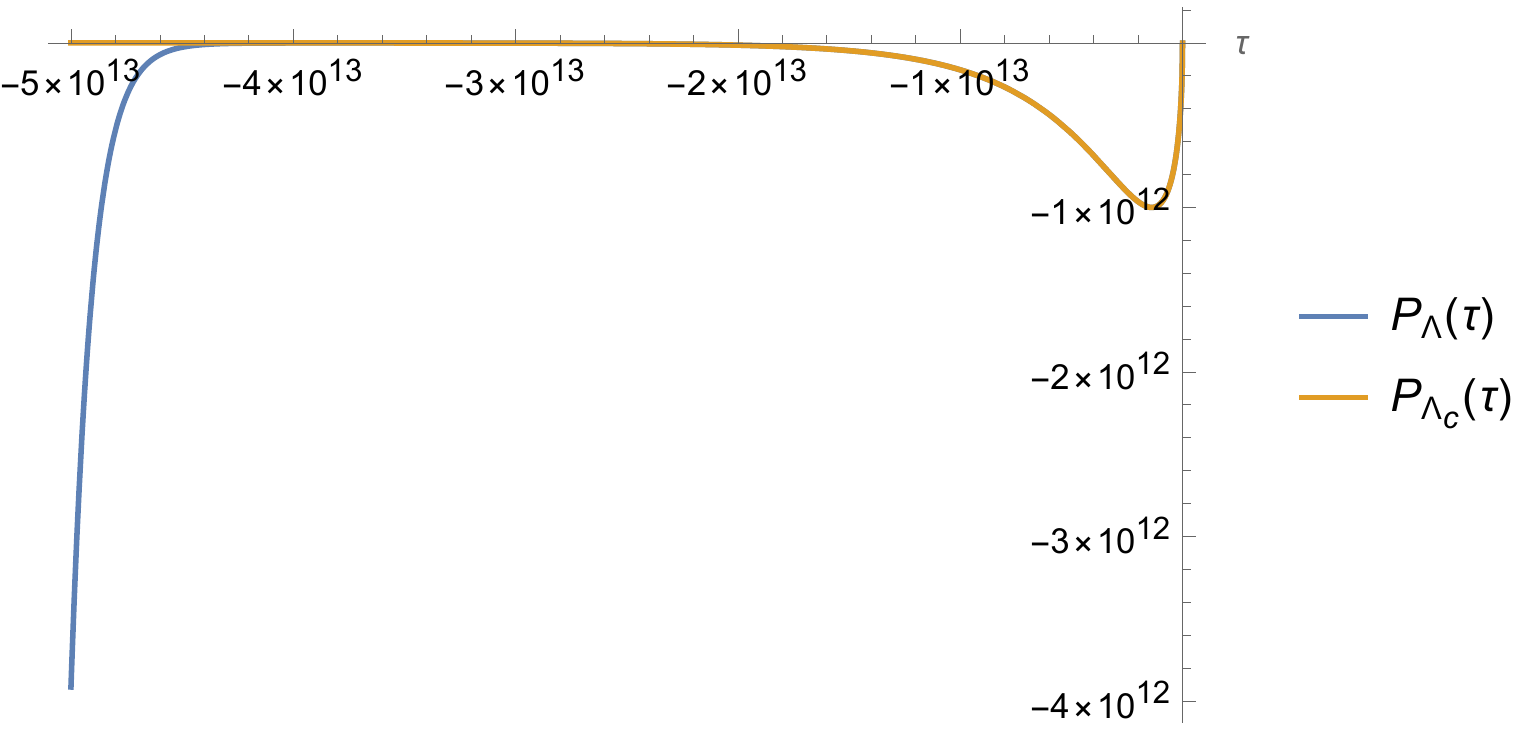}&
\includegraphics[height=4cm]{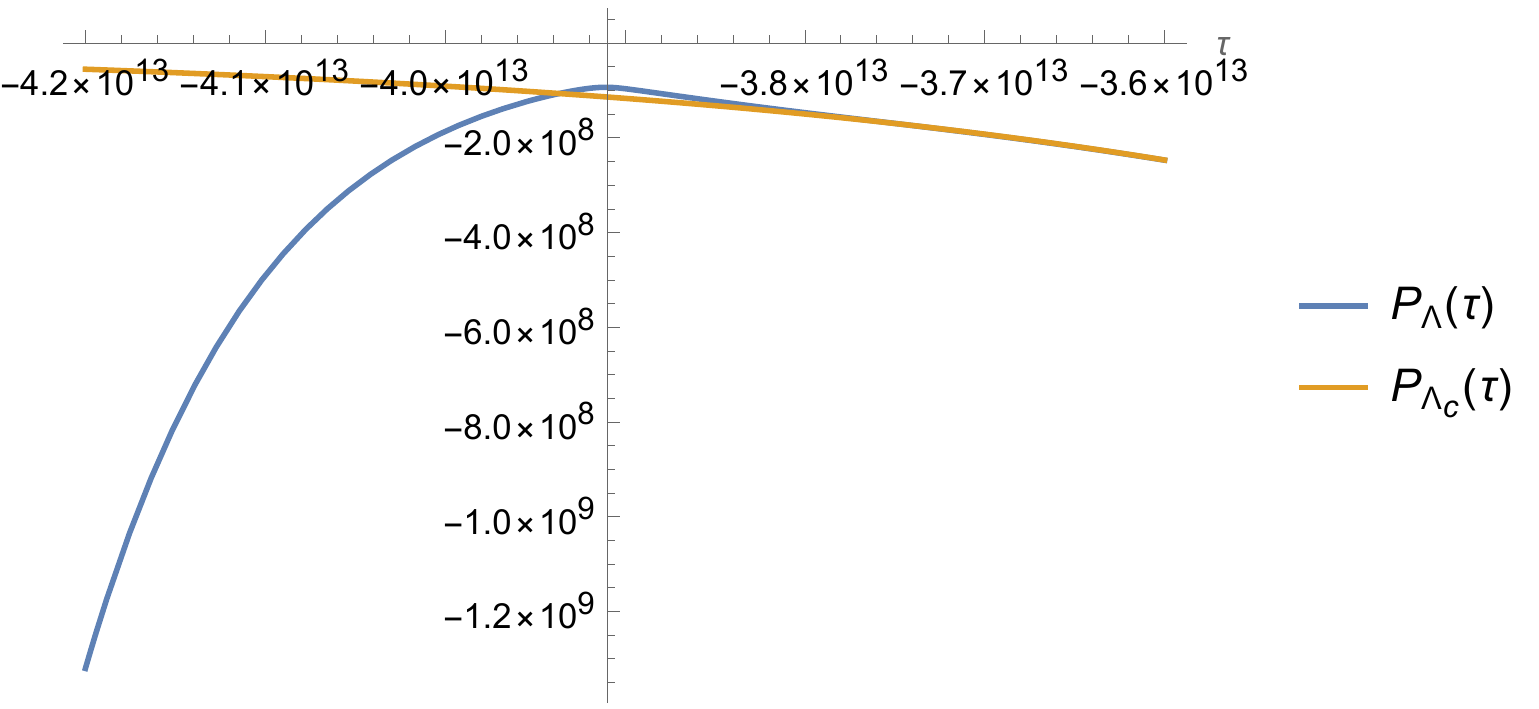}\\
	(g) & (h)  \\[6pt]
	\end{tabular}
\caption{Plots of the physical variables $\left(R, \Lambda, P_R, P_{\Lambda}\right)$ and their classical correspondences $\left(R_c, \Lambda_c, P_{R_c}, P_{\Lambda_c}\right)$.
Particular attention are paid to the region near the throat $\tau=-3.91 \times 10^{13}$. 
Graphs are plotted with $m=10^{12} m_p , \; j=j_0=10$.
} 
\lb{fig3}
\end{figure}

     \begin{figure}[h!]
 \begin{tabular}{cc} 
 		\includegraphics[height=4cm]{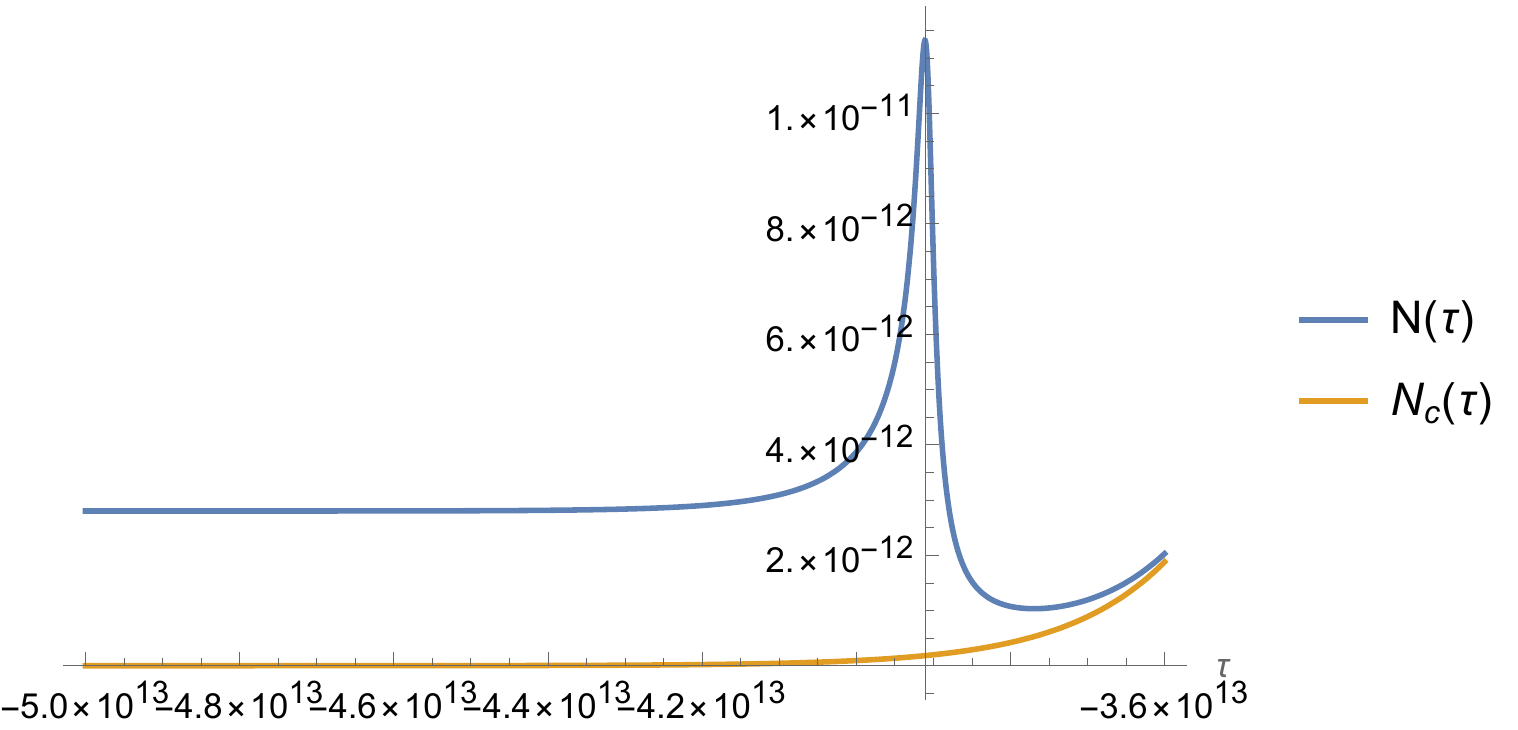}&
\includegraphics[height=4cm]{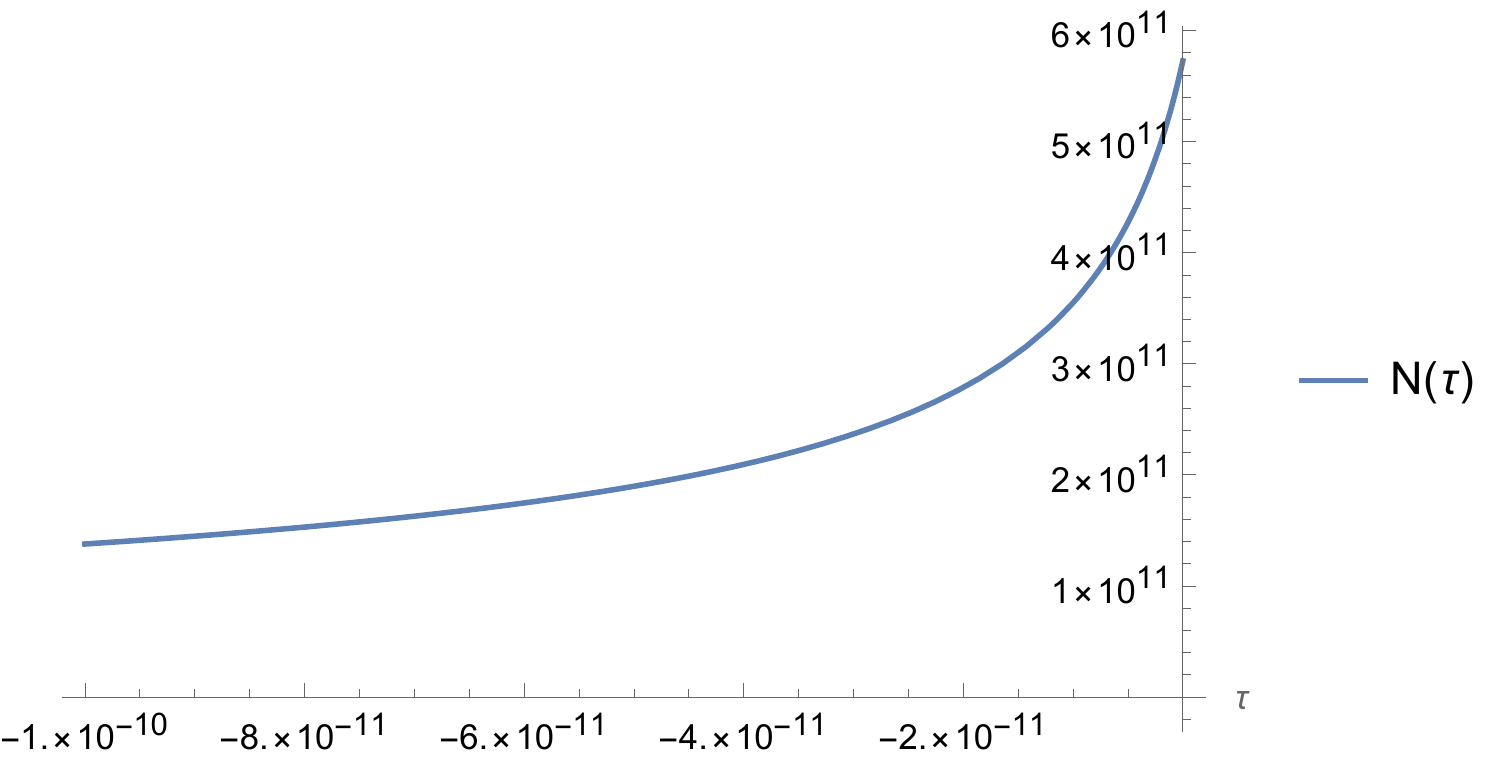}\\		
	(a) & (b)  \\[6pt]
	\includegraphics[height=4cm]{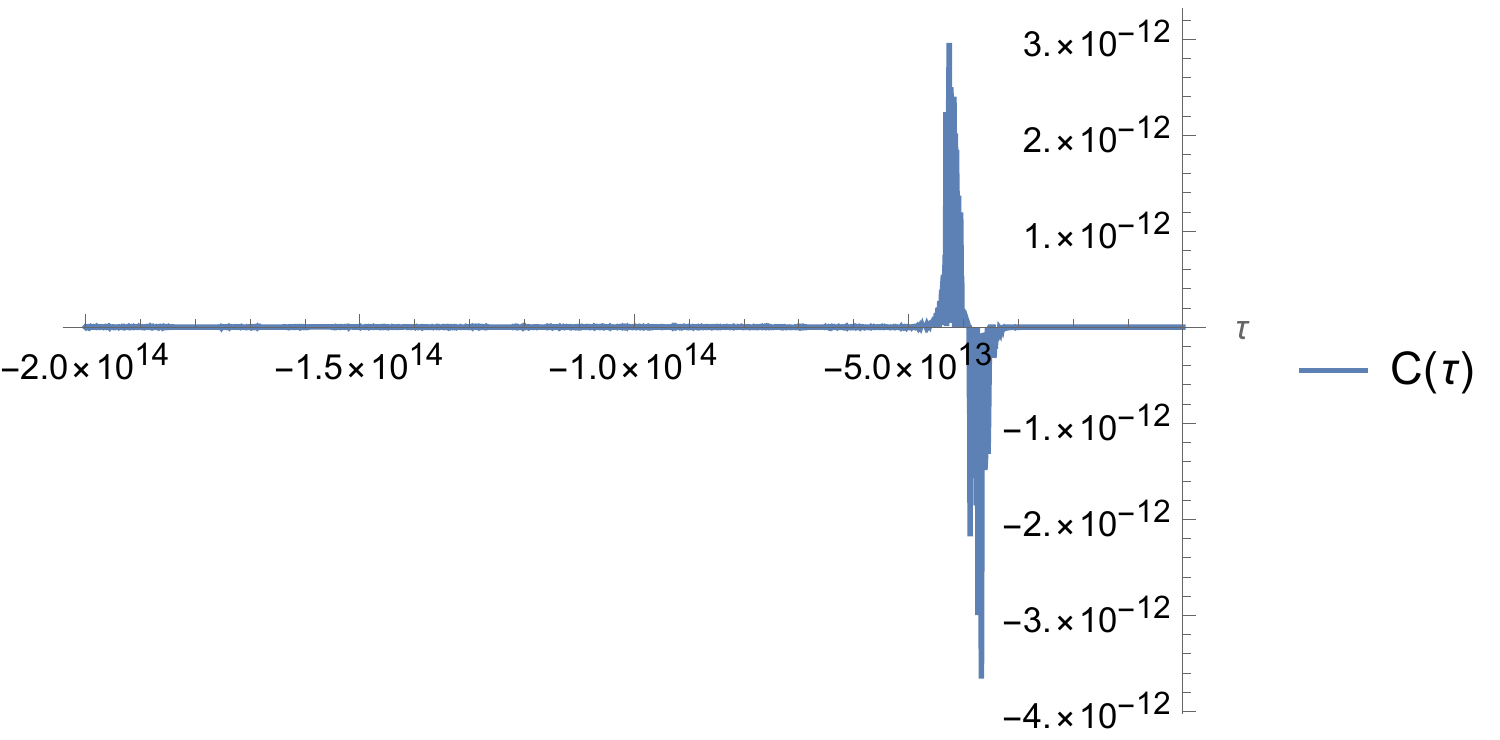}&
\includegraphics[height=4cm]{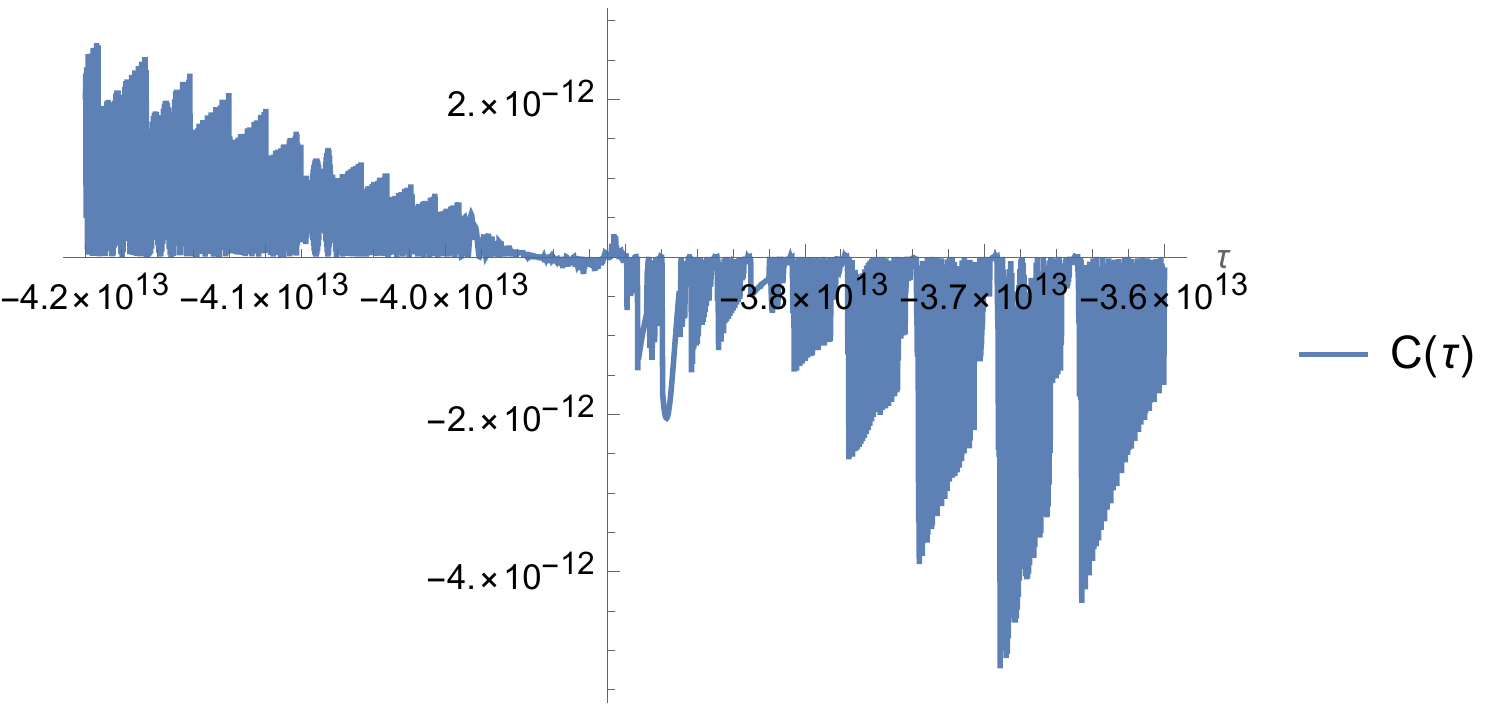}\\
	(c) & (d)  \\[6pt]
		\end{tabular}
\caption{Plots of $\mathcal{C}(\tau)$ and the lapse function $N(\tau)$ for 
 $m=10^{12} m_p ,\; j=j_0=10$.
} 
\lb{fig4}
\end{figure}

\begin{figure}[h!]
 \begin{tabular}{cc}
		\includegraphics[height=4cm]{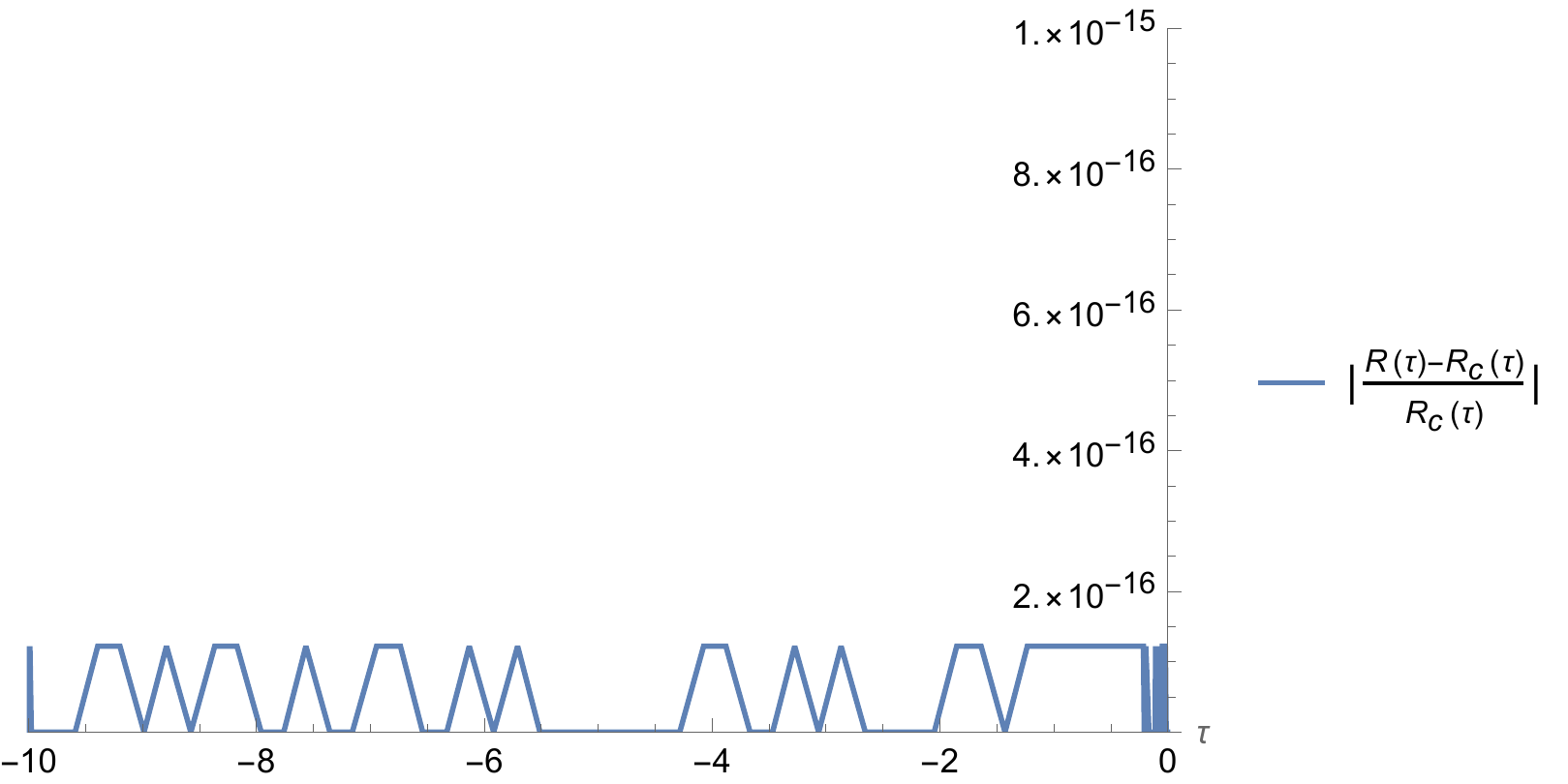}&
\includegraphics[height=4cm]{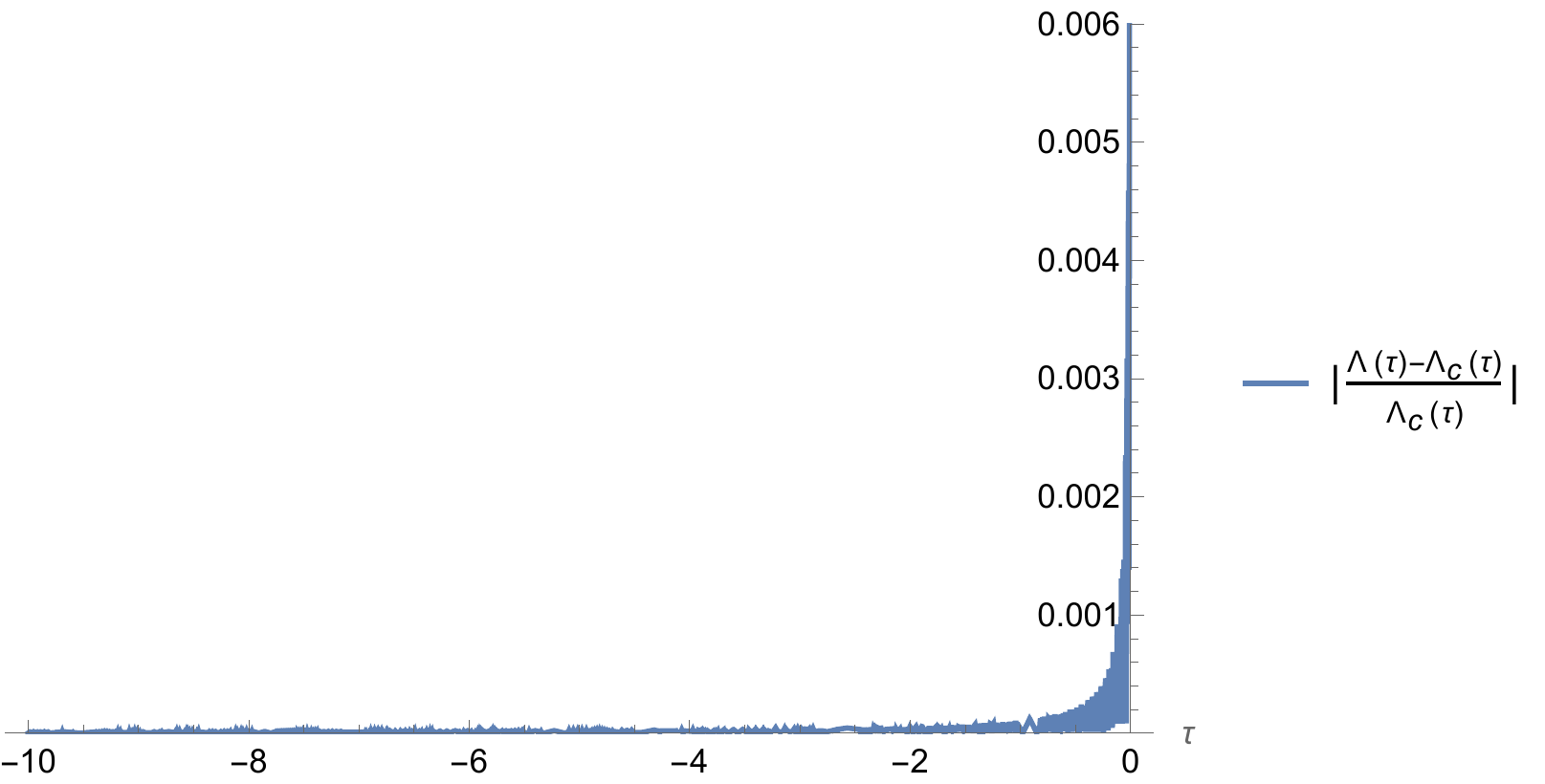}\\
	(a) & (b)  \\[6pt]
		\includegraphics[height=4cm]{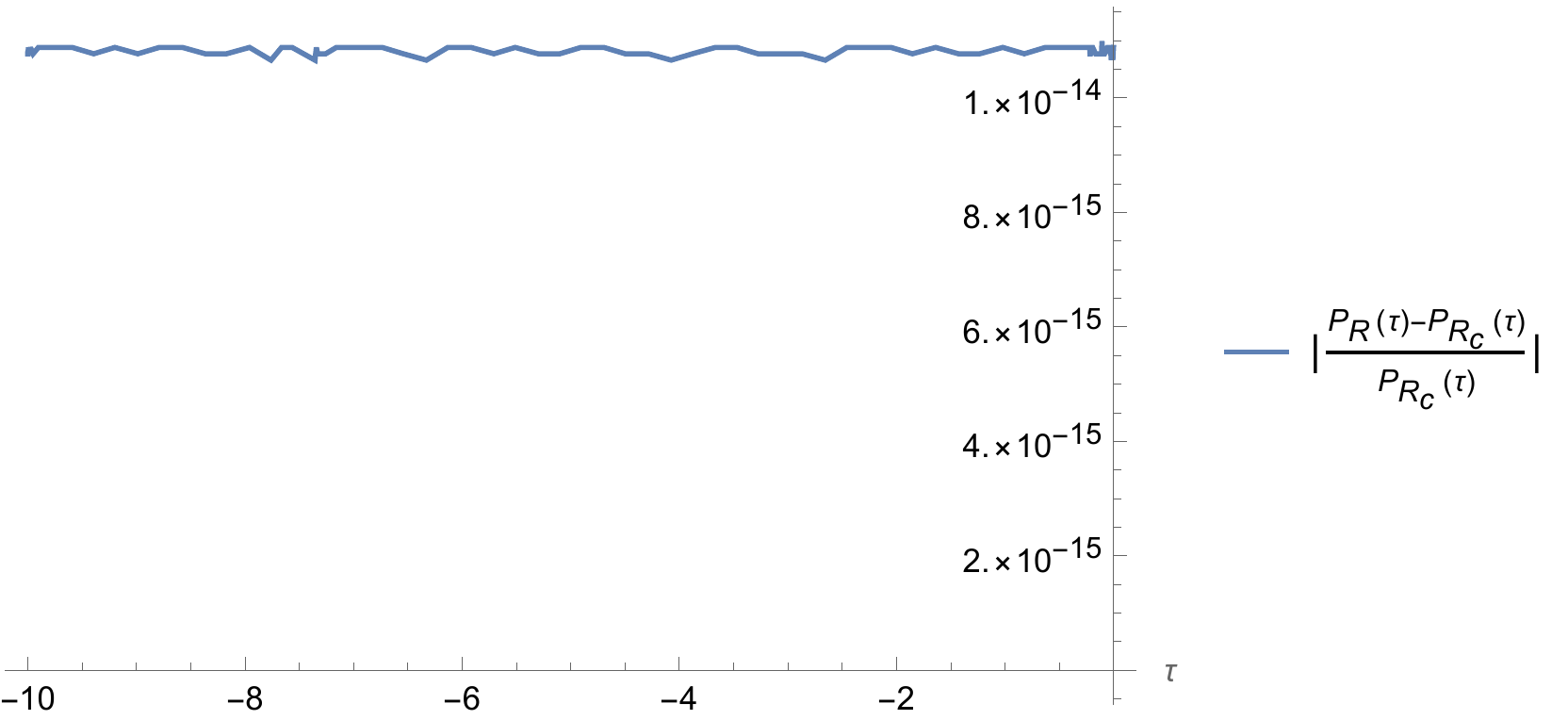}&
\includegraphics[height=4cm]{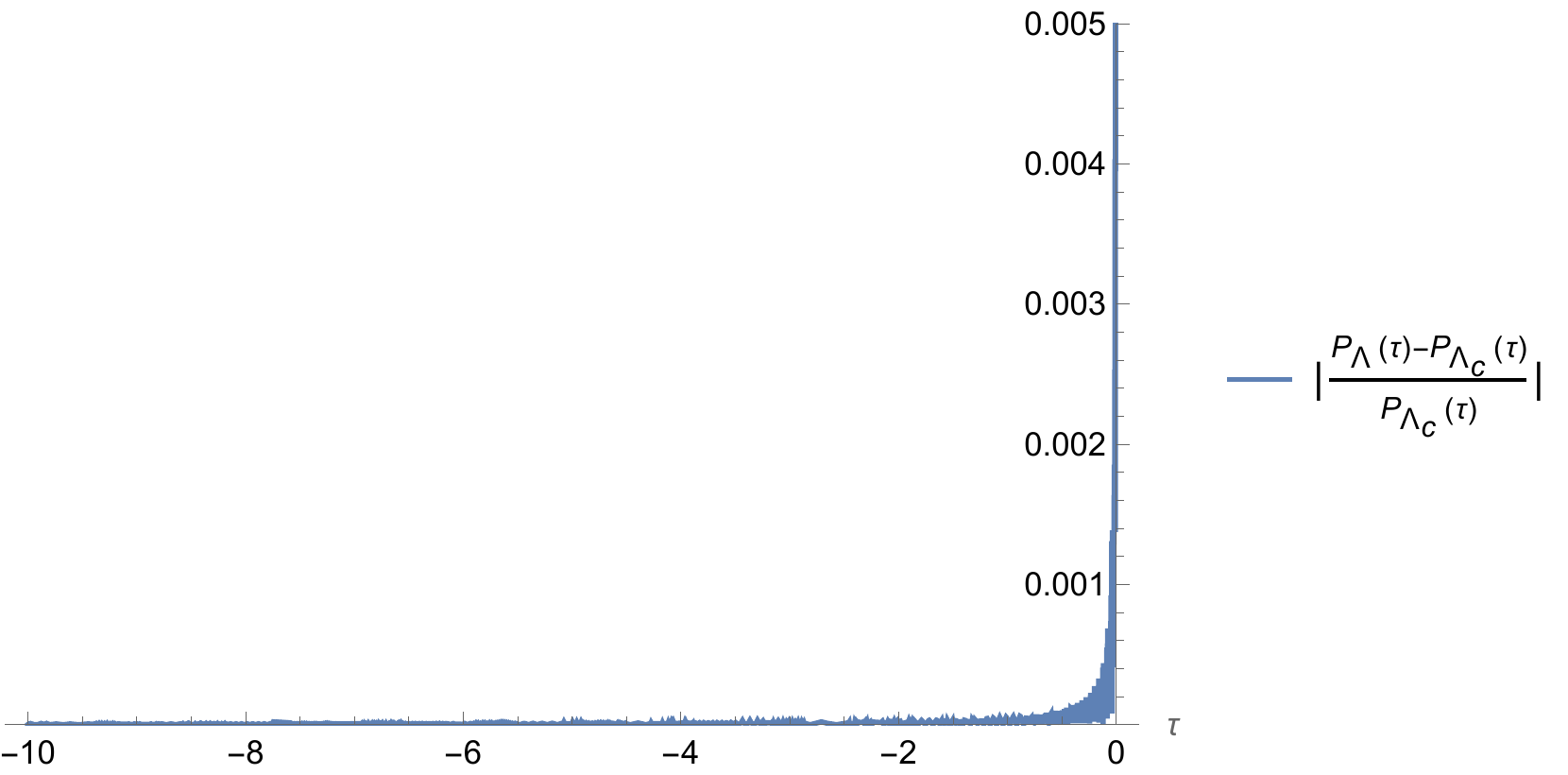}\\
	(c) & (d)  \\[6pt]  
	\includegraphics[height=4cm]{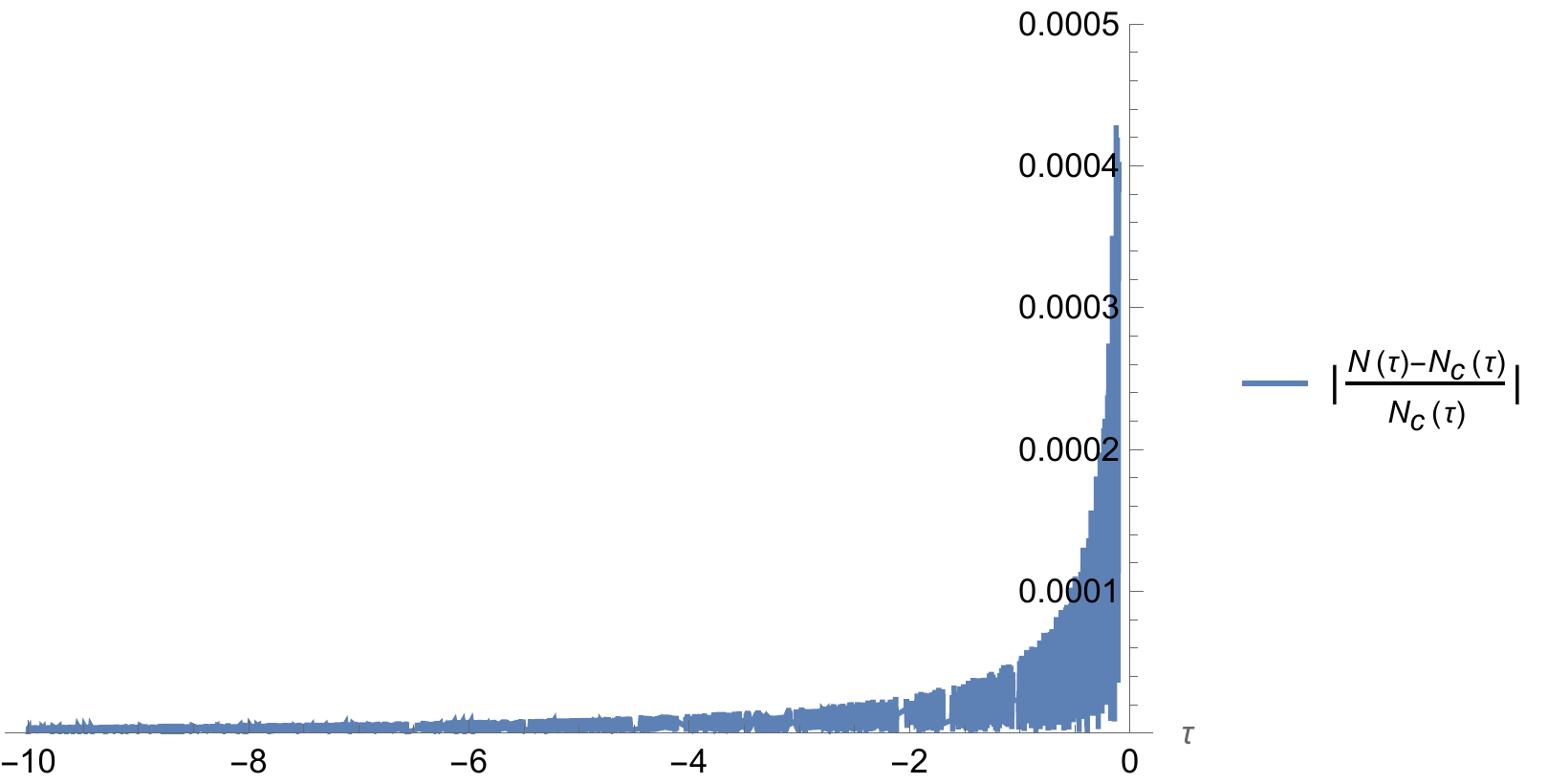} &
	\includegraphics[height=4cm]{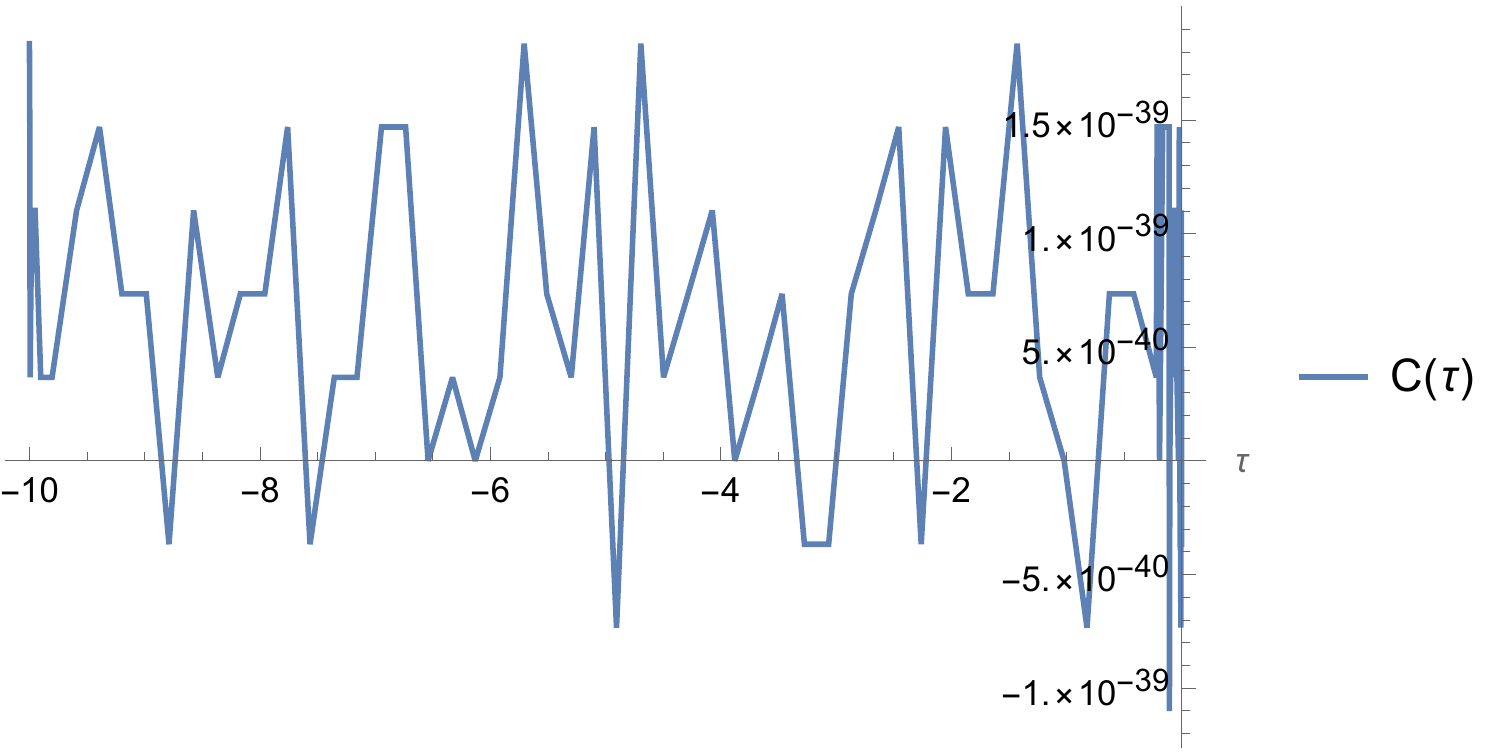}\\
	(e) & (f)  \\[6pt] 
		\end{tabular}
\caption{Plots of the relative  differences of the functions $\left(R, \Lambda, P_R, P_{\Lambda}, N\right)$ and $\mathcal{C}(\tau)$  near the black hole horizon with the same choice of the parameters $m$ and $j$, as those specified in Figs. \ref{fig3} and \ref{fig4}, that is, $m=10^{12} m_p ,\; j=j_0=10$. 
} 
\lb{fig-horizon}
\end{figure} 	

\end{widetext}

\begin{widetext}

\begin{figure}[h!]
 \begin{tabular}{cc}
\includegraphics[height=4cm]{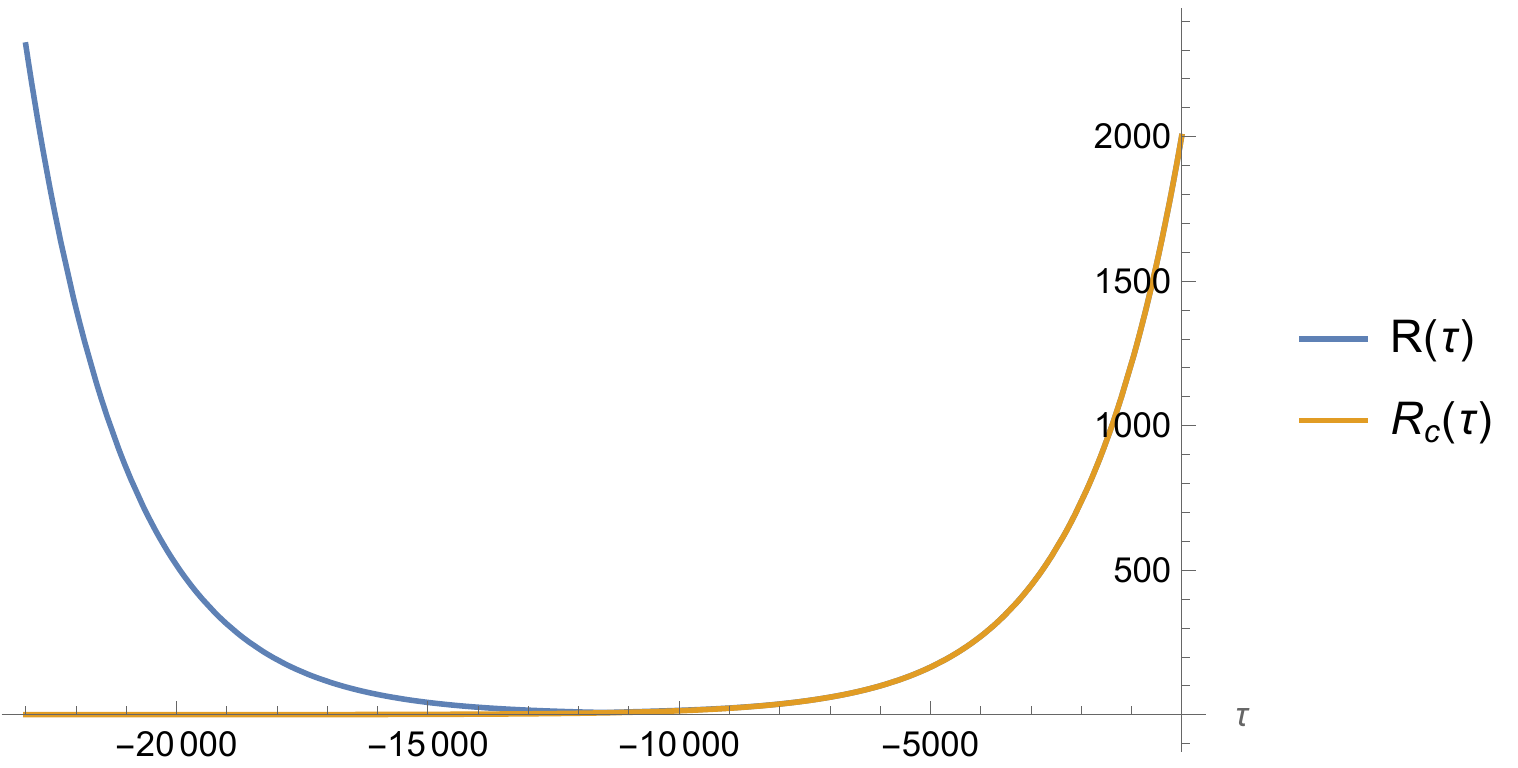}&
\includegraphics[height=4cm]{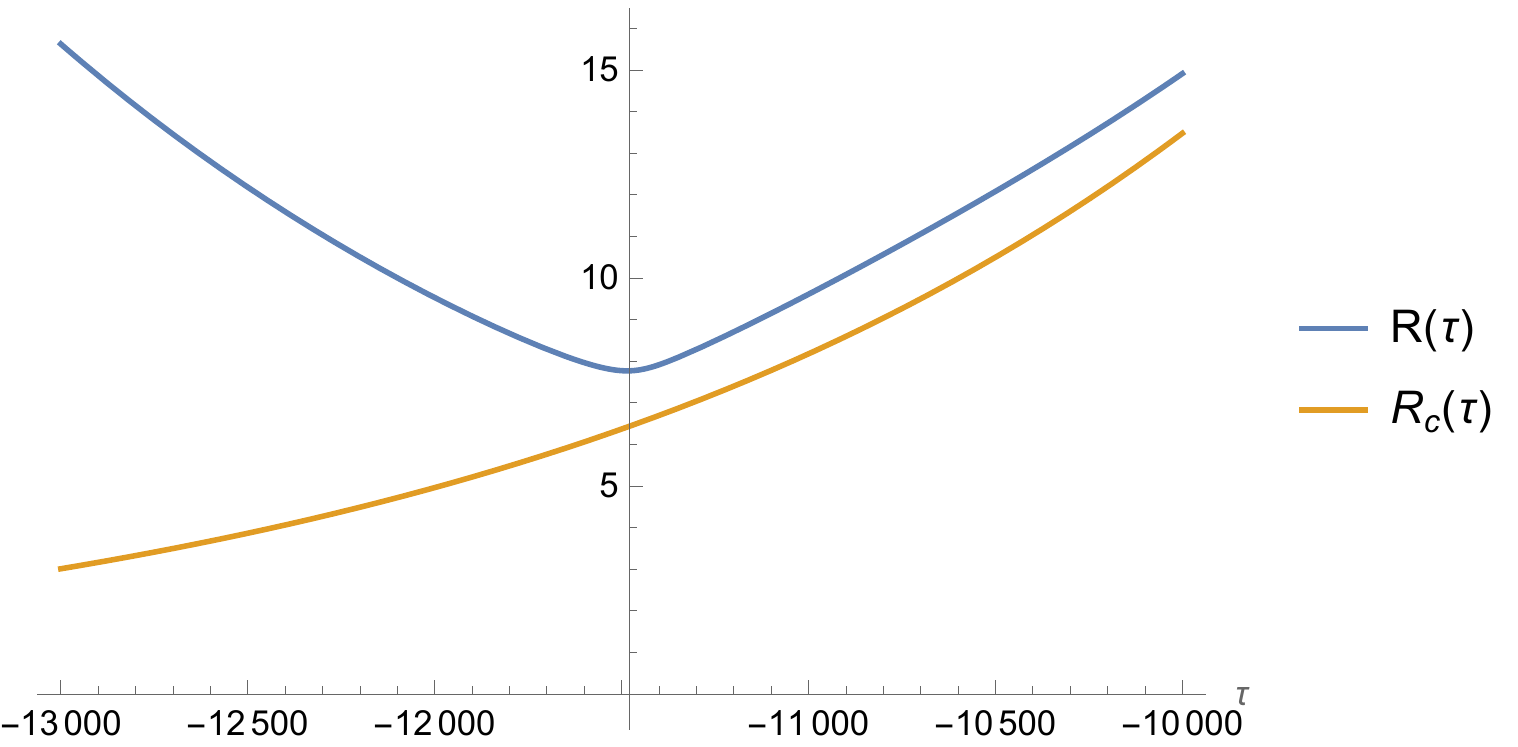}\\
	(a) & (b)  \\[6pt]
\includegraphics[height=4cm]{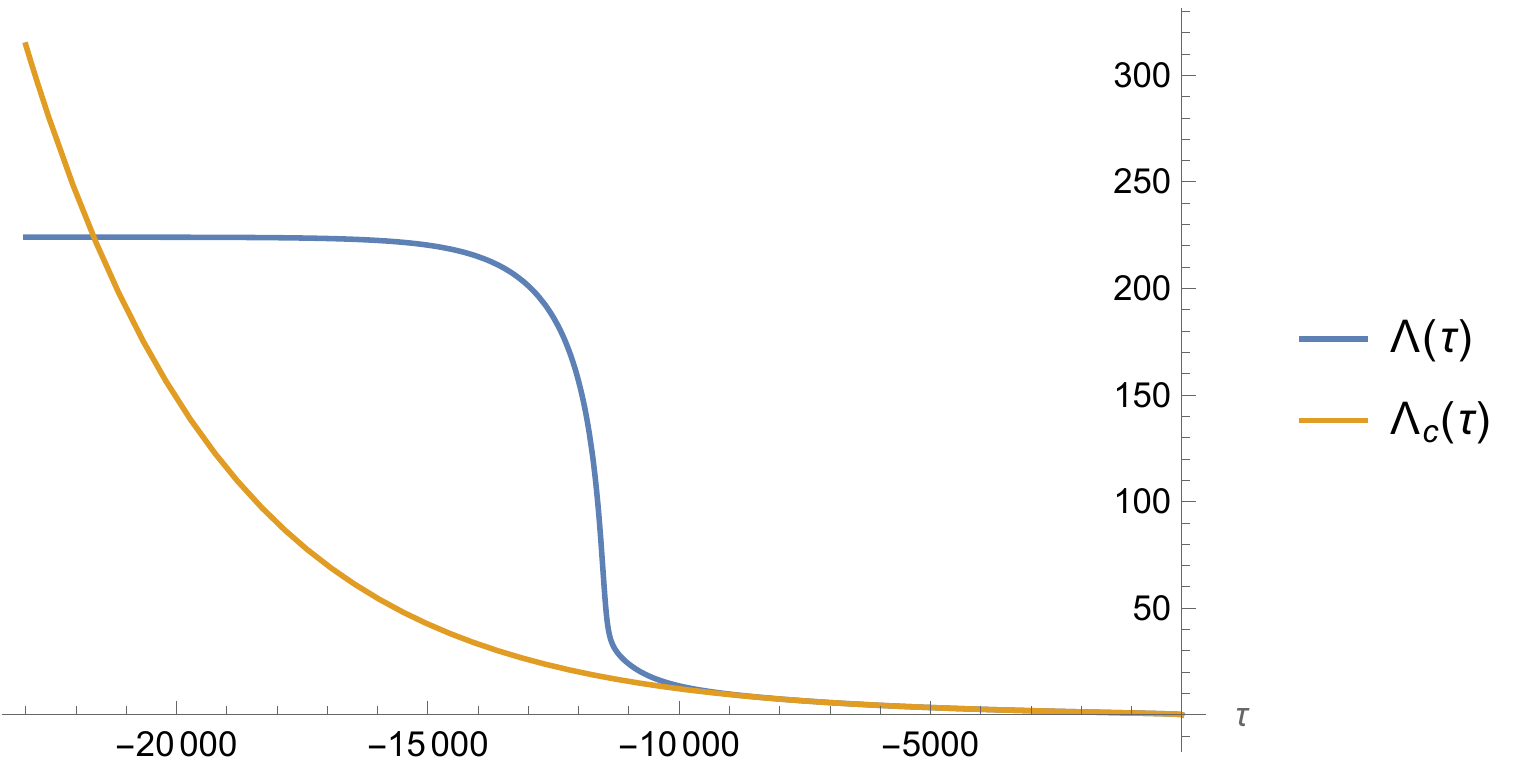}&
\includegraphics[height=4cm]{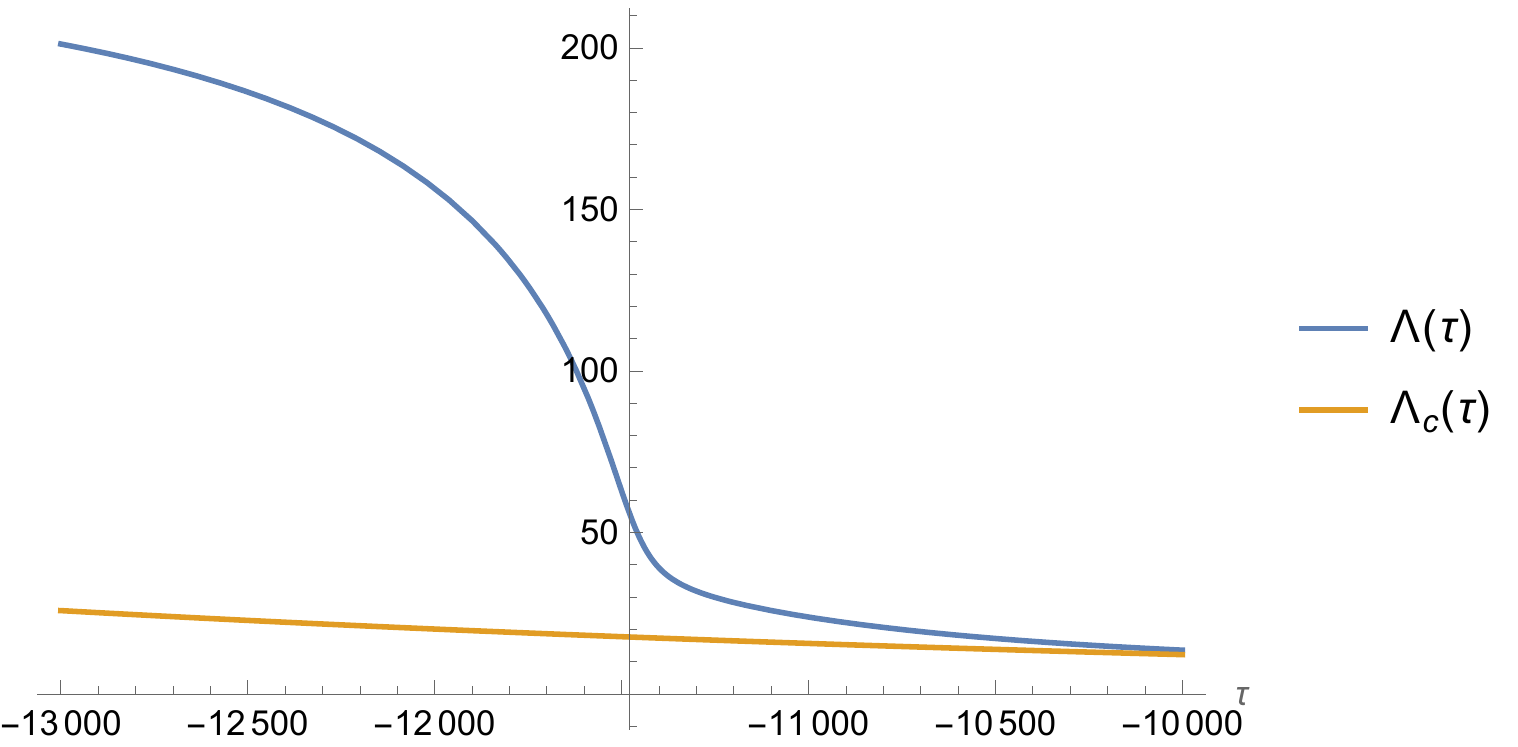}\\
	(c) & (d)  \\[6pt]
	\includegraphics[height=4cm]{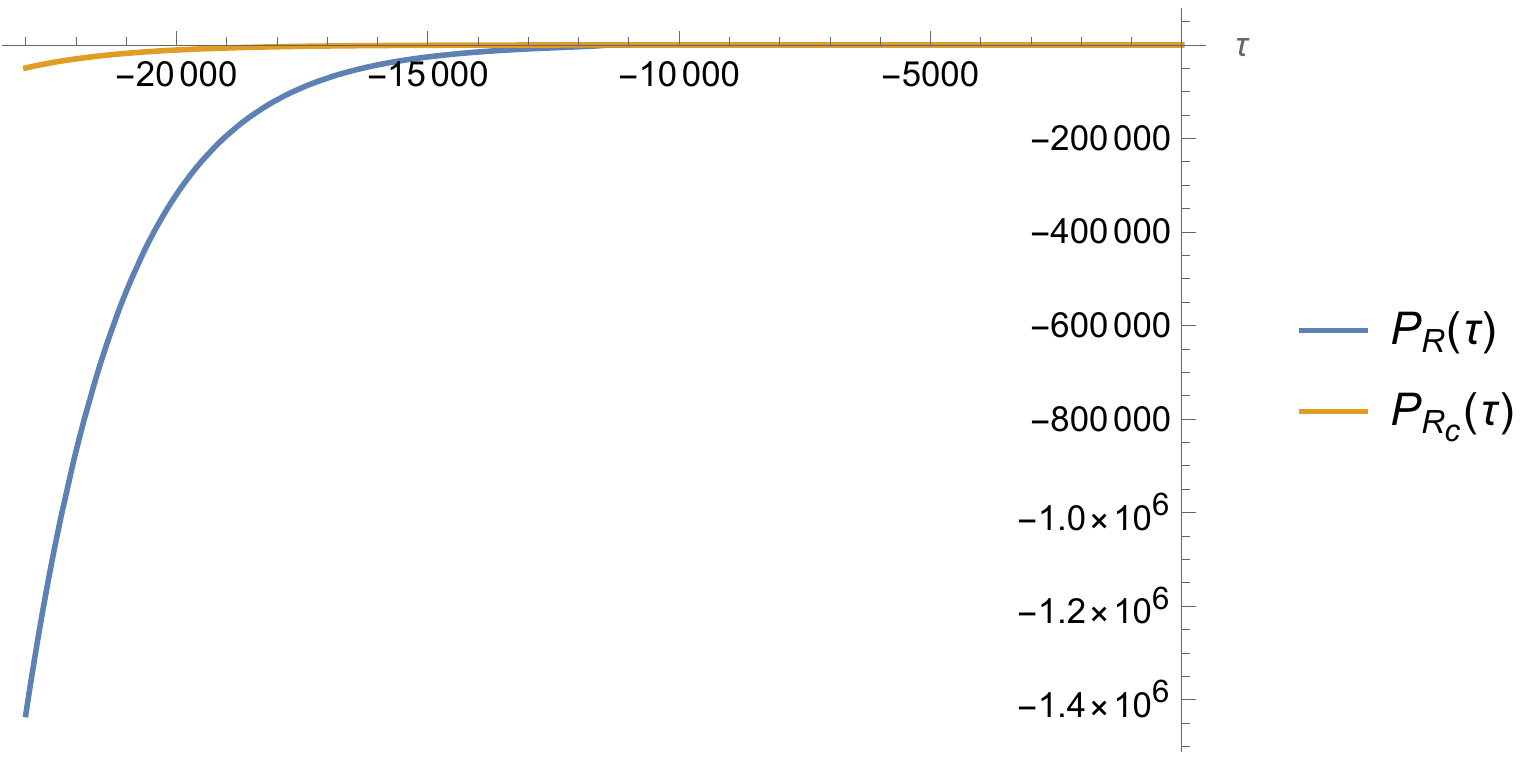}&
\includegraphics[height=4cm]{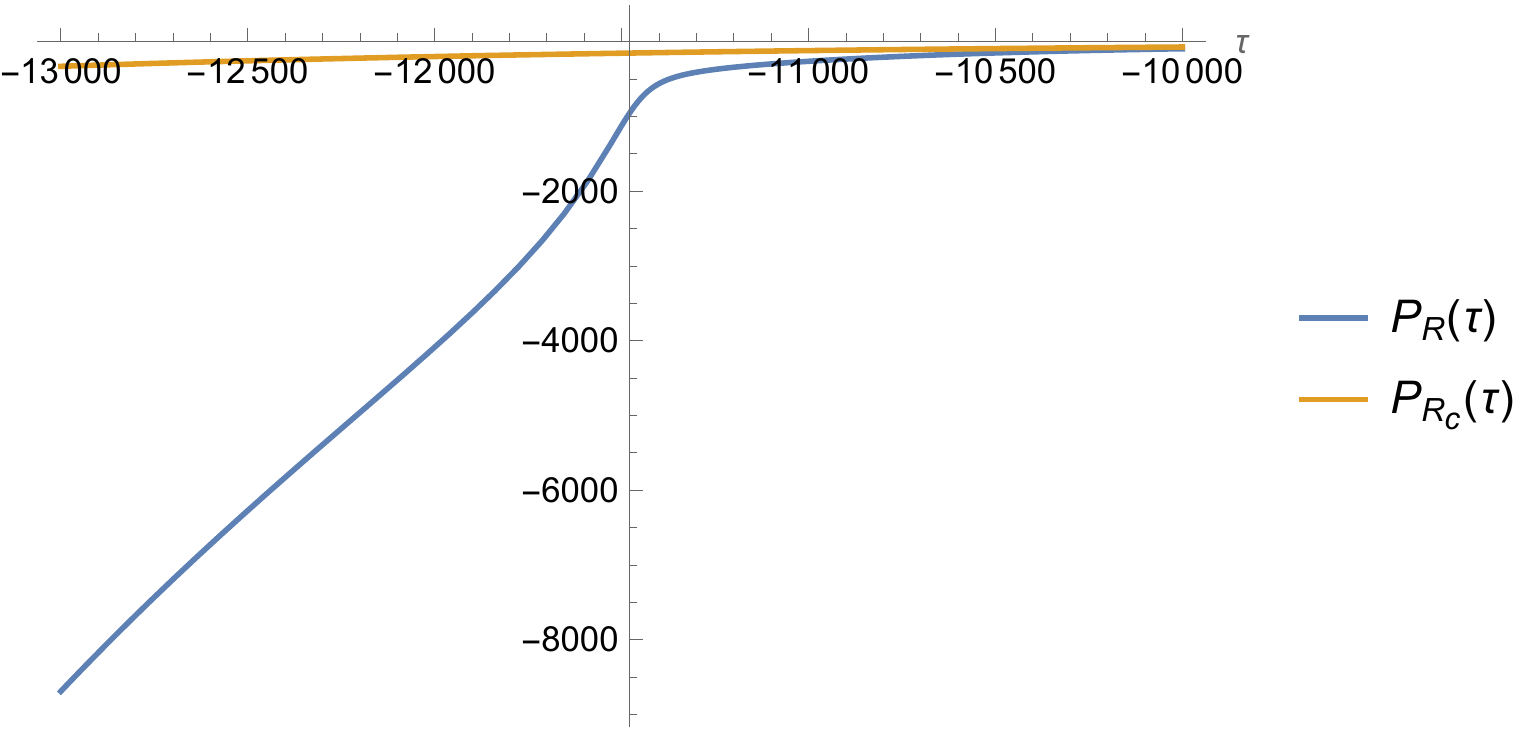}\\
	(e) & (f)  \\[6pt]
	 	\includegraphics[height=4cm]{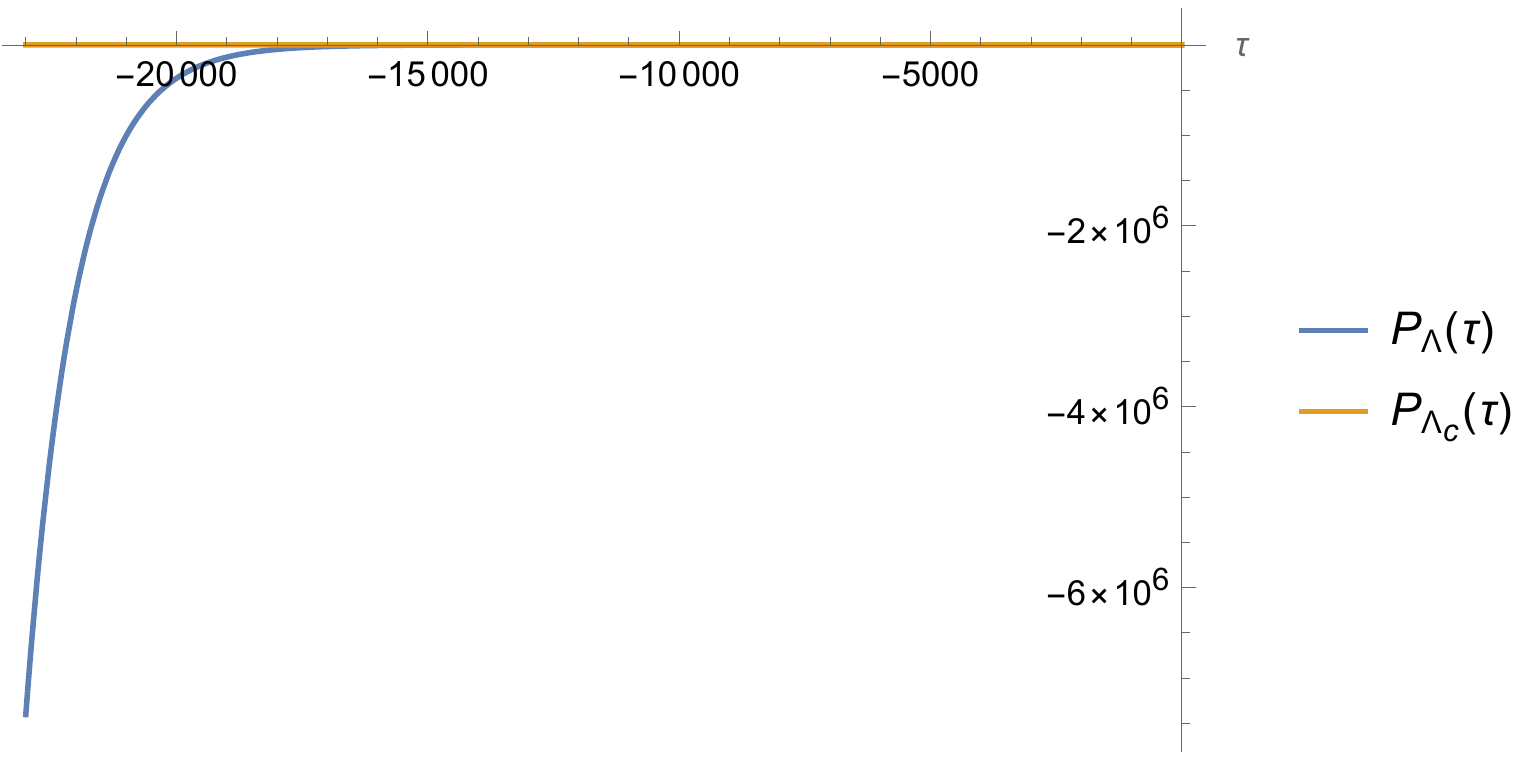}&
\includegraphics[height=4cm]{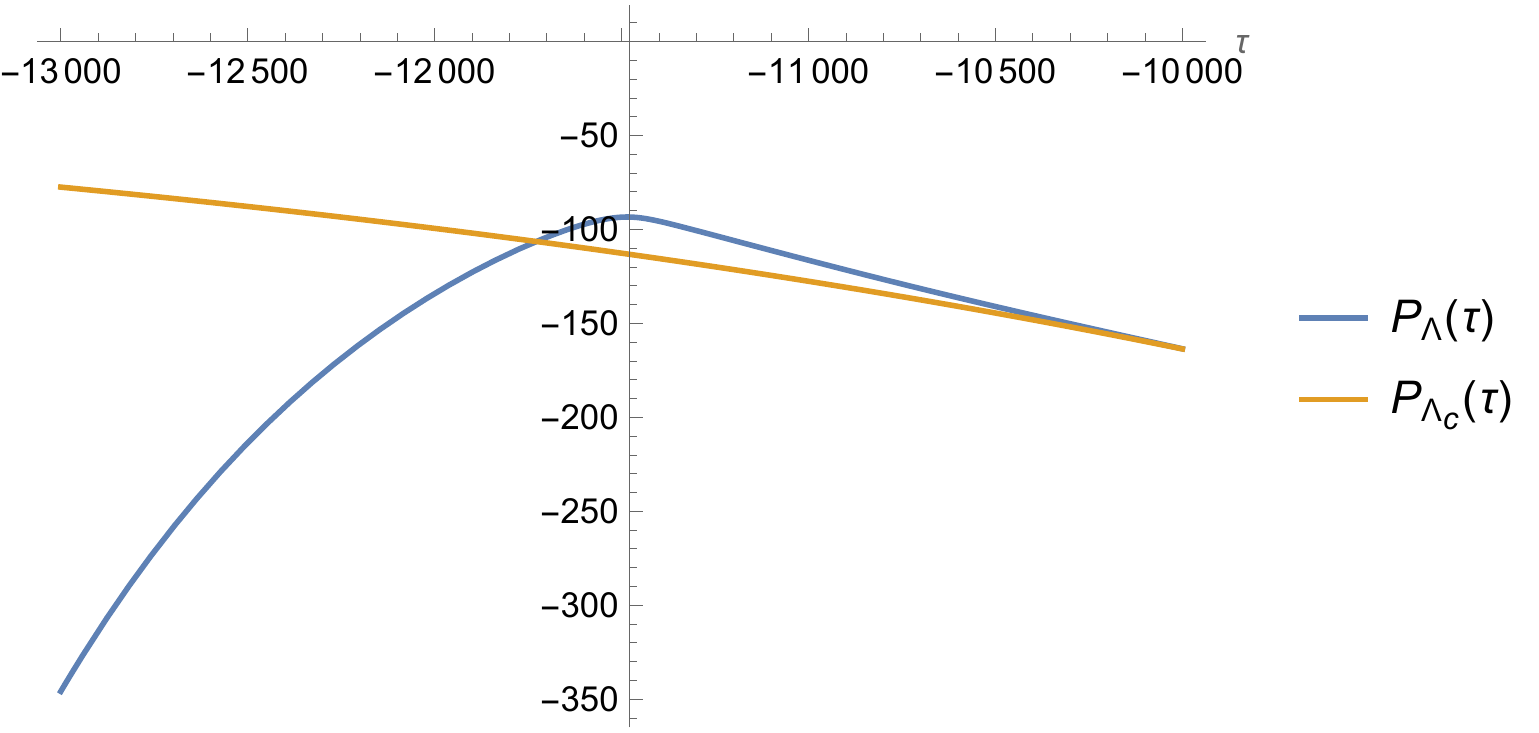}\\
	(g) & (h)  \\[6pt]
	\end{tabular}
\caption{Plots of the physical variables $\left(R, \Lambda, P_R, P_{\Lambda}\right)$ and their classical correspondences $\left(R_c, \Lambda_c, P_{R_c}, P_{\Lambda_c}\right)$.
Particular attention is paid to the region near the throat $\tau_{\text{min}} =-1.148 \times 10^{4}$. 
Graphs are plotted with $m=10^{3} m_p , \; j=j_0=10$.
} 
\lb{fig-m=10^{3}}
\end{figure}

     \begin{figure}[h!]
 \begin{tabular}{cc} 
 		\includegraphics[height=4cm]{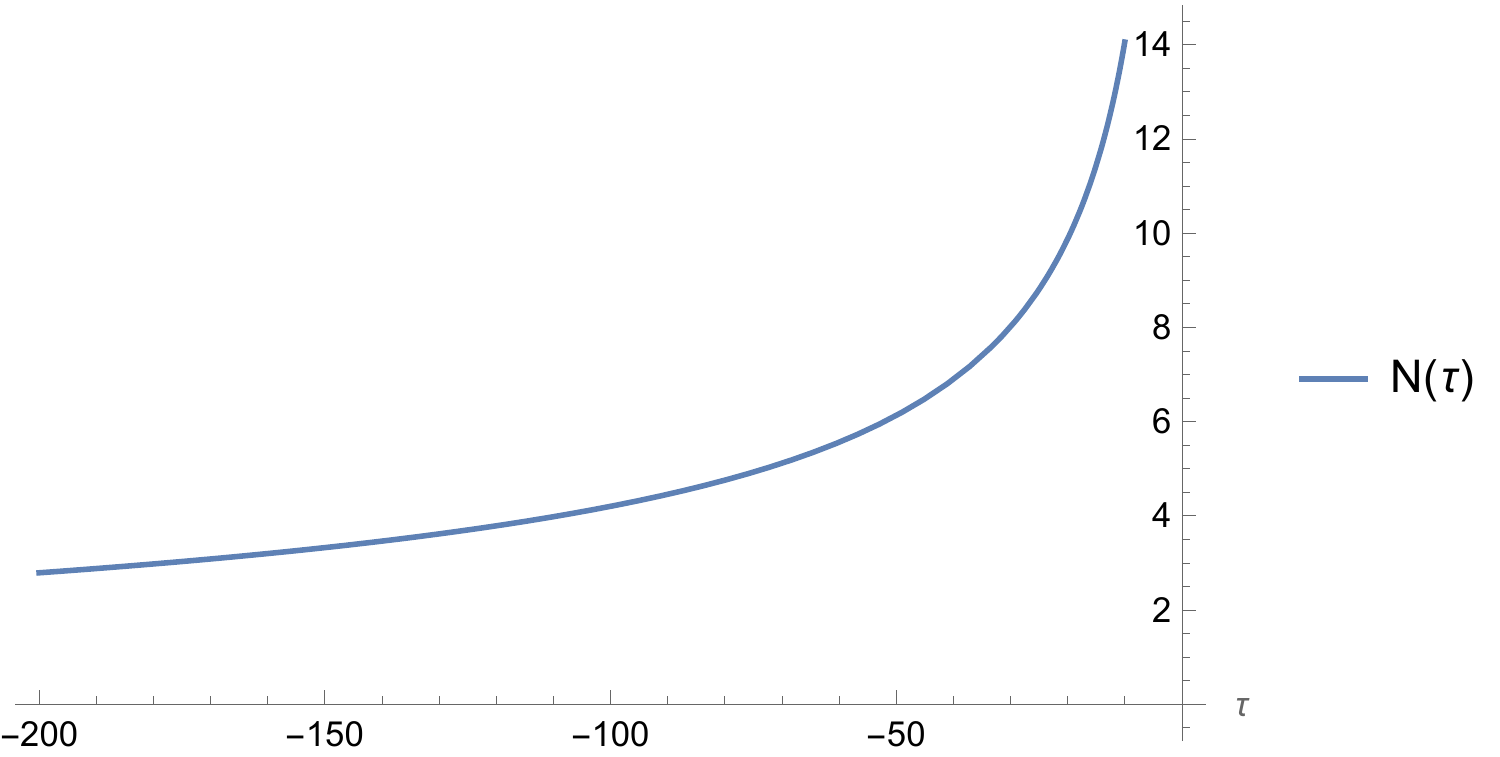}&
\includegraphics[height=4cm]{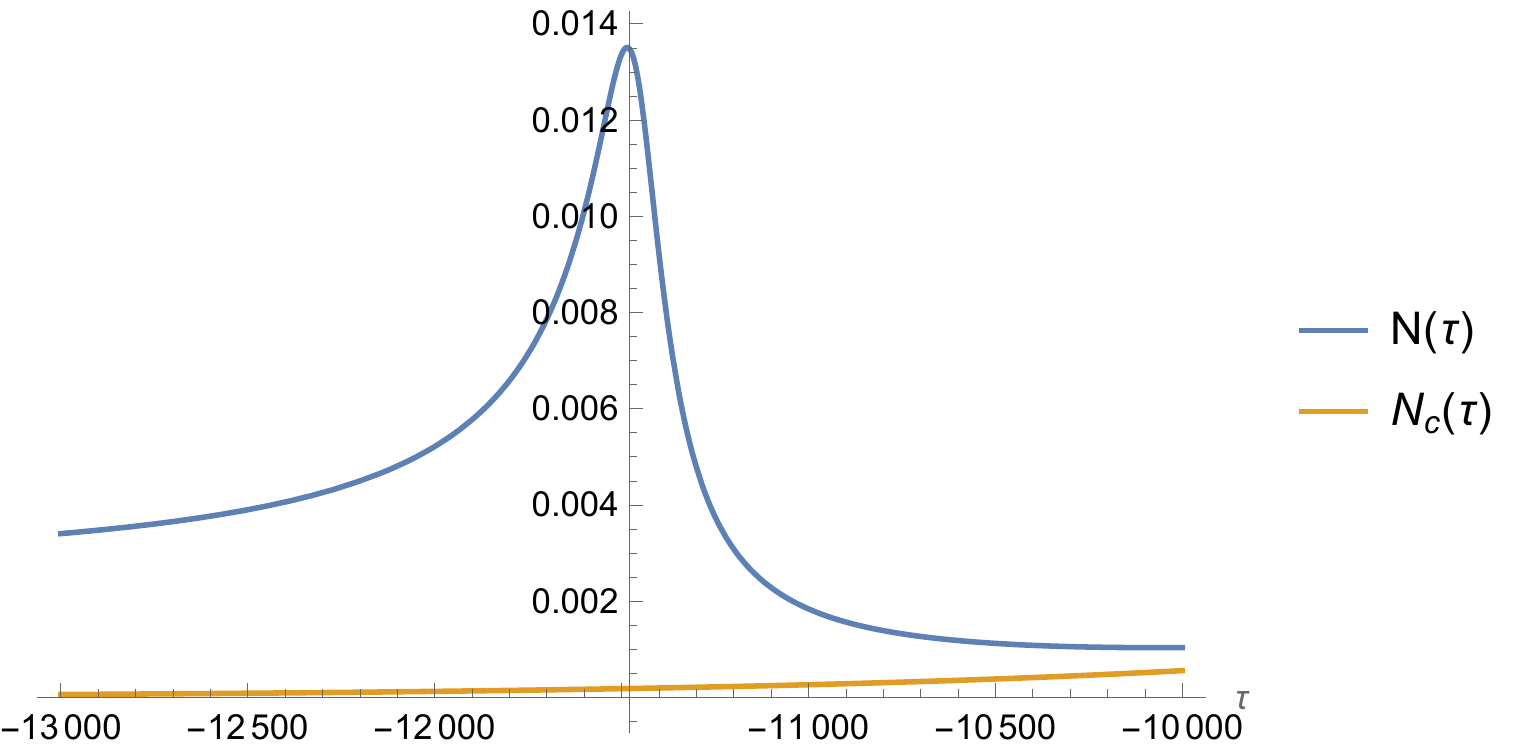}\\		
	(a) & (b)  \\[6pt]
	\includegraphics[height=4cm]{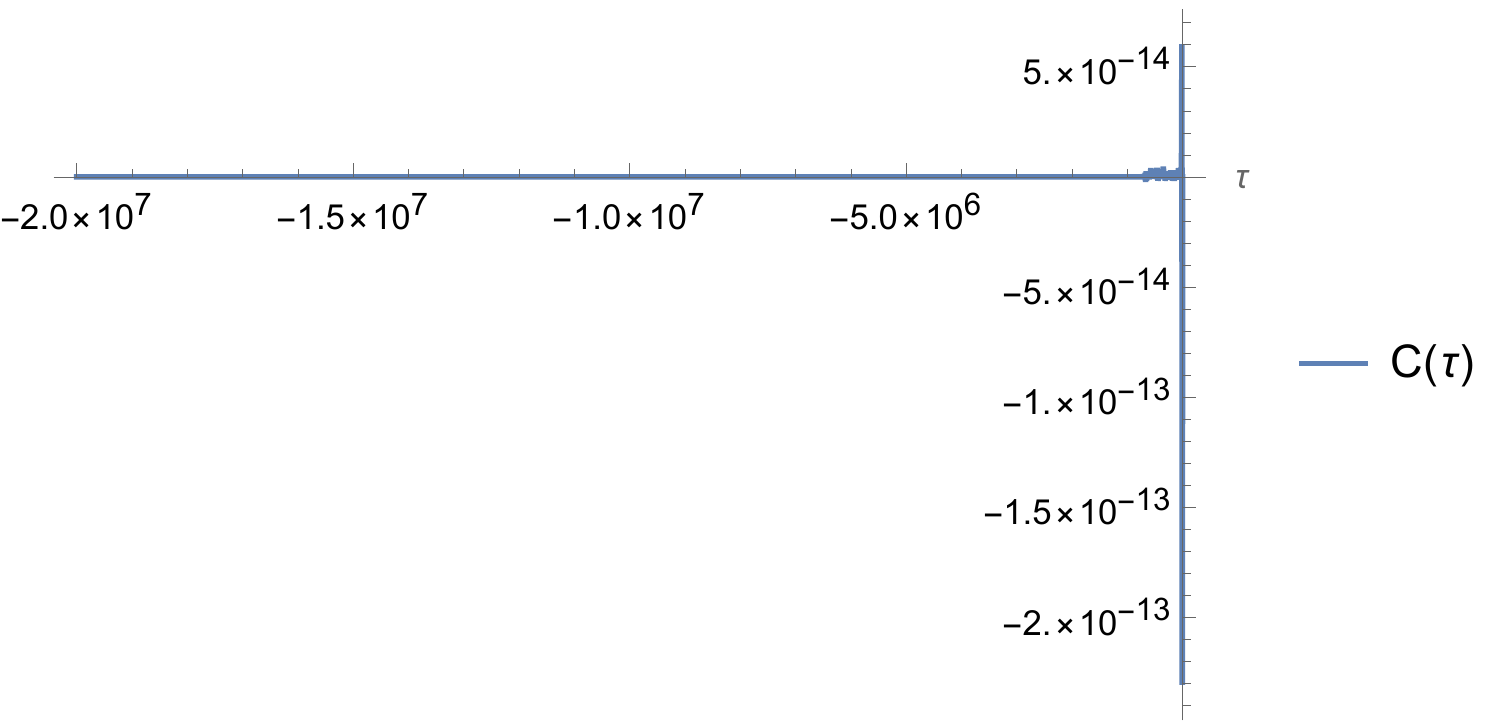}&
\includegraphics[height=4cm]{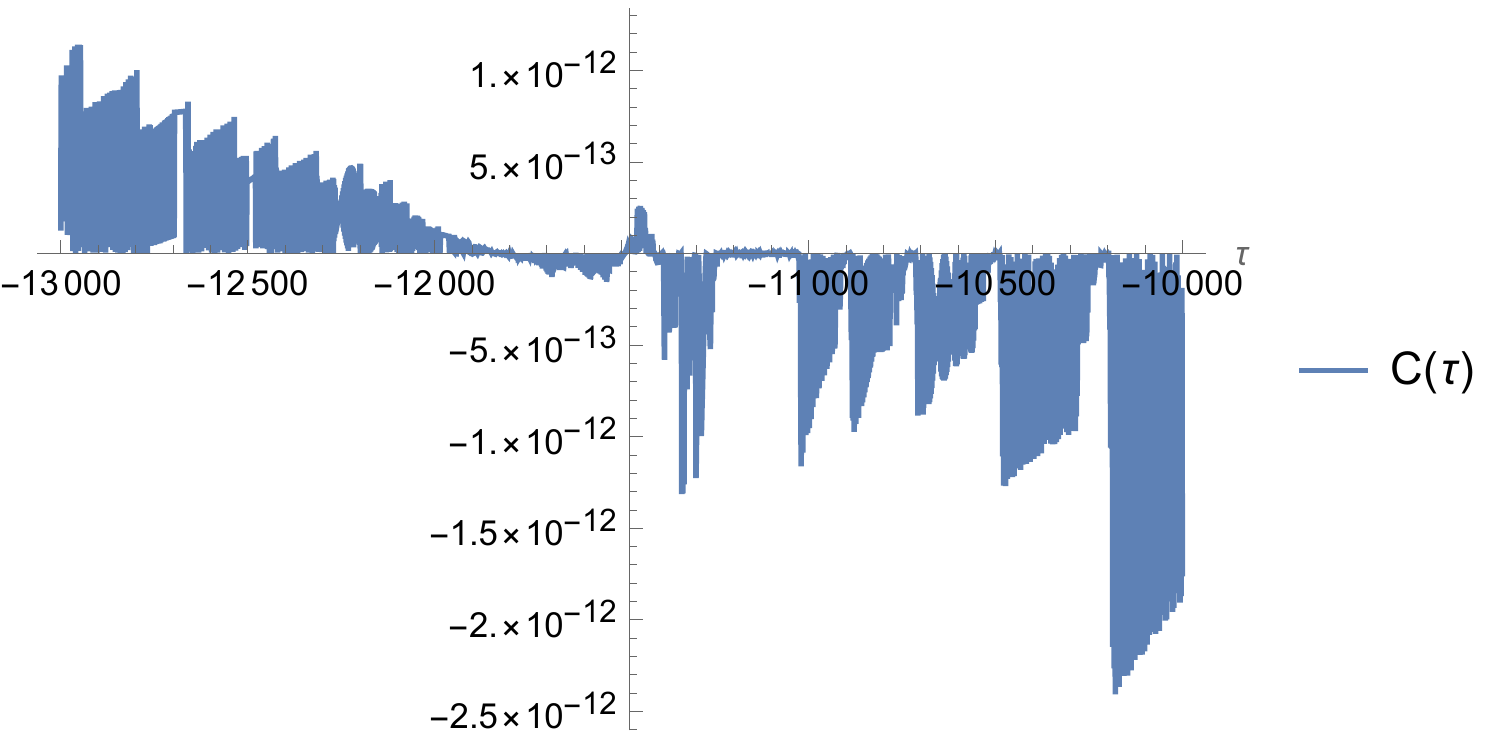}\\
	(c) & (d)  \\[6pt]
		\end{tabular}
\caption{Plots of  the lapse function $N(\tau)$ and $\mathcal{C}(\tau)$ for 
 $m=10^{3} m_p$ and $j=j_0=10$.
} 
\lb{fig-m=10^{3}-lapse}
\end{figure} 	

\begin{figure}[h!]
 \begin{tabular}{cc}
		\includegraphics[height=4cm]{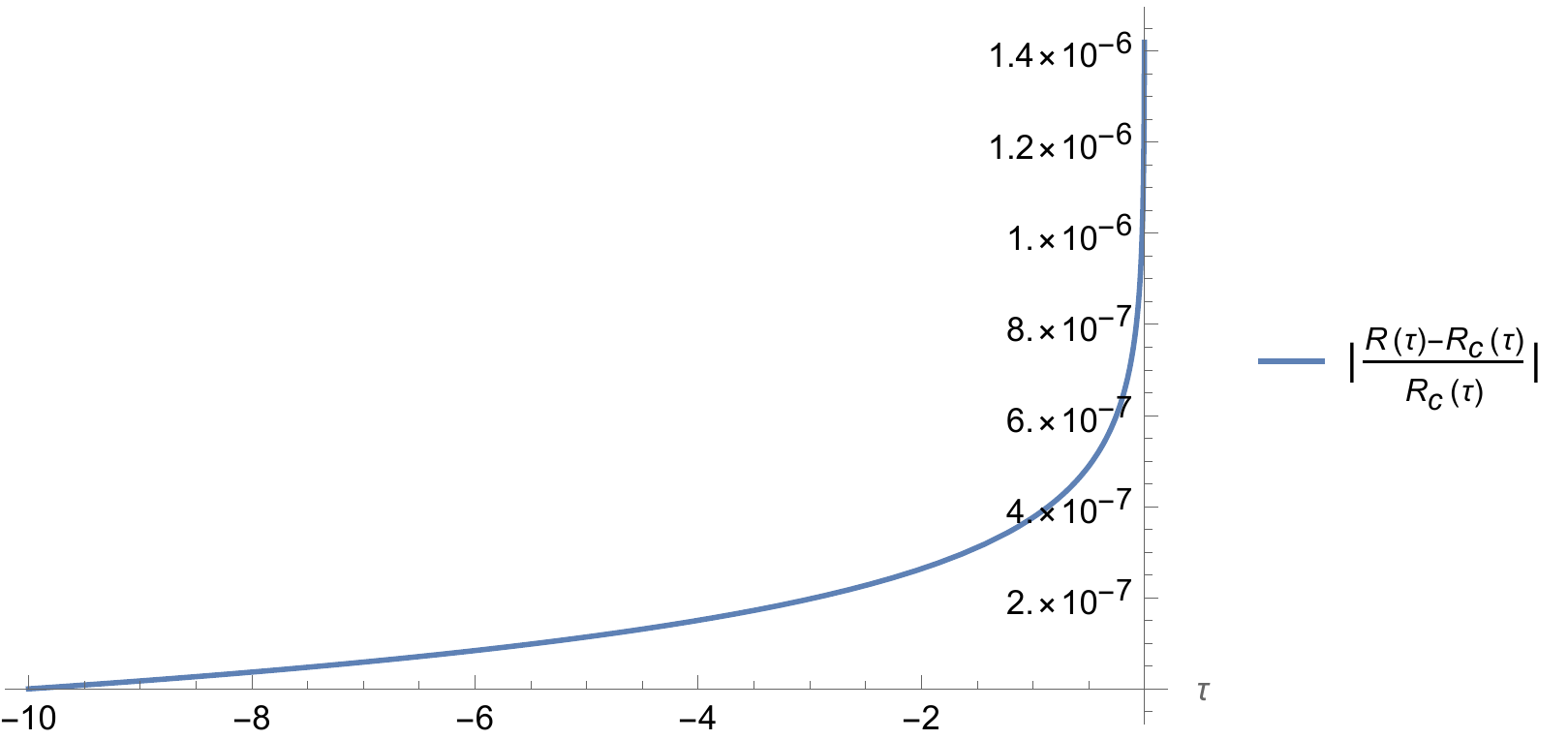}&
\includegraphics[height=4cm]{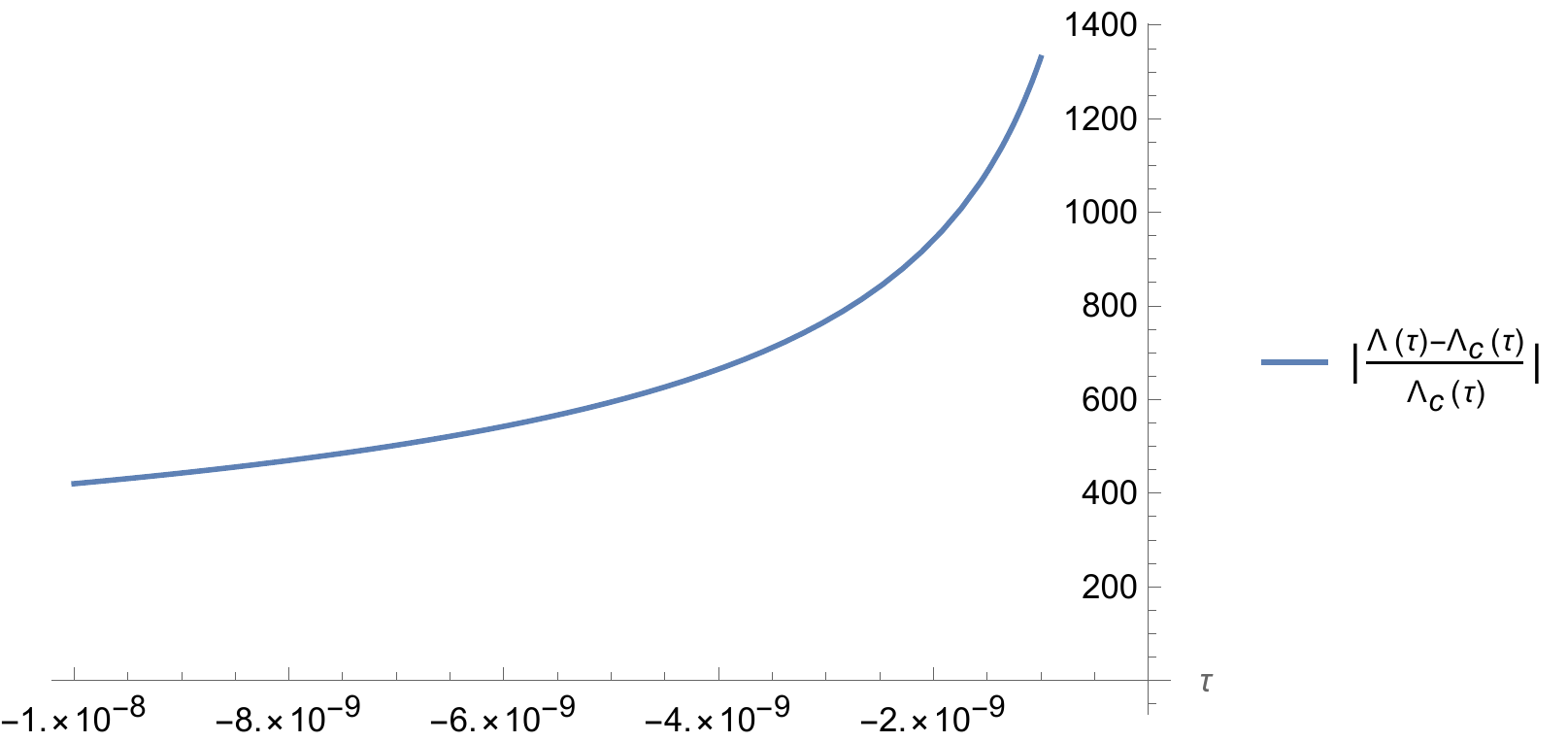}\\
	(a) & (b)  \\[6pt]
		\includegraphics[height=4cm]{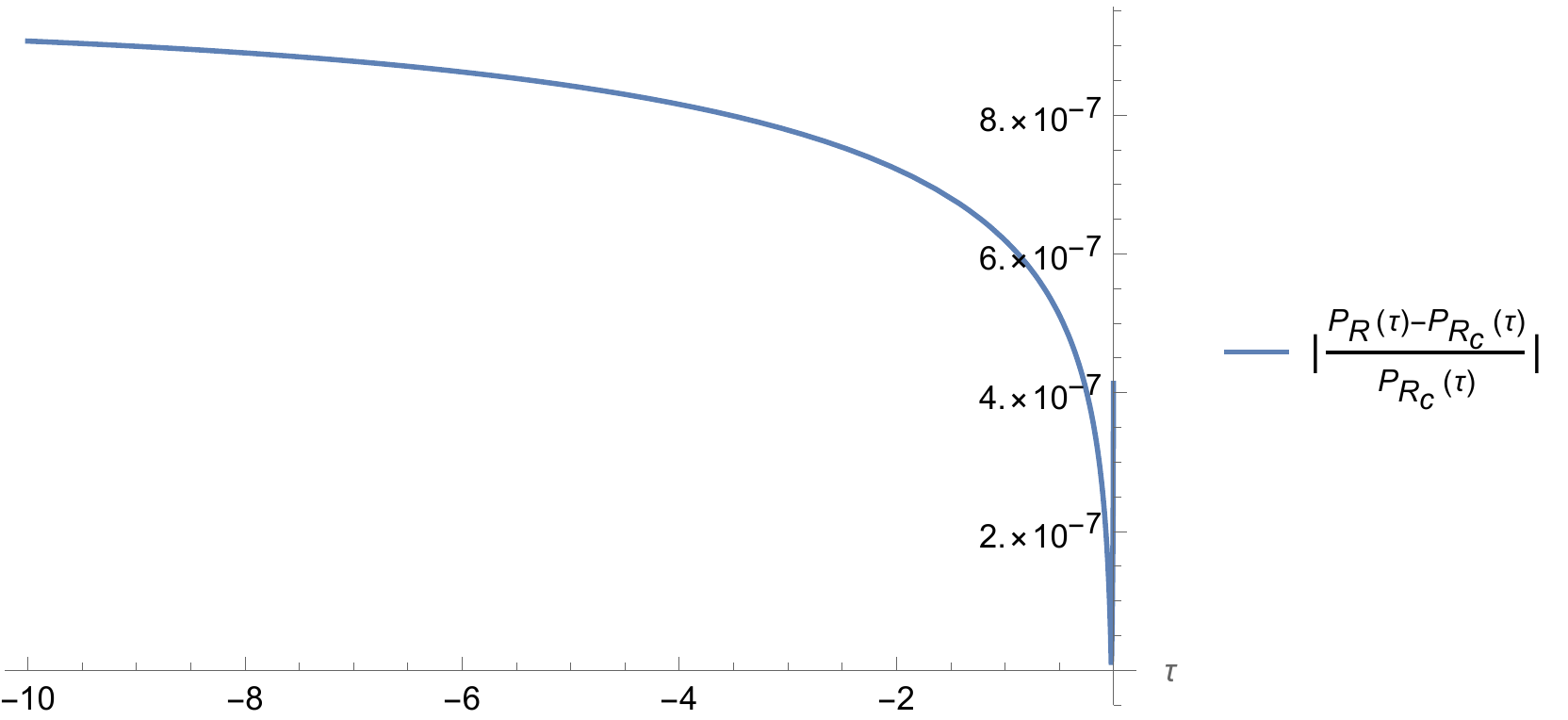}&
\includegraphics[height=4cm]{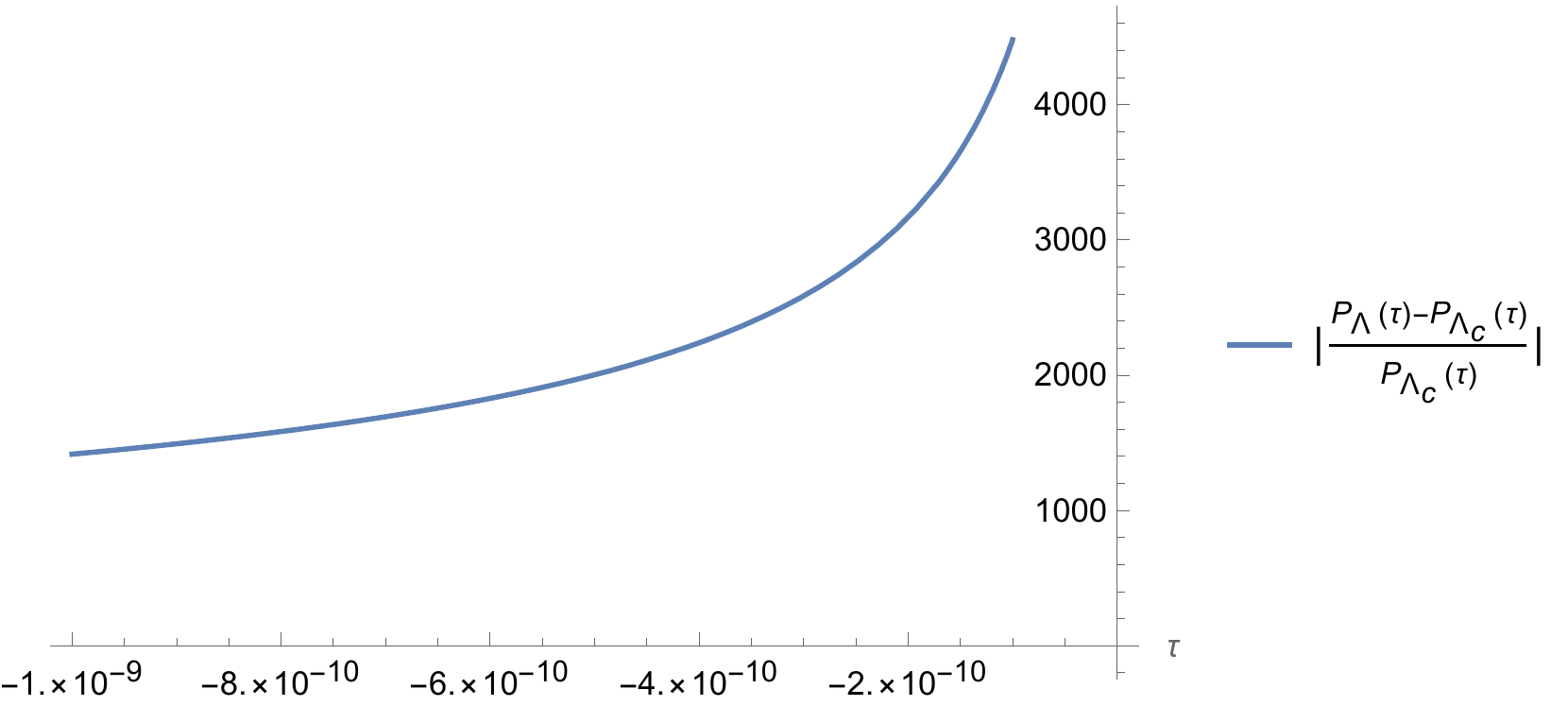}\\
	(c) & (d)  \\[6pt]
\includegraphics[height=4cm]{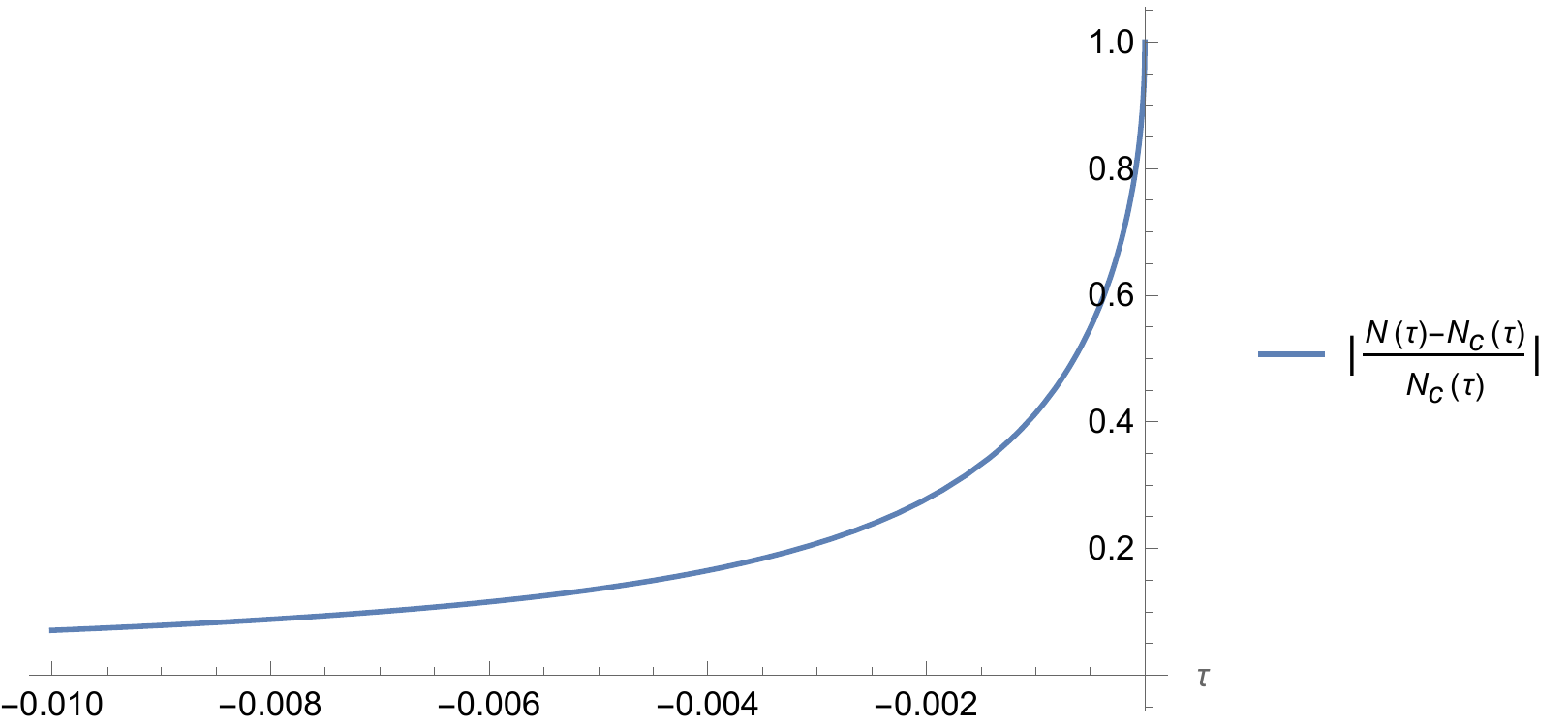}&
\includegraphics[height=4cm]{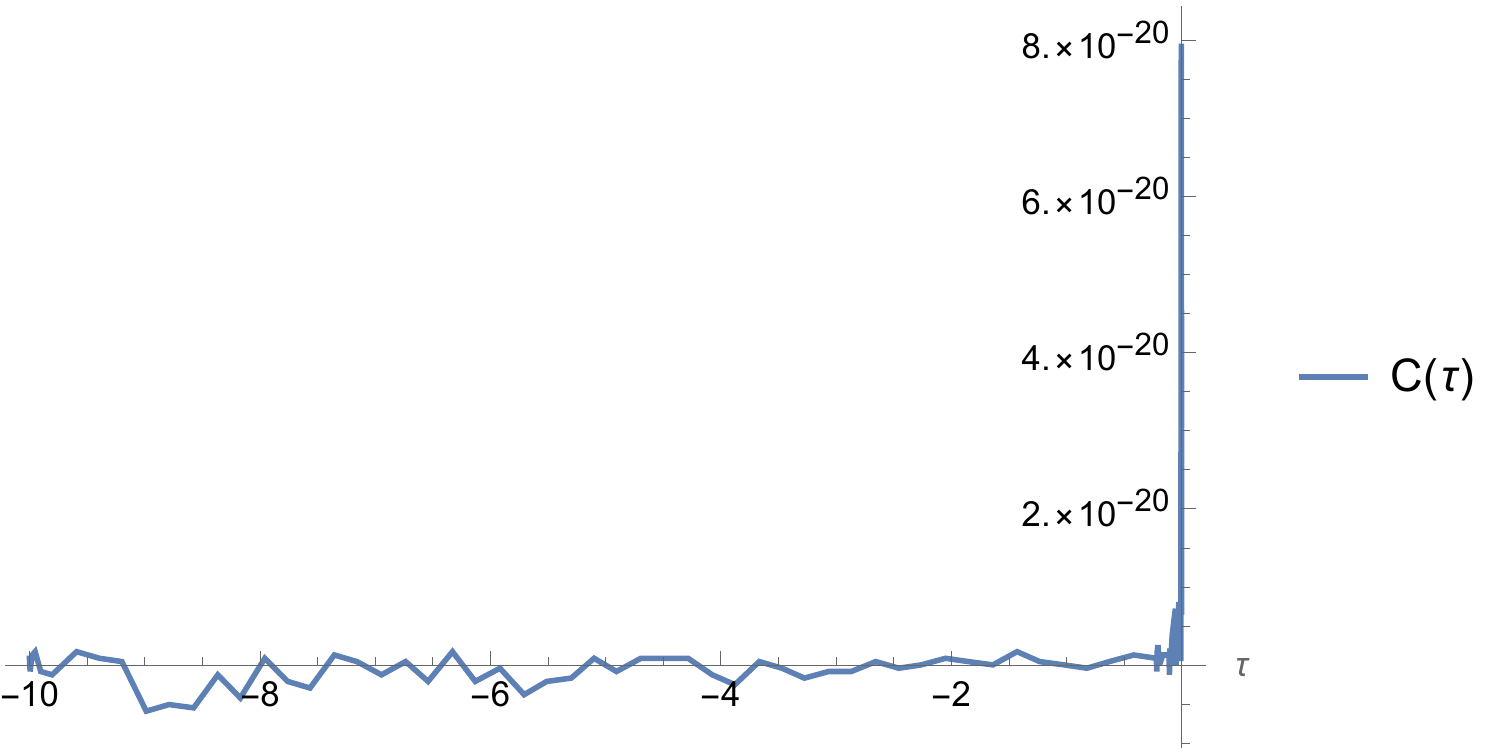}\\
	(e) & (f)  \\[6pt]	
		\end{tabular}
\caption{Plots of the relative  differences of the functions $\left(R, \Lambda, P_R, P_{\Lambda}\right)$, the lapse function  $N(\tau)$ and $\mathcal{C}(\tau)$ near the black hole horizon ($\tau = 0$) with 
 $m=10^{3} m_p$ and $j=j_0=10$, the same choice  as those specified in Figs. \ref{fig-m=10^{3}} and \ref{fig-m=10^{3}-lapse}. 
} 
\lb{fig-horizon-m=10^{3}}
\end{figure} 	

\end{widetext}

 For the ABP model, we shall choose these three variables as $\left(R, \Lambda, P_{\Lambda}\right)$, so that
 \bqn
 \lb{eq2.5j}
 \Lambda(\tau_i) &=& \Lambda_c(\tau_i),   \quad P_{\Lambda}(\tau_i) = P_{\Lambda_c}(\tau_i), \nb\\
  R(\tau_i) &=& R_c(\tau_i),  
  \eqn
while $P_{R}(\tau_i)$ is obtained from the effective Hamiltonian constraint 
\bq
 \lb{eq2.5k}
{\cal{H}}^{\text{IV+CS}}_{\text{int}}(\tau_i) = 0, \quad {\text{or}} \quad {\cal{C}}(\tau_i) = 0,
\eq
where ${\cal{C}}(\tau)$ is defined by Eq.(\ref{Cfactor}). This reduced parameter space will be referred to as $\hat{\cal{D}}$. It is clear that this reduced space is much smaller than the whole phase space ${\cal{D}}$.
However, for our current purpose, this is enough.  With such chosen initial conditions, the Hamiltonian equations will uniquely determine the evolutions of the four variables ($\Lambda, P_{\Lambda}$) and ($R, P_{R}$) at any other time $\tau$. Once these four variables are known,
 from Eq.(\ref{eq2.7b}) we can find the lapse function $N(\tau)$.

With the above prescription, we can see that the initial values of the four variables will depend not only on the choice of the initial moment $\tau_i$ but also on the values of $j_0, \; j$ and $m$. In particular, if the quantum effects are not negligible at the moment
$\tau_i$, it is expected that such obtained $P_{R}(\tau_i)$ should be significantly different from its corresponding relativistic value $P_{R_c}(\tau_i)$.

To see this clearly, 
 in Tables \ref{tab:initial} - \ref{tab:initial-3} we show such differences. In particular, in Table \ref{tab:initial} we show the dependence of $P_{R}(\tau_i)$ on the choice of the initial time $\tau_i$ for $m=10^{12} m_p , \; j=j_0=10$. From this table we can see that 
 $\Delta P_{R}(\tau_i) \equiv  P_{R}(\tau_i) - P_{R_c}(\tau_i) \simeq 0$ for  $\tau_i/\tau_p \lesssim -0.1$. As $\tau_i \rightarrow 0$, the difference becomes larger.

In Table \ref{tab:initial-2}, we show the dependence of  $P_{R}(\tau_i)$ on the choices of $j$ with $m=10^{12} m_p $ and $\tau_i=-10.0\; \tau_{p}$. Physically, the lager the parameter $j$ is, the closer to the relativistic value
 of $P_R$ should be. However, due to the accuracy of the numerical computations, it is difficult to obtain precisely the values 
 of $P_R$ from the  effective Hamiltonian constraint (\ref{eq2.5k}). So, in Table \ref{tab:initial-2} we only consider the initial values of  $P_{R}(\tau_i)$ for $ j \lesssim 10^{12}$.

     In Table \ref{tab:initial-3}, we show the dependence of  $P_{R}(\tau_i)$ on the choices of $m$ with $j=10$ and $\tau_i=-10.0\; \tau_{p} $, from which it can be seen that the deviations becomes larger for
     $m \lesssim  10^3 \; m_p$.  
      It should be also noted  that for very large masses, the initial time 
$\tau_i$ must be chosen very negative. Otherwise,  the term $e^{\tau/(Gm)}$, appearing in the   effective Hamiltonian constraint [cf. Eqs.(\ref{eq2.12a}) - \ref{eq2.12d})], becomes extremely small, and numerical errors 
can be introduced.  So, in Table \ref{tab:initial-3} for the choice of $\tau_i = -10 \tau_{p} $, we only consider the cases where $m$ is up to $10^{14}\; m_p$, although physically the larger $m$ is,  the closer   $P_{R}(\tau_i)$ is to its relativistic values.

In Fig. \ref{fig3},  we plot the four functions  
$ \left(R, \Lambda, P_R, P_{\Lambda}\right)$,
  and their classical correspondences  for $m=10^{12} m_p,\;  j=j_0=10,\; \tau_i = -10\; \tau_{p} $. With such initial conditions, we find that the location of throat (transition surface) is around $\tau_{\text{min}} \simeq-3.9108 \times 10^{13} \tau_{p} $, at which  $R(\tau)$ reaches its minimum value,   $R_{\text{min}} \simeq 7779.35\; \ell_p$. 
It is interesting to note that near the throat the four functions all change dramatically, especially  $\Lambda(\tau)$, which behaves like a step function. 
In addition, even at the transition surface, we find that the conditions of Eq.(\ref{eq2.5d1}) are well satisfied.

To closely monitor the numerical errors, we also plot out the  effective Hamiltonian (${\cal{C}}(\tau) \simeq 0$) in Fig.  \ref{fig4} together with the lapse function $N(\tau)$, from which we can see that in the region near the throat the numerical errors indeed become large. But out of this region, the numerical errors soon become negligible. From Fig. \ref{fig3} and  \ref{fig4}  we also find that our numerical solutions match well
with their asymptotic behaviors given by Eq.(\ref{eq2.5d}), as $\tau \rightarrow - \infty$.

To consider the quantum effects near the horizons, in Fig. \ref{fig-horizon}  we plot out the relative differences between functions $\left(R, \Lambda, P_R, P_{\Lambda}, N\right)$ and their classical value. To monitor the numerical errors, we also plot out the  effective Hamiltonian constraint 
${\cal{C}}(\tau) \simeq 0$.  From these plots, we can see clearly that the quantum effects indeed become negligible near the horizons
\footnote{Note that at the horizon  $N(\tau)$ diverges. So, in the region very near the horizon $N(\tau)$   becomes extremely  large, and the accurate numerical calculations become difficult,
so it is unclear whether the sudden growth of $\Delta N/N_c$, as shown in Fig. \ref{fig-horizon}    is due to  numerical errors or not. In fact, similar growths can be also noticed from  the plots of $\Delta \Lambda/\Lambda_c$ and $\Delta P_{\Lambda}/P_{\Lambda_c}$. Such sudden growths happen also in the cases $\eta >1$ and $\eta < 1$, as to be seen below.}.

On the other hand, when the mass of the black hole is near the Planck scale, such effects are not  negligible even near the horizon. To show this, in Figs. \ref{fig-m=10^{3}} - \ref{fig-horizon-m=10^{3}}  we plot various  physical variables for  $m = 10^3 \; m_p ,\; j = j_0 = 10$, for which we find that the  location of throat  is around $\tau_{\text{min}} \simeq-1.148 \times 10^{4} \tau_{p}$, at which  $R(\tau)$ reaches its minimum value,  $R_{\text{min}} \simeq 7.76\; \ell_p$. From these figures it is clear that now the quantum effects become large near the horizons, and cannot be negligible. It should be noted that for such small black hole, the semi-classical limit conditions (\ref{eq2.5d}) are not well satisfied at the throat, and as a result,  the corresponding  effective Hamiltonian may no longer describe the real quantum dynamics well. For more details, we refer readers to \cite{ABP19,ABP20}.

  \subsection{$\eta \gtrsim 1$}

  In this case,    we find
  \bqn
  \lb{eq4.19}
  X  &\simeq& \eta_0, \quad Y  \simeq \frac{\eta_0}{\eta}, \nb\\
   W  &\simeq& \pi h_0[\eta_0]+2\sin[\eta_0],\nb\\
  \frac{P_\Lambda}{R^2}  &\simeq&  \frac{\eta_0}{\alpha \gamma G},\quad
  \frac{P_R}{R \Lambda}  \simeq  \frac{2\eta_0}{\alpha \gamma G}, \nb\\
  \eqn
  as $\tau \rightarrow -\infty$. Then, 
  the metric coefficients have the following  asymptotical behavior,
\bqn
\lb{eq2.5jb}
 N(\tau)&\simeq& N_0= - \frac{2\gamma \sqrt{8\pi \gamma} \; \ell_p \; \sqrt{ j_0},}{mG \left(\pi h_0[\eta_0]+2\sin[\eta_0]\right)},  \nb\\ 
\Lambda(\tau) &\simeq& \Lambda_0 \exp\left\{\frac{{\cal{F}}(\eta)}{2 mG} \tau\right\}, \nb\\  
R(\tau) &\simeq& R_0 \exp\left\{\frac{\cos\left(\frac{\eta_0}{\eta}\right)}{2 mG} \tau\right\}, 
\eqn
where $\Lambda_0$ and $R_0$ are constants,  
and 
\bqn
\lb{eq2.5kb}
{\cal{F}}(\eta) &=&  \frac{1}{{\mathcal{D}(\eta_0)}^2}\Big[2\pi h_{-1}(\eta_0)\sin^2(\eta_0) + \pi^2 \cos(\eta_0) h_0^2(\eta_0)\Big]\nb\\
&& - \cos\left(\frac{\eta_0}{\eta}\right),
\eqn
where ${\mathcal{D}(\eta_0)}$ is defined by Eq.(\ref{eq2.9}) but now with $A = B = 0$, and the constant  $\eta_0$ is implicitly determined by 
\bq
\lb{eq2.5l}
\eta \sin\left(\frac{\eta_0}{\eta}\right) + \frac{\pi}{{\cal{D}}(\eta_0)} \sin(\eta_0) h_0(\eta_0) = 0.
\eq

In \cite{ABP19}, it was shown that ${\cal{F}}(\eta) < 0$ and $\eta_0 < - \pi$ when $\eta > 1$, so that both $R$ and $\Lambda$ grow exponentially as
$\tau  \rightarrow -\infty$. 
Setting 
\bq
\lb{eq2.5m}
a \equiv \frac{\left|{\cal{F}}(\eta)\right|}{2mG} > 0, \quad d \equiv \frac{\left|\cos\left(\frac{\eta_0}{\eta}\right)\right|}{2mG} > 0, 
\eq
we find that 
\bq
\lb{eq2.5n}
\Lambda = \Lambda_0 e^{-a\tau}, \quad R = R_0 e^{-d\tau}.
\eq 
Then, the metric takes the following asymptotical form 
\bq
\lb{eq2.5o}
ds^2 \simeq - \left(\frac{\hat{N}_0}{R}\right)^2dR^2 + R^{\frac{2a}{d}} d\bar x^2 +  R^2d\Omega^2,
\eq
where $\hat{N}_0 \equiv N_0/d$,  but now with $\bar{x} \equiv \left(\Lambda_0/R_0^{a/d}\right)x$. Similar to the last case, the corresponding spacetime is not vacuum, and the effective
energy-momentum tensor takes the same form as that given by Eq.(\ref{eq2.5g}), but now with $u_{\mu} = (\hat{N}_0/R)\delta^{R}_{\mu}$, $\bar x_{\mu} =  R^{a/d}\delta^{\bar x}_{\mu}$,   and
\bqn
\lb{eq2.5p}
\rho &\simeq&  \frac{2 a+d}{d \hat N_0^2}+\frac{1}{R^2}, \nb\\
p_{\bar x} &\simeq& -\frac{3}{\hat{N}_0^2}-\frac{1}{R^2}, \nb\\
 p_{\bot} &\simeq& -\frac{a^2+a d+d^2}{d^2 \hat{N}_0^2},
 \eqn
 from which we find that
 \bqn
\lb{eq2.5pa}
\rho + p_{\bar x} &\simeq&  \frac{2(a -d)}{d \hat N_0^2}+{\cal{O}}\left(\frac{1}{R^2}\right), \nb\\
\rho + p_{\bot} &\simeq& -  \frac{a(a -d)}{d^2 \hat N_0^2}+{\cal{O}}\left(\frac{1}{R^2}\right).
 \eqn
Therefore, in this case none of the three energy conditions is satisfied either, provided that $a \not= d$. 
When $a = d$, the spacetime is asymptotically de Sitter, as shown below. 
 In particular, we find that
\bqn
\lb{eq2.5ib}
&& {\cal{R}}  \simeq  2 \left(\frac{a^2+2 a d+3 d^2}{d^2 \hat{N}_0^2}+\frac{1}{R^2}\right), \nb\\
&& R_{\mu\nu} R^{\mu\nu} \simeq    2\frac{a^4+2 a^3 d+5 a^2 d^2+4 a d^3+6 d^4}{d^4 \hat{N}_0^4}\nb\\
&& ~~~~~~~~~~~~~~~~~~ +\frac{4 (a+2 d)}{d \hat{N}_0^2 R^2}+\frac{2}{R^4}, \nb\\
 && R_{\mu\nu\alpha\beta} R^{\mu\nu\alpha\beta} \simeq  4 \frac{a^4+2 a^2d^2+3d^4}{d^4\hat{N}_0^4}+\frac{8}{\hat{N}_0^2 R^2}+\frac{4}{R^4}, \nb\\
&& C_{\mu\nu\alpha\beta} C^{\mu\nu\alpha\beta} \simeq \frac{4 \left(a R^2 (a-d)+d^2 \hat{N}_0^2\right){}^2}{3 d^4 \hat{N}_0^4 R^4}.
\eqn
Therefore, different from   the last case,  asymptotically  the spacetime is conformally flat only when $a = d$. Otherwise, we have
$C_{\mu\nu\alpha\beta} C^{\mu\nu\alpha\beta} \simeq 4a^2(a-d)^2/(3d^4 \hat N_0^4) + \mathcal{O} \left(1/R^{2}\right)$.

 On the other hand, introducing the quantity $\bar t$ via the relation
 \bq
\lb{eq2.5q}
\bar t = - \frac{d \hat N_0}{a R^{a/d}_0} \left(\frac{R_0}{R}\right)^{a/d} \equiv - \bar t_0 \left(\frac{R_0}{R}\right)^{a/d},
\eq
we find that  the metric (\ref{eq2.5o}) takes the form
\bqn
ds^2 \simeq  R_{0}^{2a/d} \left(\frac{\bar t_0}{\bar t}\right)^2\left(-d\bar t^2 +  d\bar x^2\right) +  R^2d\Omega^2. \lb{eq2.5r}
\eqn
When $a=d$, Eq.\eqref{eq2.5r} reduces to
\bqn
\lb{eq2.5r1}
ds^2 \simeq R_{0}^{2} \left(\frac{\bar t_0}{\bar t}\right)^2\left(-d\bar t^2 +  d\bar x^2+d\Omega^2\right),\; (a = d), ~~~~
\eqn
which is the same as the de Sitter spacetime for $R \gg R_{\Lambda}$, where $R_{\Lambda}$ is the de Sitter radius. In fact, when  $R \gg R_{\Lambda}$ we have
that the de Sitter spacetime is given by   
\bqn
\lb{eq2.5s}
ds_{\Lambda}^2 &=&  -\left(1- \left(\frac{R}{R_{\Lambda}}\right)^2\right) d\bar x^2 + \left(1- \left(\frac{R}{R_{\Lambda}}\right)^2\right)^{-1}d R^2\nb\\
&&  +     R^2d\Omega^2 \nb\\
&\simeq& \left(\frac{R_{\Lambda}}{\bar t} \right)^2\Big(- d\bar t^2 + d\bar x^2 +  d\Omega^2\Big),
\eqn
but now with the rescaling $\bar x \rightarrow \bar x/R_{\Lambda}$ and  
\bq
\lb{eq2.5t}
\bar t  \equiv - \frac{R_{\Lambda}}{R}.  
\eq

\begin{figure}[h!]
\includegraphics[height=4.5cm]{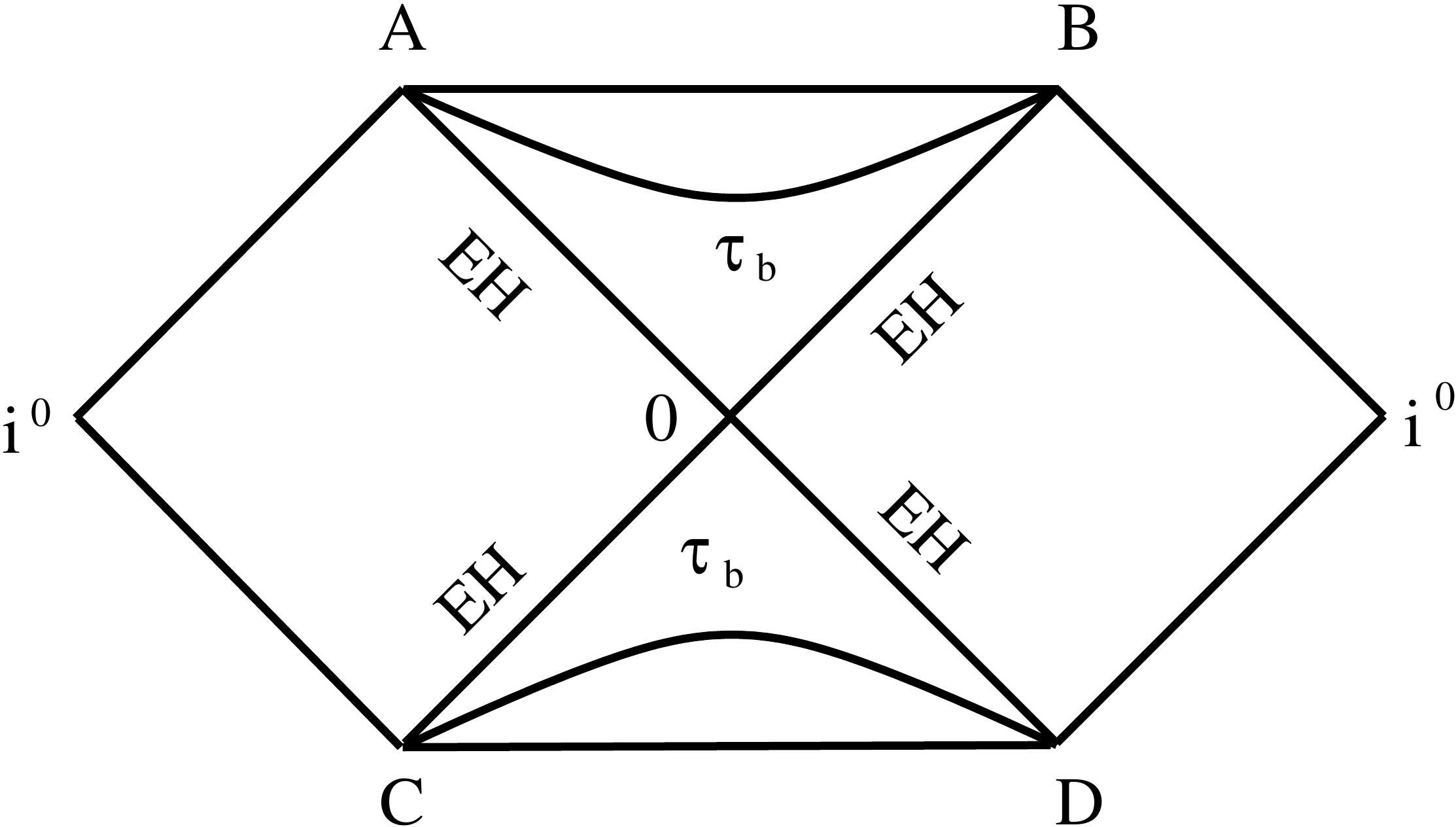}
\caption{The  Penrose diagram for the loop quantum spacetimes without the inverse volume corrections in the case $\eta > 1$ (As to be shown below, the 
corresponding Penrose diagram for the case $\eta < 1$ is also given by this figure). The curved lines denoted by $\tau_b$ are the transition surfaces (throats), and
the straight lines AD and BC  are the locations of the black hole horizons, while the   straight  lines AB and CD are the spacelike infinities, which correspond to $\bar t = 0$ and  form the future/past boundaries. The whole spacetime is free of 
 singularities.  } 
\label{fig7}
\end{figure}

Note that the angular sectors of the two metrics (\ref{eq2.5r}) and (\ref{eq2.5s})  are different in terms of $\bar t$.
In particular, in the metric (\ref{eq2.5r}) we have $R^2 \propto (-\bar t)^{-2d/a}$, while in the de Sitter spacetime we have $R^2 \propto (-\bar t)^{-2}$. Therefore, they are 
equal  only when $a =d$. However,   the sectors of the $(\bar t, \bar x$)-planes are quite similar even when $a \not= d$. 
As a result, in both cases the surfaces $\bar t = 0$ represent  spacelike
hypersurfaces and  form the boundaries of the spacetimes. Then, the corresponding Penrose diagram in the current case is given by Fig. \ref{fig7}.

When $a=d$, since $\mathcal{F}(\eta)<0$ and $\cos\left(\frac{\eta_0}{\eta}\right)<0$, from Eq.(\ref{eq2.5m}) we find
\bqn
\lb{eq2.5l1}
\mathcal{F}(\eta)=\cos\left(\frac{\eta_0}{\eta}\right).
\eqn
On the other hand, $\eta$ and $\eta_0$ must satisfy  Eq.\eqref{eq2.5l}, too. So, these two equations uniquely determine  $\eta$ and $\eta_0$.
For  $\eta_0 \lesssim - \pi$, we find that Eqs.\eqref{eq2.5l} and \eqref{eq2.5l1} have the solution,  
\bqn
\lb{eq2.5ua}
&&\left(\eta, \eta_0\right)  \approx \left(1.142,  -3.329\right),  
 \eqn 
for which, from Eqs.(\ref{eq2.5}) and (\ref{eq2.5b}) we find that 
\bqn
\lb{eq2.5v1}
\gamma = \frac{\sqrt{2\pi}}{8\eta}  \simeq
0.274.
\eqn
It is remarkable to note that this value   is precisely the one found from the analysis of black hole entropy \cite{ABBDV10}.  
It should be also noted that Eqs.\eqref{eq2.5l} and \eqref{eq2.5l1} have multi-valued solutions, as these two equations are involved with periodic  functions. 
In this paper, we consider only the case  $\eta_0 \lesssim - \pi$  \cite{ABP19}. 
 
 \begin{widetext}

\begin{figure}[h!]
 \begin{tabular}{cc}
\includegraphics[height=4cm]{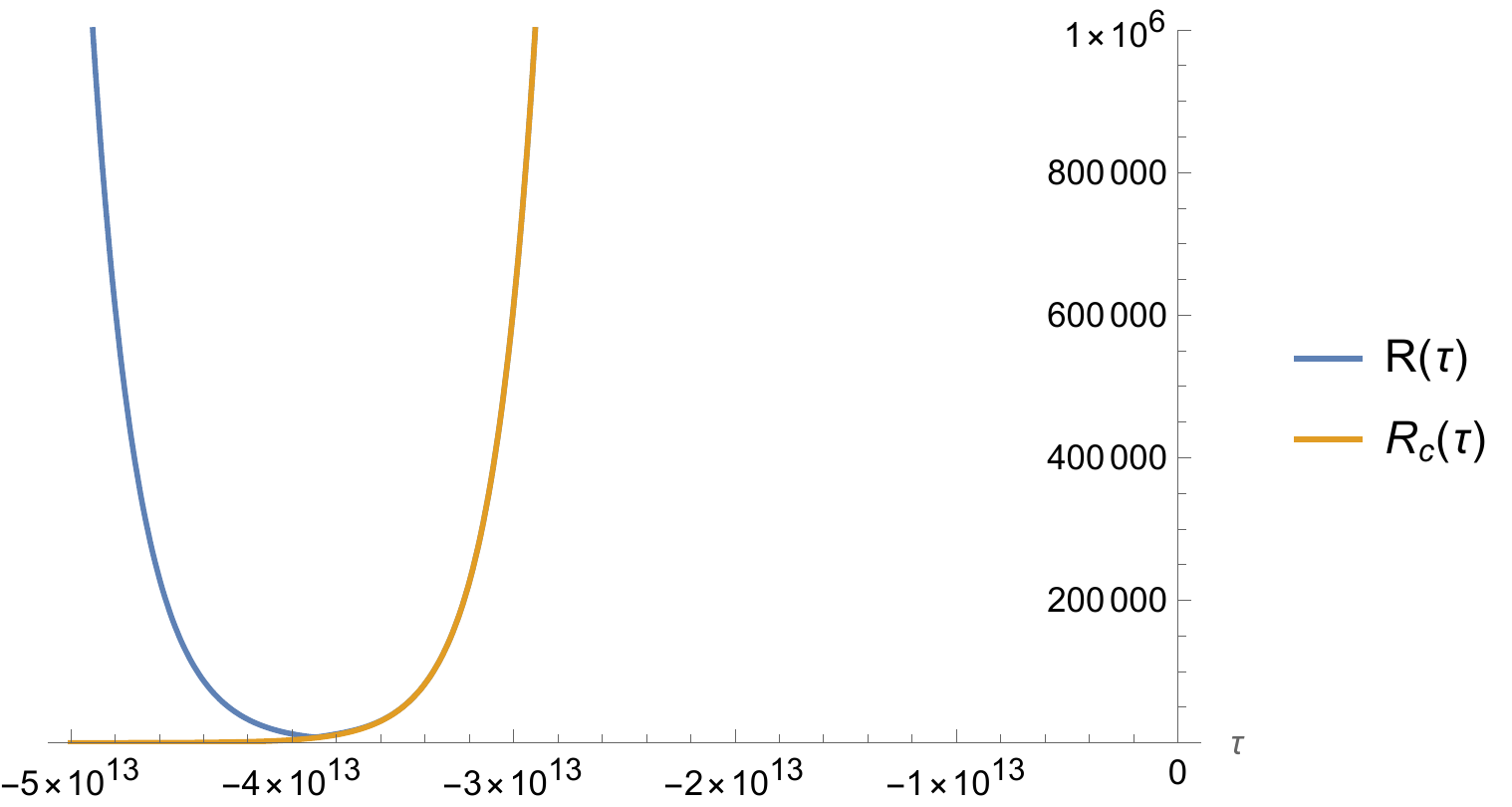}&
\includegraphics[height=4cm]{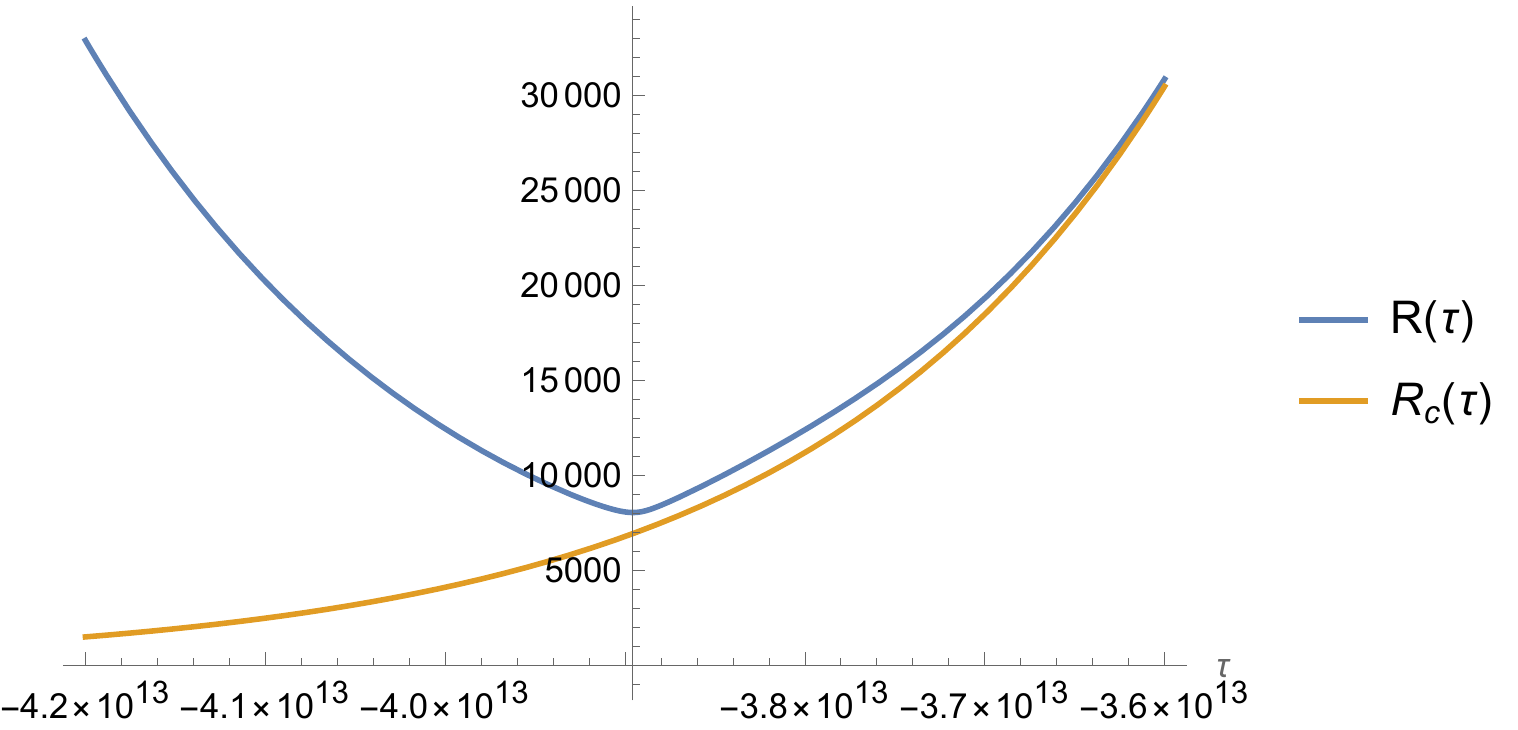}\\
	(a) & (b)  \\[6pt]
\includegraphics[height=4cm]{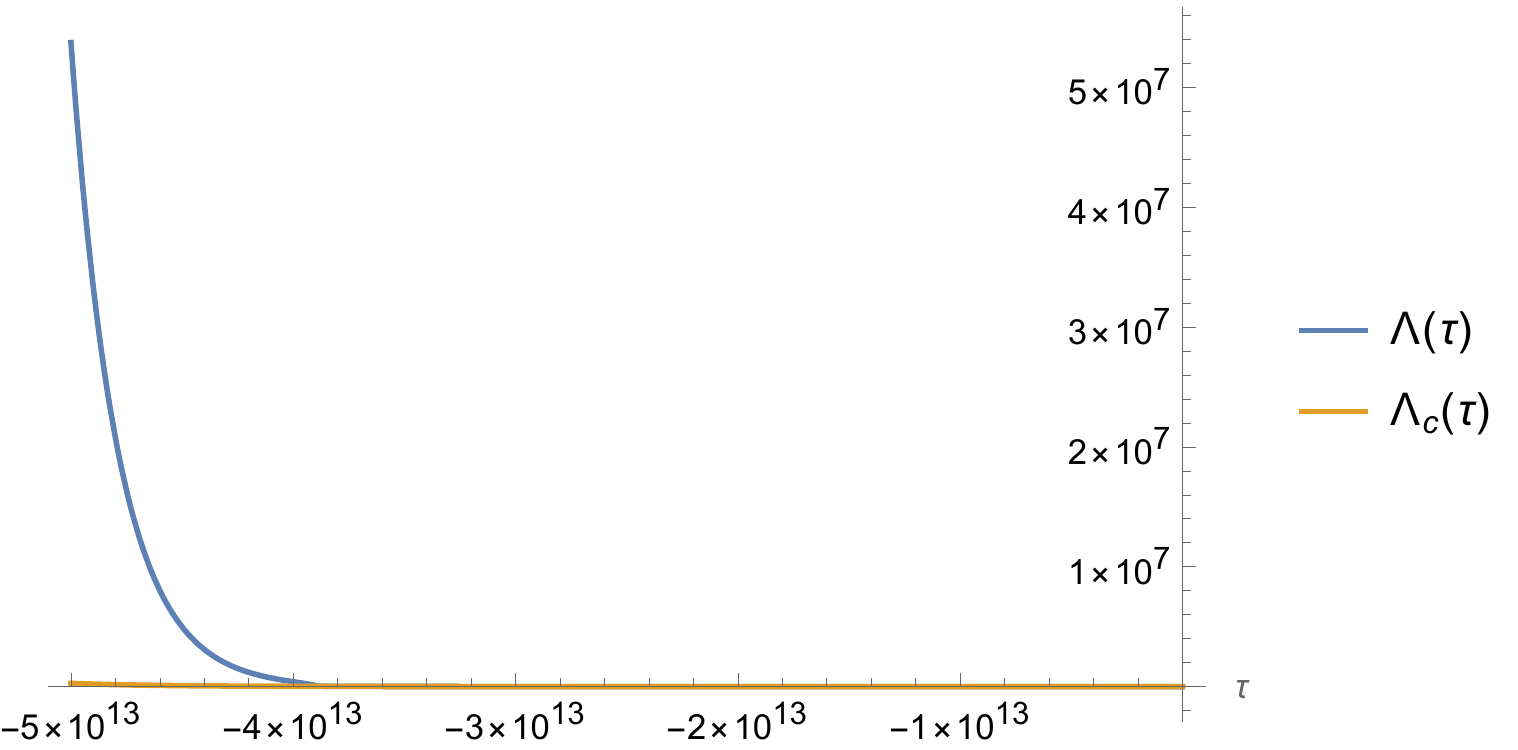}&
\includegraphics[height=4cm]{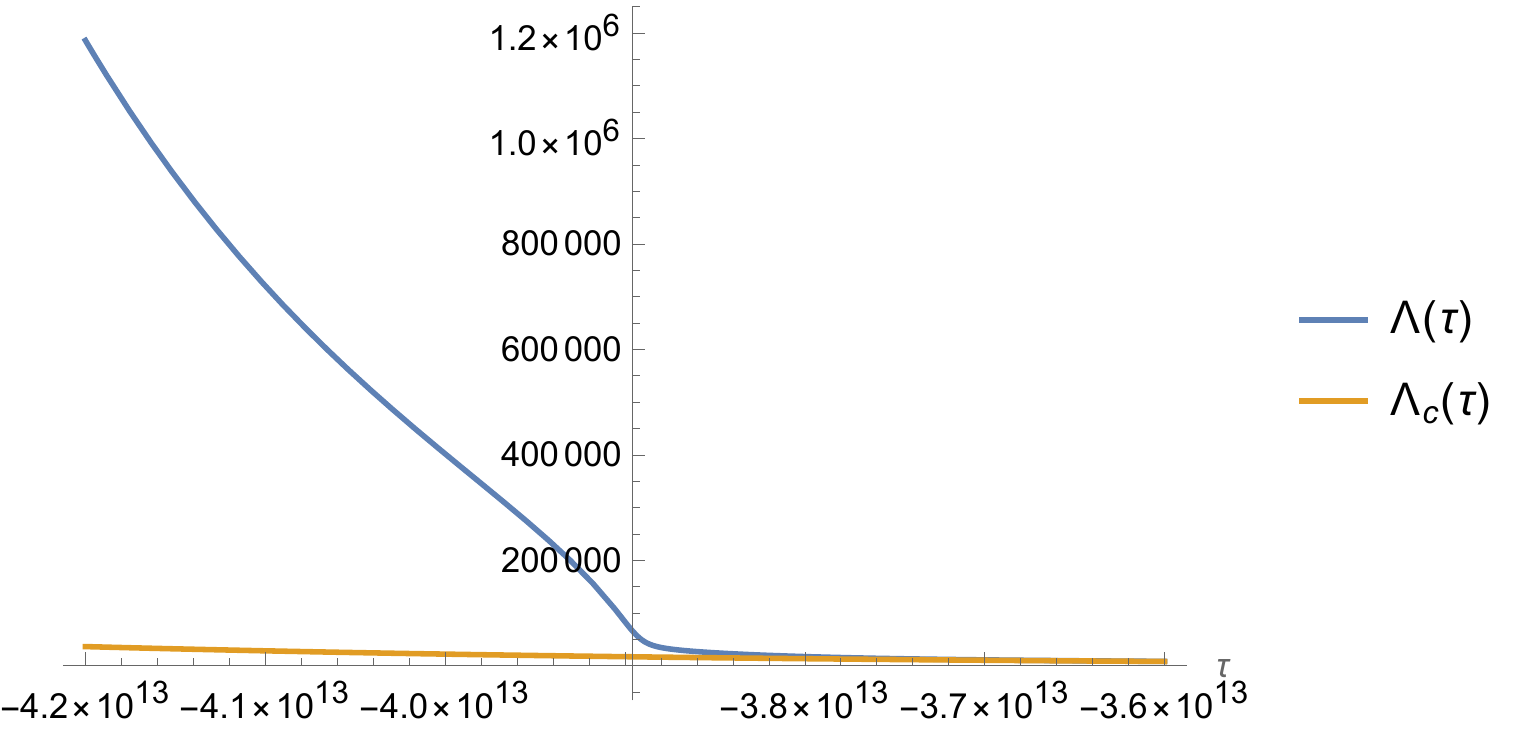}\\
	(c) & (d)  \\[6pt]
	\includegraphics[height=4cm]{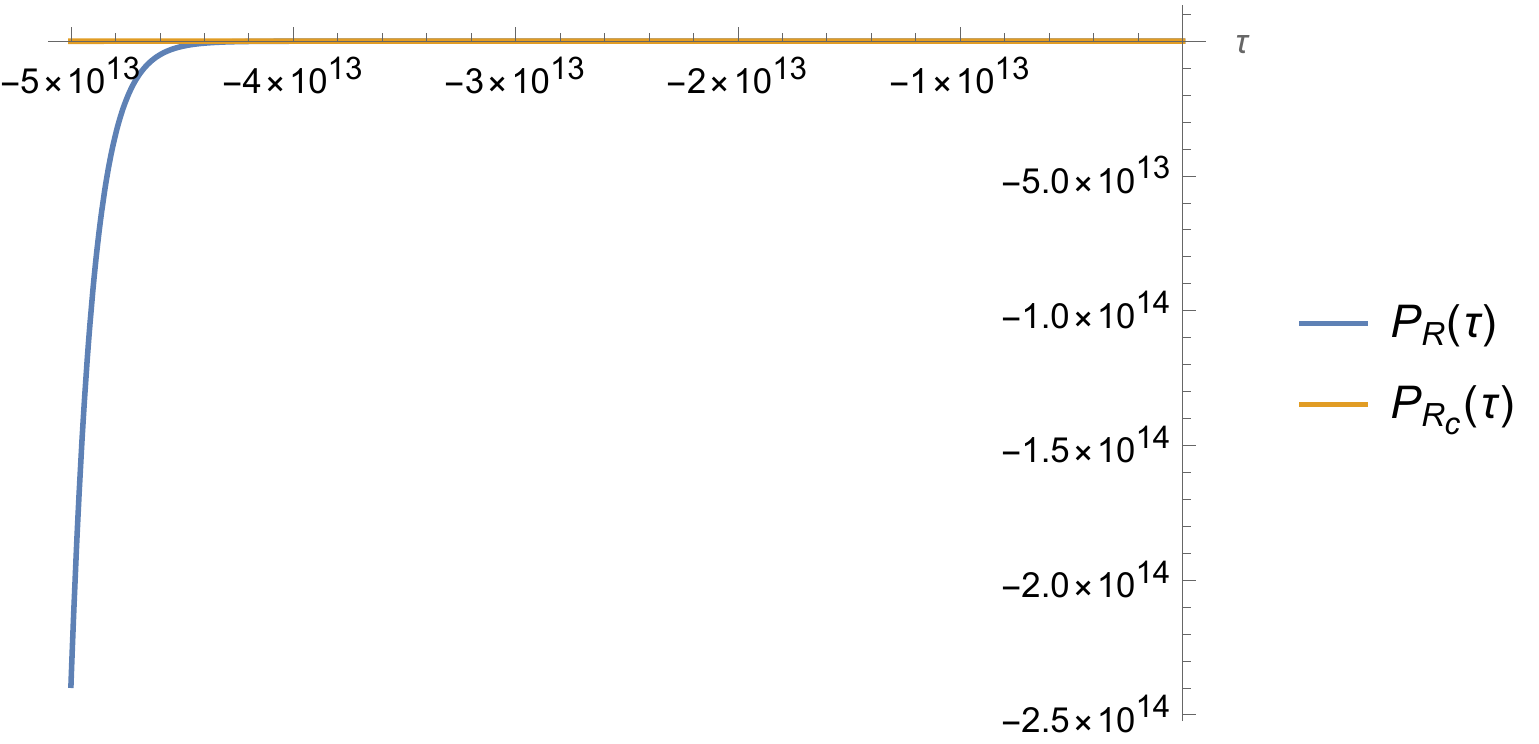}&
\includegraphics[height=4cm]{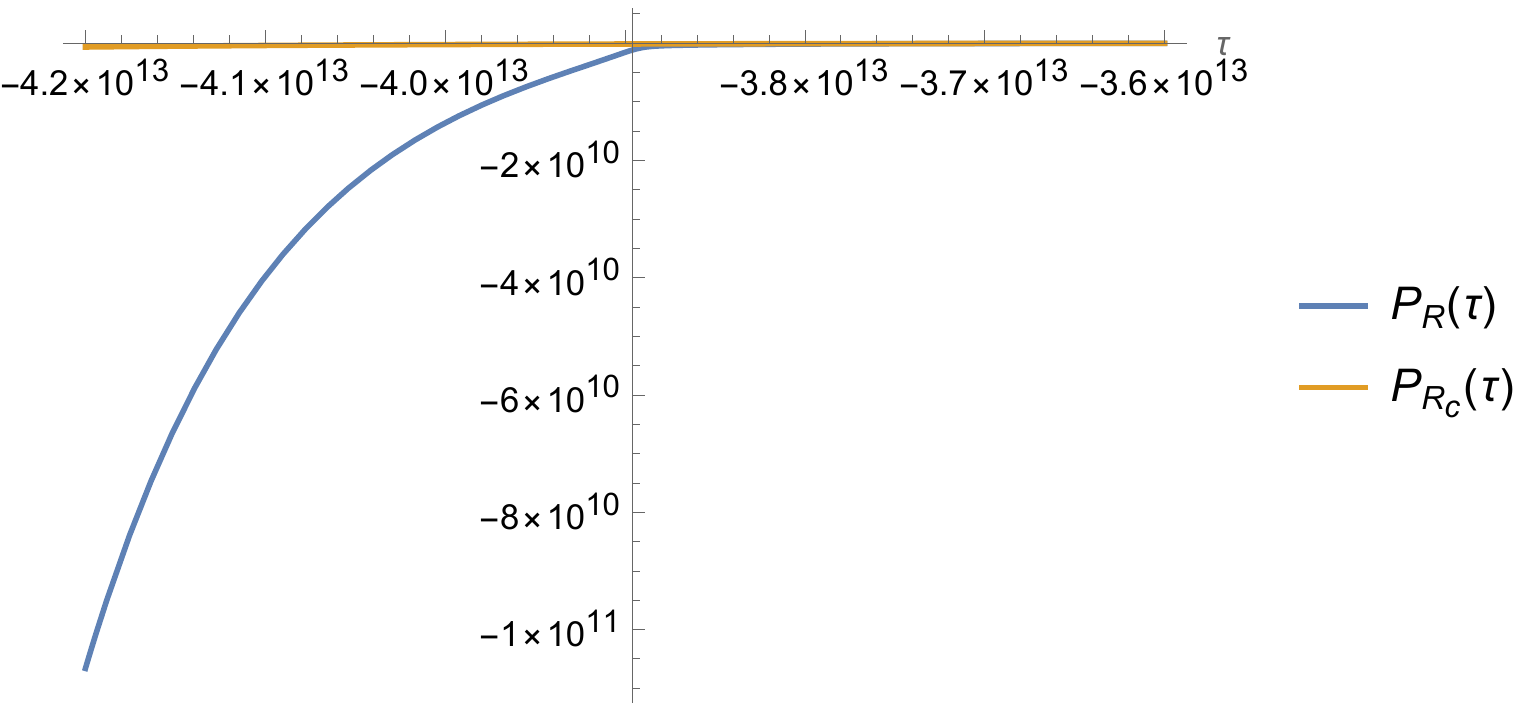}\\
	(e) & (f)  \\[6pt]
	 	\includegraphics[height=4cm]{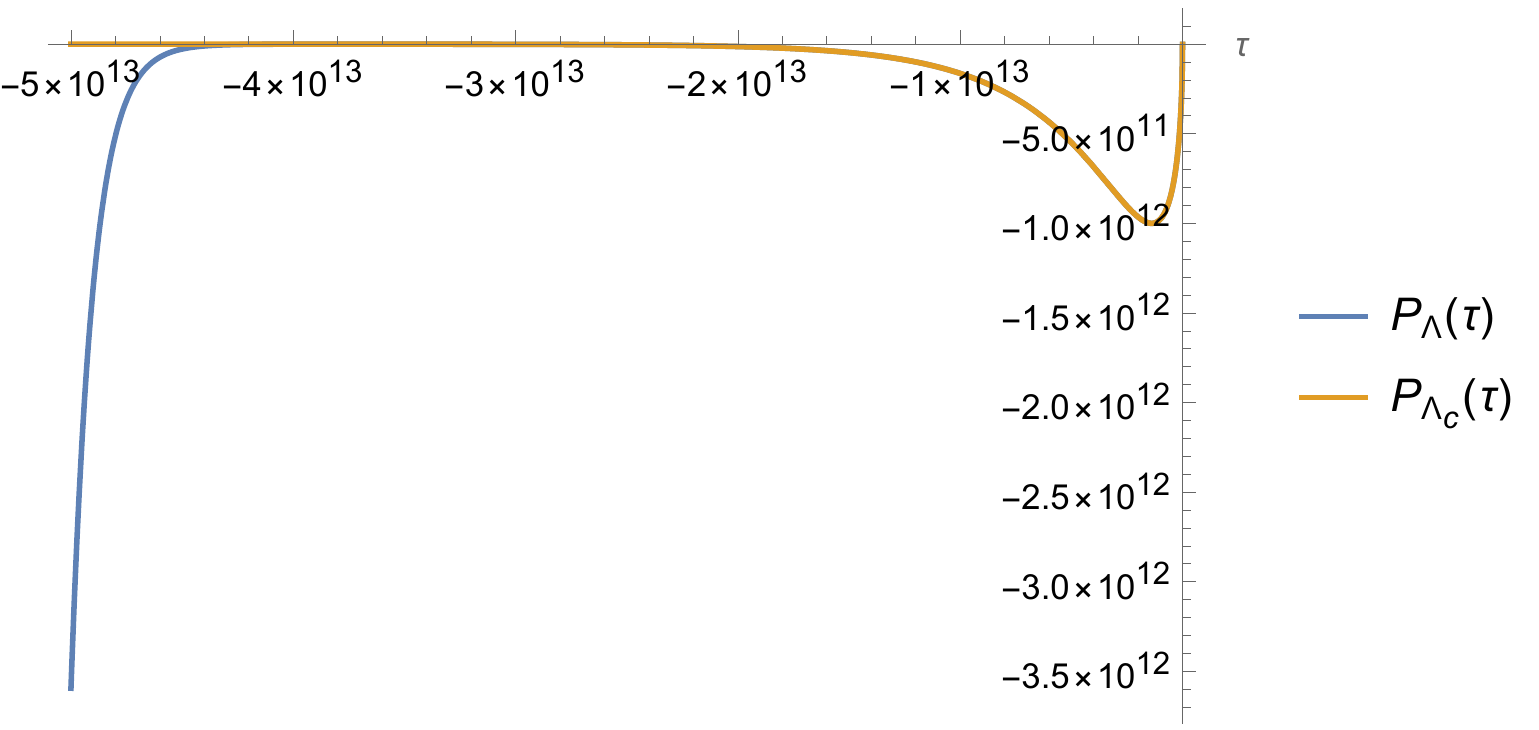}&
\includegraphics[height=4cm]{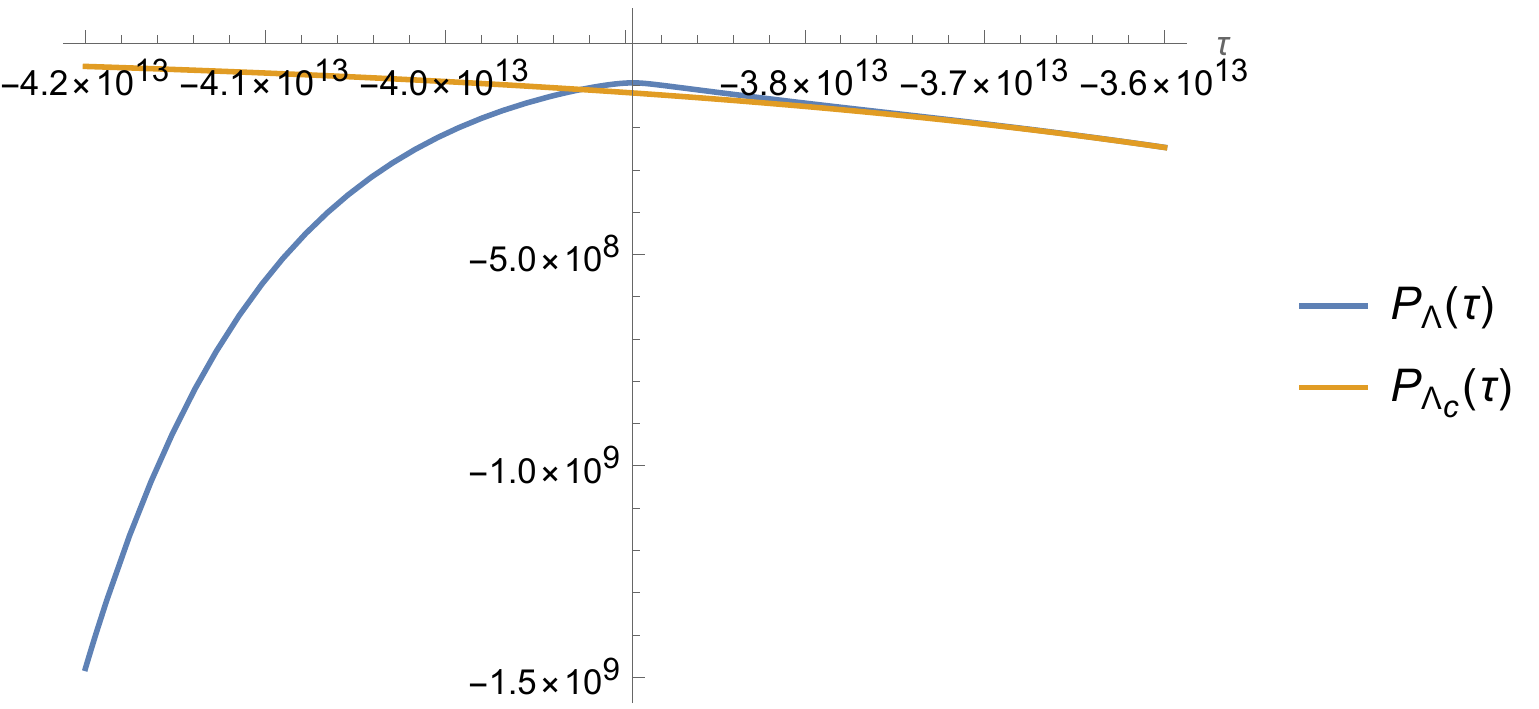}\\
	(g) & (h)  \\[6pt]
	\end{tabular}
\caption{Plots of the physical variables $\left(R, \Lambda, P_R, P_{\Lambda}\right)$ and their classical correspondences $\left(R_c, \Lambda_c, P_{R_c}, P_{\Lambda_c}\right)$.
Particular attention is paid to the region near the throat $\tau_{\text{min}} =-3.896 \times 10^{13}$, at which $R(\tau)=8059.95$. 
Graphs are plotted with $m=10^{12} m_p , \; j_0=11.42, \; j=10, \; \eta=1.142$.
} 
\lb{fig12}
\end{figure} 

     \begin{figure}[h!]
 \begin{tabular}{cc} 
 		\includegraphics[height=4cm]{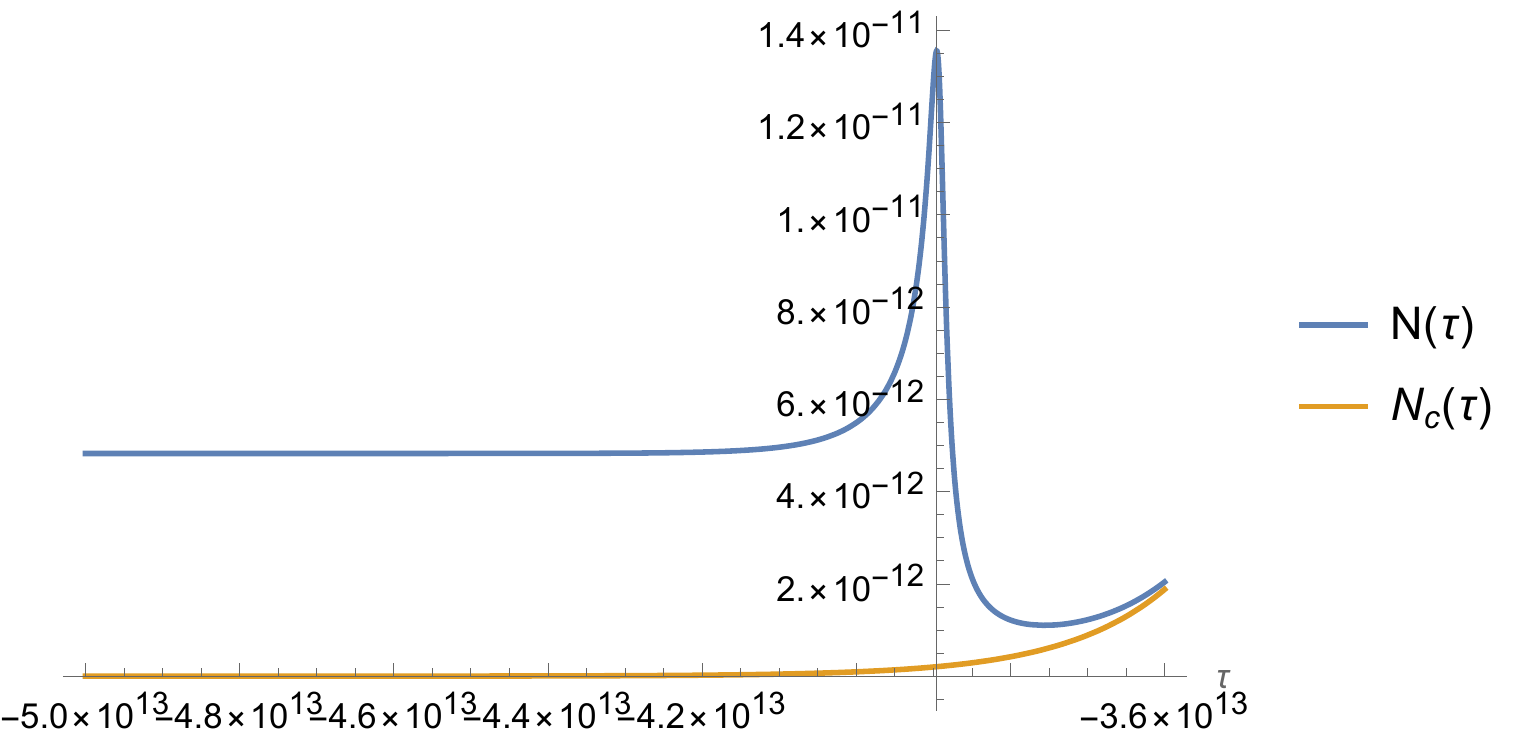}&
\includegraphics[height=4cm]{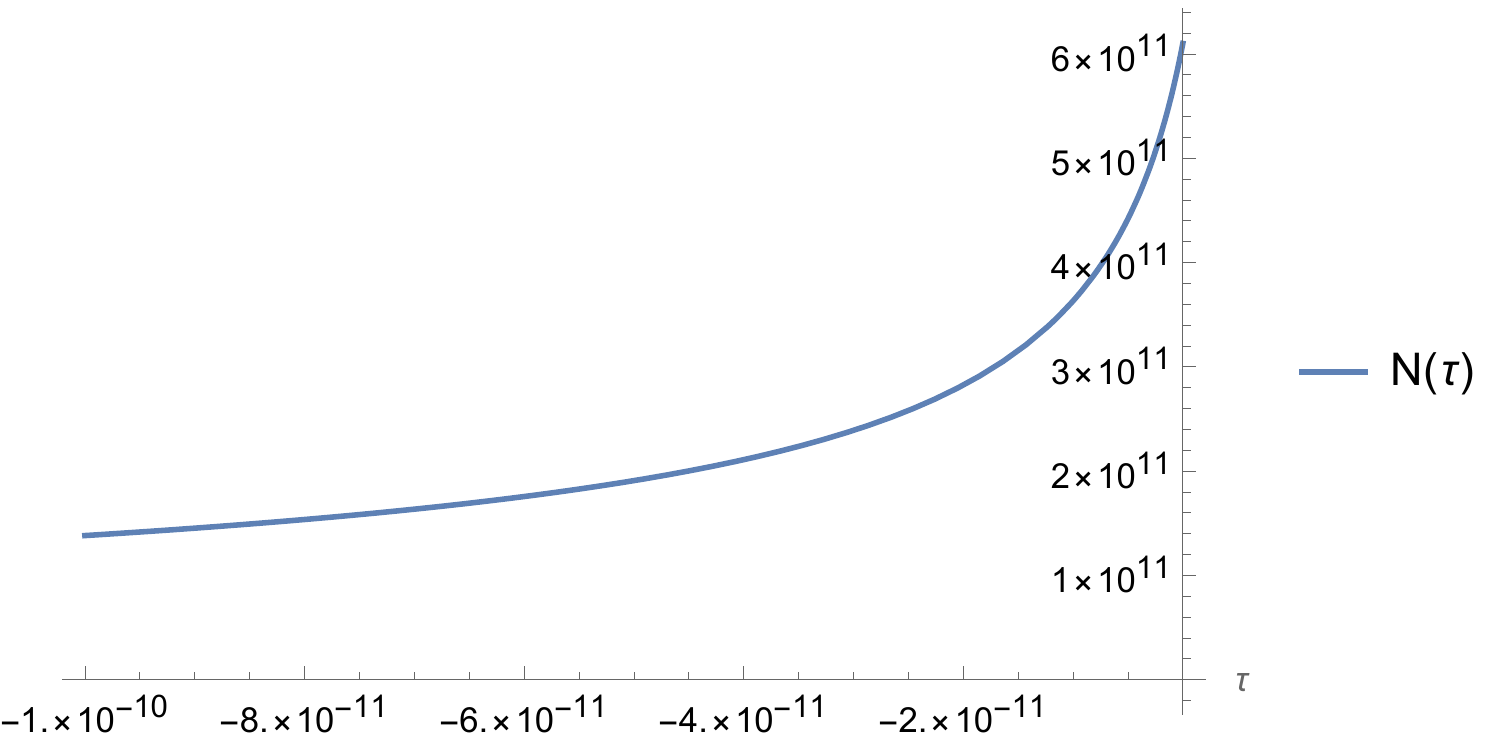}\\		
	(a) & (b)  \\[6pt]
	\includegraphics[height=4cm]{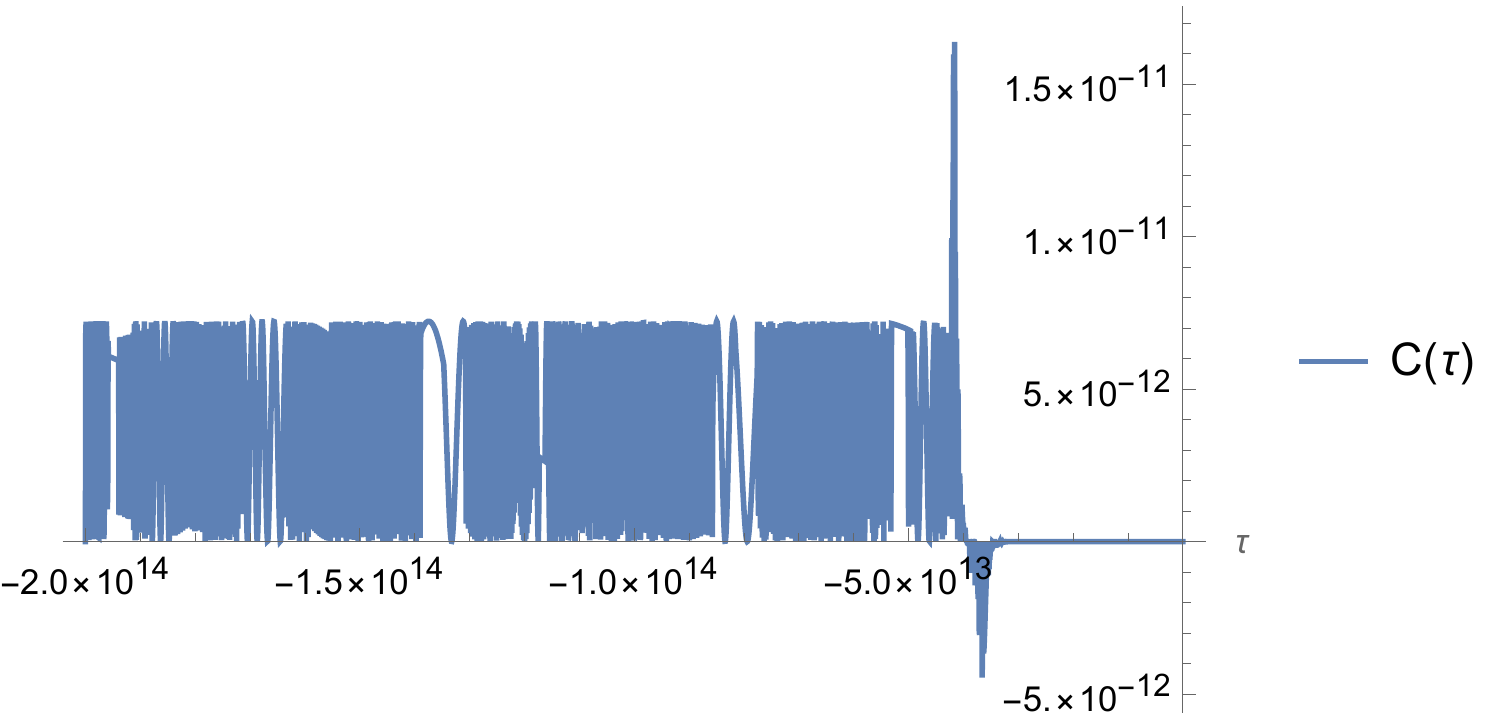}&
\includegraphics[height=4cm]{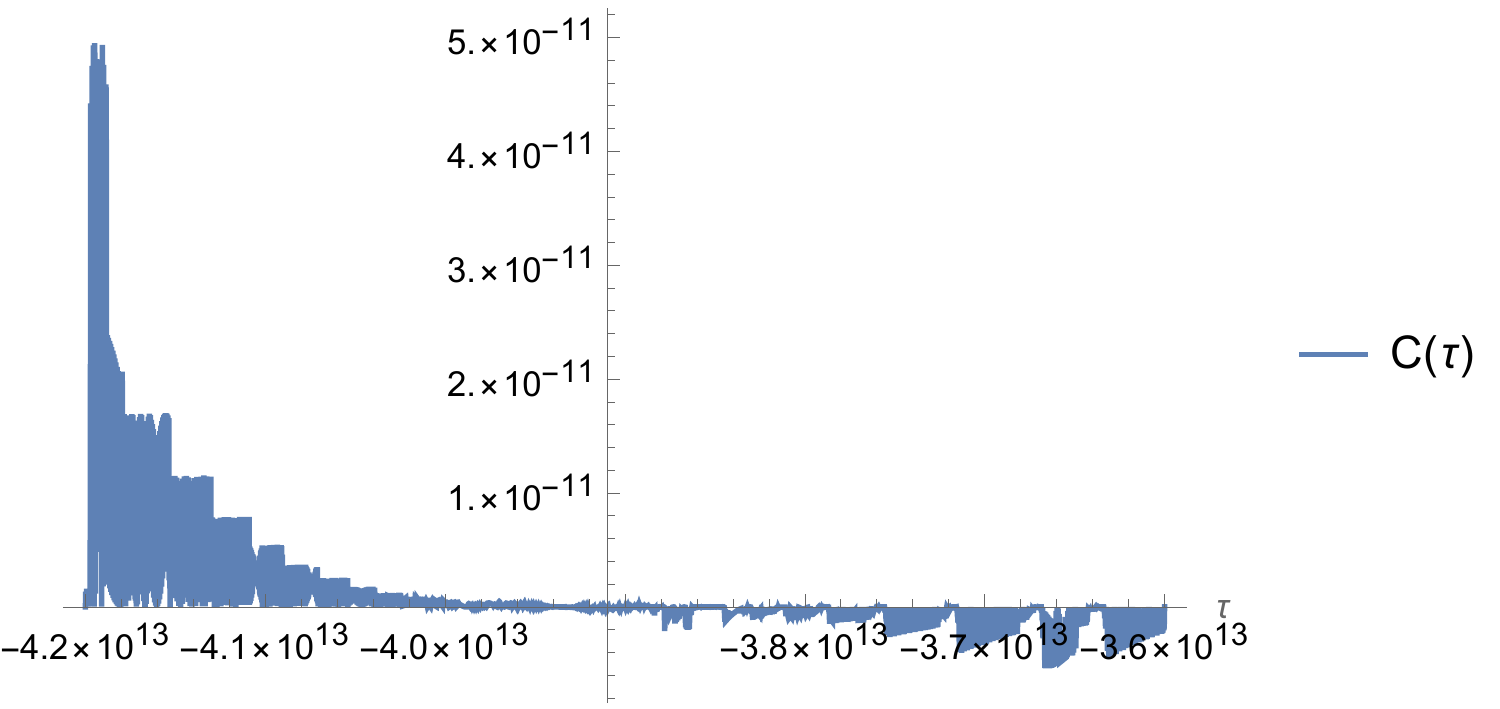}\\
	(c) & (d)  \\[6pt]
		\end{tabular}
\caption{Plots of $\mathcal{C}(\tau)$ and the lapse function $N(\tau)$ for 
 $m=10^{12} m_p , \; j_0=11.42, \; j=10, \; \eta=1.142$.
} 
\lb{fig13}
\end{figure} 

\begin{figure}[h!]
 \begin{tabular}{cc}
		\includegraphics[height=4cm]{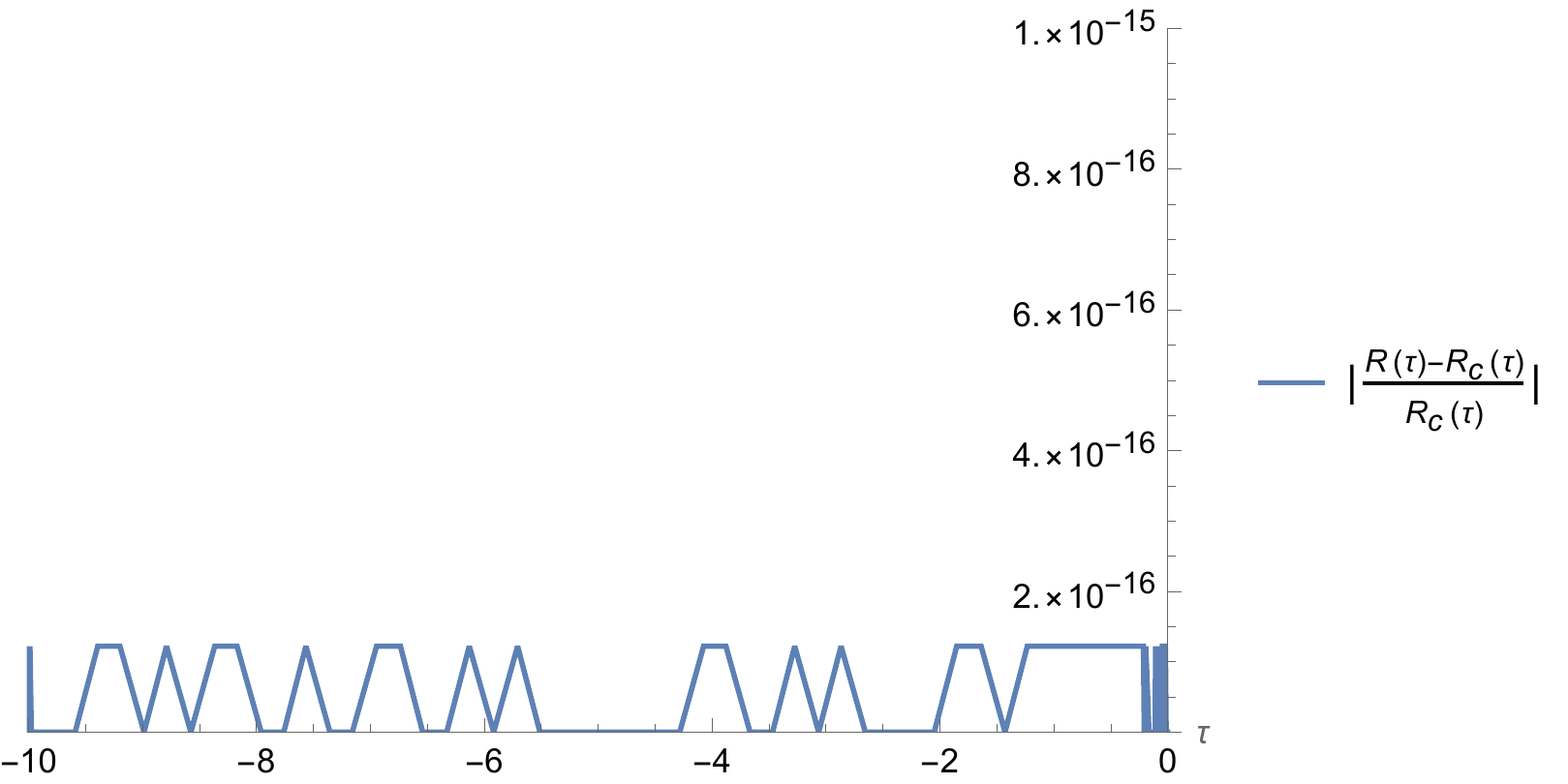}&
\includegraphics[height=4cm]{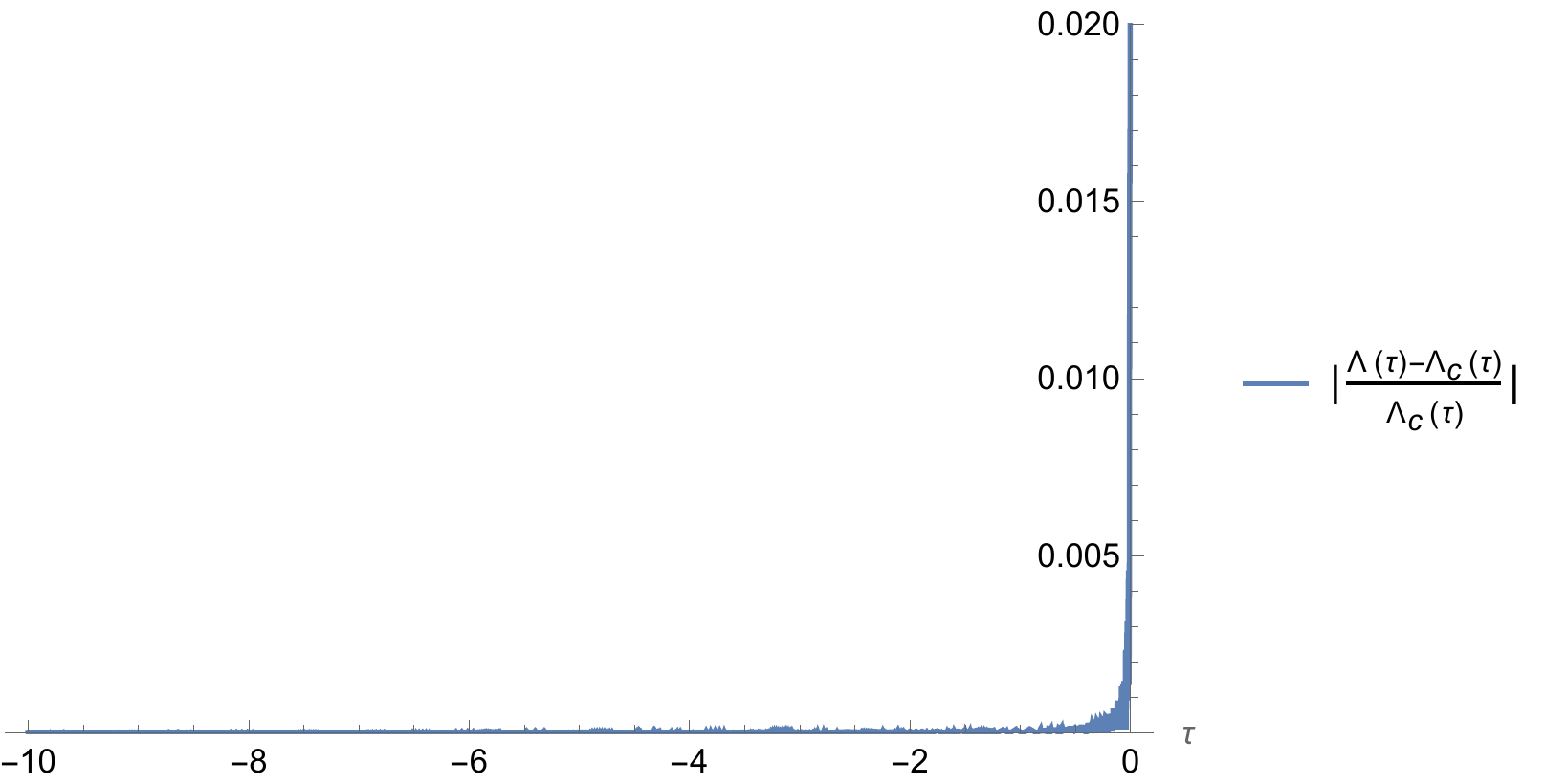}\\
	(a) & (b)  \\[6pt]
		\includegraphics[height=4cm]{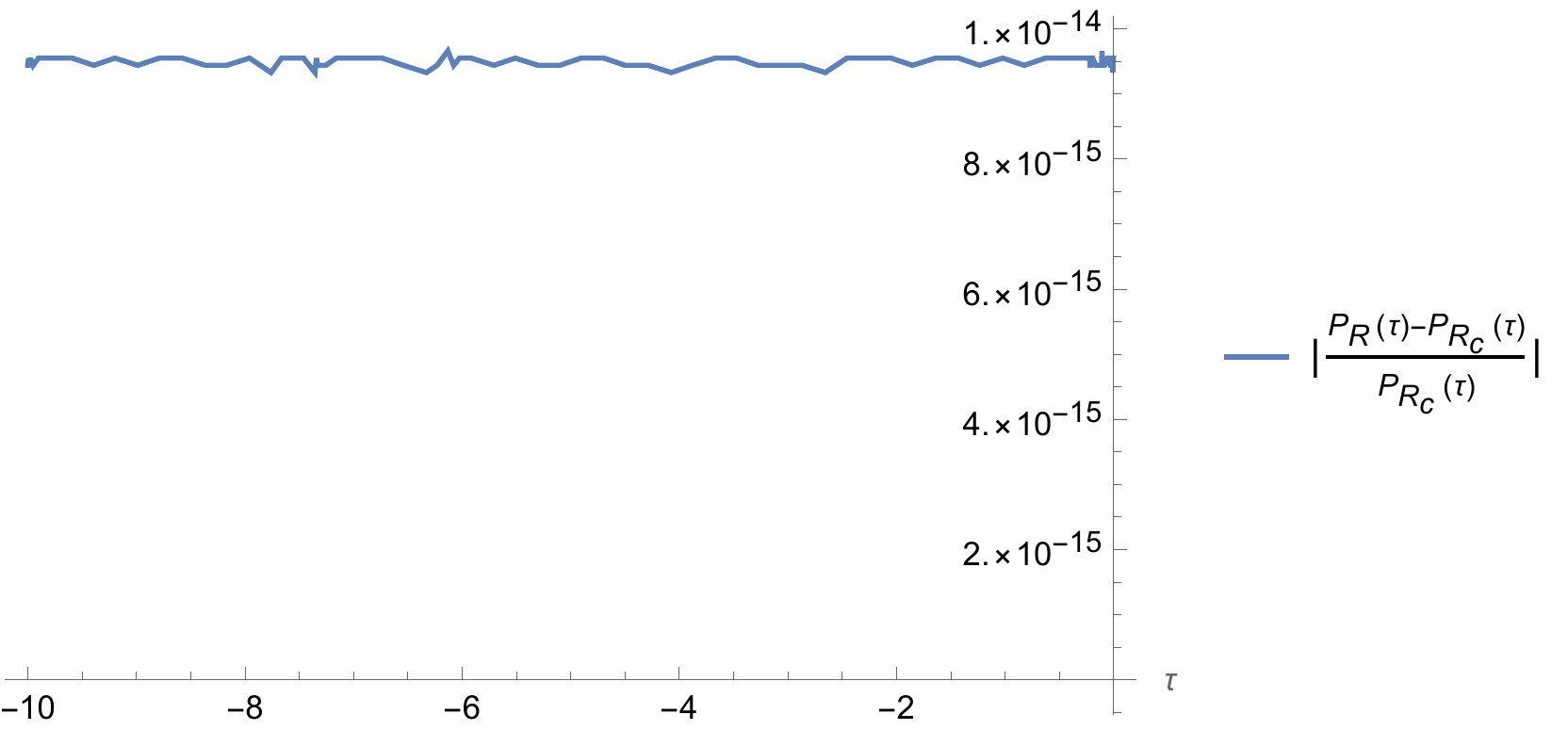}&
\includegraphics[height=4cm]{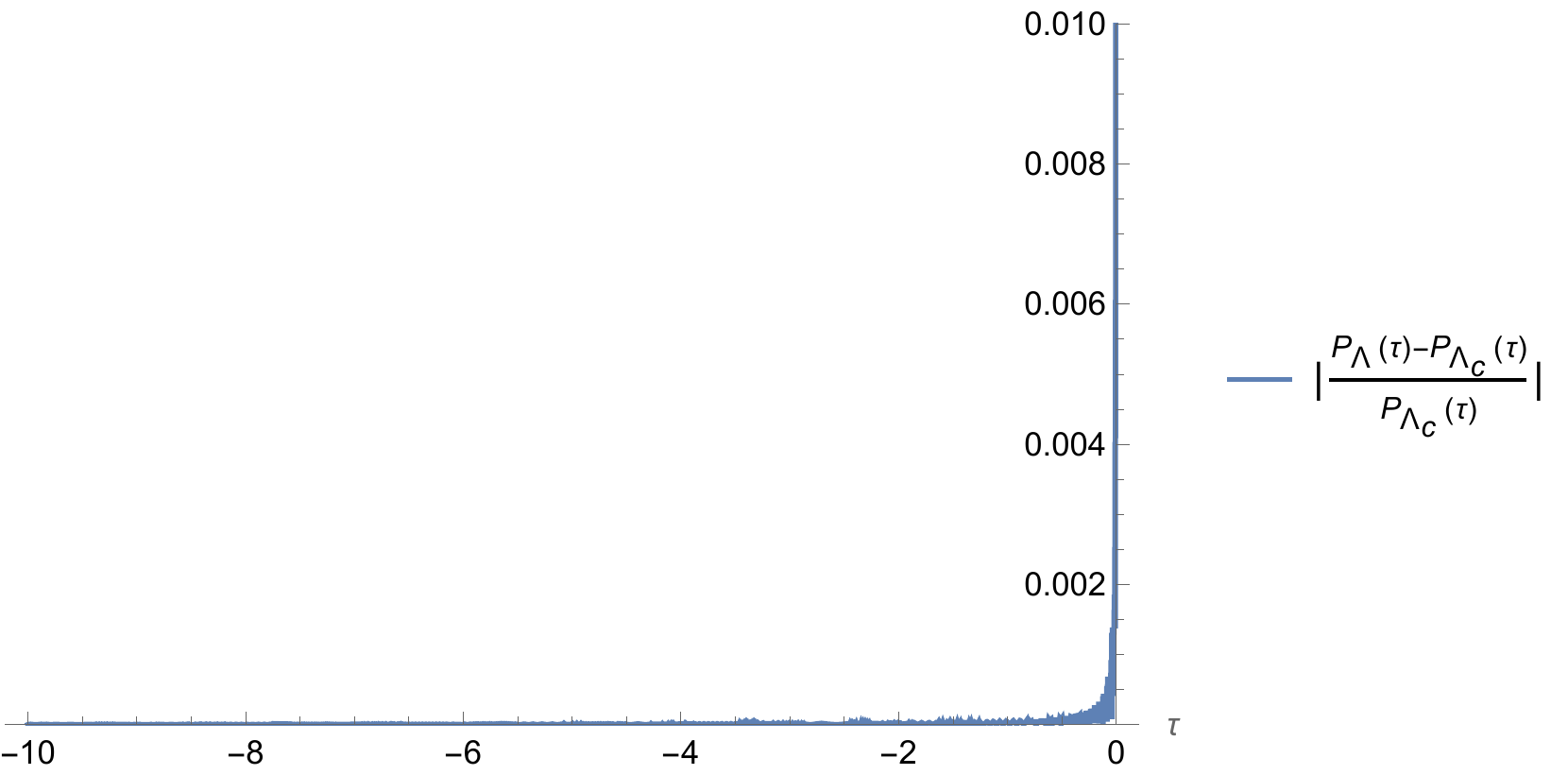}\\
	(c) & (d)  \\[6pt]  
	\includegraphics[height=4cm]{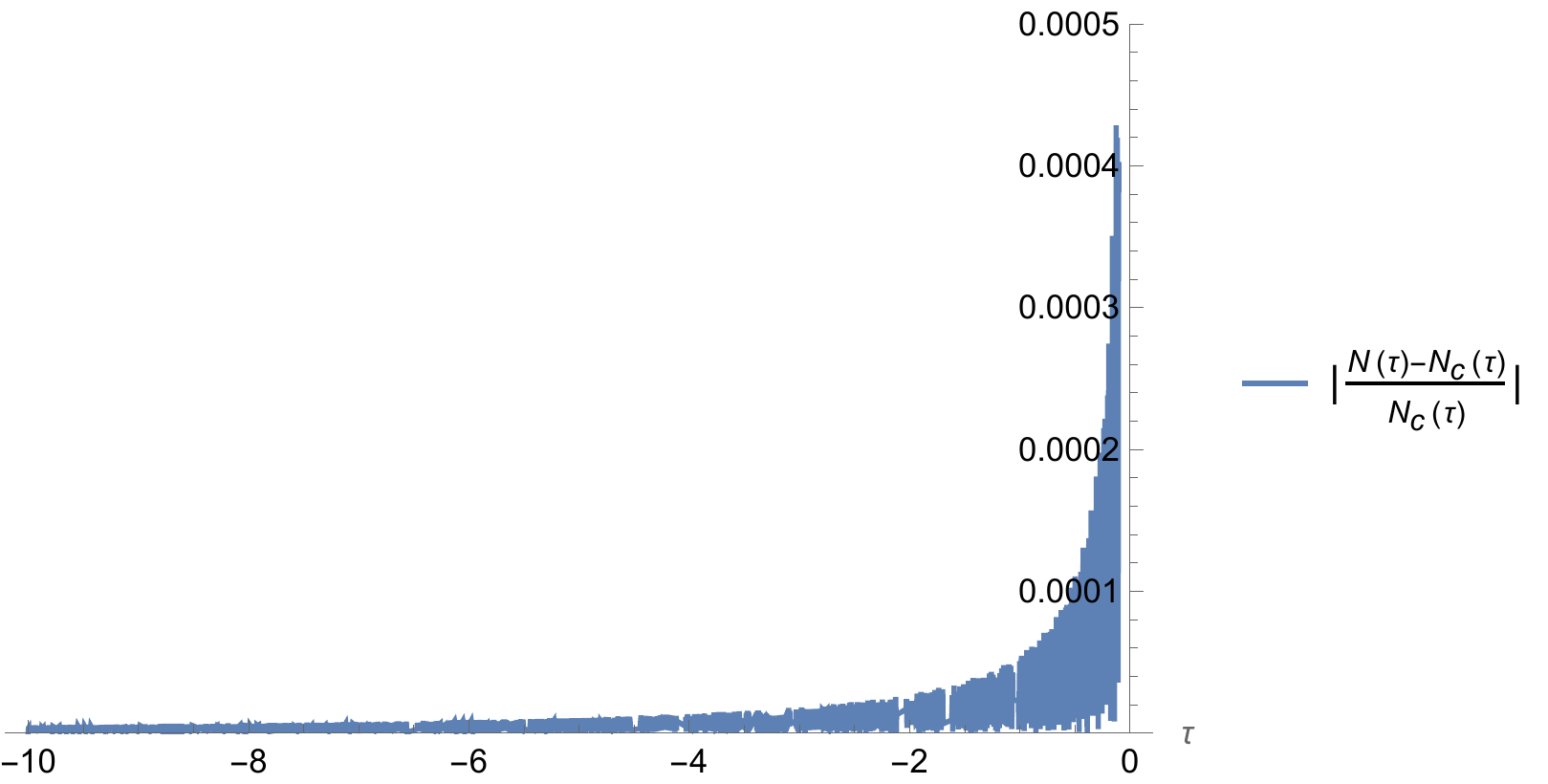}&
	\includegraphics[height=4cm]{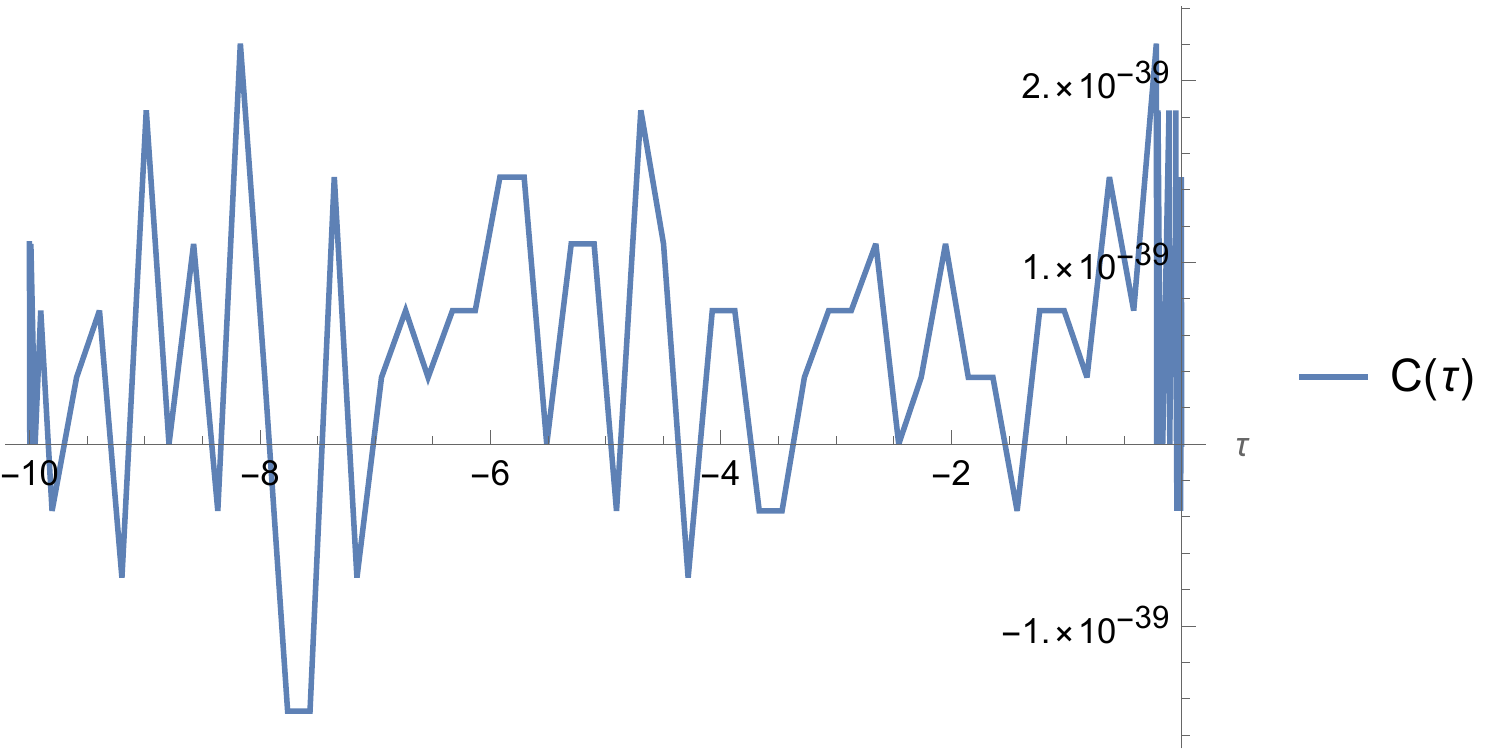}\\
	(e) & (f)  \\[6pt] 
		\end{tabular}
\caption{Plots of the relative  differences of the functions $\left(R, \Lambda, P_R, P_{\Lambda}, N(\tau)\right)$ and $\mathcal{C}(\tau)$ near the black hole horizon with the same choice of the parameters $m$ and $j$, as those specified in Figs. \ref{fig12} and \ref{fig13}, that is,  $m=10^{12} m_p , \; j_0=11.42, \; j=10, \; \eta=1.142$.
} 
\lb{fig14}
\end{figure}

\end{widetext}

\begin{table}[h]
\caption{The dependence of the constants $N_0$, $R_0$, $\Lambda_0$ of Eq.\eqref{eq2.5j} on $m$ with $\eta \approx 1.142,\; \gamma \approx 0.274,\; j_x = 10^5$. 
The corresponding transition times $\tau_{\text{min}}$ and radii  $R_{\text{min}}$ are also given.}
\begin{tabular}{|l|l|l|l|l|l|}
\hline
                $\frac{m}{m_{p}}$   & $~~~~~~\frac{\tau_{\text{min}}}{\tau_p}$     & $\frac{R_{\text{min}}}{\ell_p}$ & $~~~~~~~N_0$          
                         & $R_0$  & $\Lambda_0$ \\ \hline
$10^{12}$ & $-3.260 \times 10^{13}$ & 193114           & $5.706 \times 10^{-10}$ & 0.0226 & 0.00725     \\ \hline
$10^{10}$ & $-2.646 \times 10^{11}$ & 41605.1          & $5.706 \times 10^{-8}$  & 0.0968 & 0.0311      \\ \hline
$10^6$    & $-1.418 \times 10^7$    & 1929.73          & $5.706 \times 10^{-4}$  & 1.787  & 0.631       \\ \hline
\end{tabular}
\lb{420}
\end{table}

In Figs. \ref{fig12} - \ref{fig14}, we plot various physical quantities for $m=10^{12} m_p , \; j_0=11.42, \; j=10$, so that $\eta \equiv j_0/j = 1.142$.  This corresponds to the case studied in \cite{ABP20}, which will be analyzed in more detail in the next section with $A B C \not= 0$. Then, we find that the transition surface 
is located at $\tau_{\text{min}}/\tau_p \simeq -3.896 \times 10^{13}$, at which we have $R(\tau_{\text{min}}) \simeq 8059.95$. Note that with these choices of 
$m,\; j$ and $j_x$, the semiclassical  limit conditions (\ref{eq2.25v1}) and (\ref{eq2.5d1}) are well satisfied. Then, from Figs. \ref{fig12} and \ref{fig13} 
we find that the asymptotical behavior of the metric coefficients given  by Eq.(\ref{eq2.5j}) is well justified, while Fig. \ref{fig14} shows that the quantum effects 
near the black hole horizon ($\tau \simeq 0$) are negligible even for $m/m_p = 10^{12}$.
 For the cases with solar mass 
$m/m_p \gtrsim 10^{38}$, it is expected that such effects are even smaller. 

It should be noted that the specific values of the factors $N_0, R_0$ and $\Lambda_0$ appearing in Eq.(\ref{eq2.5j}) depend on the choice of $m$, although 
the asymptotic behavior of $N, R$ and $\Lambda$ all take the form of  Eq.(\ref{eq2.5j}). As a result, the corresponding Penrose diagram is the same
and  given by Fig. \ref{fig7} for any given $\eta > 1$. In Table \ref{420} we present their values for several choices of $m$.

We also study the effects of $\eta$, and find that the quality behaviors of the spacetimes
are quite similar to the above  even when $\eta = 2$, as long as the semiclassical  limit conditions (\ref{eq2.25v1}) and (\ref{eq2.5d1}) are satisfied and $m$ is not too small ($m/m_p  \gtrsim 10^{6}$).

  \begin{widetext}

\begin{figure}[h!]
 \begin{tabular}{cc}
\includegraphics[height=4cm]{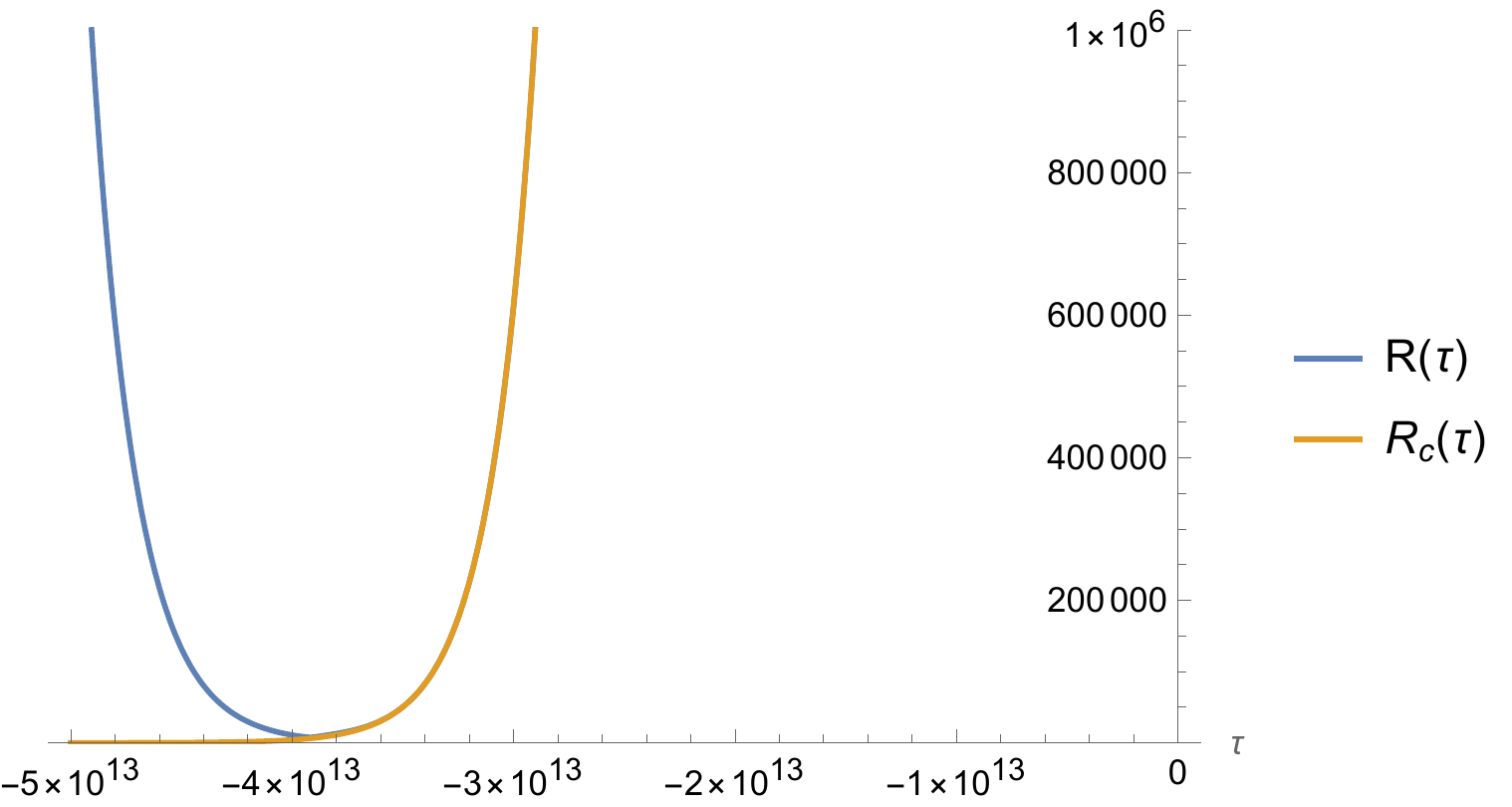}&
\includegraphics[height=4cm]{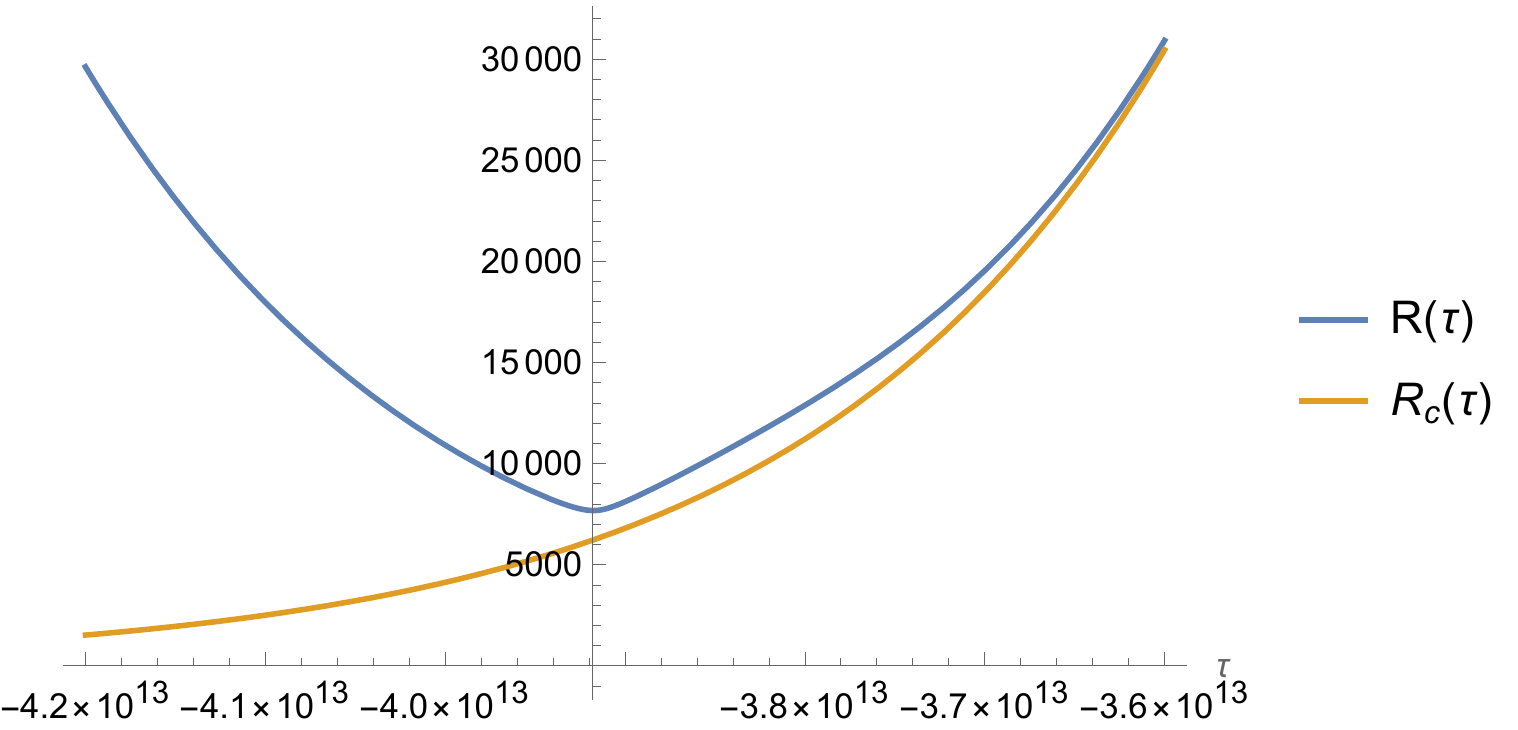}\\
	(a) & (b)  \\[6pt]
\includegraphics[height=4cm]{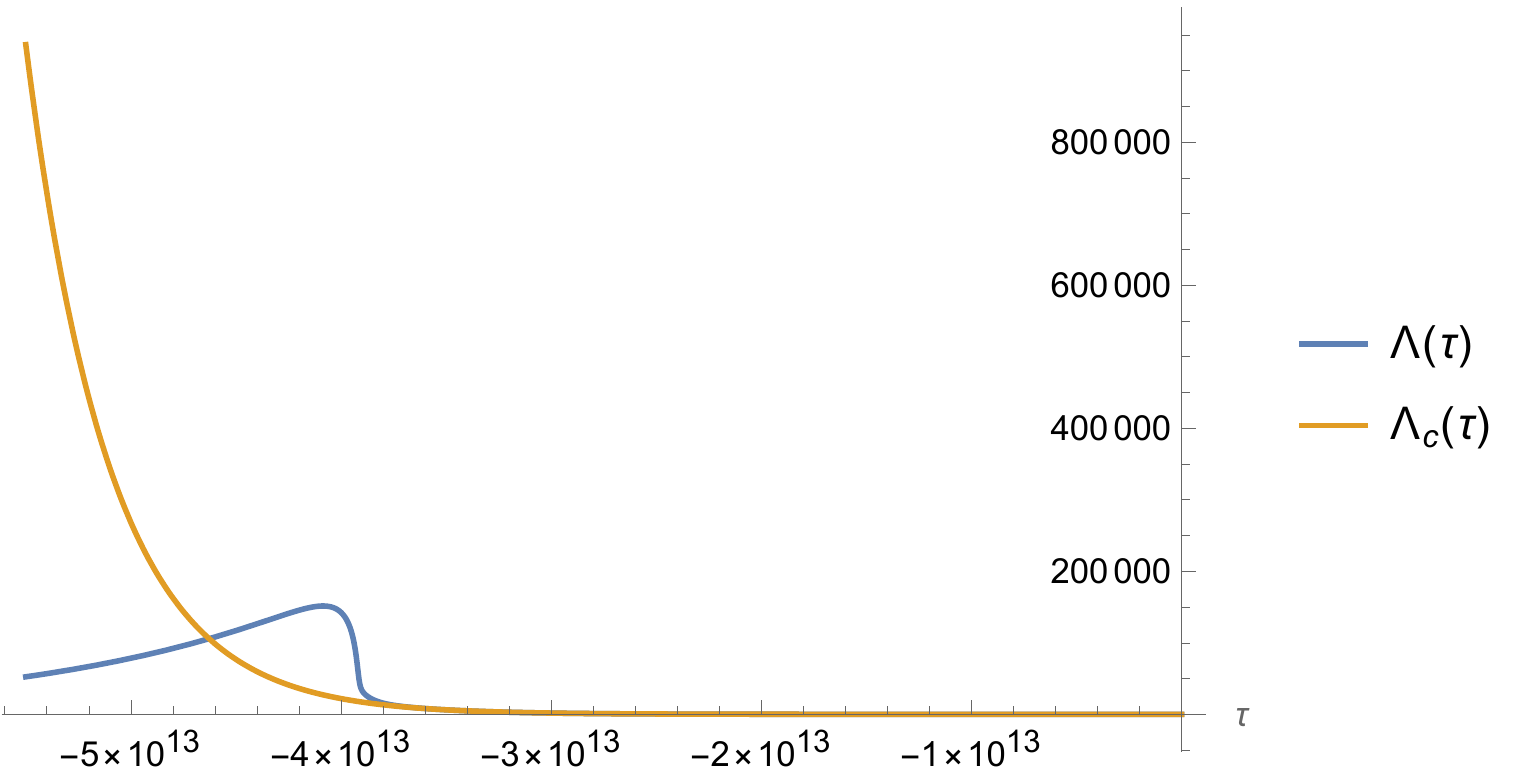}&
\includegraphics[height=4cm]{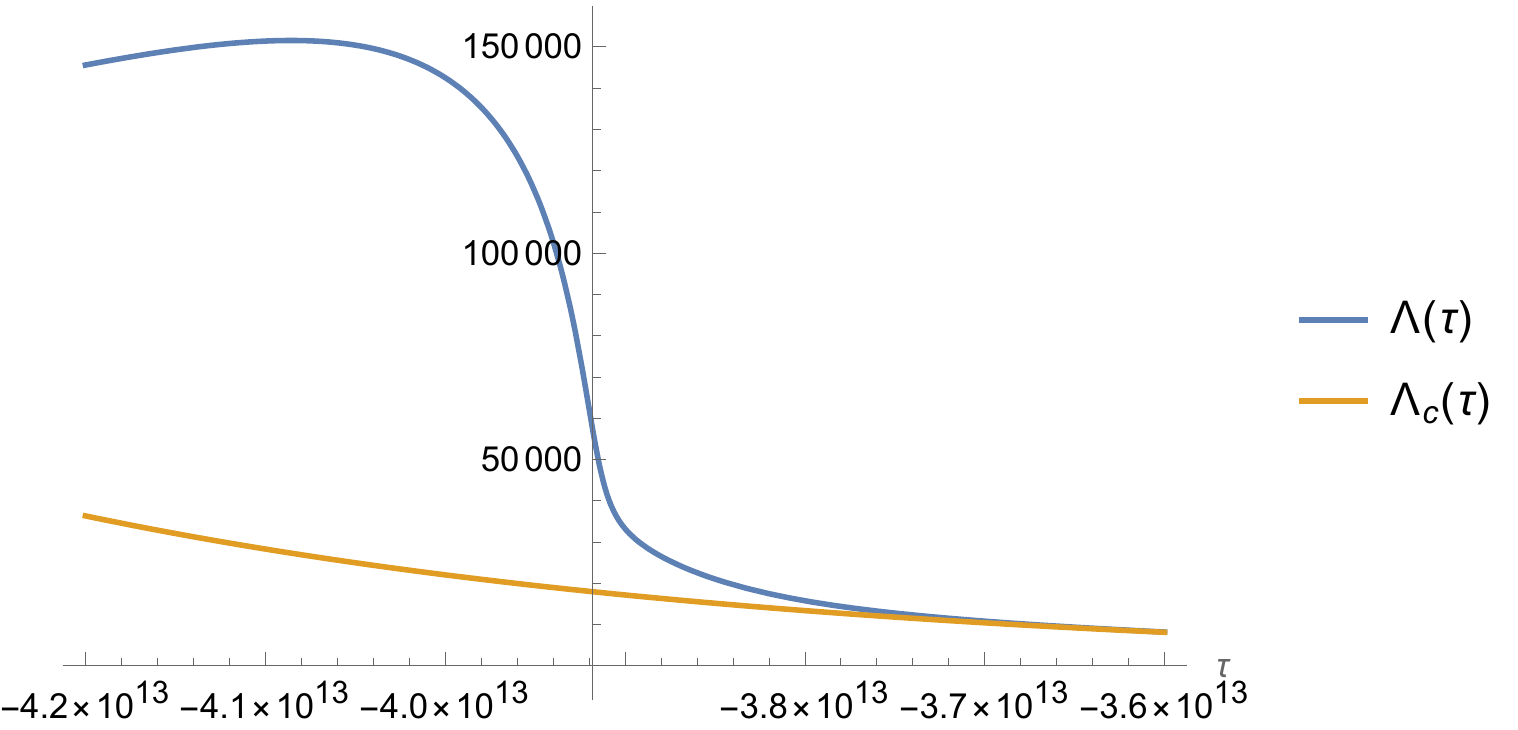}\\
	(c) & (d)  \\[6pt]
	\includegraphics[height=4cm]{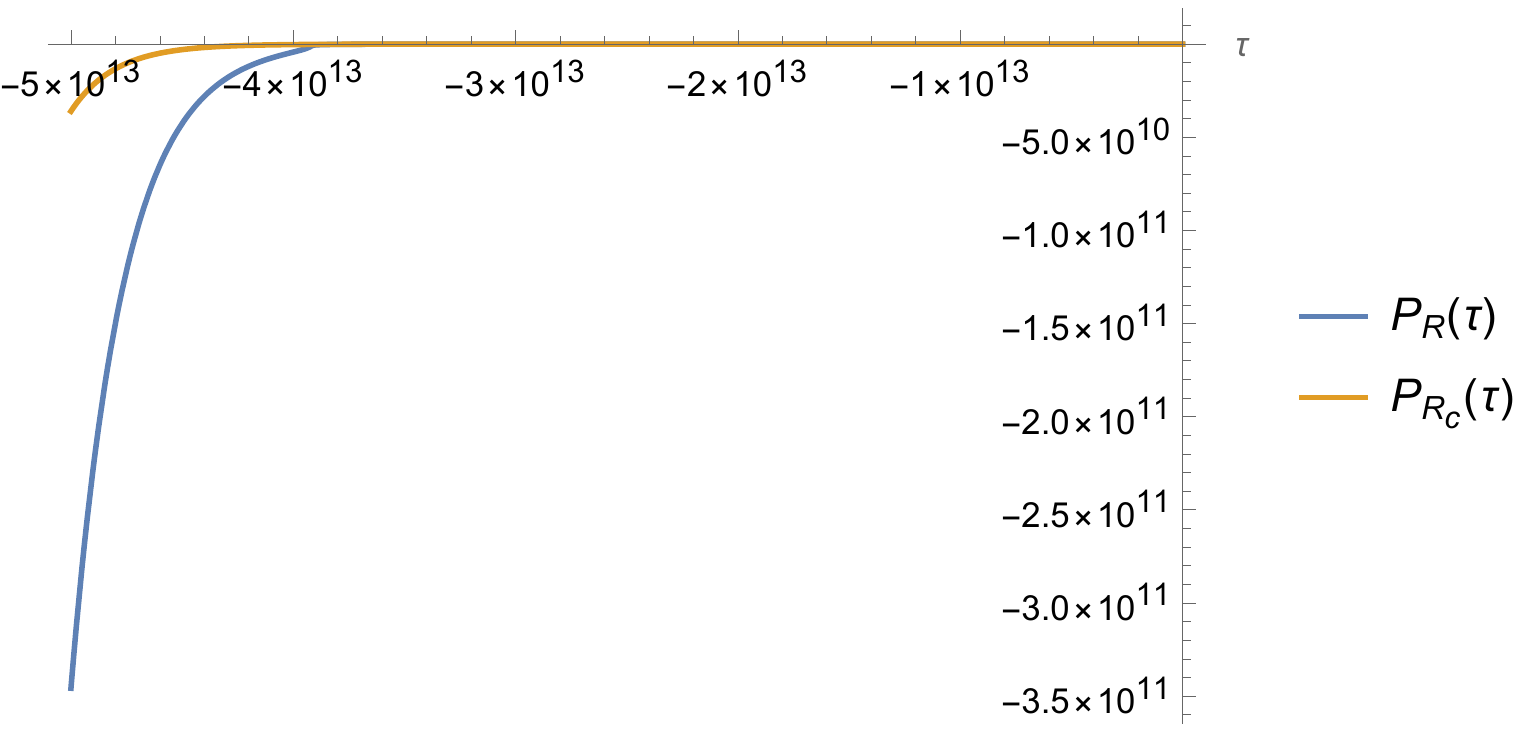}&
\includegraphics[height=4cm]{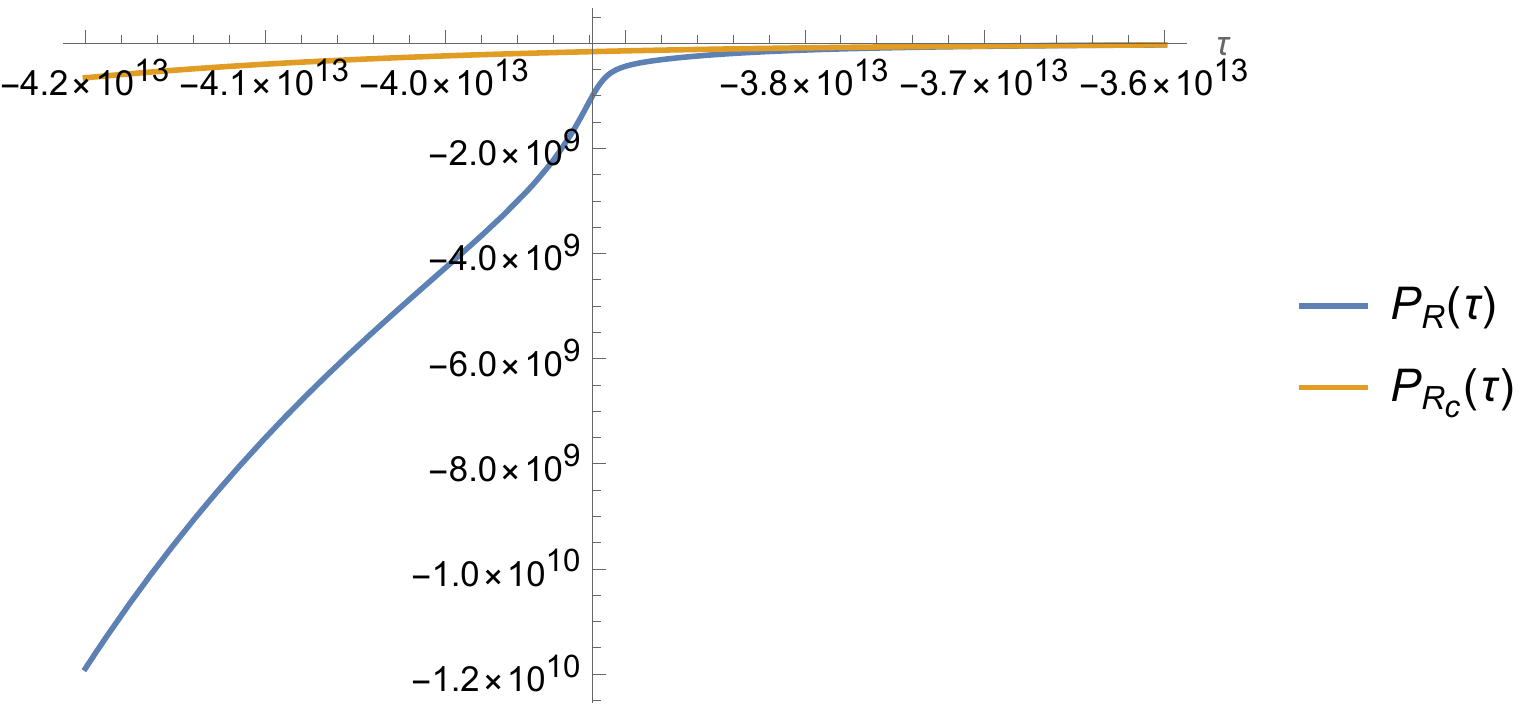}\\
	(e) & (f)  \\[6pt]
	 	\includegraphics[height=4cm]{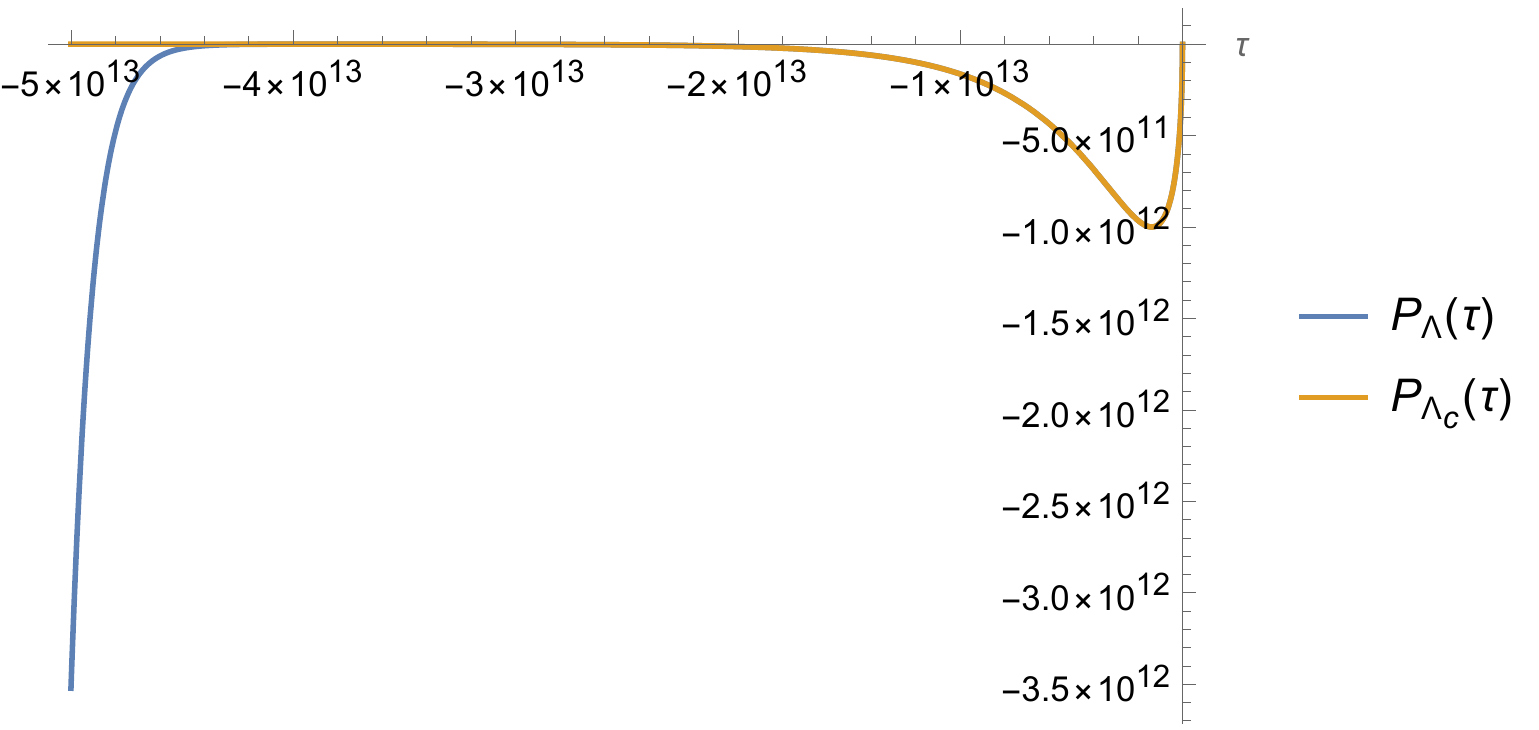}&
\includegraphics[height=4cm]{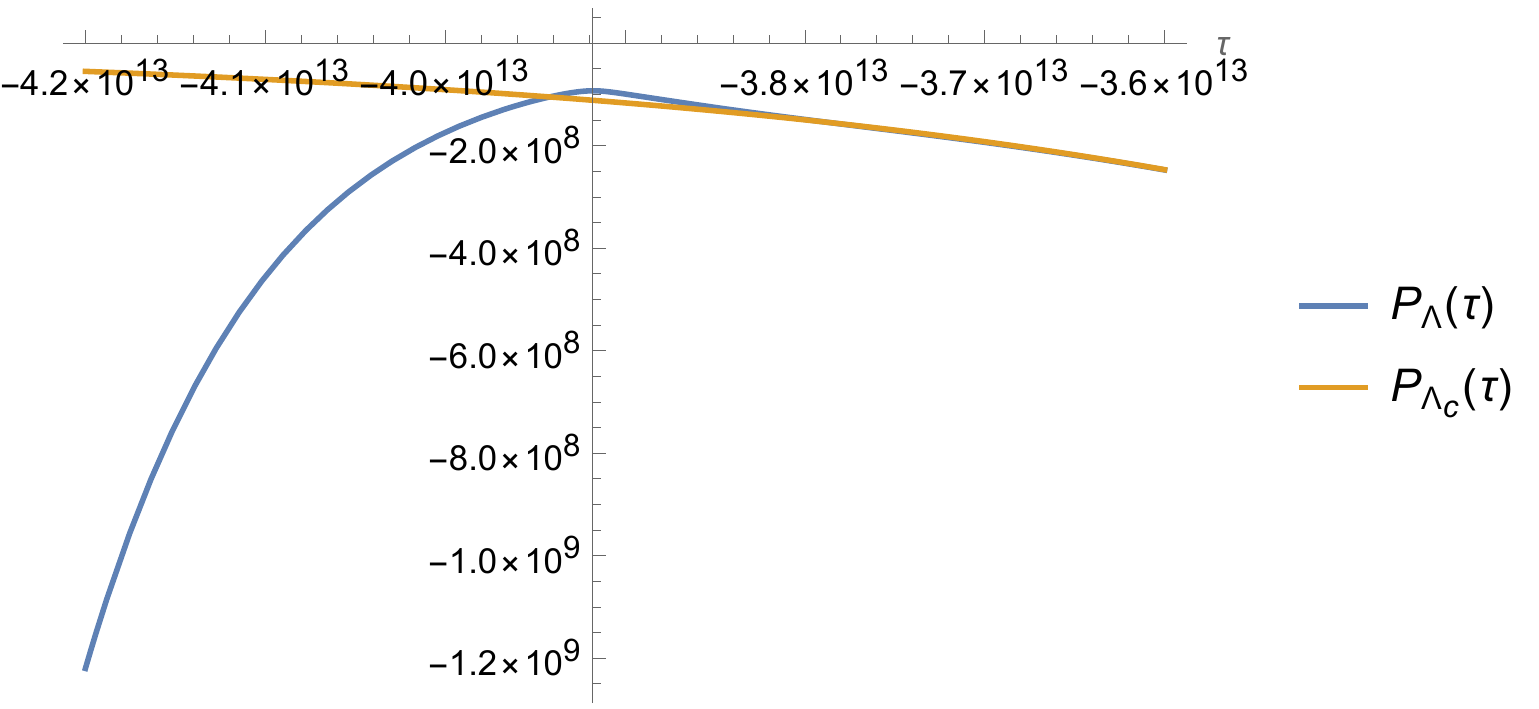}\\
	(g) & (h)  \\[6pt]
	\end{tabular}
\caption{Plots of the physical variables $\left(R, \Lambda, P_R, P_{\Lambda}\right)$ and their classical correspondences $\left(R_c, \Lambda_c, P_{R_c}, P_{\Lambda_c}\right)$.
Particular attention is paid to the region near the throat $\tau_{\text{min}} =-3.918 \times 10^{13}$, at which $R(\tau_{\text{min}})=7676.1$. 
Graphs are plotted with $m=10^{12} m_p , \; j_0=9.5, \; j=10, \; \eta=0.95$.
} 
\lb{fig24}
\end{figure} 

     \begin{figure}[h!]
 \begin{tabular}{cc} 
 		\includegraphics[height=4cm]{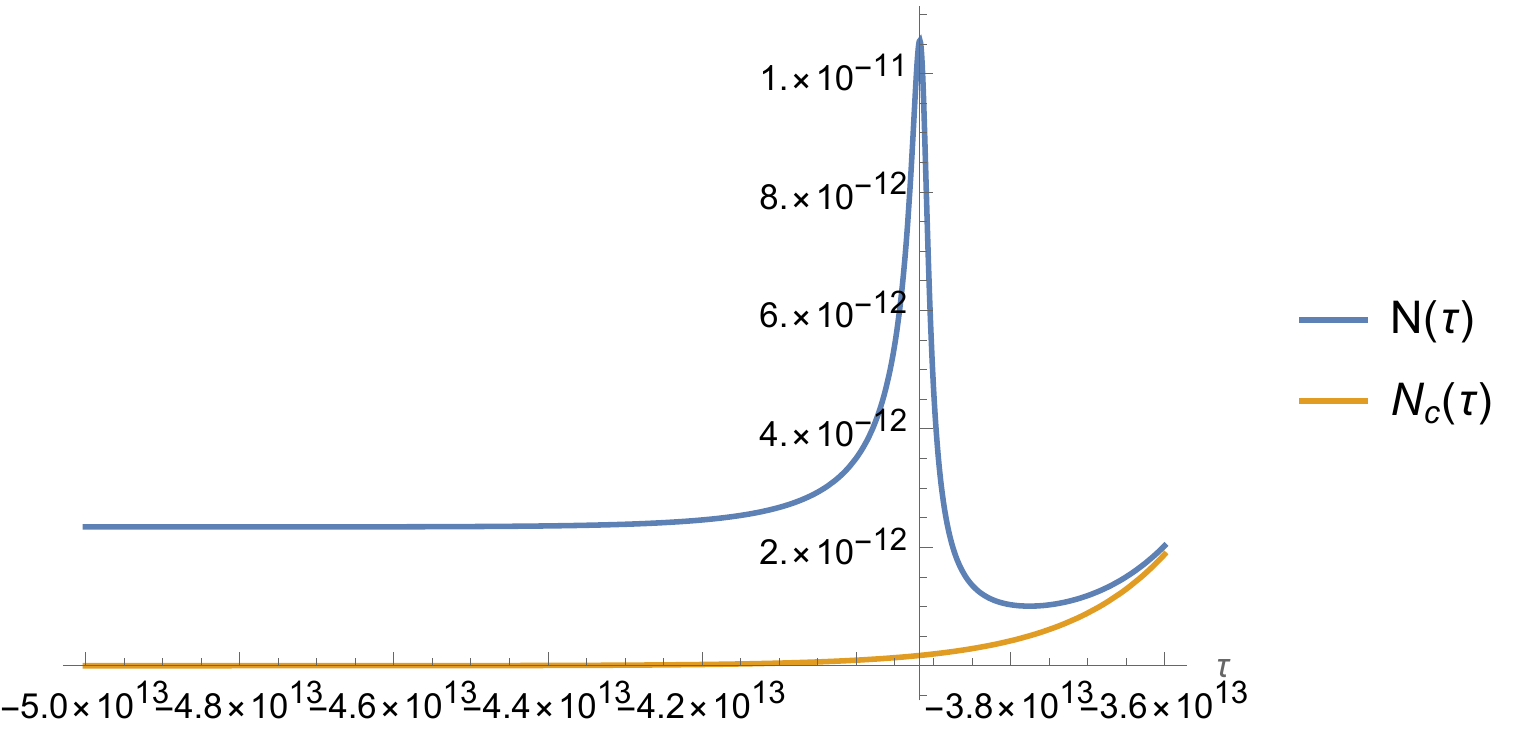}&
\includegraphics[height=4cm]{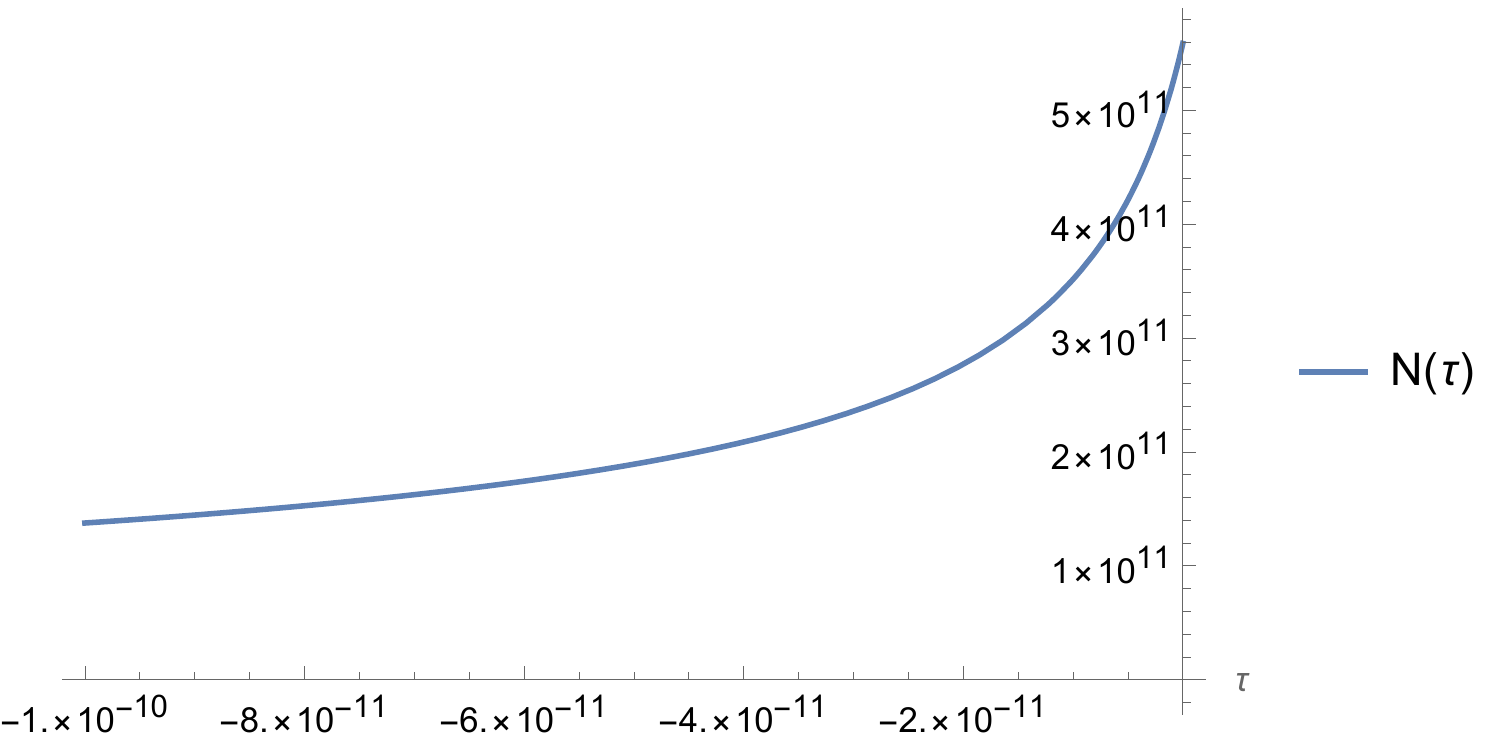}\\		
	(a) & (b)  \\[6pt]
	\includegraphics[height=4cm]{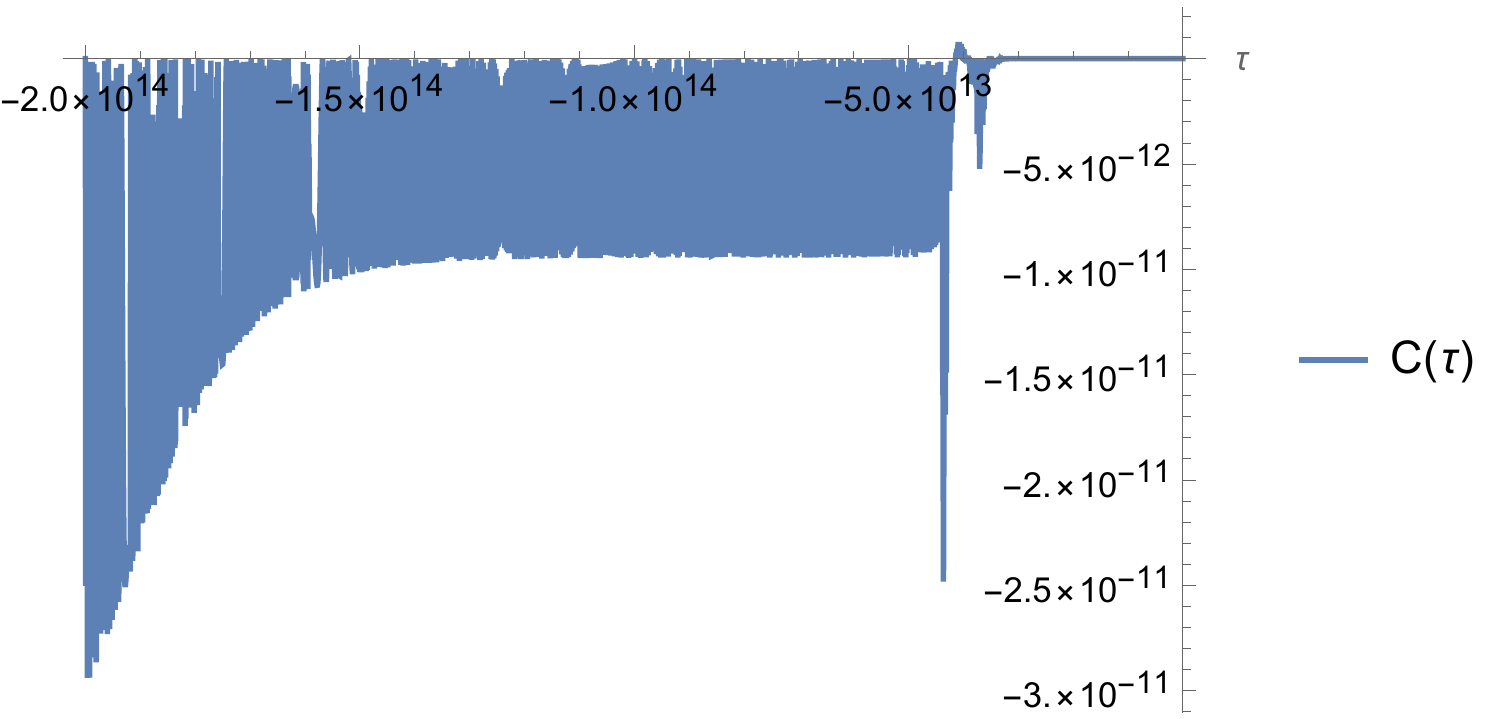}&
\includegraphics[height=4cm]{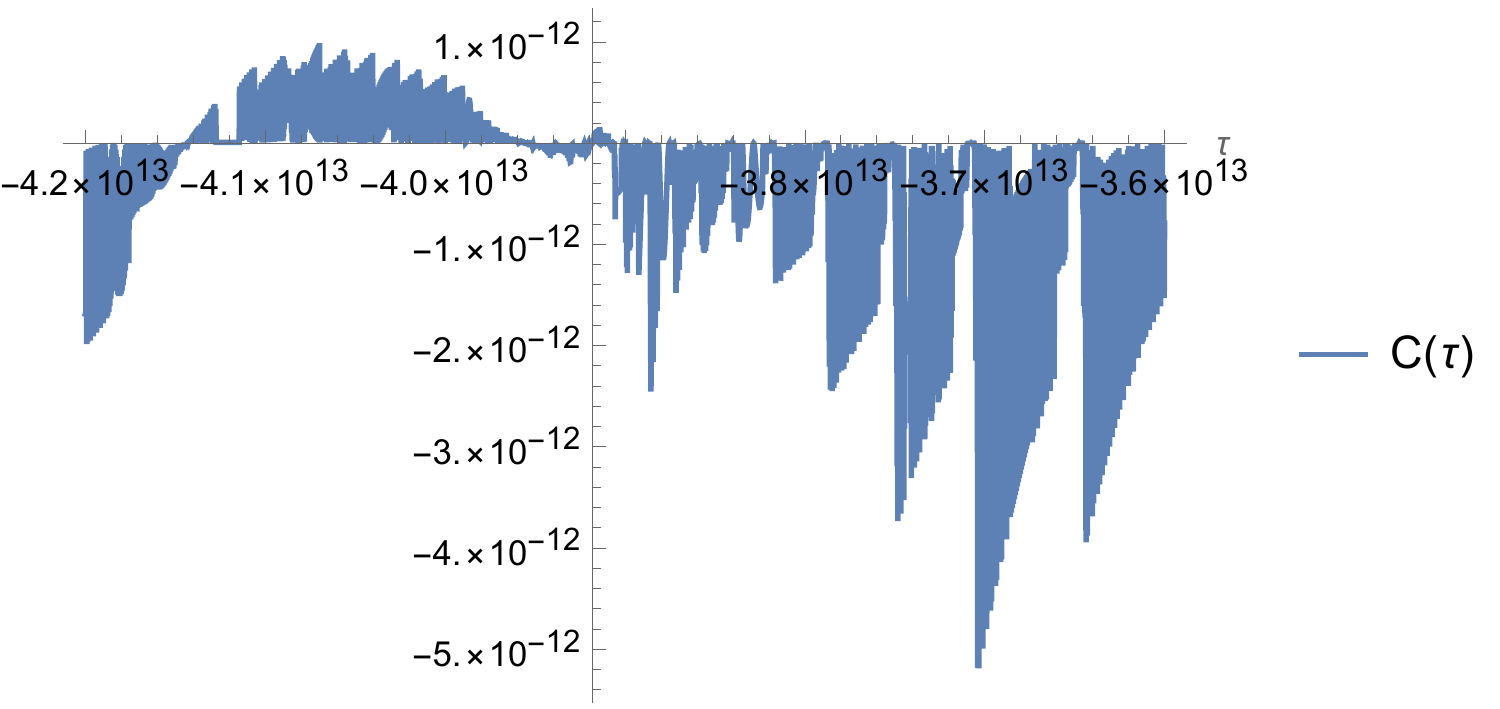}\\
	(c) & (d)  \\[6pt]
		\end{tabular}
\caption{Plots of $\mathcal{C}(\tau)$ and the lapse function $N(\tau)$ for 
 $m=10^{12} m_p , \; j_0=9.5, \; j=10, \; \eta=0.95$.
} 
\lb{fig25}
\end{figure} 

\begin{figure}[h!]
 \begin{tabular}{cc}
		\includegraphics[height=4cm]{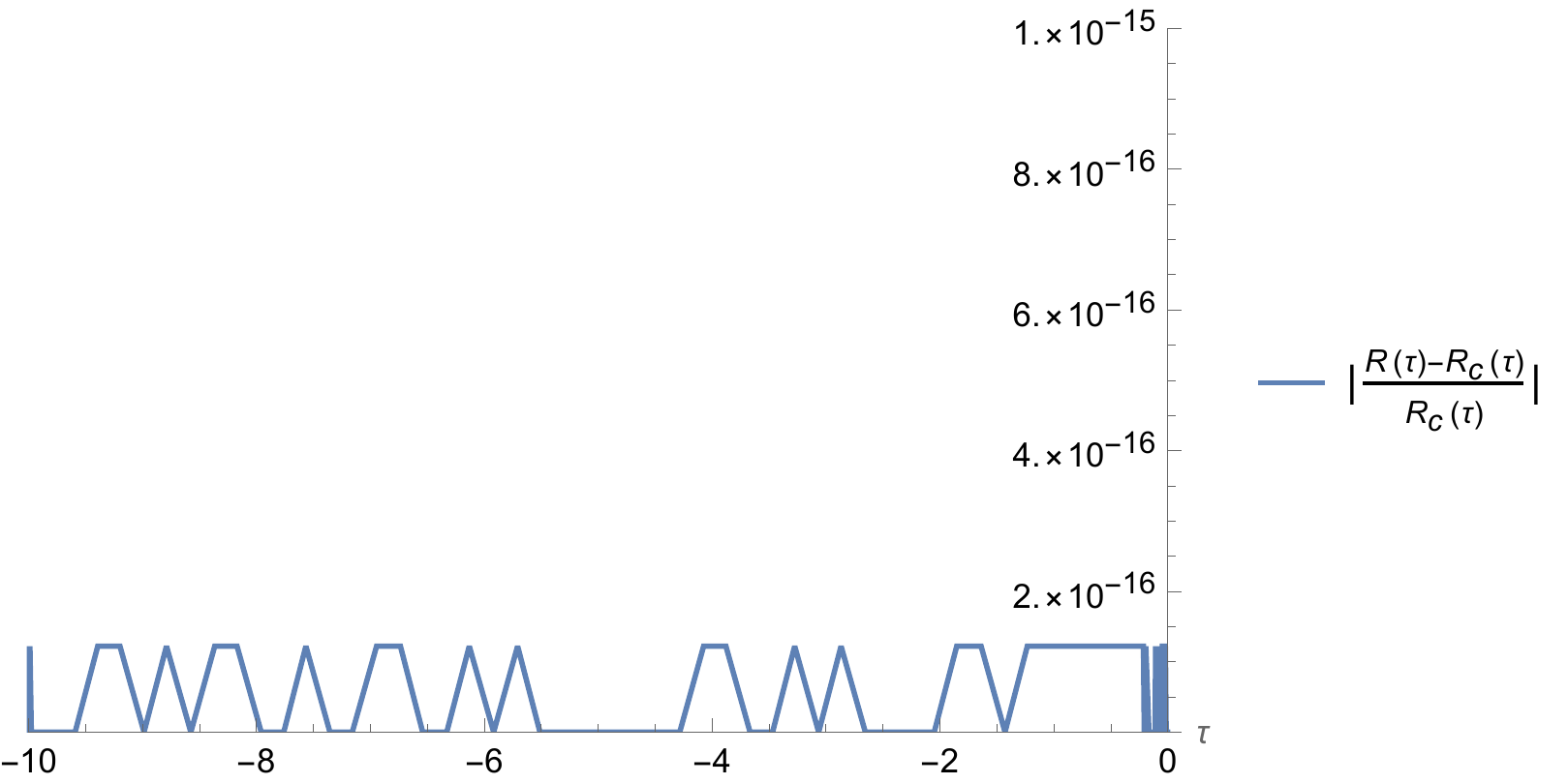}&
\includegraphics[height=4cm]{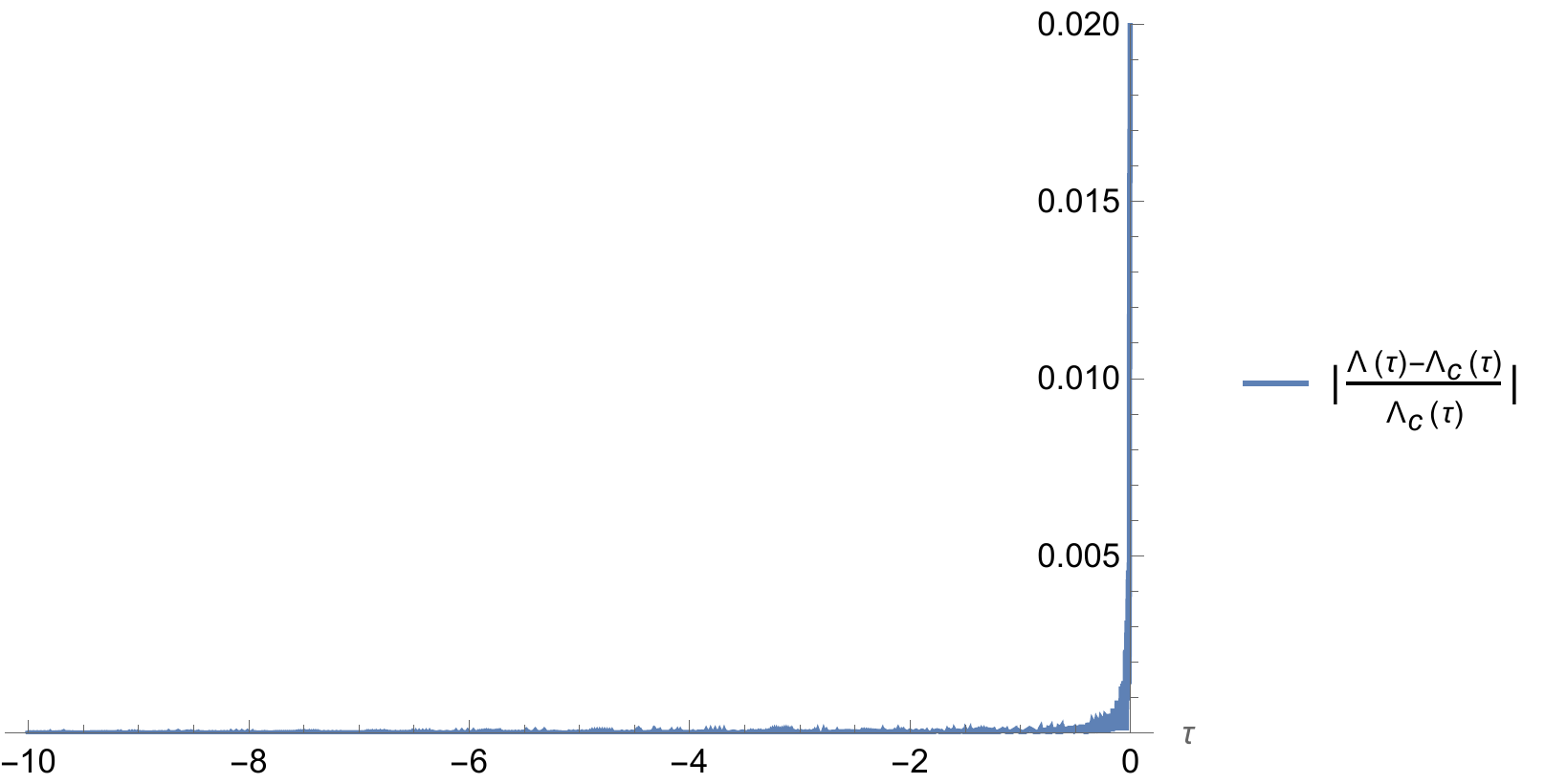}\\
	(a) & (b)  \\[6pt]
		\includegraphics[height=4cm]{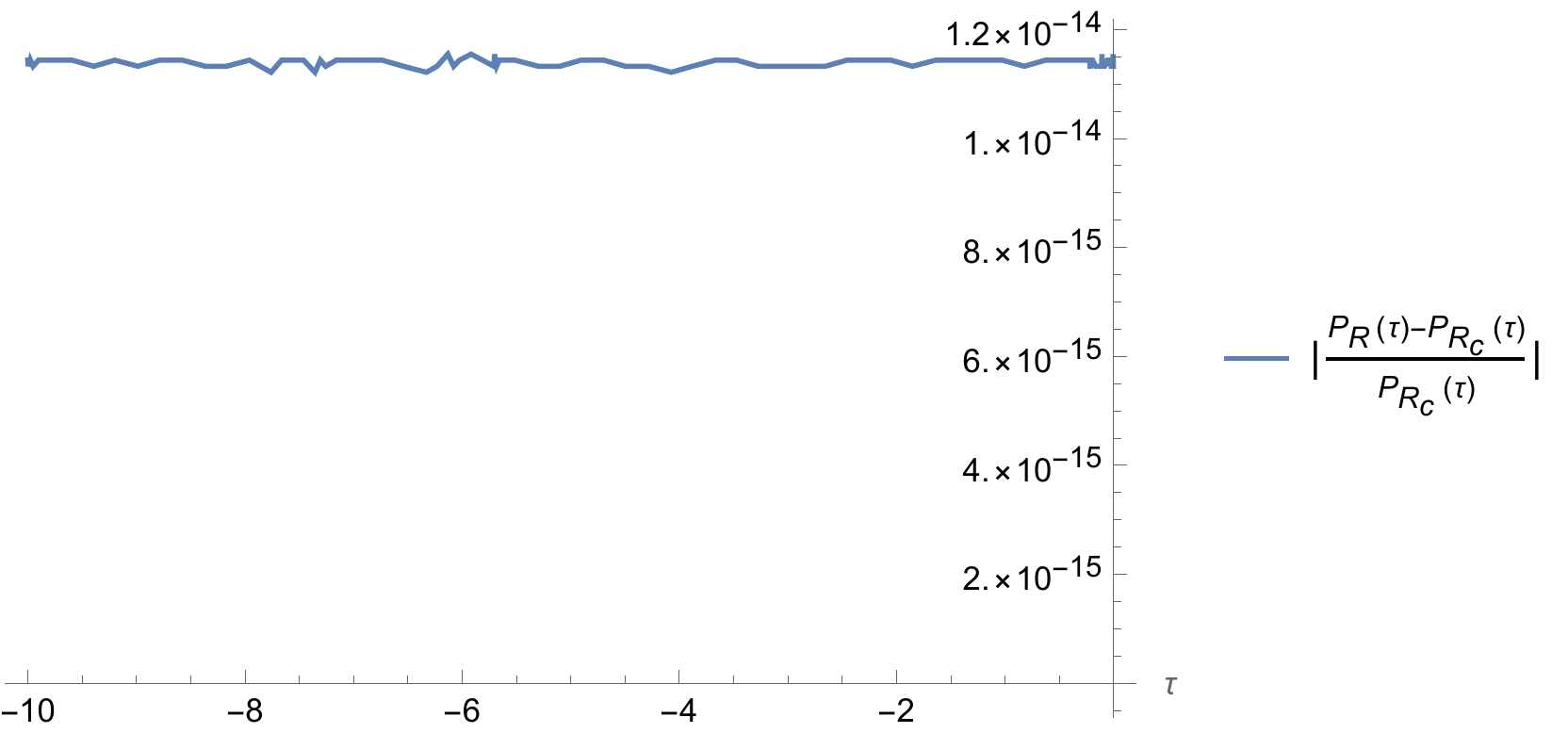}&
\includegraphics[height=4cm]{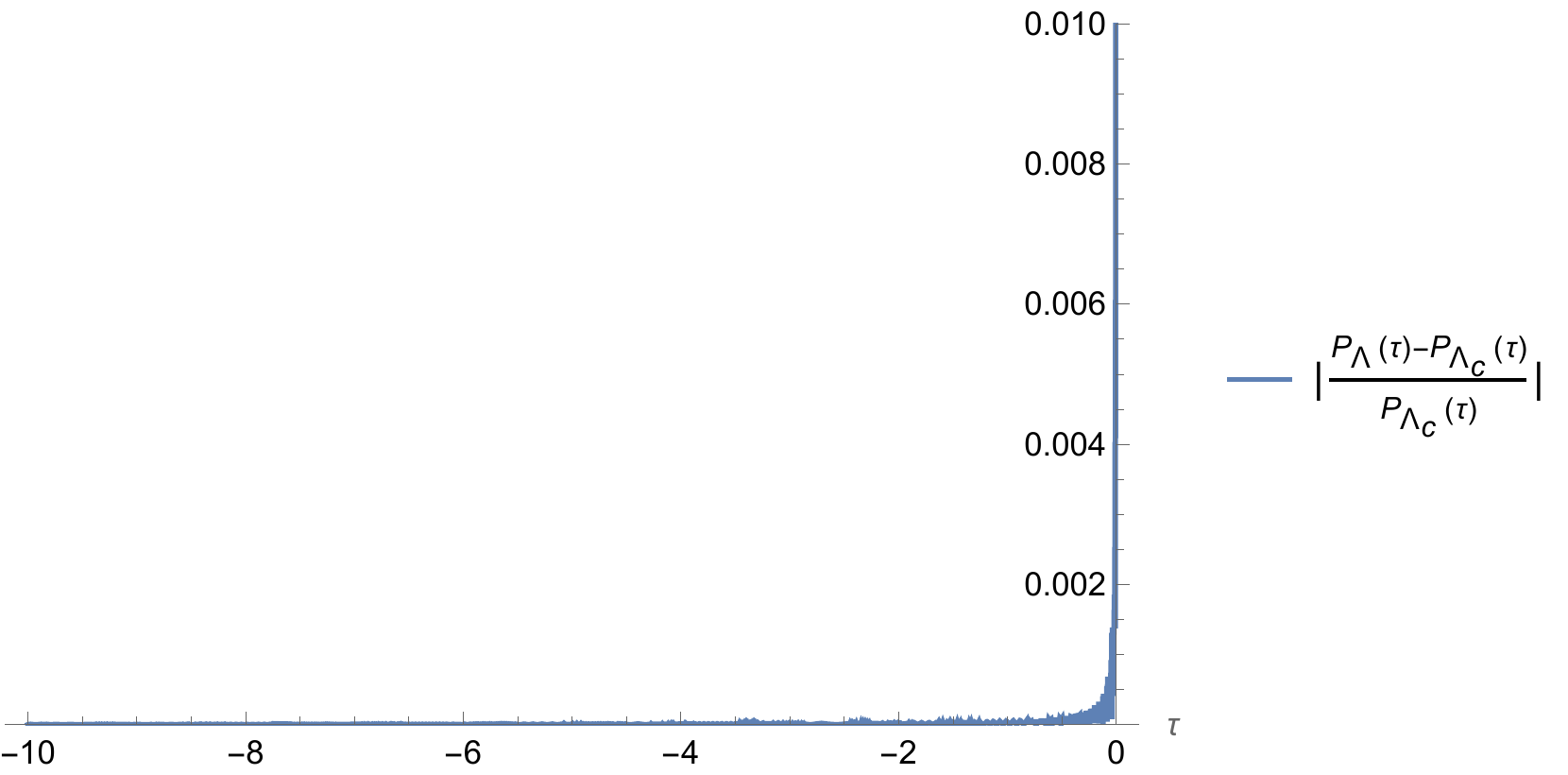}\\
	(c) & (d)  \\[6pt]  
	\includegraphics[height=4cm]{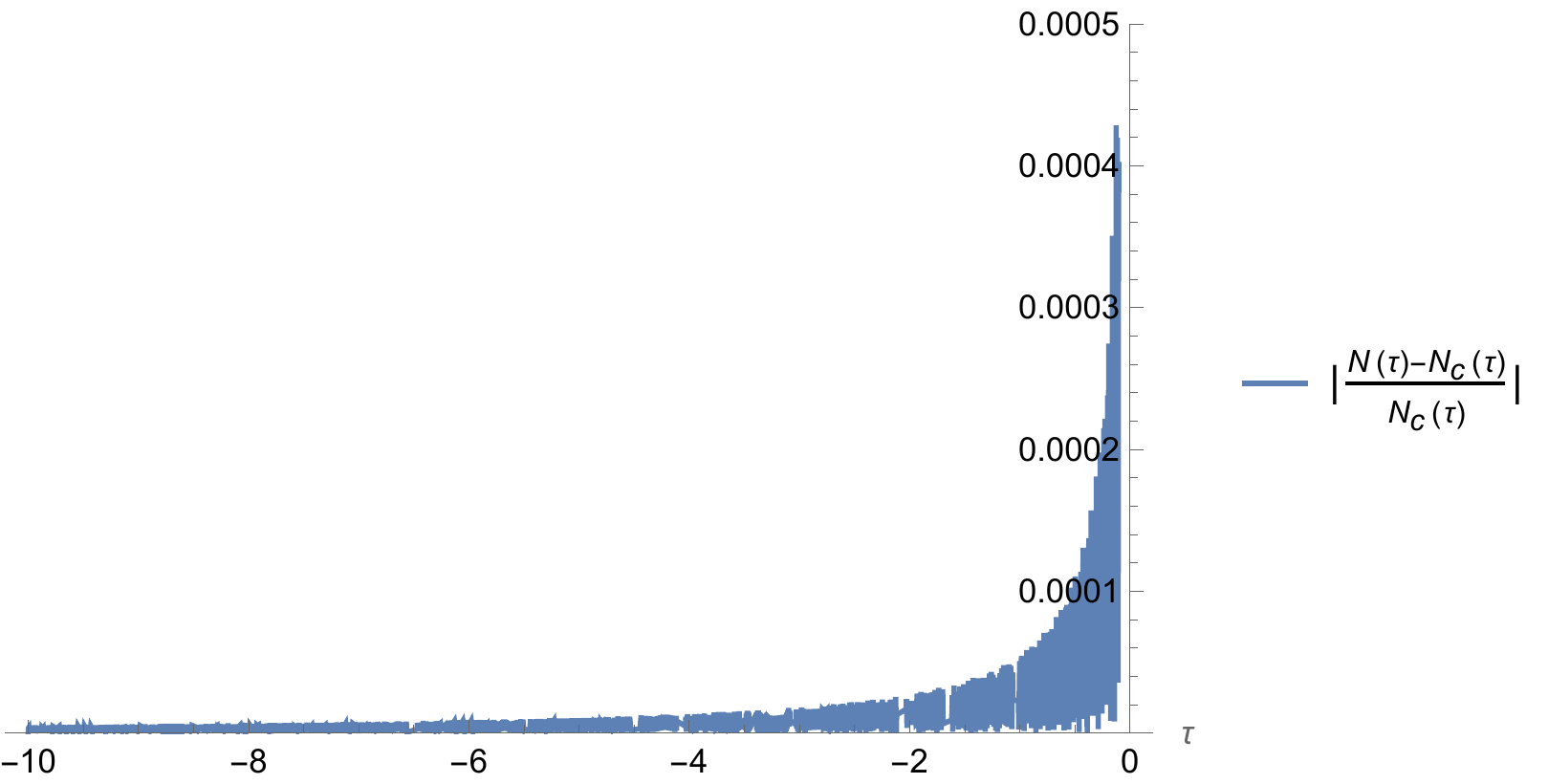}&
	\includegraphics[height=4cm]{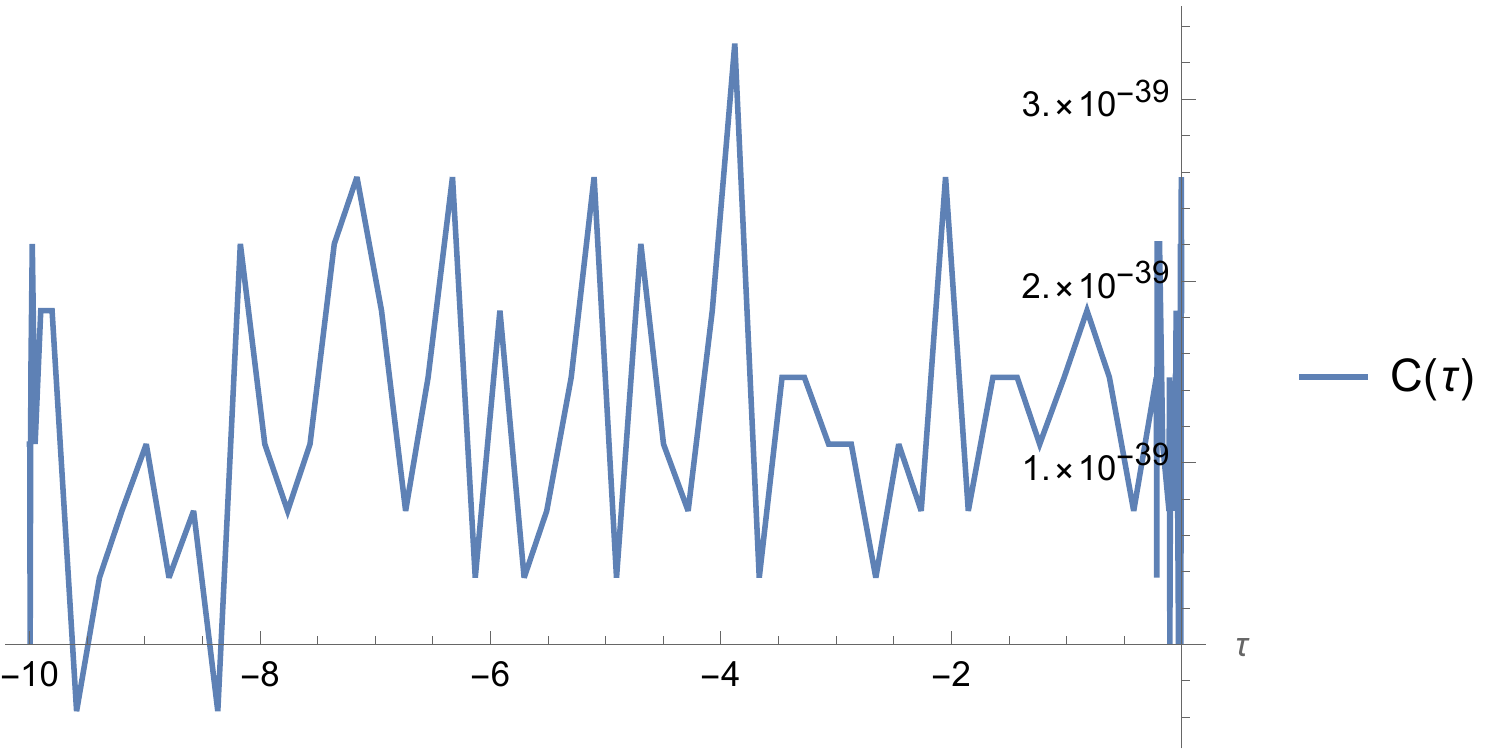}\\
	(e) & (f)  \\[6pt] 
		\end{tabular}
\caption{Plots of the relative  differences of the functions $\left(R, \Lambda, P_R, P_{\Lambda}, N(\tau)\right)$ and $\mathcal{C}(\tau)$ near the black hole horizon with the same choice of the parameters $m$ and $j$, as those specified in Figs. \ref{fig24} and \ref{fig25}, that is,   $m=10^{12} m_p , \; j_0=9.5, \; j=10, \; \eta=0.95$.
} 
\lb{fig26}
\end{figure} 		

 \end{widetext}

  \subsection{$\eta \lesssim 1$}

  When   $\eta \lesssim 1$, the metric coefficients take the same  asymptotical forms as  {those} given by Eqs.(\ref{eq2.5j}) - (\ref{eq2.5l}), but now with ${\cal{F}}(\eta) > 0$ and $\eta_0 > - \pi$   \cite{ABP19}.
Therefore, now  $\Lambda$ decreases exponentially as
$\tau  \rightarrow -\infty$, while $R$ still keeps increasing exponentially, i.e.
\bqn
\lb{eq2.5u}
N  &\simeq& - \frac{2\gamma \sqrt{8\pi \gamma} \; \ell_p \; \sqrt{ j_0},}{mG \left(\pi h_0[\eta_0]+2\sin[\eta_0]\right)},  \nb\\
 \Lambda &=& \Lambda_0 e^{ a\tau}, \quad R = R_0 e^{-d\tau}.
\eqn
 Then, the metric takes the following asymptotical form 
\bq
\lb{eq2.5v}
ds^2 \simeq - \left(\frac{\hat{N}_0}{R}\right)^2dR^2 +  \frac{d\bar x^2}{R^{{2a}/{d}}} +  R^2d\Omega^2.
\eq
The corresponding   effective
energy-momentum tensor also takes the same form as that given by Eq.(\ref{eq2.5g}), but now with $u_{\mu} = (\hat{N}_0/R)\delta^{R}_{\mu}$, $\bar x_{\mu} =  R^{-a/b}\delta^{\bar x}_{\mu}$,   and
\bqn
\lb{eq2.5w}
\rho &\simeq& \frac{ d - 2 a}{d \hat{N}_0^2} - \frac{1}{R^2}, \nb\\
p_{\bar x}  &\simeq&  -\frac{3}{\hat{N}_0^2}-\frac{1}{R^2}, \nb\\
 p_{\bot} &\simeq& -\frac{a^2-a d+d^2}{d^2 \hat{N}_0^2},
 \eqn
 from which we can see that none of the three energy conditions are satisfied for any given $a$ and $d$. In particular, when $a = d$ we have
 $\rho \simeq p_{\bar x}/3 \simeq p_{\bot} < 0$. In addition, we also have
\begin{widetext}
\bqn
\lb{eq2.5x}
&& {\cal{R}}  \simeq  2 \left(\frac{a^2-2 a d+3 d^2}{d^2 \hat{N}_0^2}+\frac{1}{R^2}\right), \nb\\
&& R_{\mu\nu} R^{\mu\nu} 
 \simeq 2\frac{a^4-2 a^3 d+5 a^2 d^2-4 a d^3+6 d^4}{d^4 \hat{N}_0^4}  -\frac{4 (a-2 d)}{d \hat{N}_0^2 R^2}+\frac{2}{R^4}, \nb\\
 && R_{\mu\nu\alpha\beta} R^{\mu\nu\alpha\beta} \simeq 4 \left(\frac{a^4+ 2 a^2d^2 +3d^4}{d^4 \hat{N}_0^4}+\frac{2}{\hat{N}_0^2 R^2}+\frac{1}{R^4}\right), \nb\\
&& C_{\mu\nu\alpha\beta} C^{\mu\nu\alpha\beta} \simeq \frac{4 \left(a R^2 (a+d)+d^2 \hat{N}_0^2\right){}^2}{3 d^4 \hat{N}_0^4 R^4},
\eqn
which can be obtained from Eq.(\ref{eq2.5i}) by the replacement $ a \rightarrow -a$, as  expected. 
 \end{widetext}

To consider the corresponding Penrose diagram, we first write the metric (\ref{eq2.5v}) in the form 
\bq
\lb{eq2.5y}
ds^2 \simeq -R_0^{-2a/d}\left(\frac{\bar t_0}{\bar t}\right)^2   \left( - d\bar t^2  +  d\bar x^2\right)  +  R^2d\Omega^2,
\eq
where   
\bqn
\lb{eq2.5z}
\bar t &=& \bar t_0 \left(\frac{R}{R_0}\right)^{a/d}, \quad \bar{x} \equiv \left(\Lambda_0R_0^{a/d}\right)x, \nb\\
R &=& R_0\left(\frac{\bar t}{\bar t_0}\right)^{d/a}, \;\; \bar t_0 \equiv \frac{d \hat N_0 R^{a/d}_0}{a}.
\eqn
Comparing Eq.(\ref{eq2.5y}) with Eq.(\ref{eq2.5r}),  we find that the ($\bar t, \bar x$)-planes in both spacetimes have the same structure, and the only difference is to replace $a$ by $-a$.
Thus, the corresponding Penrose diagram is also given by Fig. \ref{fig7}. It is interesting to note that now the spacetime is not asymptotically de Sitter, even when $a = d$. In fact,  now  it is even not asymptotically conformally flat as can be seen from Eq.(\ref{eq2.5x}). In addition, in the current case none of the three energy conditions are satisfied.

  In Figs. \ref{fig24} - \ref{fig26}, we plot various physical quantities for $m/m_p = 10^{12}, \; j_0 = 9.5, \; j = 10$ so that $\eta \equiv j_0/j = 0.95 < 1$. In this case, the transition surface is located at $\tau_{\text{min}} =-3.918 \times 10^{13}$, at which we find $R(\tau_{\text{min}})=7676.1$. Then, it can be shown that both of the conditions 
  (\ref{eq2.25v1}) and (\ref{eq2.5d1}) are satisfied. Therefore, the corresponding semiclassical description of the quantum black holes is well justified. In particular, from Figs. \ref{fig24} 
  and  \ref{fig25} we find that the asymptotic behavior of the metric coefficients are well approximated by Eq.(\ref{eq2.5u}), while Fig. \ref{fig26} shows that near the horizon
  ($\tau \simeq 0$) the quantum geometric effects become negligible, possibly except  the region very near to the horizon [cf. Fig. \ref{fig26}].
  
  It is interesting to note that the asymptotic behavior in the current case is very sensitive to the choice of $\eta$. In particular, we find that when $\eta = 0.5$ the asymptotic behavior of the spacetime is already quite different from the one described by  Eq.(\ref{eq2.5u}), although   the semiclassical  conditions 
  (\ref{eq2.25v1}) and (\ref{eq2.5d1}) are still well justified.

 \begin{widetext}
  
 \begin{figure}[h!]
 \begin{tabular}{cc}
		\includegraphics[height=4cm]{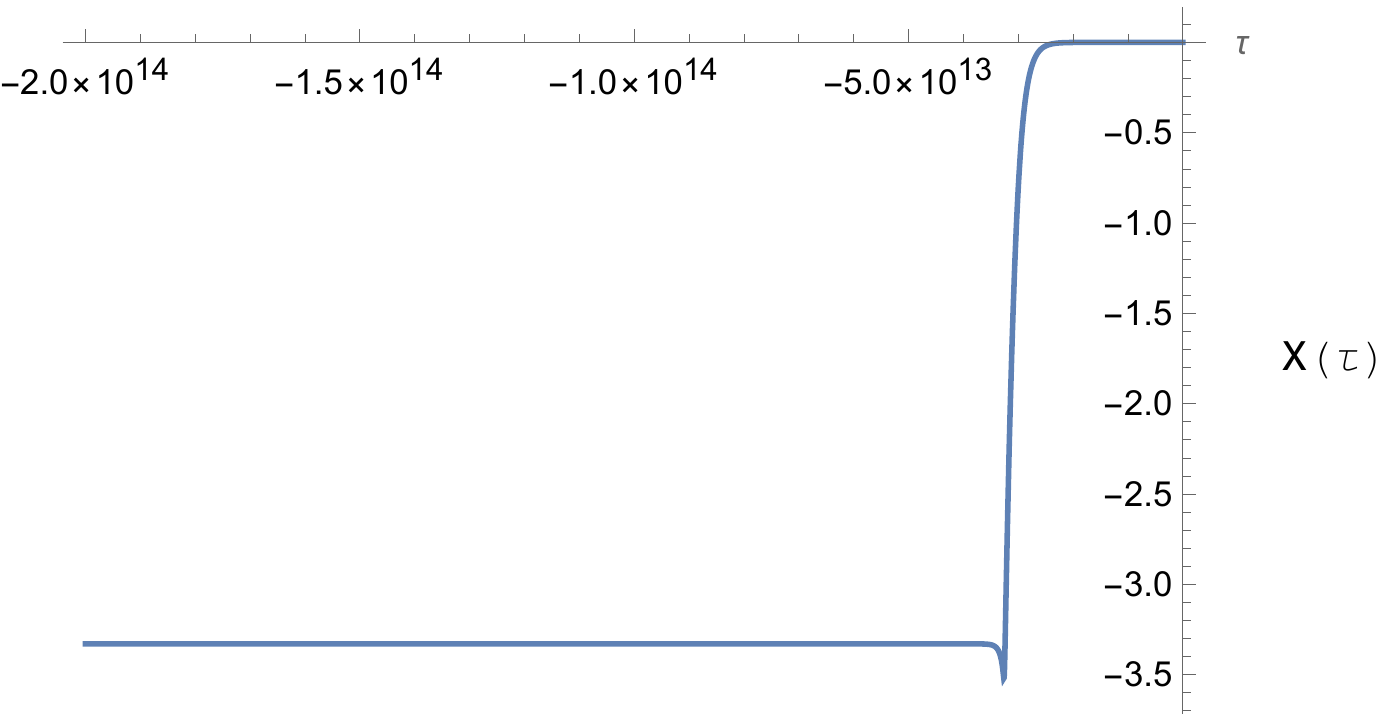}&
\includegraphics[height=4cm]{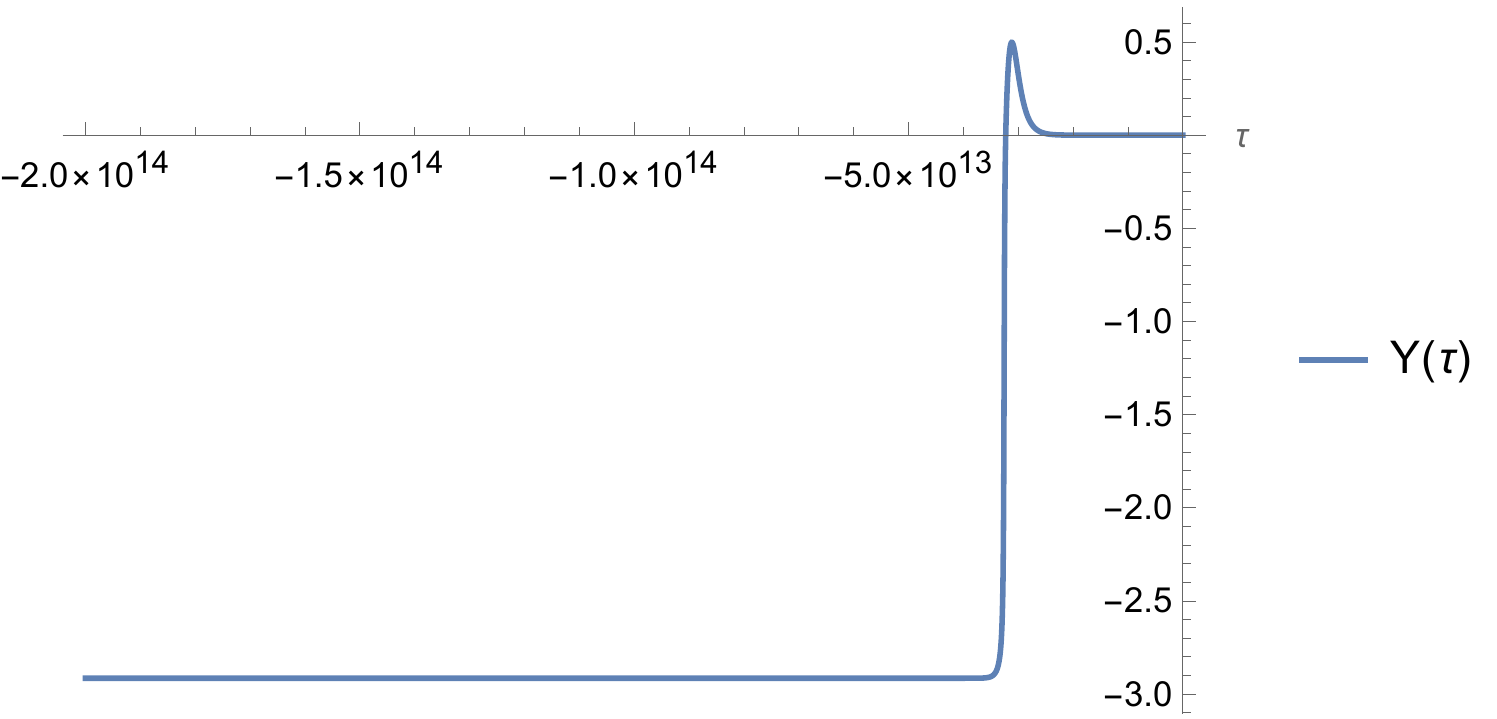}\\
	(a) & (b)  \\[6pt]
		\includegraphics[height=4cm]{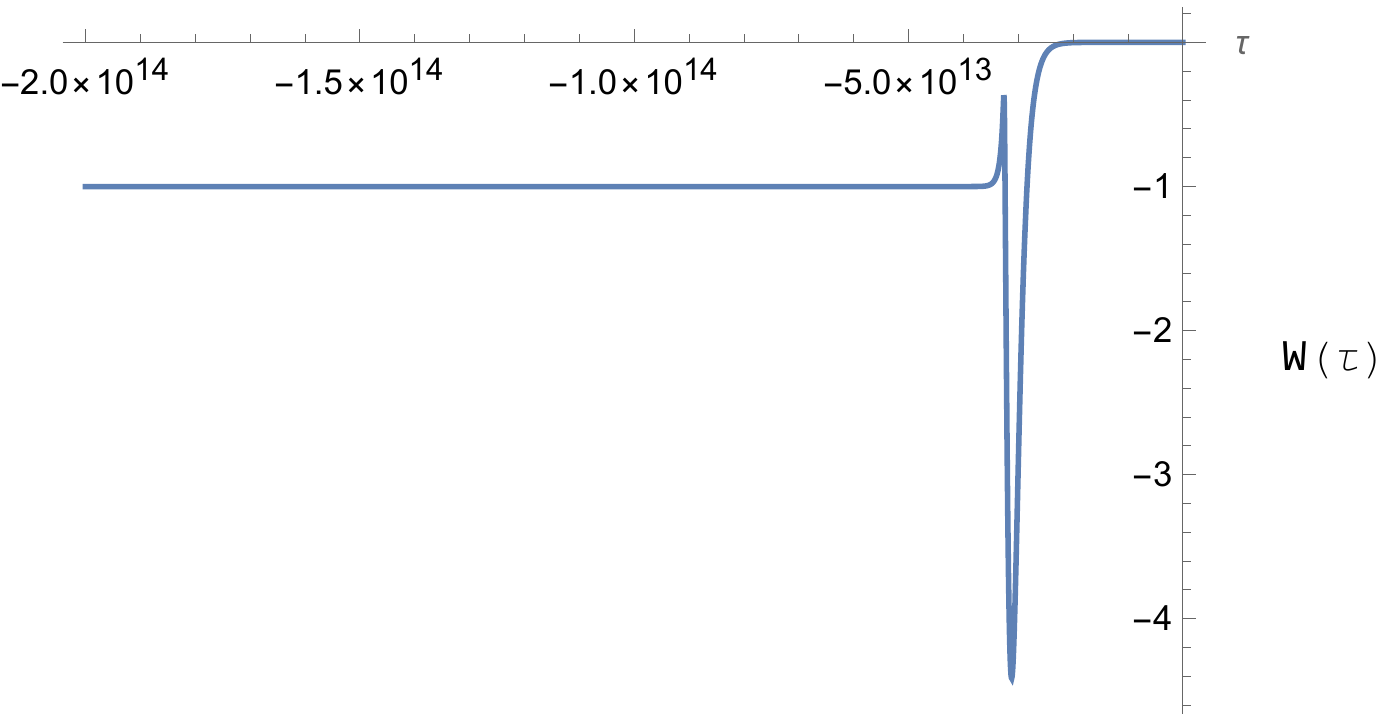}&
\includegraphics[height=4cm]{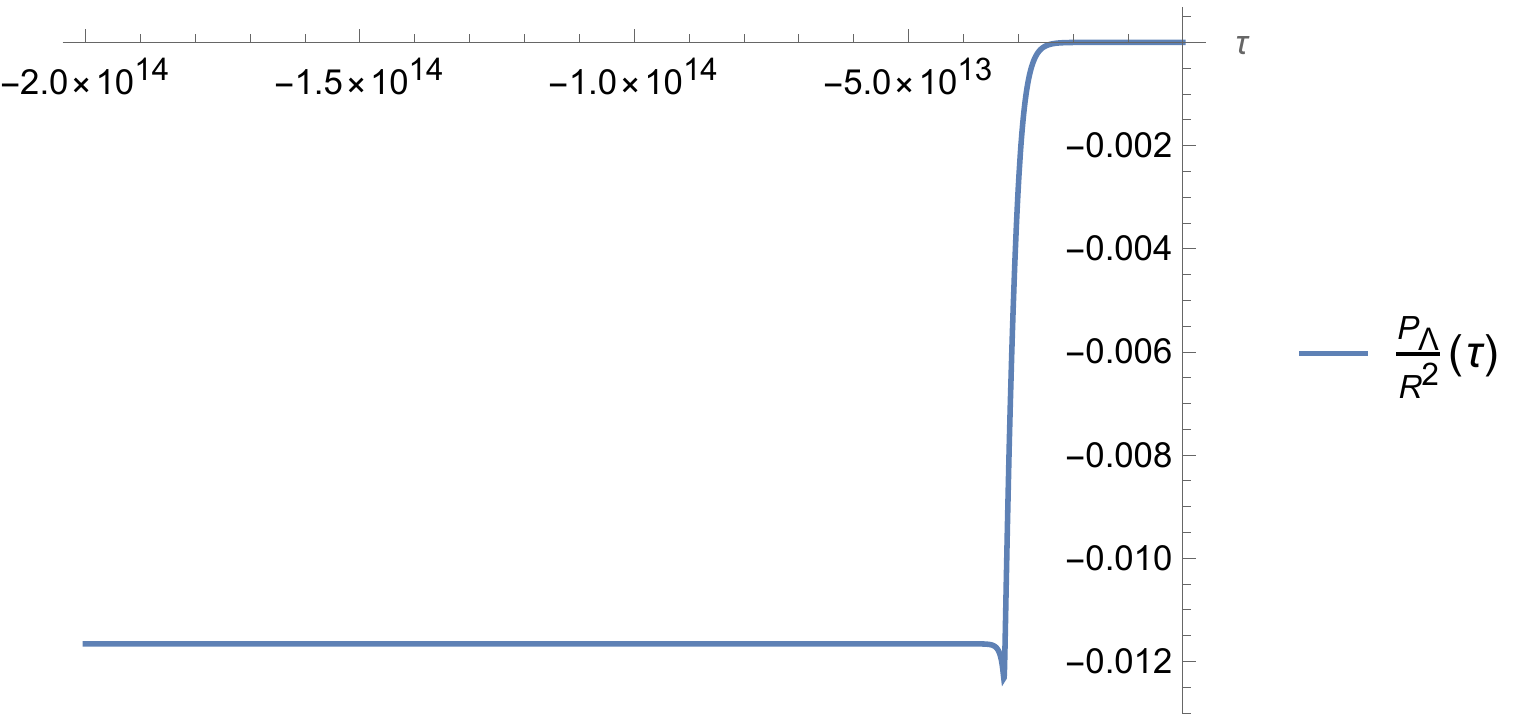}\\
	(c) & (d)  \\[6pt]  
	\includegraphics[height=4cm]{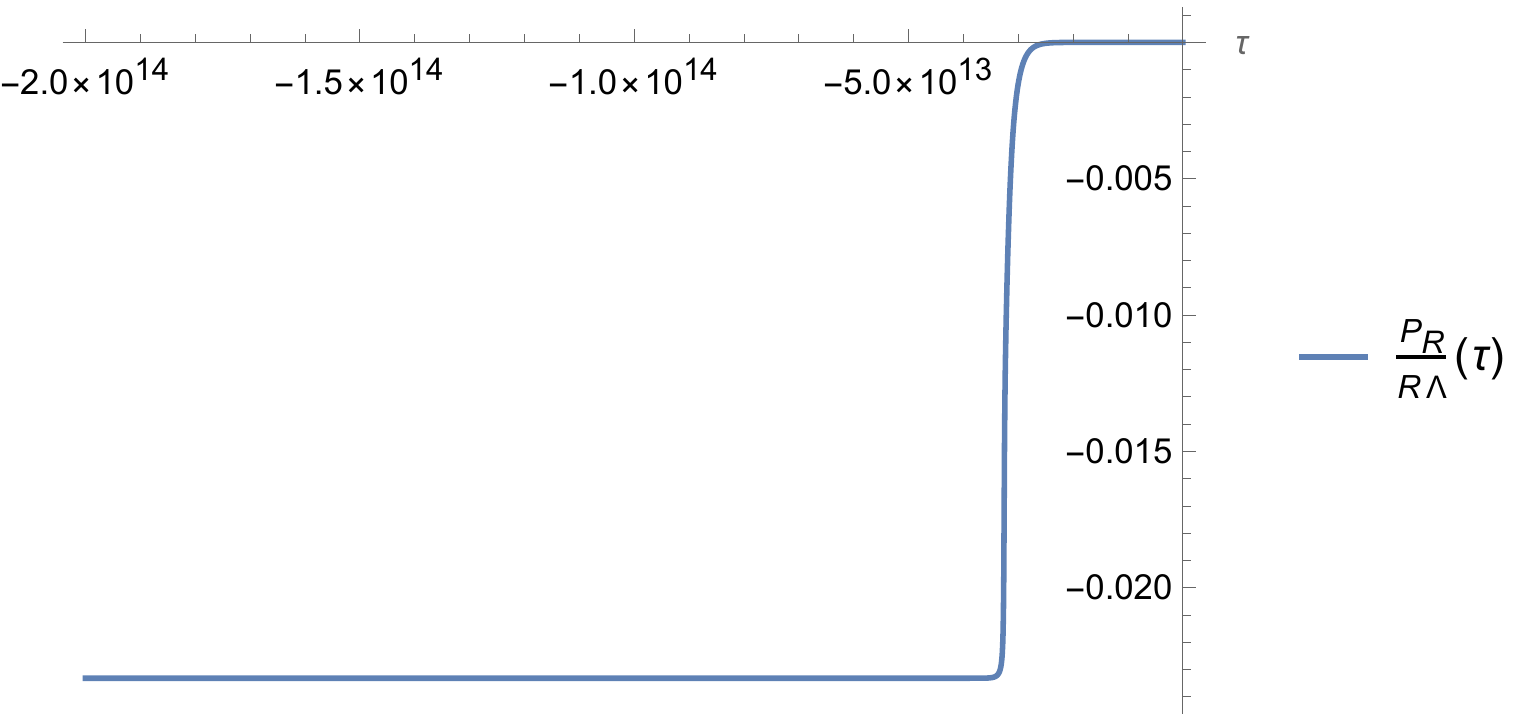}\\
	(e)   \\[6pt] 
		\end{tabular}
\caption{Plots of  the functions $\left(X, Y, W, \frac{P_\Lambda}{R^2}, \frac{P_R}{R \Lambda}\right)$. The throat is lcated at $\tau_{\text{min}} =-3.260 \times 10^{13}$, at which $R(\tau_{\text{min}})=193115$. 
Curves are plotted with $\gamma \approx 0.274, \; m=10^{12} m_p , \; j_x=10^5, \; \eta \approx 1.142$.
} 
\lb{fig36}
\end{figure} 	

 \begin{figure}[h!]
 \begin{tabular}{cc}
		\includegraphics[height=4cm]{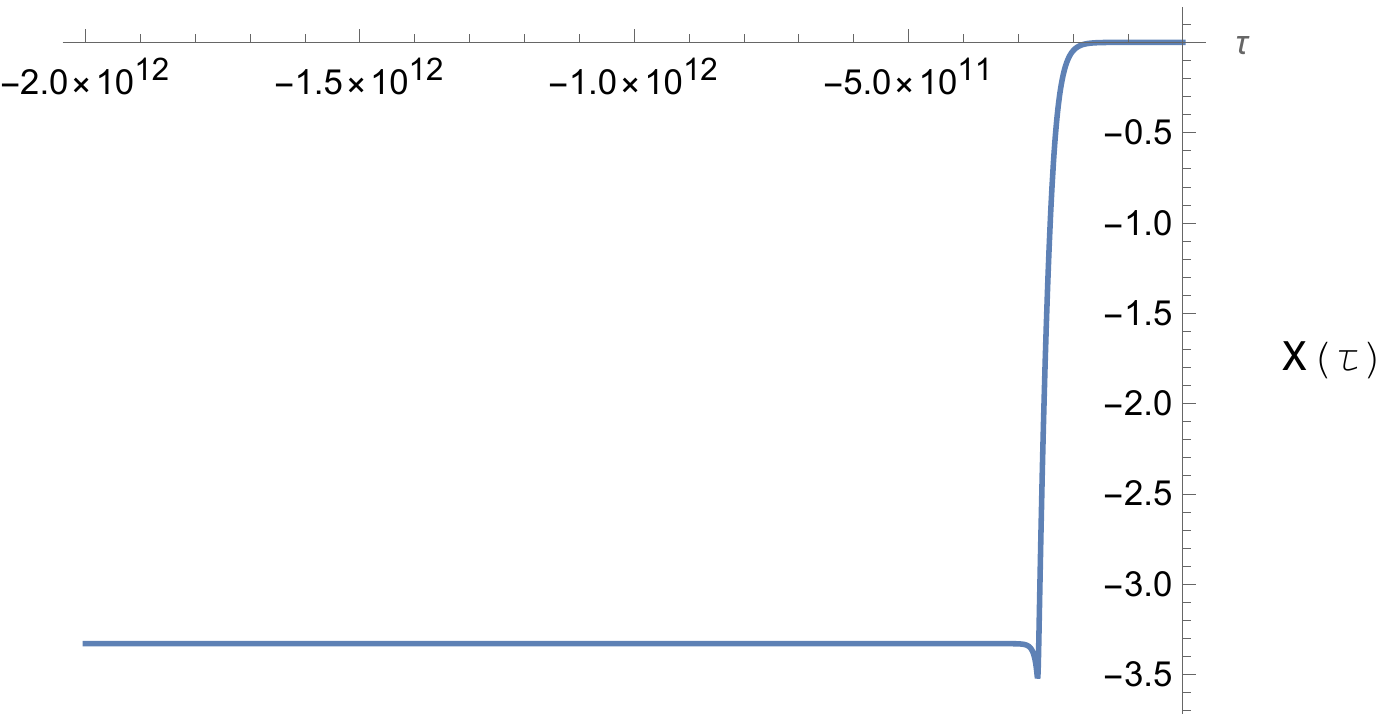}&
\includegraphics[height=4cm]{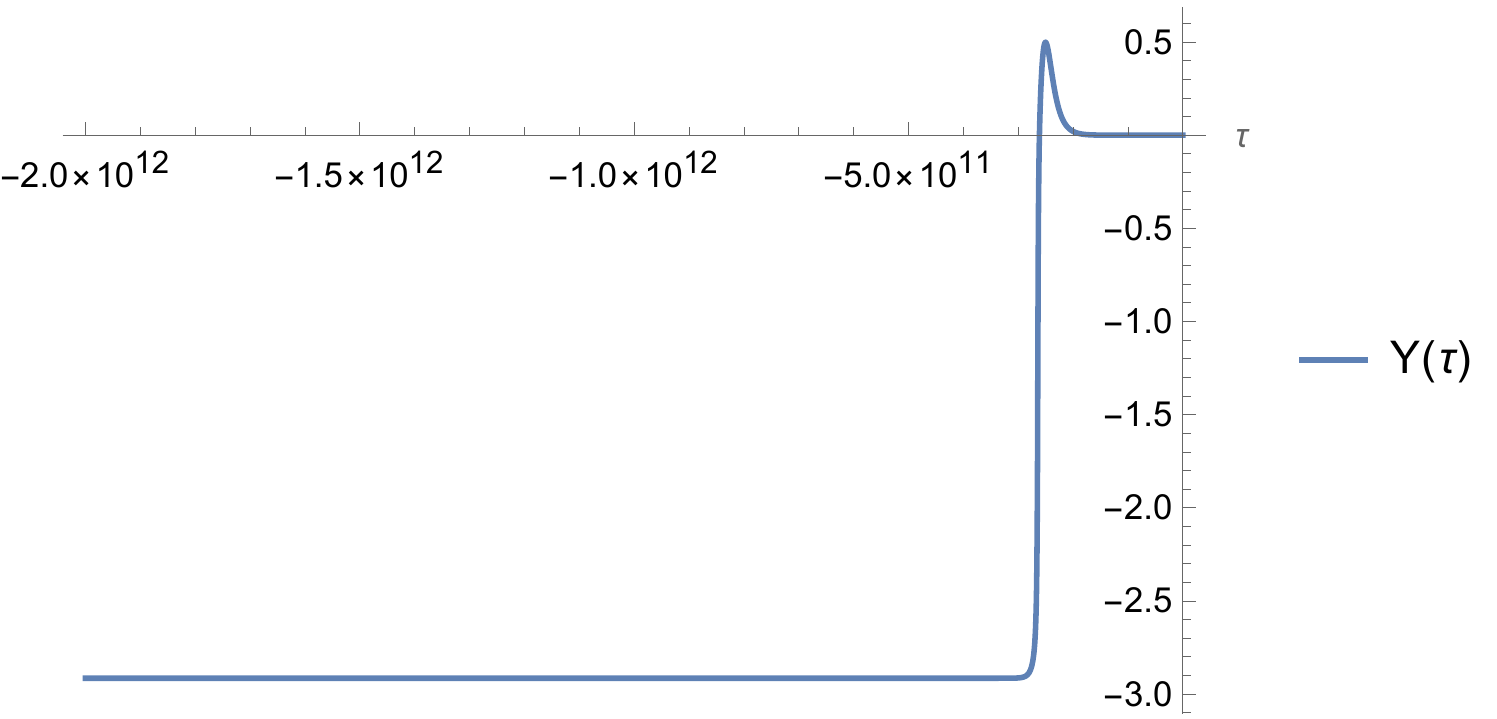}\\
	(a) & (b)  \\[6pt]
		\includegraphics[height=4cm]{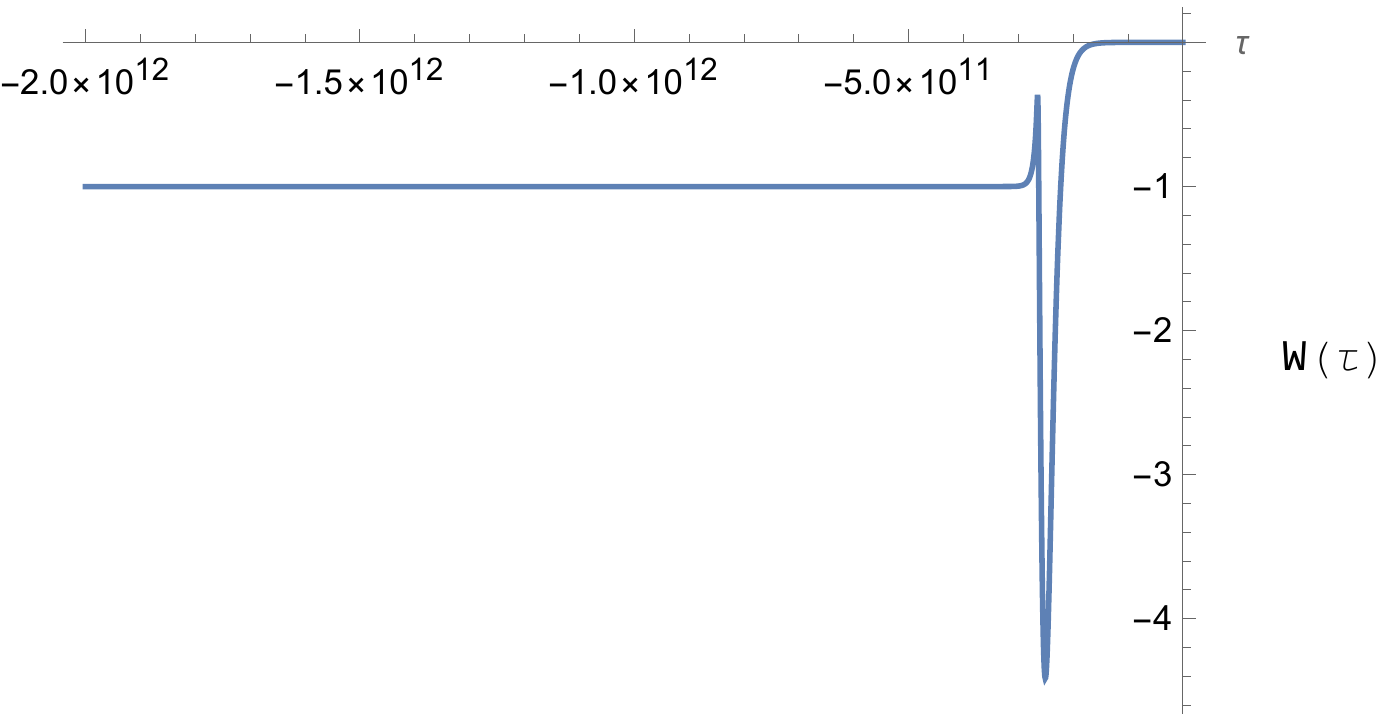}&
\includegraphics[height=4cm]{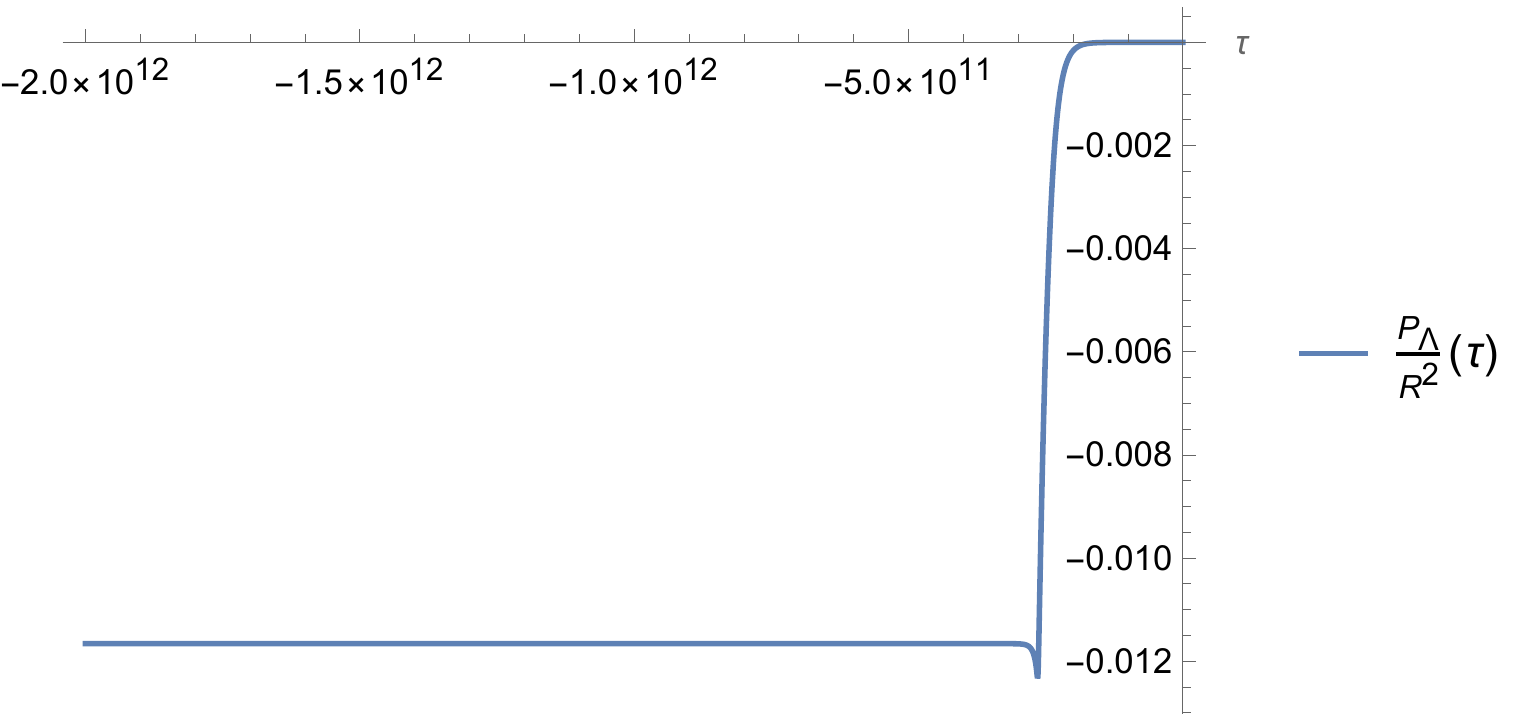}\\
	(c) & (d)  \\[6pt]  
	\includegraphics[height=4cm]{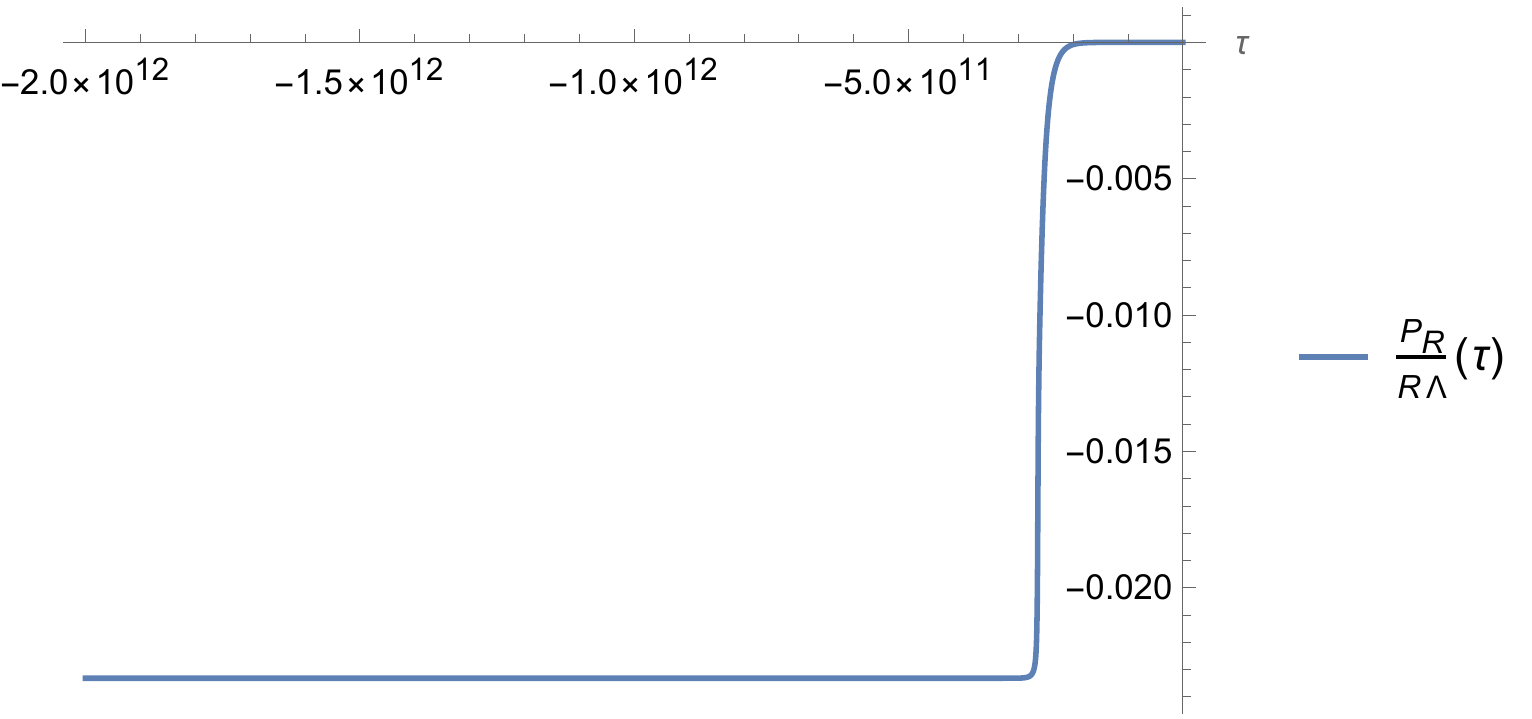}\\
	(e)   \\[6pt] 
		\end{tabular}
\caption{Plots of  the functions $\left(X, Y, W, \frac{P_\Lambda}{R^2}, \frac{P_R}{R \Lambda}\right)$. The throat is at $\tau_{\text{min}} =-2.646 \times 10^{11}$, at which $R(\tau_{\text{min}})=41609.4$. 
Graphs are plotted with $\gamma \approx 0.274, \; m=10^{10} m_p , \; j_x=10^5, \; \eta \approx 1.142$.
} 
\lb{fig37}
\end{figure} 		

 \begin{figure}[h!]
 \begin{tabular}{cc}
		\includegraphics[height=4cm]{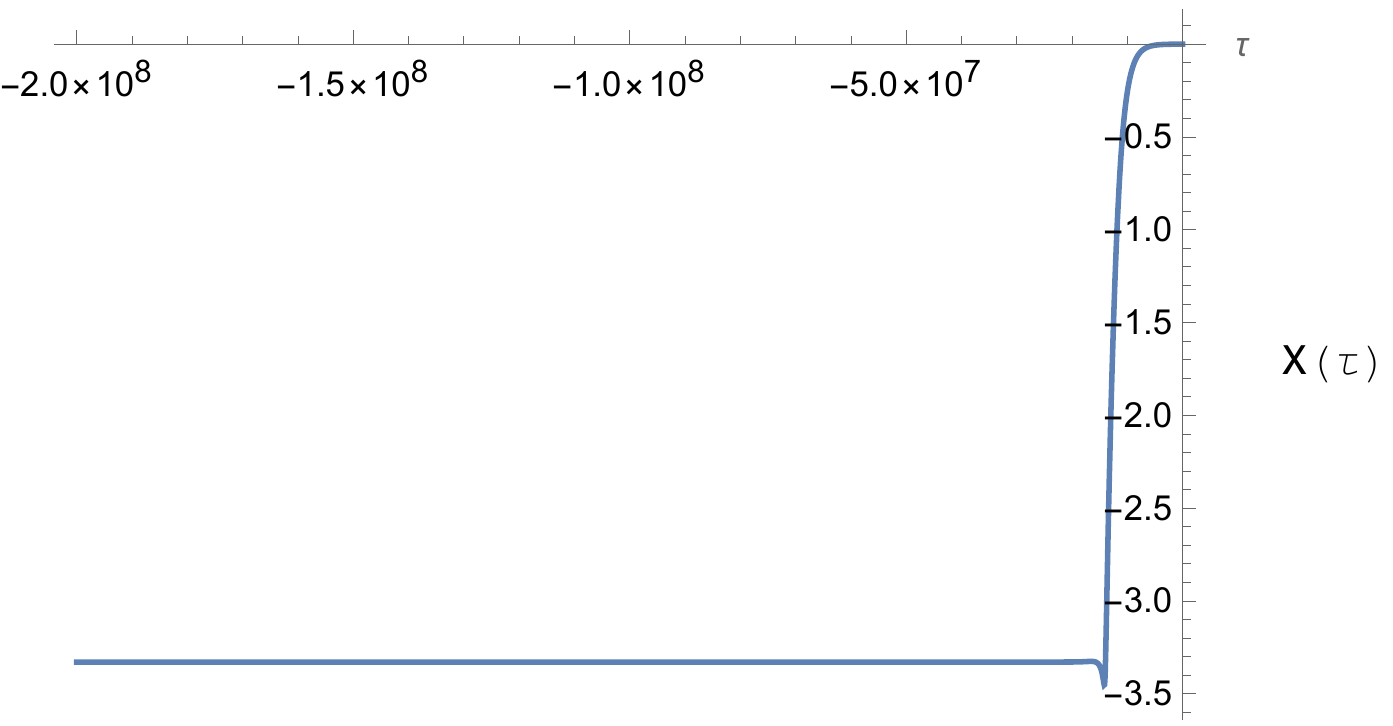}&
\includegraphics[height=4cm]{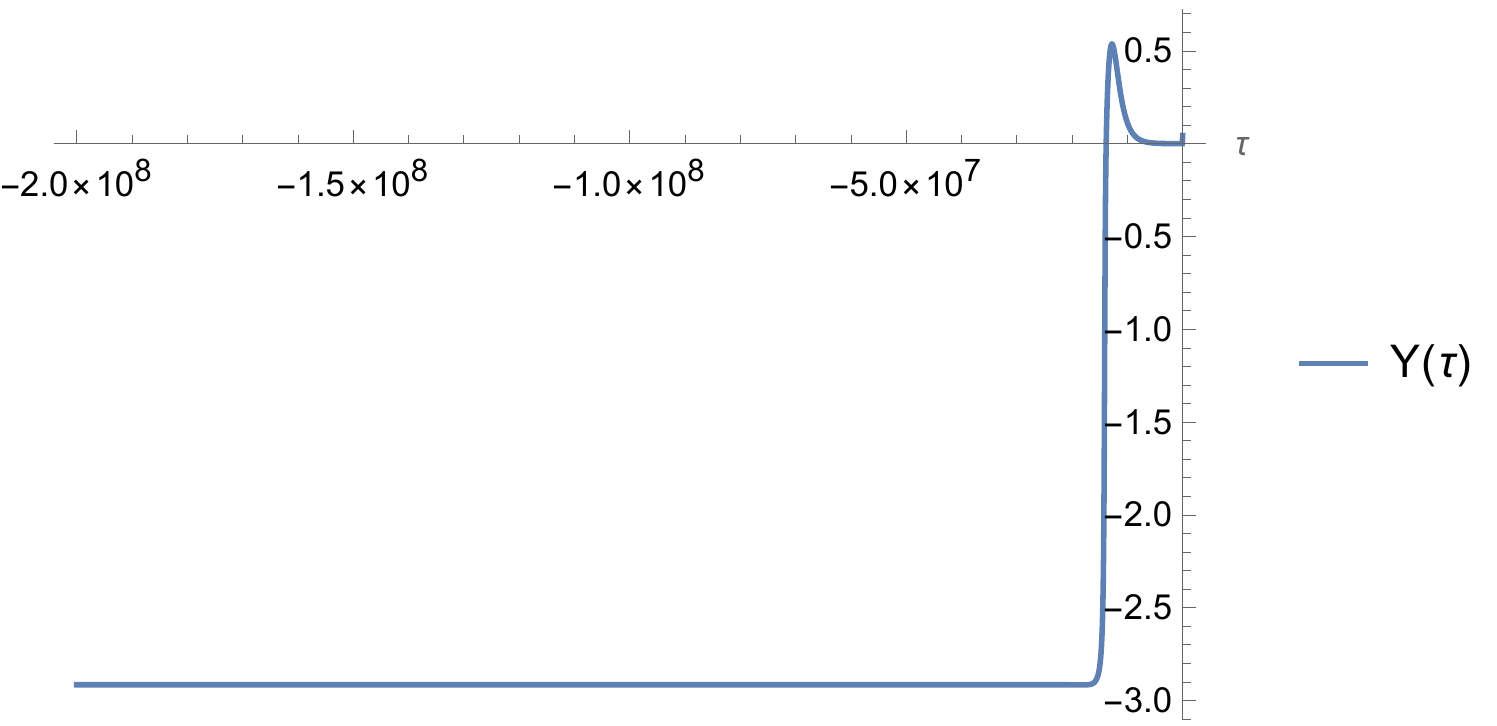}\\
	(a) & (b)  \\[6pt]
		\includegraphics[height=4cm]{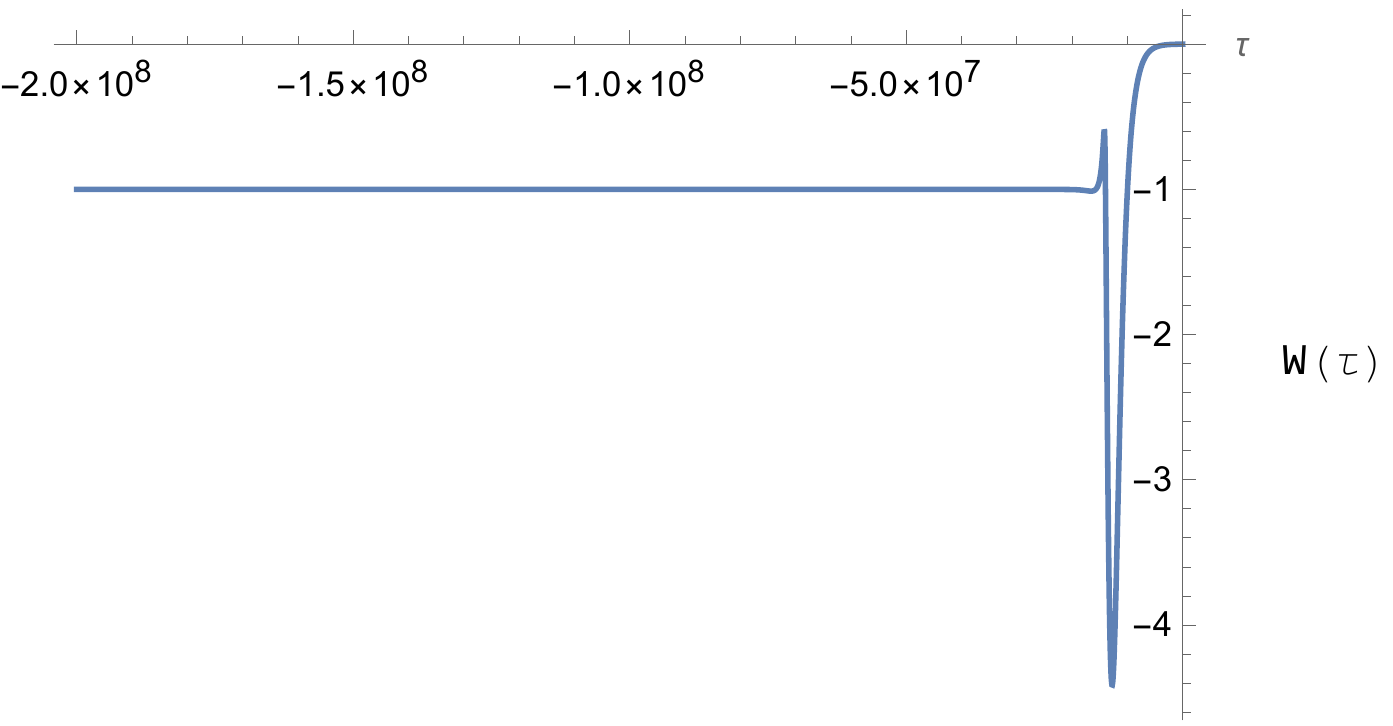}&
\includegraphics[height=4cm]{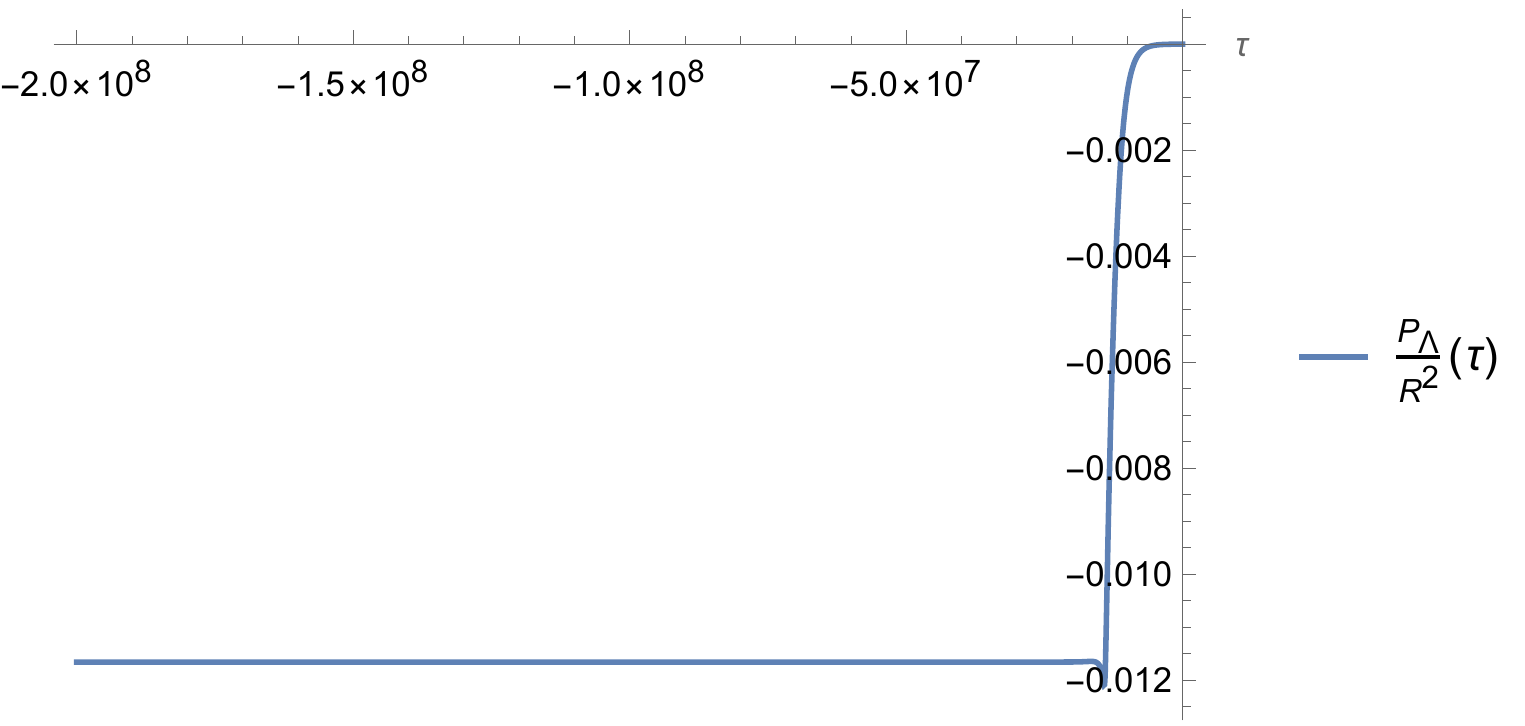}\\
	(c) & (d)  \\[6pt]  
	\includegraphics[height=4cm]{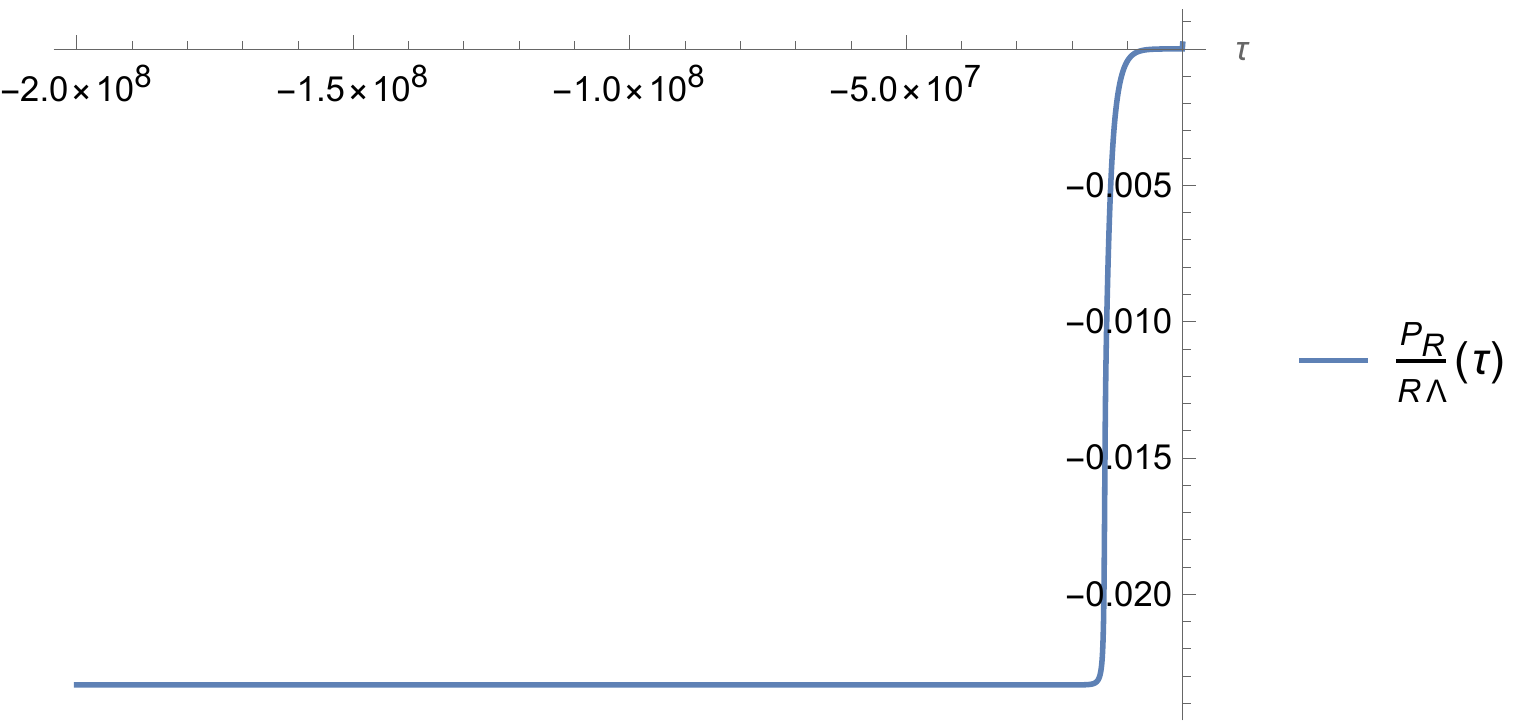}\\
	(e)   \\[6pt] 
		\end{tabular}
\caption{Plots of  the functions $\left(X, Y, W, \frac{P_\Lambda}{R^2}, \frac{P_R}{R \Lambda}\right)$. The throat is at $\tau_{\text{min}} =-1.416 \times 10^{7}$, at which $R(\tau_{\text{min}})=2012.19$. 
Graphs are plotted with $\gamma \approx 0.274, \; m=10^{6} m_p , \; j_x=10^5, \; \eta \approx 1.142$.
} 
\lb{fig38}
\end{figure} 		

 \end{widetext}

 \section{Main Properties of the quantum reduced loop Black Holes with the Inverse Volume Corrections} \lb{rlqg}
 \renewcommand{\theequation}{5.\arabic{equation}}\setcounter{equation}{0}
 
As shown in \cite{ABP20}, the  inverse volume corrections, represented by terms proportional to the constants $A, B$ and $C$ in the effective Hamiltonian given by Eqs.(\ref{hamiltonian})
and (\ref{Cfactor}),  are sub-leading. This  can be also seen clearly from the analysis 
given in the beginning of the last section.  Therefore, the inverse volume corrections should not change the main properties of the solutions with $\eta = 1,\; \eta  > 1,\; \eta < 1$, respectively. 
However,   demanding that the spatial manifold triangulation remain consistent on both sides of the black hole horizons, ABP found \cite{ABP20}
\bq
\lb{eq5.1}
j = \gamma j_x,  
\eq
which immediately leads to
\bq
\lb{eq5.2a}
\eta \equiv \frac{\alpha }{\beta} =\frac{\sqrt{2\pi}}{8\gamma},
\eq
 as can be seen from Eq.(\ref{eq2.5}).  On the other hand, the considerations of black hole entropy in LQG showed that  \cite{ABBDV10}
 \bq
\lb{eq5.3a}
\gamma \simeq 0.274,
\eq
which is precisely the solution obtained by requiring $a = d$ in Section IV.B for the case $\eta > 1$, in order to have the spacetime on the other side of the transition surface to be de Sitter,
where $a$ and $b$ are the constants defined in Eq.(\ref{eq2.5m}). This ``surprising coincidence" was first noted in \cite{ABP20} with a different approach, but in this paper we obtained it
simply by requiring that the transition surface connect two regions, one is asymptotically the Schwarzschild and the other is de Sitter. Therefore, following  \cite{ABP20} in this section we consider only
the case $\gamma \simeq 0.274$ \footnote{It should be noted that a second solution in \cite{ABP20} was also found with $\gamma \simeq 0.227$. However, we find that this solution
does not satisfy the Hamiltonian constraint ${\cal{H}}^{\text{IV+CS}}_{\text{int}} \simeq 0$,  so it must be discarded.}, for which we have $\eta \simeq 1.142$. 

Once $\gamma$ and $\eta$ are fixed, the five-parameter solutions of ABP are uniquely determined, after the inverse value correction parameters $\nu, \delta$ and $\delta_x$ are 
given. In the following, we adopt the values given by ABP \cite{ABP20},
 \bqn
\lb{eq5.4}
\nu &=& 1.802, \quad \delta= \frac{1.458}{\beta^2} + {\cal{O}}\left(\beta^{-6}\right),\nb\\
\delta_x&=&  \frac{0.729}{\beta^2}  + {\cal{O}}\left(\beta^{-6}\right).
\eqn
 
  In Figs. \ref{fig36} - \ref{fig38}, we plot out the functions 
  $  \left(X,\; Y, \; W,\;  \frac{P_\Lambda}{R^2},\;  \frac{P_R}{R \Lambda}\right)$,
   for different $m$. From these figures we find
  \bqn
  X  &\simeq& -\iota \approx -3.329, \quad Y  \simeq - \frac{\iota}{\eta} \approx -2.915, \nb\\
   W  &\simeq& -(\pi h_0[\iota]+2\sin[\iota]) \approx -1.001,\nb\\
  \frac{P_\Lambda}{R^2}  &\simeq& - \frac{\iota}{\alpha \gamma G} \approx -0.012,\nb\\
  \frac{P_R}{R \Lambda}  &\simeq & - \frac{2 \iota}{\alpha \gamma G} \approx -0.023, 
  \eqn
  as $\tau \rightarrow -\infty$, where $  \iota \equiv -\eta_0 \simeq  3.329$ \cite{ABP20}.
  With the above expressions, we find that the asymptotical behavior of $N(\tau), R(\tau)$ and $\Lambda(\tau)$ is precisely given by Eq.(\ref{eq2.5jb}), with
  the dependence of the three constants $N_0, \; R_0$ and $\Lambda_0$ being given by Table \ref{420}.

  As shown in Sec.  IV.B for the case $\eta > 1$, the inverse volume corrections become important only when the geometric radius $R$ is in the order of the Planck scale, 
  $R \simeq \ell_p$. However, for macroscopic black holes, the radius of the transition surface $R_{\text{min}}$ is always much larger than $\ell_p$. For example, when
  $m/m_p = 10^{12}$,   $R_{\text{min}}/\ell_p \simeq 8059.95 \gg 1$ [cf. Fig. \ref{fig12}]. Therefore, for macroscopic black holes the inverse volume corrections can be safely neglected. 
  This is true not only for the case $\eta = 1.142$, but also true for all the cases considered in Sec. IV for macroscopic black holes. Therefore, in this section we shall not repeat 
  our analyses carried out in that
  section.

  \section{Concluding Remarks} 
  \lb{SecV}
 \renewcommand{\theequation}{6.\arabic{equation}}\setcounter{equation}{0}

In this paper, we systematically study quantum black holes in  the framework of QRLG, proposed recently by ABP \cite{ABP18,ABP19,ABP20}.  
Starting from the full theory of LQG, ABP derived the effective Hamiltonian with respect to coherent 
states peaked around spherically symmetric geometry, by including both the holonomy and inverse volume corrections.
Then, they showed that the classical  singularity used to appear inside the Schwarzschild black hole is replaced by a regular transition surface with a finite and non-zero radius. 

To understand such obtained effective Hamiltonian well and shed light on the relations to models obtained by the  bottom-up approach, in Sec. \ref{cla limit} we first consider its classical limit, and obtained the
 desired Schwarzschild black hole solution, whereby the physical and geometric interpretation  of the quantities used in the effective  Hamiltonian are made clear. Then, 
  in Sec. \ref{bv limit} and Sec. \ref{aos-limit} by  taking proper limits 
 we re-derive respectively the BV \cite{BV07} and AOS \cite{AOS18a,AOS18b,AO20} solutions, all obtained by the bottom-up approach. In doing so, we can see  clearly the relation between models obtained by the two different  approaches, top-down and bottom-up. 
 
 In particular, the BV effective Hamiltonian   was originally obtained from the classical Hamiltonian (\ref{eq2.14b}) with the polymerization,  
\bq
\lb{eq5.2}
b \rightarrow \frac{\sin(\delta_b b)}{\delta_b}, \quad
c \rightarrow \frac{\sin(\delta_c c)}{\delta_c}.
\eq
However, instead of taking the parameters $\delta_b$ and $\delta_c$ as constants,   following the $\bar\mu$-scheme first proposed in LQC \cite{Ashtekar:2006wn} \footnote{This   is known to be the only possible choice in LQC, and results
 in physics that is independent from underlying fiducial structures used during quantization, and meanwhile yields a consistent infrared behavior for all matter obeying the weak energy condition \cite{cs08}.}, BV
 took them as
 \bq
\lb{eq5.3}
 \delta_b^{\text{(BV)}} = \sqrt{\frac{\Delta}{p_c}}, \quad \delta_c^{\text{(BV)}}  = \frac{\sqrt{\Delta p_c}}{p_b}.    
\eq
 In Sec. III.B, 
 we show explicitly that the BV effective Hamiltonian can be obtained from the ABP  Hamiltonian by taking the following replacement and limit,
 \bqn
 \lb{eq5.4a}
 && (i) \; h_0[X] \rightarrow \frac{2}{\pi}\sin[X], \quad h_{-1}[X] \rightarrow \frac{2}{\pi}\cos[X],\\
  \lb{eq5.4b}
 && (ii)\;   \frac{A}{R^2}, \; \frac{B}{R^2},\;  \frac{C}{R^2} \ll 1.
 \eqn
 It should be noted that with the choice of Eq.(\ref{eq5.3}), the corresponding values of $j_x$ and $j$ are given by Eq.(\ref{eq2.25iii}), from which we can see that they all violate the semi-classical limit (\ref{eq2.25v1}),
 with which the ABP effective Hamiltonian (\ref{hamiltonian}) was derived. As a result, the BV model cannot be physically realized in the framework of QRLG, although formally they can be obtained from the ABP effective Hamiltonian by the above  replacement and limit.

On the other hand, in addition to the replacement and limit given respectively by Eqs.(\ref{eq5.4a}) and (\ref{eq5.4b}), if we further assume that 
\bqn
 \lb{eq5.5}
 \delta^{\text{(AOS)}}_b, \;\;\;  \delta^{\text{(AOS)}}_c  = \text{Constants},
 \eqn
and are determined by Eqs.(\ref{eqAOS3a}) and (\ref{eqAOS3b}), the ABP effective Hamiltonian (\ref{hamiltonian}) reduces precisely to the AOS
one \cite{AOS18a,AOS18b,AO20}. However, as shown explicitly by Eq.(\ref{eqAOS6}), such choices are also out of the semi-classical limit
(\ref{eq2.25iii}). Therefore, the AOS model cannot be realized in the framework of QRLG either.

It must be noted that the above conclusions do not imply that the BV and AOS models are unphysical, but rather than the fact that they must be realized in a different top-down approach.

With the above in mind, in Sec. IV we study the ABP effective Hamiltonian without the inverse volume corrections, represented by the $A, B, C$ terms in Eq.(\ref{hamiltonian}) in detail, by first
confirming the main conclusions obtained in \cite{ABP19} and then clarifying some silent points. In particular, we find that the spacetime on the other side of the transition surface (throat) indeed sensitively 
depends on the ratio $\eta \equiv \alpha/\beta$, where
$\alpha$ and $\beta$ are defined by Eq.(\ref{eq2.5}) in terms of $(j_x,\;  j)$, or Eq.(\ref{eq2.5c}) in terms of  $(\hat j_0,\;  \hat j)$, where the parameters  $(j_x,\;  j)$ were introduced in \cite{ABP20}, while 
$(\hat j_0,\;  \hat j)$ were used in \cite{ABP19}, and related one to the other through Eq.(\ref{eq2.5g2}). As noticed previously, in Sec. IV we drop the hats from $(\hat j_0,\;  \hat j) \rightarrow
(j_0,\;  j)$, for the  {sake} of simplicity. 

When $\eta = 1$,  the spacetime  {on}  the other side of the transition surface is conformally flat, and the non-vanishing curvatures are all of the order of the Planck scale, as can be seen from Eq.(\ref{eq2.5i}). Then, the corresponding Penrose diagram is given by Fig. \ref{fig2}. At this point, we find that it is very helpful to make a closer comparison of the ABP model with the BV one, as for the BV choice of Eq.(\ref{eq2.25ii}),
we have $\eta^{\text{(BV)}} = 1$. In particular, we find the following:

 \begin{itemize}
 
 \item  In both models, the spacetime singularity used to appear at the center is replaced by a transition surface with a finite non-zero radius.

 \item   In both models, the spacetime on one side of the  transition surface  is quite similar to the internal region
 of a Schwarzschild black hole with a black hole like horizon located at a finite distance from the  transition surface
 (but with the removal of the black hole singularity used to occur at  the center).

 \item    In both models, the spacetime is asymmetric with respect to the   transition surface, and model-dependent. In particular, 
  in the BV model, the spacetime on the other side of the  black hole like internal region   approaches asymptotically to a charged Nariai space \cite{Bousso97,Nariai99,Bousso02}, of which the radius of the two-sphere $S^2$ approaches to a Planck scale constant, $R \rightarrow R_0 \simeq {\cal{O}}(\ell_p)$. 
In contrast,  in the ABP model the radius grows exponentially without limits, $R \rightarrow \exp\left(- \frac{\tau}{2mG}\right)$ as $\tau \rightarrow - \infty$, 
and a macroscopic universe is obtained. The corresponding global structure can be seen clearly from its Penrose diagram given by Fig. \ref{fig2}.

\item   In the BV model, there exists multiple transition surfaces at which we have $dp_c/d\tau = 0$. When passing each transition surface, $p_c$ decreases. 
As a result, $p_c$ will soon decreases to a value at which the two-spheres $S^2$ have  areas smaller than $\Delta$, whereby the effective Hamiltonian is no longer 
valid.  On the other hand,  in the ABP model, only one such transition surface exists, and the above mentioned problem is absent. As a matter of fact, 
the two-planes spanned by $\tau$ and $x$ are asymptotically flat, as shown explicitly by Eq.(\ref{eq2.5e}), although the four-dimensional spacetime is not [cf. 
  Eq.\eqref{eq2.5i}].

\end{itemize}

When  $\eta \gtrsim 1$, the spacetime 
  in general  { does not become} conformally flat, as can seen from Eq.(\ref{eq2.5ib}), unless $a = d$, where $a$ and $d$ are two constants defined by Eq.(\ref{eq2.5m}). 
  Then, the corresponding Penrose diagram is given by Fig. \ref{fig7}. When 
  \bq
  \lb{eq5.6}
  a = d, 
  \eq
  the spacetime is conformally flat and asymptotically de Sitter. It is remarkable that the condition (\ref{eq5.6}) together with the one (\ref{eq2.5a}) leads to 
  \bq
  \lb{eq2.7}
  \gamma = \frac{\sqrt{2\pi}}{8\eta} \simeq 0.274,
  \eq
  which is precisely the value obtained from the consideration of loop quantum black hole entropy obtained in  \cite{ABBDV10}. As emphasized in \cite{ABP20}, this coincidence should not be 
  underestimated, and may provide some profound physics. In particular,   the above picture is also consistent with the recently emerging picture in modified LQC models \cite{LSW21}, in which the quantum bounce,
  which corresponds to the current transition surface, connects two regions, one is asymptotically de Sitter, and the other is  asymptotically relativistic,
after considering the expectation values of the Hamiltonian operator in LQG \cite{DL17,DL18,adlp}, by using complexifier coherent states \cite{states}, as shown explicitly in \cite{lsw2018,lsw2018b,lsw2019}. 
In addition,   a similar structure of the spacetime of a spherical black hole also emerges in the framework of string \cite{BHR21}, but now the transition surface is replaced by an S-Brane.

When  $\eta \lesssim 1$, the spacetime cannot be conformally flat for any given values of $a$ and $d$, as it can be seen from Eq.(\ref{eq2.5x}). However, the corresponding Penrose diagram is the same as
that of  the case with $\eta \gtrsim 1$, and  given precisely by Fig. \ref{fig7}.

In review of all the above three cases, it is clear that the spacetime on the other side of the transition surface is no longer a white hole structure without spacetime singularities, as obtained from most of the bottom-up models \cite{Bodendorfer:2019nvy,Ashtekar20,GOP21}, so that the corresponding Penrose diagram is extended repeatedly along the vertical line to include infinite identical  universes of black holes and white holes
(without spacetime singularities). Instead, the white hole region is replaced by either a conformally flat spacetime or a non-conformally flat one, given respectively by Figs. \ref{fig2} and \ref{fig7}. But, in any case the
spacetime is already geodesically complete, and no extensions are needed beyond their boundaries, so that in this framework  {multiple} identical universes do not exist. 

In addition, the undesirable feature in the BV model that multiple horizons exist on the other side of the transition surface disappears in the ABP model. In this model, the large quantum gravitational effects near the black hole horizons   seemingly do not exist either, despite the fact that our numerical computations show that deviations may exist when very near to the black hole horizons, as shown explicitly in Figs. \ref{fig-horizon},
 \ref{fig14}, and \ref{fig26}. However,  more careful analysis is required, as the metric becomes singular when crossing the horizons, and our numerical simulations may become unreliable.  We wish to come back to this important question in another occasion. 

 When inverse volume corrections, represented by terms proportional to the constants $A, B, C$ in the effective Hamiltonian (\ref{hamiltonian}),
 are taken into account, the effects are always sub-leading, as these terms become important only when the radius of the  {two-sphere}  $\tau, x =$ Constant is of the order of the Planck scale. For macroscopic black holes, we find that
 the corresponding radii   of the transition surfaces are always much larger than  the Planck scale, so their effects will be always sub-leading even when across the transition surface. Such analysis was carried out in Sec. V, in which we mainly focus 
 on the case in which the conditions (\ref{eq5.6}) and (\ref{eq2.7}) hold. In \cite{ABP20} it was shown that these sub-leading terms precisely make up all the requirement for a spacetime to be asymptotically de Sitter,
 defined in \cite{ABK15}, even to the sub-leading order.

 \begin{acknowledgments}
 
 W-C.G. is supported by Baylor University through the Baylor Physics graduate program.   This work is  also partially supported by the National Natural Science
  Foundation of China with the Grants No. 11975116, and No. 11975203,
  and Jiangxi Science Foundation for Distinguished Young Scientists under the Grant No. 20192BCB23007.

  \end{acknowledgments}
 %%%%%%%%%%%%%%%%%%%%%%%%%%%%%%%%%%%

  \appendix
  
\section{Some properties of the Struve functions}\lb{app-a}
 \renewcommand{\theequation}{A.\arabic{equation}}\setcounter{equation}{0}

In general, the  $\nu$-th order Struve function   $h_{\nu}[X]$ is defined as \cite{AS72}, 
\bq
\lb{A.1}
h_{\nu}[z] \equiv \left(\frac{1}{2}z\right)^{\nu + 1}  \sum_{k =0}^{\infty}{ \frac{(-1)^k \left(\frac{1}{2}z\right)^{2k}}{\Gamma\left(k+\frac{3}{2}\right)\Gamma\left(k+\nu \frac{3}{2}\right)}},
\eq
which satisfies the differential equation,
\bq
\lb{A.2}
z^2\frac{d^2w}{dz^2} + z \frac{dw}{dz} + \left(z^2 - \nu^2\right) w = \frac{4\left(\frac{1}{2}z\right)^{\nu+1}}{\sqrt{\pi}\; \Gamma\left(\nu + \frac{1}{2}\right)}.
\eq
The general solution of the above equation is 
\bq
\lb{A.3}
 w = a J_{\nu}(z) + bY_{\nu}(z) + h_{\nu}(z),  
\eq
where $a$ and $b$ are two integration constants,  $J_{\nu}(z)$ and $Y_{\nu}(z)$ are the Bessel functions of the first and second kind, respectively, and satisfy the associated homogeneous 
differential equation. 

Some useful properties of $ h_{\nu}(z)$ are,
\bqn
\lb{A.2a}
\frac{d\left(z^{\nu}h_{\nu}\right)}{dz} &=& z^{\nu} h_{\nu -1}, \nb\\
\frac{d\left(z^{-\nu}h_{\nu}\right)}{dz} &=&  \frac{1}{\sqrt{\pi} \; 2^{\nu}\; \Gamma\left(\nu+\frac{3}{2}\right)} - z^{-\nu} h_{\nu +1},
\eqn
 while their asymptotic behaviors   are given by
\begin{widetext}
\bqn
\lb{A.3a}
h_0[X] &\simeq& \begin{cases}
\frac{2}{\pi  X} +\frac{1}{\sqrt{\pi X}}\left(\sin X - \cos X\right) + {\cal{O}}\left(X^{-3/2}\right), & X  \rightarrow  \infty,\cr
\frac{2 X}{\pi }-\frac{2 X^3}{9 \pi }+ {\cal{O}}\left(X^4\right), & X  \rightarrow  0,\cr
\end{cases}
\eqn
and 
\bqn
\lb{A.3b}
h_{-1}[X]&\simeq&  \begin{cases}
\frac{2}{\pi  X} +\frac{1}{\sqrt{\pi X}}\left(\sin X + \cos X\right) + {\cal{O}}\left(X^{-3/2}\right), & X  \rightarrow  \infty,\cr
\frac{2}{\pi }-\frac{2 X^2}{3 \pi }+ {\cal{O}}\left(X^4\right), & X  \rightarrow  0.\cr
\end{cases}
\eqn
\end{widetext}

 In Fig. \ref{fig1}, we plot out the  Struve function $h_0$ together with $h_{-1}$.
For other properties of the Struve functions, we refer readers to \cite{AS72}.

\end{document}